\documentclass[12pt]{iopart}

\usepackage{graphicx}
\usepackage{xstring}
\usepackage{xparse}
\usepackage{booktabs}
\usepackage{xcolor}
\usepackage[sorting=none,style=numeric-comp]{biblatex}
\DeclareRobustCommand{\totalmasssourcelower}[1]{\IfEqCase{#1}{{GW151205_195525}{-25.2}{GW151216_092416}{-11.8}{GW170121_212536}{-7.6}{GW170202_135657}{-8.4}{GW170304_163753}{-13.1}{GW170403_230611}{-15.2}{GW170425_055334}{-15.4}{GW170727_010430}{-10.6}{GW190426_082124}{-18.2}{GW190427_180650}{-2.3}{GW190511_125545}{-17.0}{GW190511_163209}{-77.5}{GW190514_065416}{-13.5}{GW190523_085933}{-34.5}{GW190524_134109}{-29.3}{GW190530_030659}{-10.2}{GW190530_133833}{-48.6}{GW190604_103812}{-74.7}{GW190605_025957}{-89.3}{GW190607_083827}{-14.7}{GW190614_134749}{-19.0}{GW190615_030234}{-23.1}{GW190705_164632}{-23.9}{GW190707_083226}{-18.0}{GW190711_030756}{-22.6}{GW190718_160159}{-2.3}{GW190725_174728}{-2.1}{GW190805_105432}{-2.9}{GW190806_033721}{-29.2}{GW190814_192009}{-30.4}{GW190818_232544}{-34.9}{GW190821_124821}{-2.8}{GW190904_104631}{-37.3}{GW190906_054335}{-17.0}{GW190910_012619}{-4.9}{GW190911_195101}{-34.1}{GW190916_200658}{-15.8}{GW190926_050336}{-15.3}{GW191113_103541}{-30.8}{GW191117_023843}{-31.9}{GW191127_050227}{-21.6}{GW191208_080334}{-10.6}{GW191224_043228}{-3.2}{GW191228_195619}{-61.3}{GW200106_134123}{-16.1}{GW200109_195634}{-25.4}{GW200129_114245}{-47.5}{GW200208_211609}{-31.6}{GW200210_005122}{-1.5}{GW200210_100022}{-24.2}{GW200220_124850}{-13.0}{GW200214_223307}{-24.6}{GW200225_075134}{-17.6}{GW200301_211019}{-6.3}{GW200304_172806}{-45.6}{GW200305_084739}{-10.4}{GW200318_191337}{-15.8}}}
\DeclareRobustCommand{\totalmasssourceupper}[1]{\IfEqCase{#1}{{GW151205_195525}{26.2}{GW151216_092416}{11.1}{GW170121_212536}{7.7}{GW170202_135657}{8.8}{GW170304_163753}{14.9}{GW170403_230611}{17.6}{GW170425_055334}{17.4}{GW170727_010430}{11.7}{GW190426_082124}{20.5}{GW190427_180650}{2.7}{GW190511_125545}{21.5}{GW190511_163209}{116.5}{GW190514_065416}{16.8}{GW190523_085933}{34.5}{GW190524_134109}{34.7}{GW190530_030659}{10.8}{GW190530_133833}{55.8}{GW190604_103812}{81.0}{GW190605_025957}{104.6}{GW190607_083827}{16.6}{GW190614_134749}{24.3}{GW190615_030234}{25.8}{GW190705_164632}{25.1}{GW190707_083226}{20.1}{GW190711_030756}{33.9}{GW190718_160159}{2.6}{GW190725_174728}{2.5}{GW190805_105432}{3.1}{GW190806_033721}{31.8}{GW190814_192009}{36.5}{GW190818_232544}{37.2}{GW190821_124821}{2.8}{GW190904_104631}{39.7}{GW190906_054335}{19.4}{GW190910_012619}{5.0}{GW190911_195101}{42.9}{GW190916_200658}{17.6}{GW190926_050336}{18.8}{GW191113_103541}{36.1}{GW191117_023843}{35.3}{GW191127_050227}{27.2}{GW191208_080334}{10.6}{GW191224_043228}{3.6}{GW191228_195619}{132.8}{GW200106_134123}{16.5}{GW200109_195634}{28.7}{GW200129_114245}{36.3}{GW200208_211609}{292.4}{GW200210_005122}{1.9}{GW200210_100022}{27.4}{GW200220_124850}{15.1}{GW200214_223307}{28.6}{GW200225_075134}{19.9}{GW200301_211019}{6.5}{GW200304_172806}{48.5}{GW200305_084739}{12.6}{GW200318_191337}{18.7}}}
\DeclareRobustCommand{\totalmasssourcemedian}[1]{\IfEqCase{#1}{{GW151205_195525}{104.9}{GW151216_092416}{53.9}{GW170121_212536}{57.3}{GW170202_135657}{43.5}{GW170304_163753}{76.0}{GW170403_230611}{82.3}{GW170425_055334}{76.8}{GW170727_010430}{70.4}{GW190426_082124}{68.7}{GW190427_180650}{18.0}{GW190511_125545}{66.4}{GW190511_163209}{116.2}{GW190514_065416}{68.2}{GW190523_085933}{64.0}{GW190524_134109}{113.1}{GW190530_030659}{52.4}{GW190530_133833}{145.8}{GW190604_103812}{154.0}{GW190605_025957}{214.8}{GW190607_083827}{74.3}{GW190614_134749}{64.3}{GW190615_030234}{118.0}{GW190705_164632}{83.9}{GW190707_083226}{87.3}{GW190711_030756}{85.3}{GW190718_160159}{16.9}{GW190725_174728}{18.2}{GW190805_105432}{21.7}{GW190806_033721}{110.5}{GW190814_192009}{120.8}{GW190818_232544}{110.7}{GW190821_124821}{14.0}{GW190904_104631}{63.8}{GW190906_054335}{62.6}{GW190910_012619}{38.5}{GW190911_195101}{142.1}{GW190916_200658}{69.6}{GW190926_050336}{62.6}{GW191113_103541}{121.6}{GW191117_023843}{115.0}{GW191127_050227}{84.4}{GW191208_080334}{56.7}{GW191224_043228}{23.2}{GW191228_195619}{239.9}{GW200106_134123}{72.2}{GW200109_195634}{106.9}{GW200129_114245}{115.9}{GW200208_211609}{51.3}{GW200210_005122}{15.2}{GW200210_100022}{100.8}{GW200220_124850}{65.3}{GW200214_223307}{81.6}{GW200225_075134}{100.6}{GW200301_211019}{35.2}{GW200304_172806}{140.0}{GW200305_084739}{56.6}{GW200318_191337}{76.6}}}
\DeclareRobustCommand{\totalmasssource}[1]{\IfEqCase{#1}{{GW151205_195525}{$\totalmasssourcemedian{GW151205_195525}^{+\totalmasssourceupper{GW151205_195525}}_{\totalmasssourcelower{GW151205_195525}}$}{GW151216_092416}{$\totalmasssourcemedian{GW151216_092416}^{+\totalmasssourceupper{GW151216_092416}}_{\totalmasssourcelower{GW151216_092416}}$}{GW170121_212536}{$\totalmasssourcemedian{GW170121_212536}^{+\totalmasssourceupper{GW170121_212536}}_{\totalmasssourcelower{GW170121_212536}}$}{GW170202_135657}{$\totalmasssourcemedian{GW170202_135657}^{+\totalmasssourceupper{GW170202_135657}}_{\totalmasssourcelower{GW170202_135657}}$}{GW170304_163753}{$\totalmasssourcemedian{GW170304_163753}^{+\totalmasssourceupper{GW170304_163753}}_{\totalmasssourcelower{GW170304_163753}}$}{GW170403_230611}{$\totalmasssourcemedian{GW170403_230611}^{+\totalmasssourceupper{GW170403_230611}}_{\totalmasssourcelower{GW170403_230611}}$}{GW170425_055334}{$\totalmasssourcemedian{GW170425_055334}^{+\totalmasssourceupper{GW170425_055334}}_{\totalmasssourcelower{GW170425_055334}}$}{GW170727_010430}{$\totalmasssourcemedian{GW170727_010430}^{+\totalmasssourceupper{GW170727_010430}}_{\totalmasssourcelower{GW170727_010430}}$}{GW190426_082124}{$\totalmasssourcemedian{GW190426_082124}^{+\totalmasssourceupper{GW190426_082124}}_{\totalmasssourcelower{GW190426_082124}}$}{GW190427_180650}{$\totalmasssourcemedian{GW190427_180650}^{+\totalmasssourceupper{GW190427_180650}}_{\totalmasssourcelower{GW190427_180650}}$}{GW190511_125545}{$\totalmasssourcemedian{GW190511_125545}^{+\totalmasssourceupper{GW190511_125545}}_{\totalmasssourcelower{GW190511_125545}}$}{GW190511_163209}{$\totalmasssourcemedian{GW190511_163209}^{+\totalmasssourceupper{GW190511_163209}}_{\totalmasssourcelower{GW190511_163209}}$}{GW190514_065416}{$\totalmasssourcemedian{GW190514_065416}^{+\totalmasssourceupper{GW190514_065416}}_{\totalmasssourcelower{GW190514_065416}}$}{GW190523_085933}{$\totalmasssourcemedian{GW190523_085933}^{+\totalmasssourceupper{GW190523_085933}}_{\totalmasssourcelower{GW190523_085933}}$}{GW190524_134109}{$\totalmasssourcemedian{GW190524_134109}^{+\totalmasssourceupper{GW190524_134109}}_{\totalmasssourcelower{GW190524_134109}}$}{GW190530_030659}{$\totalmasssourcemedian{GW190530_030659}^{+\totalmasssourceupper{GW190530_030659}}_{\totalmasssourcelower{GW190530_030659}}$}{GW190530_133833}{$\totalmasssourcemedian{GW190530_133833}^{+\totalmasssourceupper{GW190530_133833}}_{\totalmasssourcelower{GW190530_133833}}$}{GW190604_103812}{$\totalmasssourcemedian{GW190604_103812}^{+\totalmasssourceupper{GW190604_103812}}_{\totalmasssourcelower{GW190604_103812}}$}{GW190605_025957}{$\totalmasssourcemedian{GW190605_025957}^{+\totalmasssourceupper{GW190605_025957}}_{\totalmasssourcelower{GW190605_025957}}$}{GW190607_083827}{$\totalmasssourcemedian{GW190607_083827}^{+\totalmasssourceupper{GW190607_083827}}_{\totalmasssourcelower{GW190607_083827}}$}{GW190614_134749}{$\totalmasssourcemedian{GW190614_134749}^{+\totalmasssourceupper{GW190614_134749}}_{\totalmasssourcelower{GW190614_134749}}$}{GW190615_030234}{$\totalmasssourcemedian{GW190615_030234}^{+\totalmasssourceupper{GW190615_030234}}_{\totalmasssourcelower{GW190615_030234}}$}{GW190705_164632}{$\totalmasssourcemedian{GW190705_164632}^{+\totalmasssourceupper{GW190705_164632}}_{\totalmasssourcelower{GW190705_164632}}$}{GW190707_083226}{$\totalmasssourcemedian{GW190707_083226}^{+\totalmasssourceupper{GW190707_083226}}_{\totalmasssourcelower{GW190707_083226}}$}{GW190711_030756}{$\totalmasssourcemedian{GW190711_030756}^{+\totalmasssourceupper{GW190711_030756}}_{\totalmasssourcelower{GW190711_030756}}$}{GW190718_160159}{$\totalmasssourcemedian{GW190718_160159}^{+\totalmasssourceupper{GW190718_160159}}_{\totalmasssourcelower{GW190718_160159}}$}{GW190725_174728}{$\totalmasssourcemedian{GW190725_174728}^{+\totalmasssourceupper{GW190725_174728}}_{\totalmasssourcelower{GW190725_174728}}$}{GW190805_105432}{$\totalmasssourcemedian{GW190805_105432}^{+\totalmasssourceupper{GW190805_105432}}_{\totalmasssourcelower{GW190805_105432}}$}{GW190806_033721}{$\totalmasssourcemedian{GW190806_033721}^{+\totalmasssourceupper{GW190806_033721}}_{\totalmasssourcelower{GW190806_033721}}$}{GW190814_192009}{$\totalmasssourcemedian{GW190814_192009}^{+\totalmasssourceupper{GW190814_192009}}_{\totalmasssourcelower{GW190814_192009}}$}{GW190818_232544}{$\totalmasssourcemedian{GW190818_232544}^{+\totalmasssourceupper{GW190818_232544}}_{\totalmasssourcelower{GW190818_232544}}$}{GW190821_124821}{$\totalmasssourcemedian{GW190821_124821}^{+\totalmasssourceupper{GW190821_124821}}_{\totalmasssourcelower{GW190821_124821}}$}{GW190904_104631}{$\totalmasssourcemedian{GW190904_104631}^{+\totalmasssourceupper{GW190904_104631}}_{\totalmasssourcelower{GW190904_104631}}$}{GW190906_054335}{$\totalmasssourcemedian{GW190906_054335}^{+\totalmasssourceupper{GW190906_054335}}_{\totalmasssourcelower{GW190906_054335}}$}{GW190910_012619}{$\totalmasssourcemedian{GW190910_012619}^{+\totalmasssourceupper{GW190910_012619}}_{\totalmasssourcelower{GW190910_012619}}$}{GW190911_195101}{$\totalmasssourcemedian{GW190911_195101}^{+\totalmasssourceupper{GW190911_195101}}_{\totalmasssourcelower{GW190911_195101}}$}{GW190916_200658}{$\totalmasssourcemedian{GW190916_200658}^{+\totalmasssourceupper{GW190916_200658}}_{\totalmasssourcelower{GW190916_200658}}$}{GW190926_050336}{$\totalmasssourcemedian{GW190926_050336}^{+\totalmasssourceupper{GW190926_050336}}_{\totalmasssourcelower{GW190926_050336}}$}{GW191113_103541}{$\totalmasssourcemedian{GW191113_103541}^{+\totalmasssourceupper{GW191113_103541}}_{\totalmasssourcelower{GW191113_103541}}$}{GW191117_023843}{$\totalmasssourcemedian{GW191117_023843}^{+\totalmasssourceupper{GW191117_023843}}_{\totalmasssourcelower{GW191117_023843}}$}{GW191127_050227}{$\totalmasssourcemedian{GW191127_050227}^{+\totalmasssourceupper{GW191127_050227}}_{\totalmasssourcelower{GW191127_050227}}$}{GW191208_080334}{$\totalmasssourcemedian{GW191208_080334}^{+\totalmasssourceupper{GW191208_080334}}_{\totalmasssourcelower{GW191208_080334}}$}{GW191224_043228}{$\totalmasssourcemedian{GW191224_043228}^{+\totalmasssourceupper{GW191224_043228}}_{\totalmasssourcelower{GW191224_043228}}$}{GW191228_195619}{$\totalmasssourcemedian{GW191228_195619}^{+\totalmasssourceupper{GW191228_195619}}_{\totalmasssourcelower{GW191228_195619}}$}{GW200106_134123}{$\totalmasssourcemedian{GW200106_134123}^{+\totalmasssourceupper{GW200106_134123}}_{\totalmasssourcelower{GW200106_134123}}$}{GW200109_195634}{$\totalmasssourcemedian{GW200109_195634}^{+\totalmasssourceupper{GW200109_195634}}_{\totalmasssourcelower{GW200109_195634}}$}{GW200129_114245}{$\totalmasssourcemedian{GW200129_114245}^{+\totalmasssourceupper{GW200129_114245}}_{\totalmasssourcelower{GW200129_114245}}$}{GW200208_211609}{$\totalmasssourcemedian{GW200208_211609}^{+\totalmasssourceupper{GW200208_211609}}_{\totalmasssourcelower{GW200208_211609}}$}{GW200210_005122}{$\totalmasssourcemedian{GW200210_005122}^{+\totalmasssourceupper{GW200210_005122}}_{\totalmasssourcelower{GW200210_005122}}$}{GW200210_100022}{$\totalmasssourcemedian{GW200210_100022}^{+\totalmasssourceupper{GW200210_100022}}_{\totalmasssourcelower{GW200210_100022}}$}{GW200220_124850}{$\totalmasssourcemedian{GW200220_124850}^{+\totalmasssourceupper{GW200220_124850}}_{\totalmasssourcelower{GW200220_124850}}$}{GW200214_223307}{$\totalmasssourcemedian{GW200214_223307}^{+\totalmasssourceupper{GW200214_223307}}_{\totalmasssourcelower{GW200214_223307}}$}{GW200225_075134}{$\totalmasssourcemedian{GW200225_075134}^{+\totalmasssourceupper{GW200225_075134}}_{\totalmasssourcelower{GW200225_075134}}$}{GW200301_211019}{$\totalmasssourcemedian{GW200301_211019}^{+\totalmasssourceupper{GW200301_211019}}_{\totalmasssourcelower{GW200301_211019}}$}{GW200304_172806}{$\totalmasssourcemedian{GW200304_172806}^{+\totalmasssourceupper{GW200304_172806}}_{\totalmasssourcelower{GW200304_172806}}$}{GW200305_084739}{$\totalmasssourcemedian{GW200305_084739}^{+\totalmasssourceupper{GW200305_084739}}_{\totalmasssourcelower{GW200305_084739}}$}{GW200318_191337}{$\totalmasssourcemedian{GW200318_191337}^{+\totalmasssourceupper{GW200318_191337}}_{\totalmasssourcelower{GW200318_191337}}$}}}

\DeclareRobustCommand{\chirpmasssourcelower}[1]{\IfEqCase{#1}{{GW151205_195525}{-11.4}{GW151216_092416}{-5.0}{GW170121_212536}{-3.2}{GW170202_135657}{-3.3}{GW170304_163753}{-5.9}{GW170403_230611}{-6.9}{GW170425_055334}{-6.4}{GW170727_010430}{-4.6}{GW190426_082124}{-7.4}{GW190427_180650}{-0.6}{GW190511_125545}{-7.6}{GW190511_163209}{-23.6}{GW190514_065416}{-5.7}{GW190523_085933}{-12.0}{GW190524_134109}{-13.8}{GW190530_030659}{-3.7}{GW190530_133833}{-20.6}{GW190604_103812}{-25.5}{GW190605_025957}{-41.2}{GW190607_083827}{-6.6}{GW190614_134749}{-7.6}{GW190615_030234}{-10.9}{GW190705_164632}{-10.3}{GW190707_083226}{-7.9}{GW190711_030756}{-6.4}{GW190718_160159}{-0.7}{GW190725_174728}{-0.5}{GW190805_105432}{-0.8}{GW190806_033721}{-12.8}{GW190814_192009}{-13.9}{GW190818_232544}{-13.5}{GW190821_124821}{-0.2}{GW190904_104631}{-15.3}{GW190906_054335}{-8.0}{GW190910_012619}{-0.9}{GW190911_195101}{-15.7}{GW190916_200658}{-6.5}{GW190926_050336}{-6.5}{GW191113_103541}{-9.7}{GW191117_023843}{-7.9}{GW191127_050227}{-8.3}{GW191208_080334}{-3.9}{GW191224_043228}{-0.8}{GW191228_195619}{-31.8}{GW200106_134123}{-7.2}{GW200109_195634}{-11.7}{GW200129_114245}{-15.6}{GW200208_211609}{-14.5}{GW200210_005122}{-0.4}{GW200210_100022}{-7.3}{GW200220_124850}{-5.7}{GW200214_223307}{-10.6}{GW200225_075134}{-7.9}{GW200301_211019}{-2.1}{GW200304_172806}{-19.6}{GW200305_084739}{-4.7}{GW200318_191337}{-6.5}}}
\DeclareRobustCommand{\chirpmasssourceupper}[1]{\IfEqCase{#1}{{GW151205_195525}{12.1}{GW151216_092416}{3.9}{GW170121_212536}{3.3}{GW170202_135657}{3.2}{GW170304_163753}{6.8}{GW170403_230611}{7.8}{GW170425_055334}{7.1}{GW170727_010430}{5.2}{GW190426_082124}{8.7}{GW190427_180650}{0.6}{GW190511_125545}{9.7}{GW190511_163209}{35.5}{GW190514_065416}{6.8}{GW190523_085933}{18.0}{GW190524_134109}{16.8}{GW190530_030659}{3.7}{GW190530_133833}{23.6}{GW190604_103812}{31.5}{GW190605_025957}{45.0}{GW190607_083827}{7.4}{GW190614_134749}{9.6}{GW190615_030234}{12.5}{GW190705_164632}{11.2}{GW190707_083226}{8.9}{GW190711_030756}{6.7}{GW190718_160159}{0.7}{GW190725_174728}{0.5}{GW190805_105432}{0.8}{GW190806_033721}{15.2}{GW190814_192009}{17.2}{GW190818_232544}{15.0}{GW190821_124821}{0.3}{GW190904_104631}{15.8}{GW190906_054335}{8.4}{GW190910_012619}{0.8}{GW190911_195101}{19.2}{GW190916_200658}{7.5}{GW190926_050336}{7.5}{GW191113_103541}{13.3}{GW191117_023843}{9.7}{GW191127_050227}{9.5}{GW191208_080334}{4.1}{GW191224_043228}{0.9}{GW191228_195619}{40.4}{GW200106_134123}{7.4}{GW200109_195634}{12.7}{GW200129_114245}{16.0}{GW200208_211609}{35.6}{GW200210_005122}{0.5}{GW200210_100022}{8.9}{GW200220_124850}{6.7}{GW200214_223307}{11.5}{GW200225_075134}{9.0}{GW200301_211019}{2.1}{GW200304_172806}{21.9}{GW200305_084739}{5.6}{GW200318_191337}{8.5}}}
\DeclareRobustCommand{\chirpmasssourcemedian}[1]{\IfEqCase{#1}{{GW151205_195525}{43.6}{GW151216_092416}{20.7}{GW170121_212536}{24.5}{GW170202_135657}{18.0}{GW170304_163753}{32.2}{GW170403_230611}{34.9}{GW170425_055334}{31.5}{GW170727_010430}{30.1}{GW190426_082124}{28.9}{GW190427_180650}{7.3}{GW190511_125545}{27.2}{GW190511_163209}{39.2}{GW190514_065416}{28.9}{GW190523_085933}{23.2}{GW190524_134109}{46.0}{GW190530_030659}{21.4}{GW190530_133833}{58.6}{GW190604_103812}{55.4}{GW190605_025957}{88.1}{GW190607_083827}{31.6}{GW190614_134749}{26.3}{GW190615_030234}{49.8}{GW190705_164632}{33.6}{GW190707_083226}{35.9}{GW190711_030756}{31.1}{GW190718_160159}{7.0}{GW190725_174728}{7.5}{GW190805_105432}{8.8}{GW190806_033721}{42.4}{GW190814_192009}{50.3}{GW190818_232544}{41.0}{GW190821_124821}{4.8}{GW190904_104631}{26.1}{GW190906_054335}{25.7}{GW190910_012619}{9.6}{GW190911_195101}{59.4}{GW190916_200658}{27.6}{GW190926_050336}{25.7}{GW191113_103541}{36.3}{GW191117_023843}{38.3}{GW191127_050227}{32.3}{GW191208_080334}{22.7}{GW191224_043228}{9.3}{GW191228_195619}{96.0}{GW200106_134123}{29.7}{GW200109_195634}{44.9}{GW200129_114245}{43.1}{GW200208_211609}{21.0}{GW200210_005122}{6.4}{GW200210_100022}{29.1}{GW200220_124850}{27.7}{GW200214_223307}{34.0}{GW200225_075134}{42.8}{GW200301_211019}{14.5}{GW200304_172806}{52.6}{GW200305_084739}{24.0}{GW200318_191337}{31.9}}}
\DeclareRobustCommand{\chirpmasssource}[1]{\IfEqCase{#1}{{GW151205_195525}{$\chirpmasssourcemedian{GW151205_195525}^{+\chirpmasssourceupper{GW151205_195525}}_{\chirpmasssourcelower{GW151205_195525}}$}{GW151216_092416}{$\chirpmasssourcemedian{GW151216_092416}^{+\chirpmasssourceupper{GW151216_092416}}_{\chirpmasssourcelower{GW151216_092416}}$}{GW170121_212536}{$\chirpmasssourcemedian{GW170121_212536}^{+\chirpmasssourceupper{GW170121_212536}}_{\chirpmasssourcelower{GW170121_212536}}$}{GW170202_135657}{$\chirpmasssourcemedian{GW170202_135657}^{+\chirpmasssourceupper{GW170202_135657}}_{\chirpmasssourcelower{GW170202_135657}}$}{GW170304_163753}{$\chirpmasssourcemedian{GW170304_163753}^{+\chirpmasssourceupper{GW170304_163753}}_{\chirpmasssourcelower{GW170304_163753}}$}{GW170403_230611}{$\chirpmasssourcemedian{GW170403_230611}^{+\chirpmasssourceupper{GW170403_230611}}_{\chirpmasssourcelower{GW170403_230611}}$}{GW170425_055334}{$\chirpmasssourcemedian{GW170425_055334}^{+\chirpmasssourceupper{GW170425_055334}}_{\chirpmasssourcelower{GW170425_055334}}$}{GW170727_010430}{$\chirpmasssourcemedian{GW170727_010430}^{+\chirpmasssourceupper{GW170727_010430}}_{\chirpmasssourcelower{GW170727_010430}}$}{GW190426_082124}{$\chirpmasssourcemedian{GW190426_082124}^{+\chirpmasssourceupper{GW190426_082124}}_{\chirpmasssourcelower{GW190426_082124}}$}{GW190427_180650}{$\chirpmasssourcemedian{GW190427_180650}^{+\chirpmasssourceupper{GW190427_180650}}_{\chirpmasssourcelower{GW190427_180650}}$}{GW190511_125545}{$\chirpmasssourcemedian{GW190511_125545}^{+\chirpmasssourceupper{GW190511_125545}}_{\chirpmasssourcelower{GW190511_125545}}$}{GW190511_163209}{$\chirpmasssourcemedian{GW190511_163209}^{+\chirpmasssourceupper{GW190511_163209}}_{\chirpmasssourcelower{GW190511_163209}}$}{GW190514_065416}{$\chirpmasssourcemedian{GW190514_065416}^{+\chirpmasssourceupper{GW190514_065416}}_{\chirpmasssourcelower{GW190514_065416}}$}{GW190523_085933}{$\chirpmasssourcemedian{GW190523_085933}^{+\chirpmasssourceupper{GW190523_085933}}_{\chirpmasssourcelower{GW190523_085933}}$}{GW190524_134109}{$\chirpmasssourcemedian{GW190524_134109}^{+\chirpmasssourceupper{GW190524_134109}}_{\chirpmasssourcelower{GW190524_134109}}$}{GW190530_030659}{$\chirpmasssourcemedian{GW190530_030659}^{+\chirpmasssourceupper{GW190530_030659}}_{\chirpmasssourcelower{GW190530_030659}}$}{GW190530_133833}{$\chirpmasssourcemedian{GW190530_133833}^{+\chirpmasssourceupper{GW190530_133833}}_{\chirpmasssourcelower{GW190530_133833}}$}{GW190604_103812}{$\chirpmasssourcemedian{GW190604_103812}^{+\chirpmasssourceupper{GW190604_103812}}_{\chirpmasssourcelower{GW190604_103812}}$}{GW190605_025957}{$\chirpmasssourcemedian{GW190605_025957}^{+\chirpmasssourceupper{GW190605_025957}}_{\chirpmasssourcelower{GW190605_025957}}$}{GW190607_083827}{$\chirpmasssourcemedian{GW190607_083827}^{+\chirpmasssourceupper{GW190607_083827}}_{\chirpmasssourcelower{GW190607_083827}}$}{GW190614_134749}{$\chirpmasssourcemedian{GW190614_134749}^{+\chirpmasssourceupper{GW190614_134749}}_{\chirpmasssourcelower{GW190614_134749}}$}{GW190615_030234}{$\chirpmasssourcemedian{GW190615_030234}^{+\chirpmasssourceupper{GW190615_030234}}_{\chirpmasssourcelower{GW190615_030234}}$}{GW190705_164632}{$\chirpmasssourcemedian{GW190705_164632}^{+\chirpmasssourceupper{GW190705_164632}}_{\chirpmasssourcelower{GW190705_164632}}$}{GW190707_083226}{$\chirpmasssourcemedian{GW190707_083226}^{+\chirpmasssourceupper{GW190707_083226}}_{\chirpmasssourcelower{GW190707_083226}}$}{GW190711_030756}{$\chirpmasssourcemedian{GW190711_030756}^{+\chirpmasssourceupper{GW190711_030756}}_{\chirpmasssourcelower{GW190711_030756}}$}{GW190718_160159}{$\chirpmasssourcemedian{GW190718_160159}^{+\chirpmasssourceupper{GW190718_160159}}_{\chirpmasssourcelower{GW190718_160159}}$}{GW190725_174728}{$\chirpmasssourcemedian{GW190725_174728}^{+\chirpmasssourceupper{GW190725_174728}}_{\chirpmasssourcelower{GW190725_174728}}$}{GW190805_105432}{$\chirpmasssourcemedian{GW190805_105432}^{+\chirpmasssourceupper{GW190805_105432}}_{\chirpmasssourcelower{GW190805_105432}}$}{GW190806_033721}{$\chirpmasssourcemedian{GW190806_033721}^{+\chirpmasssourceupper{GW190806_033721}}_{\chirpmasssourcelower{GW190806_033721}}$}{GW190814_192009}{$\chirpmasssourcemedian{GW190814_192009}^{+\chirpmasssourceupper{GW190814_192009}}_{\chirpmasssourcelower{GW190814_192009}}$}{GW190818_232544}{$\chirpmasssourcemedian{GW190818_232544}^{+\chirpmasssourceupper{GW190818_232544}}_{\chirpmasssourcelower{GW190818_232544}}$}{GW190821_124821}{$\chirpmasssourcemedian{GW190821_124821}^{+\chirpmasssourceupper{GW190821_124821}}_{\chirpmasssourcelower{GW190821_124821}}$}{GW190904_104631}{$\chirpmasssourcemedian{GW190904_104631}^{+\chirpmasssourceupper{GW190904_104631}}_{\chirpmasssourcelower{GW190904_104631}}$}{GW190906_054335}{$\chirpmasssourcemedian{GW190906_054335}^{+\chirpmasssourceupper{GW190906_054335}}_{\chirpmasssourcelower{GW190906_054335}}$}{GW190910_012619}{$\chirpmasssourcemedian{GW190910_012619}^{+\chirpmasssourceupper{GW190910_012619}}_{\chirpmasssourcelower{GW190910_012619}}$}{GW190911_195101}{$\chirpmasssourcemedian{GW190911_195101}^{+\chirpmasssourceupper{GW190911_195101}}_{\chirpmasssourcelower{GW190911_195101}}$}{GW190916_200658}{$\chirpmasssourcemedian{GW190916_200658}^{+\chirpmasssourceupper{GW190916_200658}}_{\chirpmasssourcelower{GW190916_200658}}$}{GW190926_050336}{$\chirpmasssourcemedian{GW190926_050336}^{+\chirpmasssourceupper{GW190926_050336}}_{\chirpmasssourcelower{GW190926_050336}}$}{GW191113_103541}{$\chirpmasssourcemedian{GW191113_103541}^{+\chirpmasssourceupper{GW191113_103541}}_{\chirpmasssourcelower{GW191113_103541}}$}{GW191117_023843}{$\chirpmasssourcemedian{GW191117_023843}^{+\chirpmasssourceupper{GW191117_023843}}_{\chirpmasssourcelower{GW191117_023843}}$}{GW191127_050227}{$\chirpmasssourcemedian{GW191127_050227}^{+\chirpmasssourceupper{GW191127_050227}}_{\chirpmasssourcelower{GW191127_050227}}$}{GW191208_080334}{$\chirpmasssourcemedian{GW191208_080334}^{+\chirpmasssourceupper{GW191208_080334}}_{\chirpmasssourcelower{GW191208_080334}}$}{GW191224_043228}{$\chirpmasssourcemedian{GW191224_043228}^{+\chirpmasssourceupper{GW191224_043228}}_{\chirpmasssourcelower{GW191224_043228}}$}{GW191228_195619}{$\chirpmasssourcemedian{GW191228_195619}^{+\chirpmasssourceupper{GW191228_195619}}_{\chirpmasssourcelower{GW191228_195619}}$}{GW200106_134123}{$\chirpmasssourcemedian{GW200106_134123}^{+\chirpmasssourceupper{GW200106_134123}}_{\chirpmasssourcelower{GW200106_134123}}$}{GW200109_195634}{$\chirpmasssourcemedian{GW200109_195634}^{+\chirpmasssourceupper{GW200109_195634}}_{\chirpmasssourcelower{GW200109_195634}}$}{GW200129_114245}{$\chirpmasssourcemedian{GW200129_114245}^{+\chirpmasssourceupper{GW200129_114245}}_{\chirpmasssourcelower{GW200129_114245}}$}{GW200208_211609}{$\chirpmasssourcemedian{GW200208_211609}^{+\chirpmasssourceupper{GW200208_211609}}_{\chirpmasssourcelower{GW200208_211609}}$}{GW200210_005122}{$\chirpmasssourcemedian{GW200210_005122}^{+\chirpmasssourceupper{GW200210_005122}}_{\chirpmasssourcelower{GW200210_005122}}$}{GW200210_100022}{$\chirpmasssourcemedian{GW200210_100022}^{+\chirpmasssourceupper{GW200210_100022}}_{\chirpmasssourcelower{GW200210_100022}}$}{GW200220_124850}{$\chirpmasssourcemedian{GW200220_124850}^{+\chirpmasssourceupper{GW200220_124850}}_{\chirpmasssourcelower{GW200220_124850}}$}{GW200214_223307}{$\chirpmasssourcemedian{GW200214_223307}^{+\chirpmasssourceupper{GW200214_223307}}_{\chirpmasssourcelower{GW200214_223307}}$}{GW200225_075134}{$\chirpmasssourcemedian{GW200225_075134}^{+\chirpmasssourceupper{GW200225_075134}}_{\chirpmasssourcelower{GW200225_075134}}$}{GW200301_211019}{$\chirpmasssourcemedian{GW200301_211019}^{+\chirpmasssourceupper{GW200301_211019}}_{\chirpmasssourcelower{GW200301_211019}}$}{GW200304_172806}{$\chirpmasssourcemedian{GW200304_172806}^{+\chirpmasssourceupper{GW200304_172806}}_{\chirpmasssourcelower{GW200304_172806}}$}{GW200305_084739}{$\chirpmasssourcemedian{GW200305_084739}^{+\chirpmasssourceupper{GW200305_084739}}_{\chirpmasssourcelower{GW200305_084739}}$}{GW200318_191337}{$\chirpmasssourcemedian{GW200318_191337}^{+\chirpmasssourceupper{GW200318_191337}}_{\chirpmasssourcelower{GW200318_191337}}$}}}

\DeclareRobustCommand{\massonesourcelower}[1]{\IfEqCase{#1}{{GW151205_195525}{-19.6}{GW151216_092416}{-12.5}{GW170121_212536}{-6.4}{GW170202_135657}{-8.0}{GW170304_163753}{-10.4}{GW170403_230611}{-11.9}{GW170425_055334}{-15.2}{GW170727_010430}{-8.3}{GW190426_082124}{-13.8}{GW190427_180650}{-3.5}{GW190511_125545}{-14.1}{GW190511_163209}{-65.8}{GW190514_065416}{-10.8}{GW190523_085933}{-27.0}{GW190524_134109}{-23.8}{GW190530_030659}{-10.2}{GW190530_133833}{-39.3}{GW190604_103812}{-70.1}{GW190605_025957}{-60.6}{GW190607_083827}{-10.3}{GW190614_134749}{-15.2}{GW190615_030234}{-16.8}{GW190705_164632}{-20.0}{GW190707_083226}{-15.9}{GW190711_030756}{-25.5}{GW190718_160159}{-2.5}{GW190725_174728}{-3.3}{GW190805_105432}{-4.2}{GW190806_033721}{-28.5}{GW190814_192009}{-23.2}{GW190818_232544}{-34.9}{GW190821_124821}{-3.8}{GW190904_104631}{-26.8}{GW190906_054335}{-13.6}{GW190910_012619}{-5.2}{GW190911_195101}{-26.6}{GW190916_200658}{-15.2}{GW190926_050336}{-13.1}{GW191113_103541}{-29.8}{GW191117_023843}{-31.9}{GW191127_050227}{-22.3}{GW191208_080334}{-12.3}{GW191224_043228}{-4.8}{GW191228_195619}{-52.4}{GW200106_134123}{-13.2}{GW200109_195634}{-18.7}{GW200129_114245}{-41.6}{GW200208_211609}{-15.7}{GW200210_005122}{-2.1}{GW200210_100022}{-25.2}{GW200220_124850}{-9.9}{GW200214_223307}{-18.1}{GW200225_075134}{-13.1}{GW200301_211019}{-6.5}{GW200304_172806}{-42.3}{GW200305_084739}{-8.4}{GW200318_191337}{-13.2}}}
\DeclareRobustCommand{\massonesourceupper}[1]{\IfEqCase{#1}{{GW151205_195525}{21.5}{GW151216_092416}{12.0}{GW170121_212536}{7.5}{GW170202_135657}{9.6}{GW170304_163753}{11.8}{GW170403_230611}{14.1}{GW170425_055334}{16.5}{GW170727_010430}{9.4}{GW190426_082124}{16.0}{GW190427_180650}{3.9}{GW190511_125545}{15.8}{GW190511_163209}{109.4}{GW190514_065416}{13.8}{GW190523_085933}{24.6}{GW190524_134109}{27.5}{GW190530_030659}{13.8}{GW190530_133833}{45.5}{GW190604_103812}{69.2}{GW190605_025957}{81.2}{GW190607_083827}{11.8}{GW190614_134749}{20.6}{GW190615_030234}{18.1}{GW190705_164632}{20.8}{GW190707_083226}{18.4}{GW190711_030756}{39.6}{GW190718_160159}{4.6}{GW190725_174728}{3.6}{GW190805_105432}{4.0}{GW190806_033721}{29.7}{GW190814_192009}{25.2}{GW190818_232544}{37.8}{GW190821_124821}{3.5}{GW190904_104631}{31.4}{GW190906_054335}{15.6}{GW190910_012619}{5.1}{GW190911_195101}{30.8}{GW190916_200658}{18.7}{GW190926_050336}{16.1}{GW191113_103541}{32.5}{GW191117_023843}{37.4}{GW191127_050227}{30.7}{GW191208_080334}{12.0}{GW191224_043228}{4.8}{GW191228_195619}{116.3}{GW200106_134123}{14.2}{GW200109_195634}{20.9}{GW200129_114245}{34.0}{GW200208_211609}{815.2}{GW200210_005122}{3.0}{GW200210_100022}{26.3}{GW200220_124850}{10.9}{GW200214_223307}{21.8}{GW200225_075134}{14.4}{GW200301_211019}{8.9}{GW200304_172806}{49.4}{GW200305_084739}{9.8}{GW200318_191337}{15.2}}}
\DeclareRobustCommand{\massonesourcemedian}[1]{\IfEqCase{#1}{{GW151205_195525}{63.3}{GW151216_092416}{37.6}{GW170121_212536}{32.3}{GW170202_135657}{26.4}{GW170304_163753}{44.1}{GW170403_230611}{47.4}{GW170425_055334}{48.3}{GW170727_010430}{39.8}{GW190426_082124}{40.6}{GW190427_180650}{11.6}{GW190511_125545}{41.9}{GW190511_163209}{87.2}{GW190514_065416}{39.4}{GW190523_085933}{46.2}{GW190524_134109}{70.0}{GW190530_030659}{32.1}{GW190530_133833}{92.7}{GW190604_103812}{114.1}{GW190605_025957}{132.4}{GW190607_083827}{42.6}{GW190614_134749}{40.3}{GW190615_030234}{68.9}{GW190705_164632}{53.8}{GW190707_083226}{53.5}{GW190711_030756}{62.8}{GW190718_160159}{9.9}{GW190725_174728}{11.6}{GW190805_105432}{14.2}{GW190806_033721}{74.3}{GW190814_192009}{72.9}{GW190818_232544}{78.1}{GW190821_124821}{10.9}{GW190904_104631}{39.7}{GW190906_054335}{39.1}{GW190910_012619}{34.3}{GW190911_195101}{84.5}{GW190916_200658}{45.0}{GW190926_050336}{38.9}{GW191113_103541}{101.4}{GW191117_023843}{90.5}{GW191127_050227}{57.0}{GW191208_080334}{37.1}{GW191224_043228}{15.3}{GW191228_195619}{164.6}{GW200106_134123}{44.6}{GW200109_195634}{63.4}{GW200129_114245}{80.5}{GW200208_211609}{31.5}{GW200210_005122}{9.0}{GW200210_100022}{85.3}{GW200220_124850}{38.0}{GW200214_223307}{48.9}{GW200225_075134}{57.7}{GW200301_211019}{21.3}{GW200304_172806}{94.2}{GW200305_084739}{32.7}{GW200318_191337}{45.4}}}
\DeclareRobustCommand{\massonesource}[1]{\IfEqCase{#1}{{GW151205_195525}{$\massonesourcemedian{GW151205_195525}^{+\massonesourceupper{GW151205_195525}}_{\massonesourcelower{GW151205_195525}}$}{GW151216_092416}{$\massonesourcemedian{GW151216_092416}^{+\massonesourceupper{GW151216_092416}}_{\massonesourcelower{GW151216_092416}}$}{GW170121_212536}{$\massonesourcemedian{GW170121_212536}^{+\massonesourceupper{GW170121_212536}}_{\massonesourcelower{GW170121_212536}}$}{GW170202_135657}{$\massonesourcemedian{GW170202_135657}^{+\massonesourceupper{GW170202_135657}}_{\massonesourcelower{GW170202_135657}}$}{GW170304_163753}{$\massonesourcemedian{GW170304_163753}^{+\massonesourceupper{GW170304_163753}}_{\massonesourcelower{GW170304_163753}}$}{GW170403_230611}{$\massonesourcemedian{GW170403_230611}^{+\massonesourceupper{GW170403_230611}}_{\massonesourcelower{GW170403_230611}}$}{GW170425_055334}{$\massonesourcemedian{GW170425_055334}^{+\massonesourceupper{GW170425_055334}}_{\massonesourcelower{GW170425_055334}}$}{GW170727_010430}{$\massonesourcemedian{GW170727_010430}^{+\massonesourceupper{GW170727_010430}}_{\massonesourcelower{GW170727_010430}}$}{GW190426_082124}{$\massonesourcemedian{GW190426_082124}^{+\massonesourceupper{GW190426_082124}}_{\massonesourcelower{GW190426_082124}}$}{GW190427_180650}{$\massonesourcemedian{GW190427_180650}^{+\massonesourceupper{GW190427_180650}}_{\massonesourcelower{GW190427_180650}}$}{GW190511_125545}{$\massonesourcemedian{GW190511_125545}^{+\massonesourceupper{GW190511_125545}}_{\massonesourcelower{GW190511_125545}}$}{GW190511_163209}{$\massonesourcemedian{GW190511_163209}^{+\massonesourceupper{GW190511_163209}}_{\massonesourcelower{GW190511_163209}}$}{GW190514_065416}{$\massonesourcemedian{GW190514_065416}^{+\massonesourceupper{GW190514_065416}}_{\massonesourcelower{GW190514_065416}}$}{GW190523_085933}{$\massonesourcemedian{GW190523_085933}^{+\massonesourceupper{GW190523_085933}}_{\massonesourcelower{GW190523_085933}}$}{GW190524_134109}{$\massonesourcemedian{GW190524_134109}^{+\massonesourceupper{GW190524_134109}}_{\massonesourcelower{GW190524_134109}}$}{GW190530_030659}{$\massonesourcemedian{GW190530_030659}^{+\massonesourceupper{GW190530_030659}}_{\massonesourcelower{GW190530_030659}}$}{GW190530_133833}{$\massonesourcemedian{GW190530_133833}^{+\massonesourceupper{GW190530_133833}}_{\massonesourcelower{GW190530_133833}}$}{GW190604_103812}{$\massonesourcemedian{GW190604_103812}^{+\massonesourceupper{GW190604_103812}}_{\massonesourcelower{GW190604_103812}}$}{GW190605_025957}{$\massonesourcemedian{GW190605_025957}^{+\massonesourceupper{GW190605_025957}}_{\massonesourcelower{GW190605_025957}}$}{GW190607_083827}{$\massonesourcemedian{GW190607_083827}^{+\massonesourceupper{GW190607_083827}}_{\massonesourcelower{GW190607_083827}}$}{GW190614_134749}{$\massonesourcemedian{GW190614_134749}^{+\massonesourceupper{GW190614_134749}}_{\massonesourcelower{GW190614_134749}}$}{GW190615_030234}{$\massonesourcemedian{GW190615_030234}^{+\massonesourceupper{GW190615_030234}}_{\massonesourcelower{GW190615_030234}}$}{GW190705_164632}{$\massonesourcemedian{GW190705_164632}^{+\massonesourceupper{GW190705_164632}}_{\massonesourcelower{GW190705_164632}}$}{GW190707_083226}{$\massonesourcemedian{GW190707_083226}^{+\massonesourceupper{GW190707_083226}}_{\massonesourcelower{GW190707_083226}}$}{GW190711_030756}{$\massonesourcemedian{GW190711_030756}^{+\massonesourceupper{GW190711_030756}}_{\massonesourcelower{GW190711_030756}}$}{GW190718_160159}{$\massonesourcemedian{GW190718_160159}^{+\massonesourceupper{GW190718_160159}}_{\massonesourcelower{GW190718_160159}}$}{GW190725_174728}{$\massonesourcemedian{GW190725_174728}^{+\massonesourceupper{GW190725_174728}}_{\massonesourcelower{GW190725_174728}}$}{GW190805_105432}{$\massonesourcemedian{GW190805_105432}^{+\massonesourceupper{GW190805_105432}}_{\massonesourcelower{GW190805_105432}}$}{GW190806_033721}{$\massonesourcemedian{GW190806_033721}^{+\massonesourceupper{GW190806_033721}}_{\massonesourcelower{GW190806_033721}}$}{GW190814_192009}{$\massonesourcemedian{GW190814_192009}^{+\massonesourceupper{GW190814_192009}}_{\massonesourcelower{GW190814_192009}}$}{GW190818_232544}{$\massonesourcemedian{GW190818_232544}^{+\massonesourceupper{GW190818_232544}}_{\massonesourcelower{GW190818_232544}}$}{GW190821_124821}{$\massonesourcemedian{GW190821_124821}^{+\massonesourceupper{GW190821_124821}}_{\massonesourcelower{GW190821_124821}}$}{GW190904_104631}{$\massonesourcemedian{GW190904_104631}^{+\massonesourceupper{GW190904_104631}}_{\massonesourcelower{GW190904_104631}}$}{GW190906_054335}{$\massonesourcemedian{GW190906_054335}^{+\massonesourceupper{GW190906_054335}}_{\massonesourcelower{GW190906_054335}}$}{GW190910_012619}{$\massonesourcemedian{GW190910_012619}^{+\massonesourceupper{GW190910_012619}}_{\massonesourcelower{GW190910_012619}}$}{GW190911_195101}{$\massonesourcemedian{GW190911_195101}^{+\massonesourceupper{GW190911_195101}}_{\massonesourcelower{GW190911_195101}}$}{GW190916_200658}{$\massonesourcemedian{GW190916_200658}^{+\massonesourceupper{GW190916_200658}}_{\massonesourcelower{GW190916_200658}}$}{GW190926_050336}{$\massonesourcemedian{GW190926_050336}^{+\massonesourceupper{GW190926_050336}}_{\massonesourcelower{GW190926_050336}}$}{GW191113_103541}{$\massonesourcemedian{GW191113_103541}^{+\massonesourceupper{GW191113_103541}}_{\massonesourcelower{GW191113_103541}}$}{GW191117_023843}{$\massonesourcemedian{GW191117_023843}^{+\massonesourceupper{GW191117_023843}}_{\massonesourcelower{GW191117_023843}}$}{GW191127_050227}{$\massonesourcemedian{GW191127_050227}^{+\massonesourceupper{GW191127_050227}}_{\massonesourcelower{GW191127_050227}}$}{GW191208_080334}{$\massonesourcemedian{GW191208_080334}^{+\massonesourceupper{GW191208_080334}}_{\massonesourcelower{GW191208_080334}}$}{GW191224_043228}{$\massonesourcemedian{GW191224_043228}^{+\massonesourceupper{GW191224_043228}}_{\massonesourcelower{GW191224_043228}}$}{GW191228_195619}{$\massonesourcemedian{GW191228_195619}^{+\massonesourceupper{GW191228_195619}}_{\massonesourcelower{GW191228_195619}}$}{GW200106_134123}{$\massonesourcemedian{GW200106_134123}^{+\massonesourceupper{GW200106_134123}}_{\massonesourcelower{GW200106_134123}}$}{GW200109_195634}{$\massonesourcemedian{GW200109_195634}^{+\massonesourceupper{GW200109_195634}}_{\massonesourcelower{GW200109_195634}}$}{GW200129_114245}{$\massonesourcemedian{GW200129_114245}^{+\massonesourceupper{GW200129_114245}}_{\massonesourcelower{GW200129_114245}}$}{GW200208_211609}{$\massonesourcemedian{GW200208_211609}^{+\massonesourceupper{GW200208_211609}}_{\massonesourcelower{GW200208_211609}}$}{GW200210_005122}{$\massonesourcemedian{GW200210_005122}^{+\massonesourceupper{GW200210_005122}}_{\massonesourcelower{GW200210_005122}}$}{GW200210_100022}{$\massonesourcemedian{GW200210_100022}^{+\massonesourceupper{GW200210_100022}}_{\massonesourcelower{GW200210_100022}}$}{GW200220_124850}{$\massonesourcemedian{GW200220_124850}^{+\massonesourceupper{GW200220_124850}}_{\massonesourcelower{GW200220_124850}}$}{GW200214_223307}{$\massonesourcemedian{GW200214_223307}^{+\massonesourceupper{GW200214_223307}}_{\massonesourcelower{GW200214_223307}}$}{GW200225_075134}{$\massonesourcemedian{GW200225_075134}^{+\massonesourceupper{GW200225_075134}}_{\massonesourcelower{GW200225_075134}}$}{GW200301_211019}{$\massonesourcemedian{GW200301_211019}^{+\massonesourceupper{GW200301_211019}}_{\massonesourcelower{GW200301_211019}}$}{GW200304_172806}{$\massonesourcemedian{GW200304_172806}^{+\massonesourceupper{GW200304_172806}}_{\massonesourcelower{GW200304_172806}}$}{GW200305_084739}{$\massonesourcemedian{GW200305_084739}^{+\massonesourceupper{GW200305_084739}}_{\massonesourcelower{GW200305_084739}}$}{GW200318_191337}{$\massonesourcemedian{GW200318_191337}^{+\massonesourceupper{GW200318_191337}}_{\massonesourcelower{GW200318_191337}}$}}}

\DeclareRobustCommand{\masstwosourcelower}[1]{\IfEqCase{#1}{{GW151205_195525}{-17.7}{GW151216_092416}{-6.5}{GW170121_212536}{-5.7}{GW170202_135657}{-5.5}{GW170304_163753}{-10.0}{GW170403_230611}{-11.0}{GW170425_055334}{-10.7}{GW170727_010430}{-7.8}{GW190426_082124}{-10.6}{GW190427_180650}{-1.7}{GW190511_125545}{-12.4}{GW190511_163209}{-16.7}{GW190514_065416}{-9.1}{GW190523_085933}{-11.6}{GW190524_134109}{-24.4}{GW190530_030659}{-7.3}{GW190530_133833}{-29.6}{GW190604_103812}{-25.8}{GW190605_025957}{-60.2}{GW190607_083827}{-10.1}{GW190614_134749}{-11.4}{GW190615_030234}{-18.9}{GW190705_164632}{-14.7}{GW190707_083226}{-13.9}{GW190711_030756}{-9.8}{GW190718_160159}{-2.3}{GW190725_174728}{-1.5}{GW190805_105432}{-1.8}{GW190806_033721}{-20.0}{GW190814_192009}{-22.6}{GW190818_232544}{-17.6}{GW190821_124821}{-0.7}{GW190904_104631}{-16.3}{GW190906_054335}{-12.1}{GW190910_012619}{-0.6}{GW190911_195101}{-26.0}{GW190916_200658}{-11.0}{GW190926_050336}{-10.1}{GW191113_103541}{-7.9}{GW191117_023843}{-8.4}{GW191127_050227}{-12.9}{GW191208_080334}{-6.6}{GW191224_043228}{-2.0}{GW191228_195619}{-35.9}{GW200106_134123}{-11.3}{GW200109_195634}{-18.6}{GW200129_114245}{-16.0}{GW200208_211609}{-15.4}{GW200210_005122}{-1.5}{GW200210_100022}{-5.6}{GW200220_124850}{-9.2}{GW200214_223307}{-15.0}{GW200225_075134}{-12.8}{GW200301_211019}{-4.6}{GW200304_172806}{-29.5}{GW200305_084739}{-8.0}{GW200318_191337}{-10.8}}}
\DeclareRobustCommand{\masstwosourceupper}[1]{\IfEqCase{#1}{{GW151205_195525}{17.4}{GW151216_092416}{6.9}{GW170121_212536}{5.4}{GW170202_135657}{5.9}{GW170304_163753}{10.0}{GW170403_230611}{11.0}{GW170425_055334}{10.3}{GW170727_010430}{7.8}{GW190426_082124}{10.8}{GW190427_180650}{1.9}{GW190511_125545}{12.6}{GW190511_163209}{26.7}{GW190514_065416}{9.0}{GW190523_085933}{24.9}{GW190524_134109}{22.1}{GW190530_030659}{6.7}{GW190530_133833}{29.2}{GW190604_103812}{29.7}{GW190605_025957}{53.5}{GW190607_083827}{10.1}{GW190614_134749}{11.2}{GW190615_030234}{18.4}{GW190705_164632}{15.3}{GW190707_083226}{13.3}{GW190711_030756}{9.9}{GW190718_160159}{1.8}{GW190725_174728}{1.9}{GW190805_105432}{2.1}{GW190806_033721}{21.6}{GW190814_192009}{23.0}{GW190818_232544}{21.3}{GW190821_124821}{1.1}{GW190904_104631}{16.9}{GW190906_054335}{11.0}{GW190910_012619}{0.7}{GW190911_195101}{25.3}{GW190916_200658}{11.8}{GW190926_050336}{9.4}{GW191113_103541}{12.6}{GW191117_023843}{8.5}{GW191127_050227}{14.7}{GW191208_080334}{7.5}{GW191224_043228}{2.4}{GW191228_195619}{37.6}{GW200106_134123}{11.8}{GW200109_195634}{17.8}{GW200129_114245}{17.4}{GW200208_211609}{34.6}{GW200210_005122}{1.5}{GW200210_100022}{8.8}{GW200220_124850}{9.0}{GW200214_223307}{14.2}{GW200225_075134}{13.2}{GW200301_211019}{4.3}{GW200304_172806}{28.9}{GW200305_084739}{7.7}{GW200318_191337}{12.1}}}
\DeclareRobustCommand{\masstwosourcemedian}[1]{\IfEqCase{#1}{{GW151205_195525}{41.6}{GW151216_092416}{15.8}{GW170121_212536}{24.6}{GW170202_135657}{16.7}{GW170304_163753}{31.9}{GW170403_230611}{34.9}{GW170425_055334}{28.5}{GW170727_010430}{30.4}{GW190426_082124}{27.7}{GW190427_180650}{6.3}{GW190511_125545}{25.1}{GW190511_163209}{24.7}{GW190514_065416}{28.7}{GW190523_085933}{16.9}{GW190524_134109}{43.6}{GW190530_030659}{19.4}{GW190530_133833}{51.9}{GW190604_103812}{40.1}{GW190605_025957}{83.3}{GW190607_083827}{31.8}{GW190614_134749}{24.0}{GW190615_030234}{49.8}{GW190705_164632}{29.4}{GW190707_083226}{33.3}{GW190711_030756}{21.5}{GW190718_160159}{6.7}{GW190725_174728}{6.4}{GW190805_105432}{7.4}{GW190806_033721}{35.1}{GW190814_192009}{48.5}{GW190818_232544}{30.7}{GW190821_124821}{3.0}{GW190904_104631}{23.6}{GW190906_054335}{23.8}{GW190910_012619}{4.2}{GW190911_195101}{58.3}{GW190916_200658}{23.8}{GW190926_050336}{23.8}{GW191113_103541}{19.5}{GW191117_023843}{23.6}{GW191127_050227}{25.3}{GW191208_080334}{19.2}{GW191224_043228}{7.7}{GW191228_195619}{78.4}{GW200106_134123}{27.2}{GW200109_195634}{43.6}{GW200129_114245}{34.2}{GW200208_211609}{19.4}{GW200210_005122}{6.0}{GW200210_100022}{15.0}{GW200220_124850}{27.5}{GW200214_223307}{32.3}{GW200225_075134}{43.1}{GW200301_211019}{13.3}{GW200304_172806}{43.4}{GW200305_084739}{23.8}{GW200318_191337}{30.8}}}
\DeclareRobustCommand{\masstwosource}[1]{\IfEqCase{#1}{{GW151205_195525}{$\masstwosourcemedian{GW151205_195525}^{+\masstwosourceupper{GW151205_195525}}_{\masstwosourcelower{GW151205_195525}}$}{GW151216_092416}{$\masstwosourcemedian{GW151216_092416}^{+\masstwosourceupper{GW151216_092416}}_{\masstwosourcelower{GW151216_092416}}$}{GW170121_212536}{$\masstwosourcemedian{GW170121_212536}^{+\masstwosourceupper{GW170121_212536}}_{\masstwosourcelower{GW170121_212536}}$}{GW170202_135657}{$\masstwosourcemedian{GW170202_135657}^{+\masstwosourceupper{GW170202_135657}}_{\masstwosourcelower{GW170202_135657}}$}{GW170304_163753}{$\masstwosourcemedian{GW170304_163753}^{+\masstwosourceupper{GW170304_163753}}_{\masstwosourcelower{GW170304_163753}}$}{GW170403_230611}{$\masstwosourcemedian{GW170403_230611}^{+\masstwosourceupper{GW170403_230611}}_{\masstwosourcelower{GW170403_230611}}$}{GW170425_055334}{$\masstwosourcemedian{GW170425_055334}^{+\masstwosourceupper{GW170425_055334}}_{\masstwosourcelower{GW170425_055334}}$}{GW170727_010430}{$\masstwosourcemedian{GW170727_010430}^{+\masstwosourceupper{GW170727_010430}}_{\masstwosourcelower{GW170727_010430}}$}{GW190426_082124}{$\masstwosourcemedian{GW190426_082124}^{+\masstwosourceupper{GW190426_082124}}_{\masstwosourcelower{GW190426_082124}}$}{GW190427_180650}{$\masstwosourcemedian{GW190427_180650}^{+\masstwosourceupper{GW190427_180650}}_{\masstwosourcelower{GW190427_180650}}$}{GW190511_125545}{$\masstwosourcemedian{GW190511_125545}^{+\masstwosourceupper{GW190511_125545}}_{\masstwosourcelower{GW190511_125545}}$}{GW190511_163209}{$\masstwosourcemedian{GW190511_163209}^{+\masstwosourceupper{GW190511_163209}}_{\masstwosourcelower{GW190511_163209}}$}{GW190514_065416}{$\masstwosourcemedian{GW190514_065416}^{+\masstwosourceupper{GW190514_065416}}_{\masstwosourcelower{GW190514_065416}}$}{GW190523_085933}{$\masstwosourcemedian{GW190523_085933}^{+\masstwosourceupper{GW190523_085933}}_{\masstwosourcelower{GW190523_085933}}$}{GW190524_134109}{$\masstwosourcemedian{GW190524_134109}^{+\masstwosourceupper{GW190524_134109}}_{\masstwosourcelower{GW190524_134109}}$}{GW190530_030659}{$\masstwosourcemedian{GW190530_030659}^{+\masstwosourceupper{GW190530_030659}}_{\masstwosourcelower{GW190530_030659}}$}{GW190530_133833}{$\masstwosourcemedian{GW190530_133833}^{+\masstwosourceupper{GW190530_133833}}_{\masstwosourcelower{GW190530_133833}}$}{GW190604_103812}{$\masstwosourcemedian{GW190604_103812}^{+\masstwosourceupper{GW190604_103812}}_{\masstwosourcelower{GW190604_103812}}$}{GW190605_025957}{$\masstwosourcemedian{GW190605_025957}^{+\masstwosourceupper{GW190605_025957}}_{\masstwosourcelower{GW190605_025957}}$}{GW190607_083827}{$\masstwosourcemedian{GW190607_083827}^{+\masstwosourceupper{GW190607_083827}}_{\masstwosourcelower{GW190607_083827}}$}{GW190614_134749}{$\masstwosourcemedian{GW190614_134749}^{+\masstwosourceupper{GW190614_134749}}_{\masstwosourcelower{GW190614_134749}}$}{GW190615_030234}{$\masstwosourcemedian{GW190615_030234}^{+\masstwosourceupper{GW190615_030234}}_{\masstwosourcelower{GW190615_030234}}$}{GW190705_164632}{$\masstwosourcemedian{GW190705_164632}^{+\masstwosourceupper{GW190705_164632}}_{\masstwosourcelower{GW190705_164632}}$}{GW190707_083226}{$\masstwosourcemedian{GW190707_083226}^{+\masstwosourceupper{GW190707_083226}}_{\masstwosourcelower{GW190707_083226}}$}{GW190711_030756}{$\masstwosourcemedian{GW190711_030756}^{+\masstwosourceupper{GW190711_030756}}_{\masstwosourcelower{GW190711_030756}}$}{GW190718_160159}{$\masstwosourcemedian{GW190718_160159}^{+\masstwosourceupper{GW190718_160159}}_{\masstwosourcelower{GW190718_160159}}$}{GW190725_174728}{$\masstwosourcemedian{GW190725_174728}^{+\masstwosourceupper{GW190725_174728}}_{\masstwosourcelower{GW190725_174728}}$}{GW190805_105432}{$\masstwosourcemedian{GW190805_105432}^{+\masstwosourceupper{GW190805_105432}}_{\masstwosourcelower{GW190805_105432}}$}{GW190806_033721}{$\masstwosourcemedian{GW190806_033721}^{+\masstwosourceupper{GW190806_033721}}_{\masstwosourcelower{GW190806_033721}}$}{GW190814_192009}{$\masstwosourcemedian{GW190814_192009}^{+\masstwosourceupper{GW190814_192009}}_{\masstwosourcelower{GW190814_192009}}$}{GW190818_232544}{$\masstwosourcemedian{GW190818_232544}^{+\masstwosourceupper{GW190818_232544}}_{\masstwosourcelower{GW190818_232544}}$}{GW190821_124821}{$\masstwosourcemedian{GW190821_124821}^{+\masstwosourceupper{GW190821_124821}}_{\masstwosourcelower{GW190821_124821}}$}{GW190904_104631}{$\masstwosourcemedian{GW190904_104631}^{+\masstwosourceupper{GW190904_104631}}_{\masstwosourcelower{GW190904_104631}}$}{GW190906_054335}{$\masstwosourcemedian{GW190906_054335}^{+\masstwosourceupper{GW190906_054335}}_{\masstwosourcelower{GW190906_054335}}$}{GW190910_012619}{$\masstwosourcemedian{GW190910_012619}^{+\masstwosourceupper{GW190910_012619}}_{\masstwosourcelower{GW190910_012619}}$}{GW190911_195101}{$\masstwosourcemedian{GW190911_195101}^{+\masstwosourceupper{GW190911_195101}}_{\masstwosourcelower{GW190911_195101}}$}{GW190916_200658}{$\masstwosourcemedian{GW190916_200658}^{+\masstwosourceupper{GW190916_200658}}_{\masstwosourcelower{GW190916_200658}}$}{GW190926_050336}{$\masstwosourcemedian{GW190926_050336}^{+\masstwosourceupper{GW190926_050336}}_{\masstwosourcelower{GW190926_050336}}$}{GW191113_103541}{$\masstwosourcemedian{GW191113_103541}^{+\masstwosourceupper{GW191113_103541}}_{\masstwosourcelower{GW191113_103541}}$}{GW191117_023843}{$\masstwosourcemedian{GW191117_023843}^{+\masstwosourceupper{GW191117_023843}}_{\masstwosourcelower{GW191117_023843}}$}{GW191127_050227}{$\masstwosourcemedian{GW191127_050227}^{+\masstwosourceupper{GW191127_050227}}_{\masstwosourcelower{GW191127_050227}}$}{GW191208_080334}{$\masstwosourcemedian{GW191208_080334}^{+\masstwosourceupper{GW191208_080334}}_{\masstwosourcelower{GW191208_080334}}$}{GW191224_043228}{$\masstwosourcemedian{GW191224_043228}^{+\masstwosourceupper{GW191224_043228}}_{\masstwosourcelower{GW191224_043228}}$}{GW191228_195619}{$\masstwosourcemedian{GW191228_195619}^{+\masstwosourceupper{GW191228_195619}}_{\masstwosourcelower{GW191228_195619}}$}{GW200106_134123}{$\masstwosourcemedian{GW200106_134123}^{+\masstwosourceupper{GW200106_134123}}_{\masstwosourcelower{GW200106_134123}}$}{GW200109_195634}{$\masstwosourcemedian{GW200109_195634}^{+\masstwosourceupper{GW200109_195634}}_{\masstwosourcelower{GW200109_195634}}$}{GW200129_114245}{$\masstwosourcemedian{GW200129_114245}^{+\masstwosourceupper{GW200129_114245}}_{\masstwosourcelower{GW200129_114245}}$}{GW200208_211609}{$\masstwosourcemedian{GW200208_211609}^{+\masstwosourceupper{GW200208_211609}}_{\masstwosourcelower{GW200208_211609}}$}{GW200210_005122}{$\masstwosourcemedian{GW200210_005122}^{+\masstwosourceupper{GW200210_005122}}_{\masstwosourcelower{GW200210_005122}}$}{GW200210_100022}{$\masstwosourcemedian{GW200210_100022}^{+\masstwosourceupper{GW200210_100022}}_{\masstwosourcelower{GW200210_100022}}$}{GW200220_124850}{$\masstwosourcemedian{GW200220_124850}^{+\masstwosourceupper{GW200220_124850}}_{\masstwosourcelower{GW200220_124850}}$}{GW200214_223307}{$\masstwosourcemedian{GW200214_223307}^{+\masstwosourceupper{GW200214_223307}}_{\masstwosourcelower{GW200214_223307}}$}{GW200225_075134}{$\masstwosourcemedian{GW200225_075134}^{+\masstwosourceupper{GW200225_075134}}_{\masstwosourcelower{GW200225_075134}}$}{GW200301_211019}{$\masstwosourcemedian{GW200301_211019}^{+\masstwosourceupper{GW200301_211019}}_{\masstwosourcelower{GW200301_211019}}$}{GW200304_172806}{$\masstwosourcemedian{GW200304_172806}^{+\masstwosourceupper{GW200304_172806}}_{\masstwosourcelower{GW200304_172806}}$}{GW200305_084739}{$\masstwosourcemedian{GW200305_084739}^{+\masstwosourceupper{GW200305_084739}}_{\masstwosourcelower{GW200305_084739}}$}{GW200318_191337}{$\masstwosourcemedian{GW200318_191337}^{+\masstwosourceupper{GW200318_191337}}_{\masstwosourcelower{GW200318_191337}}$}}}

\DeclareRobustCommand{\chiefflower}[1]{\IfEqCase{#1}{{GW151205_195525}{-0.36}{GW151216_092416}{-0.36}{GW170121_212536}{-0.27}{GW170202_135657}{-0.31}{GW170304_163753}{-0.26}{GW170403_230611}{-0.34}{GW170425_055334}{-0.31}{GW170727_010430}{-0.27}{GW190426_082124}{-0.35}{GW190427_180650}{-0.13}{GW190511_125545}{-0.29}{GW190511_163209}{-0.40}{GW190514_065416}{-0.32}{GW190523_085933}{-0.46}{GW190524_134109}{-0.39}{GW190530_030659}{-0.23}{GW190530_133833}{-0.43}{GW190604_103812}{-0.61}{GW190605_025957}{-0.38}{GW190607_083827}{-0.30}{GW190614_134749}{-0.32}{GW190615_030234}{-0.31}{GW190705_164632}{-0.32}{GW190707_083226}{-0.34}{GW190711_030756}{-0.28}{GW190718_160159}{-0.47}{GW190725_174728}{-0.14}{GW190805_105432}{-0.16}{GW190806_033721}{-0.39}{GW190814_192009}{-0.37}{GW190818_232544}{-0.45}{GW190821_124821}{-0.31}{GW190904_104631}{-0.34}{GW190906_054335}{-0.32}{GW190910_012619}{-0.22}{GW190911_195101}{-0.38}{GW190916_200658}{-0.29}{GW190926_050336}{-0.33}{GW191113_103541}{-0.20}{GW191117_023843}{-0.35}{GW191127_050227}{-0.32}{GW191208_080334}{-0.25}{GW191224_043228}{-0.13}{GW191228_195619}{-0.38}{GW200106_134123}{-0.30}{GW200109_195634}{-0.38}{GW200129_114245}{-0.41}{GW200208_211609}{-0.34}{GW200210_005122}{-0.10}{GW200210_100022}{-0.89}{GW200220_124850}{-0.32}{GW200214_223307}{-0.33}{GW200225_075134}{-0.36}{GW200301_211019}{-0.27}{GW200304_172806}{-0.51}{GW200305_084739}{-0.32}{GW200318_191337}{-0.33}}}
\DeclareRobustCommand{\chieffupper}[1]{\IfEqCase{#1}{{GW151205_195525}{0.40}{GW151216_092416}{0.28}{GW170121_212536}{0.26}{GW170202_135657}{0.32}{GW170304_163753}{0.27}{GW170403_230611}{0.33}{GW170425_055334}{0.28}{GW170727_010430}{0.25}{GW190426_082124}{0.31}{GW190427_180650}{0.18}{GW190511_125545}{0.29}{GW190511_163209}{0.49}{GW190514_065416}{0.29}{GW190523_085933}{0.35}{GW190524_134109}{0.37}{GW190530_030659}{0.22}{GW190530_133833}{0.44}{GW190604_103812}{0.33}{GW190605_025957}{0.45}{GW190607_083827}{0.29}{GW190614_134749}{0.31}{GW190615_030234}{0.31}{GW190705_164632}{0.31}{GW190707_083226}{0.31}{GW190711_030756}{0.38}{GW190718_160159}{0.21}{GW190725_174728}{0.17}{GW190805_105432}{0.16}{GW190806_033721}{0.29}{GW190814_192009}{0.38}{GW190818_232544}{0.32}{GW190821_124821}{0.23}{GW190904_104631}{0.33}{GW190906_054335}{0.35}{GW190910_012619}{0.20}{GW190911_195101}{0.41}{GW190916_200658}{0.33}{GW190926_050336}{0.29}{GW191113_103541}{0.19}{GW191117_023843}{0.27}{GW191127_050227}{0.32}{GW191208_080334}{0.29}{GW191224_043228}{0.15}{GW191228_195619}{0.35}{GW200106_134123}{0.33}{GW200109_195634}{0.40}{GW200129_114245}{0.46}{GW200208_211609}{0.35}{GW200210_005122}{0.15}{GW200210_100022}{0.30}{GW200220_124850}{0.28}{GW200214_223307}{0.32}{GW200225_075134}{0.32}{GW200301_211019}{0.23}{GW200304_172806}{0.44}{GW200305_084739}{0.30}{GW200318_191337}{0.32}}}
\DeclareRobustCommand{\chieffmedian}[1]{\IfEqCase{#1}{{GW151205_195525}{0.13}{GW151216_092416}{0.54}{GW170121_212536}{-0.25}{GW170202_135657}{-0.11}{GW170304_163753}{0.13}{GW170403_230611}{-0.24}{GW170425_055334}{-0.05}{GW170727_010430}{-0.04}{GW190426_082124}{-0.13}{GW190427_180650}{0.00}{GW190511_125545}{0.23}{GW190511_163209}{0.01}{GW190514_065416}{-0.17}{GW190523_085933}{0.40}{GW190524_134109}{0.27}{GW190530_030659}{0.36}{GW190530_133833}{0.05}{GW190604_103812}{0.54}{GW190605_025957}{0.10}{GW190607_083827}{-0.00}{GW190614_134749}{-0.00}{GW190615_030234}{0.00}{GW190705_164632}{0.26}{GW190707_083226}{-0.05}{GW190711_030756}{0.05}{GW190718_160159}{0.61}{GW190725_174728}{-0.05}{GW190805_105432}{-0.10}{GW190806_033721}{0.52}{GW190814_192009}{0.19}{GW190818_232544}{0.41}{GW190821_124821}{-0.16}{GW190904_104631}{0.00}{GW190906_054335}{0.09}{GW190910_012619}{-0.55}{GW190911_195101}{0.11}{GW190916_200658}{0.17}{GW190926_050336}{-0.05}{GW191113_103541}{0.57}{GW191117_023843}{-0.18}{GW191127_050227}{0.24}{GW191208_080334}{0.13}{GW191224_043228}{0.09}{GW191228_195619}{-0.01}{GW200106_134123}{0.11}{GW200109_195634}{0.07}{GW200129_114245}{0.17}{GW200208_211609}{0.03}{GW200210_005122}{0.05}{GW200210_100022}{0.63}{GW200220_124850}{-0.06}{GW200214_223307}{0.03}{GW200225_075134}{-0.07}{GW200301_211019}{-0.12}{GW200304_172806}{0.39}{GW200305_084739}{-0.07}{GW200318_191337}{0.10}}}
\DeclareRobustCommand{\chieff}[1]{\IfEqCase{#1}{{GW151205_195525}{$\chieffmedian{GW151205_195525}^{+\chieffupper{GW151205_195525}}_{\chiefflower{GW151205_195525}}$}{GW151216_092416}{$\chieffmedian{GW151216_092416}^{+\chieffupper{GW151216_092416}}_{\chiefflower{GW151216_092416}}$}{GW170121_212536}{$\chieffmedian{GW170121_212536}^{+\chieffupper{GW170121_212536}}_{\chiefflower{GW170121_212536}}$}{GW170202_135657}{$\chieffmedian{GW170202_135657}^{+\chieffupper{GW170202_135657}}_{\chiefflower{GW170202_135657}}$}{GW170304_163753}{$\chieffmedian{GW170304_163753}^{+\chieffupper{GW170304_163753}}_{\chiefflower{GW170304_163753}}$}{GW170403_230611}{$\chieffmedian{GW170403_230611}^{+\chieffupper{GW170403_230611}}_{\chiefflower{GW170403_230611}}$}{GW170425_055334}{$\chieffmedian{GW170425_055334}^{+\chieffupper{GW170425_055334}}_{\chiefflower{GW170425_055334}}$}{GW170727_010430}{$\chieffmedian{GW170727_010430}^{+\chieffupper{GW170727_010430}}_{\chiefflower{GW170727_010430}}$}{GW190426_082124}{$\chieffmedian{GW190426_082124}^{+\chieffupper{GW190426_082124}}_{\chiefflower{GW190426_082124}}$}{GW190427_180650}{$\chieffmedian{GW190427_180650}^{+\chieffupper{GW190427_180650}}_{\chiefflower{GW190427_180650}}$}{GW190511_125545}{$\chieffmedian{GW190511_125545}^{+\chieffupper{GW190511_125545}}_{\chiefflower{GW190511_125545}}$}{GW190511_163209}{$\chieffmedian{GW190511_163209}^{+\chieffupper{GW190511_163209}}_{\chiefflower{GW190511_163209}}$}{GW190514_065416}{$\chieffmedian{GW190514_065416}^{+\chieffupper{GW190514_065416}}_{\chiefflower{GW190514_065416}}$}{GW190523_085933}{$\chieffmedian{GW190523_085933}^{+\chieffupper{GW190523_085933}}_{\chiefflower{GW190523_085933}}$}{GW190524_134109}{$\chieffmedian{GW190524_134109}^{+\chieffupper{GW190524_134109}}_{\chiefflower{GW190524_134109}}$}{GW190530_030659}{$\chieffmedian{GW190530_030659}^{+\chieffupper{GW190530_030659}}_{\chiefflower{GW190530_030659}}$}{GW190530_133833}{$\chieffmedian{GW190530_133833}^{+\chieffupper{GW190530_133833}}_{\chiefflower{GW190530_133833}}$}{GW190604_103812}{$\chieffmedian{GW190604_103812}^{+\chieffupper{GW190604_103812}}_{\chiefflower{GW190604_103812}}$}{GW190605_025957}{$\chieffmedian{GW190605_025957}^{+\chieffupper{GW190605_025957}}_{\chiefflower{GW190605_025957}}$}{GW190607_083827}{$\chieffmedian{GW190607_083827}^{+\chieffupper{GW190607_083827}}_{\chiefflower{GW190607_083827}}$}{GW190614_134749}{$\chieffmedian{GW190614_134749}^{+\chieffupper{GW190614_134749}}_{\chiefflower{GW190614_134749}}$}{GW190615_030234}{$\chieffmedian{GW190615_030234}^{+\chieffupper{GW190615_030234}}_{\chiefflower{GW190615_030234}}$}{GW190705_164632}{$\chieffmedian{GW190705_164632}^{+\chieffupper{GW190705_164632}}_{\chiefflower{GW190705_164632}}$}{GW190707_083226}{$\chieffmedian{GW190707_083226}^{+\chieffupper{GW190707_083226}}_{\chiefflower{GW190707_083226}}$}{GW190711_030756}{$\chieffmedian{GW190711_030756}^{+\chieffupper{GW190711_030756}}_{\chiefflower{GW190711_030756}}$}{GW190718_160159}{$\chieffmedian{GW190718_160159}^{+\chieffupper{GW190718_160159}}_{\chiefflower{GW190718_160159}}$}{GW190725_174728}{$\chieffmedian{GW190725_174728}^{+\chieffupper{GW190725_174728}}_{\chiefflower{GW190725_174728}}$}{GW190805_105432}{$\chieffmedian{GW190805_105432}^{+\chieffupper{GW190805_105432}}_{\chiefflower{GW190805_105432}}$}{GW190806_033721}{$\chieffmedian{GW190806_033721}^{+\chieffupper{GW190806_033721}}_{\chiefflower{GW190806_033721}}$}{GW190814_192009}{$\chieffmedian{GW190814_192009}^{+\chieffupper{GW190814_192009}}_{\chiefflower{GW190814_192009}}$}{GW190818_232544}{$\chieffmedian{GW190818_232544}^{+\chieffupper{GW190818_232544}}_{\chiefflower{GW190818_232544}}$}{GW190821_124821}{$\chieffmedian{GW190821_124821}^{+\chieffupper{GW190821_124821}}_{\chiefflower{GW190821_124821}}$}{GW190904_104631}{$\chieffmedian{GW190904_104631}^{+\chieffupper{GW190904_104631}}_{\chiefflower{GW190904_104631}}$}{GW190906_054335}{$\chieffmedian{GW190906_054335}^{+\chieffupper{GW190906_054335}}_{\chiefflower{GW190906_054335}}$}{GW190910_012619}{$\chieffmedian{GW190910_012619}^{+\chieffupper{GW190910_012619}}_{\chiefflower{GW190910_012619}}$}{GW190911_195101}{$\chieffmedian{GW190911_195101}^{+\chieffupper{GW190911_195101}}_{\chiefflower{GW190911_195101}}$}{GW190916_200658}{$\chieffmedian{GW190916_200658}^{+\chieffupper{GW190916_200658}}_{\chiefflower{GW190916_200658}}$}{GW190926_050336}{$\chieffmedian{GW190926_050336}^{+\chieffupper{GW190926_050336}}_{\chiefflower{GW190926_050336}}$}{GW191113_103541}{$\chieffmedian{GW191113_103541}^{+\chieffupper{GW191113_103541}}_{\chiefflower{GW191113_103541}}$}{GW191117_023843}{$\chieffmedian{GW191117_023843}^{+\chieffupper{GW191117_023843}}_{\chiefflower{GW191117_023843}}$}{GW191127_050227}{$\chieffmedian{GW191127_050227}^{+\chieffupper{GW191127_050227}}_{\chiefflower{GW191127_050227}}$}{GW191208_080334}{$\chieffmedian{GW191208_080334}^{+\chieffupper{GW191208_080334}}_{\chiefflower{GW191208_080334}}$}{GW191224_043228}{$\chieffmedian{GW191224_043228}^{+\chieffupper{GW191224_043228}}_{\chiefflower{GW191224_043228}}$}{GW191228_195619}{$\chieffmedian{GW191228_195619}^{+\chieffupper{GW191228_195619}}_{\chiefflower{GW191228_195619}}$}{GW200106_134123}{$\chieffmedian{GW200106_134123}^{+\chieffupper{GW200106_134123}}_{\chiefflower{GW200106_134123}}$}{GW200109_195634}{$\chieffmedian{GW200109_195634}^{+\chieffupper{GW200109_195634}}_{\chiefflower{GW200109_195634}}$}{GW200129_114245}{$\chieffmedian{GW200129_114245}^{+\chieffupper{GW200129_114245}}_{\chiefflower{GW200129_114245}}$}{GW200208_211609}{$\chieffmedian{GW200208_211609}^{+\chieffupper{GW200208_211609}}_{\chiefflower{GW200208_211609}}$}{GW200210_005122}{$\chieffmedian{GW200210_005122}^{+\chieffupper{GW200210_005122}}_{\chiefflower{GW200210_005122}}$}{GW200210_100022}{$\chieffmedian{GW200210_100022}^{+\chieffupper{GW200210_100022}}_{\chiefflower{GW200210_100022}}$}{GW200220_124850}{$\chieffmedian{GW200220_124850}^{+\chieffupper{GW200220_124850}}_{\chiefflower{GW200220_124850}}$}{GW200214_223307}{$\chieffmedian{GW200214_223307}^{+\chieffupper{GW200214_223307}}_{\chiefflower{GW200214_223307}}$}{GW200225_075134}{$\chieffmedian{GW200225_075134}^{+\chieffupper{GW200225_075134}}_{\chiefflower{GW200225_075134}}$}{GW200301_211019}{$\chieffmedian{GW200301_211019}^{+\chieffupper{GW200301_211019}}_{\chiefflower{GW200301_211019}}$}{GW200304_172806}{$\chieffmedian{GW200304_172806}^{+\chieffupper{GW200304_172806}}_{\chiefflower{GW200304_172806}}$}{GW200305_084739}{$\chieffmedian{GW200305_084739}^{+\chieffupper{GW200305_084739}}_{\chiefflower{GW200305_084739}}$}{GW200318_191337}{$\chieffmedian{GW200318_191337}^{+\chieffupper{GW200318_191337}}_{\chiefflower{GW200318_191337}}$}}}

\DeclareRobustCommand{\luminositydistancelower}[1]{\IfEqCase{#1}{{GW151205_195525}{-2190}{GW151216_092416}{-1070}{GW170121_212536}{-710}{GW170202_135657}{-830}{GW170304_163753}{-1570}{GW170403_230611}{-1930}{GW170425_055334}{-1430}{GW170727_010430}{-1380}{GW190426_082124}{-3080}{GW190427_180650}{-520}{GW190511_125545}{-2000}{GW190511_163209}{-4440}{GW190514_065416}{-2280}{GW190523_085933}{-3050}{GW190524_134109}{-4680}{GW190530_030659}{-1580}{GW190530_133833}{-5040}{GW190604_103812}{-5470}{GW190605_025957}{-4310}{GW190607_083827}{-2300}{GW190614_134749}{-2730}{GW190615_030234}{-2480}{GW190705_164632}{-2670}{GW190707_083226}{-2330}{GW190711_030756}{-1320}{GW190718_160159}{-660}{GW190725_174728}{-460}{GW190805_105432}{-690}{GW190806_033721}{-5110}{GW190814_192009}{-4700}{GW190818_232544}{-3150}{GW190821_124821}{-350}{GW190904_104631}{-3210}{GW190906_054335}{-3140}{GW190910_012619}{-530}{GW190911_195101}{-4440}{GW190916_200658}{-2780}{GW190926_050336}{-2470}{GW191113_103541}{-2660}{GW191117_023843}{-2340}{GW191127_050227}{-2280}{GW191208_080334}{-1730}{GW191224_043228}{-740}{GW191228_195619}{-1970}{GW200106_134123}{-2280}{GW200109_195634}{-3820}{GW200129_114245}{-3560}{GW200208_211609}{-2230}{GW200210_005122}{-560}{GW200210_100022}{-2930}{GW200220_124850}{-2570}{GW200214_223307}{-3580}{GW200225_075134}{-2060}{GW200301_211019}{-1080}{GW200304_172806}{-4810}{GW200305_084739}{-2800}{GW200318_191337}{-3750}}}
\DeclareRobustCommand{\luminositydistanceupper}[1]{\IfEqCase{#1}{{GW151205_195525}{2510}{GW151216_092416}{1090}{GW170121_212536}{890}{GW170202_135657}{900}{GW170304_163753}{1550}{GW170403_230611}{2000}{GW170425_055334}{1700}{GW170727_010430}{1410}{GW190426_082124}{3530}{GW190427_180650}{540}{GW190511_125545}{2420}{GW190511_163209}{8910}{GW190514_065416}{2370}{GW190523_085933}{3870}{GW190524_134109}{5150}{GW190530_030659}{1620}{GW190530_133833}{6070}{GW190604_103812}{6320}{GW190605_025957}{4880}{GW190607_083827}{2540}{GW190614_134749}{3140}{GW190615_030234}{2650}{GW190705_164632}{3560}{GW190707_083226}{2740}{GW190711_030756}{1520}{GW190718_160159}{670}{GW190725_174728}{460}{GW190805_105432}{680}{GW190806_033721}{5970}{GW190814_192009}{5130}{GW190818_232544}{3980}{GW190821_124821}{310}{GW190904_104631}{4620}{GW190906_054335}{3620}{GW190910_012619}{550}{GW190911_195101}{4380}{GW190916_200658}{3110}{GW190926_050336}{2870}{GW191113_103541}{3120}{GW191117_023843}{2460}{GW191127_050227}{2610}{GW191208_080334}{1970}{GW191224_043228}{680}{GW191228_195619}{2380}{GW200106_134123}{2940}{GW200109_195634}{4300}{GW200129_114245}{4400}{GW200208_211609}{3360}{GW200210_005122}{500}{GW200210_100022}{3440}{GW200220_124850}{2700}{GW200214_223307}{4210}{GW200225_075134}{2330}{GW200301_211019}{1190}{GW200304_172806}{6040}{GW200305_084739}{2510}{GW200318_191337}{4140}}}
\DeclareRobustCommand{\luminositydistancemedian}[1]{\IfEqCase{#1}{{GW151205_195525}{3390}{GW151216_092416}{1950}{GW170121_212536}{1060}{GW170202_135657}{1470}{GW170304_163753}{2590}{GW170403_230611}{3180}{GW170425_055334}{2430}{GW170727_010430}{2450}{GW190426_082124}{4920}{GW190427_180650}{1000}{GW190511_125545}{3090}{GW190511_163209}{4800}{GW190514_065416}{4080}{GW190523_085933}{4580}{GW190524_134109}{7340}{GW190530_030659}{2830}{GW190530_133833}{6810}{GW190604_103812}{7730}{GW190605_025957}{5290}{GW190607_083827}{4010}{GW190614_134749}{4140}{GW190615_030234}{4540}{GW190705_164632}{3610}{GW190707_083226}{3660}{GW190711_030756}{2260}{GW190718_160159}{1370}{GW190725_174728}{1020}{GW190805_105432}{1440}{GW190806_033721}{8120}{GW190814_192009}{7290}{GW190818_232544}{4570}{GW190821_124821}{800}{GW190904_104631}{4580}{GW190906_054335}{4900}{GW190910_012619}{990}{GW190911_195101}{6940}{GW190916_200658}{4300}{GW190926_050336}{3910}{GW191113_103541}{3890}{GW191117_023843}{3510}{GW191127_050227}{4200}{GW191208_080334}{3020}{GW191224_043228}{1670}{GW191228_195619}{2930}{GW200106_134123}{3940}{GW200109_195634}{6360}{GW200129_114245}{5640}{GW200208_211609}{2910}{GW200210_005122}{1210}{GW200210_100022}{4950}{GW200220_124850}{4250}{GW200214_223307}{5330}{GW200225_075134}{3350}{GW200301_211019}{1980}{GW200304_172806}{6610}{GW200305_084739}{4880}{GW200318_191337}{6440}}}
\DeclareRobustCommand{\luminositydistance}[1]{\IfEqCase{#1}{{GW151205_195525}{$\luminositydistancemedian{GW151205_195525}^{+\luminositydistanceupper{GW151205_195525}}_{\luminositydistancelower{GW151205_195525}}$}{GW151216_092416}{$\luminositydistancemedian{GW151216_092416}^{+\luminositydistanceupper{GW151216_092416}}_{\luminositydistancelower{GW151216_092416}}$}{GW170121_212536}{$\luminositydistancemedian{GW170121_212536}^{+\luminositydistanceupper{GW170121_212536}}_{\luminositydistancelower{GW170121_212536}}$}{GW170202_135657}{$\luminositydistancemedian{GW170202_135657}^{+\luminositydistanceupper{GW170202_135657}}_{\luminositydistancelower{GW170202_135657}}$}{GW170304_163753}{$\luminositydistancemedian{GW170304_163753}^{+\luminositydistanceupper{GW170304_163753}}_{\luminositydistancelower{GW170304_163753}}$}{GW170403_230611}{$\luminositydistancemedian{GW170403_230611}^{+\luminositydistanceupper{GW170403_230611}}_{\luminositydistancelower{GW170403_230611}}$}{GW170425_055334}{$\luminositydistancemedian{GW170425_055334}^{+\luminositydistanceupper{GW170425_055334}}_{\luminositydistancelower{GW170425_055334}}$}{GW170727_010430}{$\luminositydistancemedian{GW170727_010430}^{+\luminositydistanceupper{GW170727_010430}}_{\luminositydistancelower{GW170727_010430}}$}{GW190426_082124}{$\luminositydistancemedian{GW190426_082124}^{+\luminositydistanceupper{GW190426_082124}}_{\luminositydistancelower{GW190426_082124}}$}{GW190427_180650}{$\luminositydistancemedian{GW190427_180650}^{+\luminositydistanceupper{GW190427_180650}}_{\luminositydistancelower{GW190427_180650}}$}{GW190511_125545}{$\luminositydistancemedian{GW190511_125545}^{+\luminositydistanceupper{GW190511_125545}}_{\luminositydistancelower{GW190511_125545}}$}{GW190511_163209}{$\luminositydistancemedian{GW190511_163209}^{+\luminositydistanceupper{GW190511_163209}}_{\luminositydistancelower{GW190511_163209}}$}{GW190514_065416}{$\luminositydistancemedian{GW190514_065416}^{+\luminositydistanceupper{GW190514_065416}}_{\luminositydistancelower{GW190514_065416}}$}{GW190523_085933}{$\luminositydistancemedian{GW190523_085933}^{+\luminositydistanceupper{GW190523_085933}}_{\luminositydistancelower{GW190523_085933}}$}{GW190524_134109}{$\luminositydistancemedian{GW190524_134109}^{+\luminositydistanceupper{GW190524_134109}}_{\luminositydistancelower{GW190524_134109}}$}{GW190530_030659}{$\luminositydistancemedian{GW190530_030659}^{+\luminositydistanceupper{GW190530_030659}}_{\luminositydistancelower{GW190530_030659}}$}{GW190530_133833}{$\luminositydistancemedian{GW190530_133833}^{+\luminositydistanceupper{GW190530_133833}}_{\luminositydistancelower{GW190530_133833}}$}{GW190604_103812}{$\luminositydistancemedian{GW190604_103812}^{+\luminositydistanceupper{GW190604_103812}}_{\luminositydistancelower{GW190604_103812}}$}{GW190605_025957}{$\luminositydistancemedian{GW190605_025957}^{+\luminositydistanceupper{GW190605_025957}}_{\luminositydistancelower{GW190605_025957}}$}{GW190607_083827}{$\luminositydistancemedian{GW190607_083827}^{+\luminositydistanceupper{GW190607_083827}}_{\luminositydistancelower{GW190607_083827}}$}{GW190614_134749}{$\luminositydistancemedian{GW190614_134749}^{+\luminositydistanceupper{GW190614_134749}}_{\luminositydistancelower{GW190614_134749}}$}{GW190615_030234}{$\luminositydistancemedian{GW190615_030234}^{+\luminositydistanceupper{GW190615_030234}}_{\luminositydistancelower{GW190615_030234}}$}{GW190705_164632}{$\luminositydistancemedian{GW190705_164632}^{+\luminositydistanceupper{GW190705_164632}}_{\luminositydistancelower{GW190705_164632}}$}{GW190707_083226}{$\luminositydistancemedian{GW190707_083226}^{+\luminositydistanceupper{GW190707_083226}}_{\luminositydistancelower{GW190707_083226}}$}{GW190711_030756}{$\luminositydistancemedian{GW190711_030756}^{+\luminositydistanceupper{GW190711_030756}}_{\luminositydistancelower{GW190711_030756}}$}{GW190718_160159}{$\luminositydistancemedian{GW190718_160159}^{+\luminositydistanceupper{GW190718_160159}}_{\luminositydistancelower{GW190718_160159}}$}{GW190725_174728}{$\luminositydistancemedian{GW190725_174728}^{+\luminositydistanceupper{GW190725_174728}}_{\luminositydistancelower{GW190725_174728}}$}{GW190805_105432}{$\luminositydistancemedian{GW190805_105432}^{+\luminositydistanceupper{GW190805_105432}}_{\luminositydistancelower{GW190805_105432}}$}{GW190806_033721}{$\luminositydistancemedian{GW190806_033721}^{+\luminositydistanceupper{GW190806_033721}}_{\luminositydistancelower{GW190806_033721}}$}{GW190814_192009}{$\luminositydistancemedian{GW190814_192009}^{+\luminositydistanceupper{GW190814_192009}}_{\luminositydistancelower{GW190814_192009}}$}{GW190818_232544}{$\luminositydistancemedian{GW190818_232544}^{+\luminositydistanceupper{GW190818_232544}}_{\luminositydistancelower{GW190818_232544}}$}{GW190821_124821}{$\luminositydistancemedian{GW190821_124821}^{+\luminositydistanceupper{GW190821_124821}}_{\luminositydistancelower{GW190821_124821}}$}{GW190904_104631}{$\luminositydistancemedian{GW190904_104631}^{+\luminositydistanceupper{GW190904_104631}}_{\luminositydistancelower{GW190904_104631}}$}{GW190906_054335}{$\luminositydistancemedian{GW190906_054335}^{+\luminositydistanceupper{GW190906_054335}}_{\luminositydistancelower{GW190906_054335}}$}{GW190910_012619}{$\luminositydistancemedian{GW190910_012619}^{+\luminositydistanceupper{GW190910_012619}}_{\luminositydistancelower{GW190910_012619}}$}{GW190911_195101}{$\luminositydistancemedian{GW190911_195101}^{+\luminositydistanceupper{GW190911_195101}}_{\luminositydistancelower{GW190911_195101}}$}{GW190916_200658}{$\luminositydistancemedian{GW190916_200658}^{+\luminositydistanceupper{GW190916_200658}}_{\luminositydistancelower{GW190916_200658}}$}{GW190926_050336}{$\luminositydistancemedian{GW190926_050336}^{+\luminositydistanceupper{GW190926_050336}}_{\luminositydistancelower{GW190926_050336}}$}{GW191113_103541}{$\luminositydistancemedian{GW191113_103541}^{+\luminositydistanceupper{GW191113_103541}}_{\luminositydistancelower{GW191113_103541}}$}{GW191117_023843}{$\luminositydistancemedian{GW191117_023843}^{+\luminositydistanceupper{GW191117_023843}}_{\luminositydistancelower{GW191117_023843}}$}{GW191127_050227}{$\luminositydistancemedian{GW191127_050227}^{+\luminositydistanceupper{GW191127_050227}}_{\luminositydistancelower{GW191127_050227}}$}{GW191208_080334}{$\luminositydistancemedian{GW191208_080334}^{+\luminositydistanceupper{GW191208_080334}}_{\luminositydistancelower{GW191208_080334}}$}{GW191224_043228}{$\luminositydistancemedian{GW191224_043228}^{+\luminositydistanceupper{GW191224_043228}}_{\luminositydistancelower{GW191224_043228}}$}{GW191228_195619}{$\luminositydistancemedian{GW191228_195619}^{+\luminositydistanceupper{GW191228_195619}}_{\luminositydistancelower{GW191228_195619}}$}{GW200106_134123}{$\luminositydistancemedian{GW200106_134123}^{+\luminositydistanceupper{GW200106_134123}}_{\luminositydistancelower{GW200106_134123}}$}{GW200109_195634}{$\luminositydistancemedian{GW200109_195634}^{+\luminositydistanceupper{GW200109_195634}}_{\luminositydistancelower{GW200109_195634}}$}{GW200129_114245}{$\luminositydistancemedian{GW200129_114245}^{+\luminositydistanceupper{GW200129_114245}}_{\luminositydistancelower{GW200129_114245}}$}{GW200208_211609}{$\luminositydistancemedian{GW200208_211609}^{+\luminositydistanceupper{GW200208_211609}}_{\luminositydistancelower{GW200208_211609}}$}{GW200210_005122}{$\luminositydistancemedian{GW200210_005122}^{+\luminositydistanceupper{GW200210_005122}}_{\luminositydistancelower{GW200210_005122}}$}{GW200210_100022}{$\luminositydistancemedian{GW200210_100022}^{+\luminositydistanceupper{GW200210_100022}}_{\luminositydistancelower{GW200210_100022}}$}{GW200220_124850}{$\luminositydistancemedian{GW200220_124850}^{+\luminositydistanceupper{GW200220_124850}}_{\luminositydistancelower{GW200220_124850}}$}{GW200214_223307}{$\luminositydistancemedian{GW200214_223307}^{+\luminositydistanceupper{GW200214_223307}}_{\luminositydistancelower{GW200214_223307}}$}{GW200225_075134}{$\luminositydistancemedian{GW200225_075134}^{+\luminositydistanceupper{GW200225_075134}}_{\luminositydistancelower{GW200225_075134}}$}{GW200301_211019}{$\luminositydistancemedian{GW200301_211019}^{+\luminositydistanceupper{GW200301_211019}}_{\luminositydistancelower{GW200301_211019}}$}{GW200304_172806}{$\luminositydistancemedian{GW200304_172806}^{+\luminositydistanceupper{GW200304_172806}}_{\luminositydistancelower{GW200304_172806}}$}{GW200305_084739}{$\luminositydistancemedian{GW200305_084739}^{+\luminositydistanceupper{GW200305_084739}}_{\luminositydistancelower{GW200305_084739}}$}{GW200318_191337}{$\luminositydistancemedian{GW200318_191337}^{+\luminositydistanceupper{GW200318_191337}}_{\luminositydistancelower{GW200318_191337}}$}}}

\DeclareRobustCommand{\redshiftlower}[1]{\IfEqCase{#1}{{GW151205_195525}{-0.3}{GW151216_092416}{-0.2}{GW170121_212536}{-0.1}{GW170202_135657}{-0.1}{GW170304_163753}{-0.2}{GW170403_230611}{-0.3}{GW170425_055334}{-0.2}{GW170727_010430}{-0.2}{GW190426_082124}{-0.4}{GW190427_180650}{-0.1}{GW190511_125545}{-0.3}{GW190511_163209}{-0.6}{GW190514_065416}{-0.3}{GW190523_085933}{-0.4}{GW190524_134109}{-0.6}{GW190530_030659}{-0.2}{GW190530_133833}{-0.6}{GW190604_103812}{-0.7}{GW190605_025957}{-0.6}{GW190607_083827}{-0.3}{GW190614_134749}{-0.4}{GW190615_030234}{-0.3}{GW190705_164632}{-0.4}{GW190707_083226}{-0.3}{GW190711_030756}{-0.2}{GW190718_160159}{-0.1}{GW190725_174728}{-0.1}{GW190805_105432}{-0.1}{GW190806_033721}{-0.6}{GW190814_192009}{-0.6}{GW190818_232544}{-0.4}{GW190821_124821}{-0.1}{GW190904_104631}{-0.4}{GW190906_054335}{-0.4}{GW190910_012619}{-0.1}{GW190911_195101}{-0.5}{GW190916_200658}{-0.4}{GW190926_050336}{-0.3}{GW191113_103541}{-0.4}{GW191117_023843}{-0.3}{GW191127_050227}{-0.3}{GW191208_080334}{-0.3}{GW191224_043228}{-0.1}{GW191228_195619}{-0.3}{GW200106_134123}{-0.3}{GW200109_195634}{-0.5}{GW200129_114245}{-0.5}{GW200208_211609}{-0.3}{GW200210_005122}{-0.1}{GW200210_100022}{-0.4}{GW200220_124850}{-0.3}{GW200214_223307}{-0.5}{GW200225_075134}{-0.3}{GW200301_211019}{-0.2}{GW200304_172806}{-0.6}{GW200305_084739}{-0.4}{GW200318_191337}{-0.5}}}
\DeclareRobustCommand{\redshiftupper}[1]{\IfEqCase{#1}{{GW151205_195525}{0.3}{GW151216_092416}{0.2}{GW170121_212536}{0.2}{GW170202_135657}{0.1}{GW170304_163753}{0.2}{GW170403_230611}{0.3}{GW170425_055334}{0.2}{GW170727_010430}{0.2}{GW190426_082124}{0.4}{GW190427_180650}{0.1}{GW190511_125545}{0.3}{GW190511_163209}{1.0}{GW190514_065416}{0.3}{GW190523_085933}{0.5}{GW190524_134109}{0.6}{GW190530_030659}{0.2}{GW190530_133833}{0.7}{GW190604_103812}{0.7}{GW190605_025957}{0.6}{GW190607_083827}{0.3}{GW190614_134749}{0.4}{GW190615_030234}{0.3}{GW190705_164632}{0.5}{GW190707_083226}{0.4}{GW190711_030756}{0.2}{GW190718_160159}{0.1}{GW190725_174728}{0.1}{GW190805_105432}{0.1}{GW190806_033721}{0.7}{GW190814_192009}{0.6}{GW190818_232544}{0.5}{GW190821_124821}{0.1}{GW190904_104631}{0.6}{GW190906_054335}{0.5}{GW190910_012619}{0.1}{GW190911_195101}{0.5}{GW190916_200658}{0.4}{GW190926_050336}{0.4}{GW191113_103541}{0.4}{GW191117_023843}{0.3}{GW191127_050227}{0.3}{GW191208_080334}{0.3}{GW191224_043228}{0.1}{GW191228_195619}{0.3}{GW200106_134123}{0.4}{GW200109_195634}{0.5}{GW200129_114245}{0.5}{GW200208_211609}{0.4}{GW200210_005122}{0.1}{GW200210_100022}{0.4}{GW200220_124850}{0.4}{GW200214_223307}{0.5}{GW200225_075134}{0.3}{GW200301_211019}{0.2}{GW200304_172806}{0.7}{GW200305_084739}{0.3}{GW200318_191337}{0.5}}}
\DeclareRobustCommand{\redshiftmedian}[1]{\IfEqCase{#1}{{GW151205_195525}{0.6}{GW151216_092416}{0.4}{GW170121_212536}{0.2}{GW170202_135657}{0.3}{GW170304_163753}{0.5}{GW170403_230611}{0.5}{GW170425_055334}{0.4}{GW170727_010430}{0.4}{GW190426_082124}{0.8}{GW190427_180650}{0.2}{GW190511_125545}{0.5}{GW190511_163209}{0.8}{GW190514_065416}{0.7}{GW190523_085933}{0.7}{GW190524_134109}{1.1}{GW190530_030659}{0.5}{GW190530_133833}{1.0}{GW190604_103812}{1.1}{GW190605_025957}{0.8}{GW190607_083827}{0.7}{GW190614_134749}{0.7}{GW190615_030234}{0.7}{GW190705_164632}{0.6}{GW190707_083226}{0.6}{GW190711_030756}{0.4}{GW190718_160159}{0.3}{GW190725_174728}{0.2}{GW190805_105432}{0.3}{GW190806_033721}{1.2}{GW190814_192009}{1.1}{GW190818_232544}{0.7}{GW190821_124821}{0.2}{GW190904_104631}{0.7}{GW190906_054335}{0.8}{GW190910_012619}{0.2}{GW190911_195101}{1.0}{GW190916_200658}{0.7}{GW190926_050336}{0.6}{GW191113_103541}{0.6}{GW191117_023843}{0.6}{GW191127_050227}{0.7}{GW191208_080334}{0.5}{GW191224_043228}{0.3}{GW191228_195619}{0.5}{GW200106_134123}{0.6}{GW200109_195634}{0.9}{GW200129_114245}{0.9}{GW200208_211609}{0.5}{GW200210_005122}{0.2}{GW200210_100022}{0.8}{GW200220_124850}{0.7}{GW200214_223307}{0.8}{GW200225_075134}{0.6}{GW200301_211019}{0.4}{GW200304_172806}{1.0}{GW200305_084739}{0.8}{GW200318_191337}{1.0}}}
\DeclareRobustCommand{\redshift}[1]{\IfEqCase{#1}{{GW151205_195525}{$\redshiftmedian{GW151205_195525}^{+\redshiftupper{GW151205_195525}}_{\redshiftlower{GW151205_195525}}$}{GW151216_092416}{$\redshiftmedian{GW151216_092416}^{+\redshiftupper{GW151216_092416}}_{\redshiftlower{GW151216_092416}}$}{GW170121_212536}{$\redshiftmedian{GW170121_212536}^{+\redshiftupper{GW170121_212536}}_{\redshiftlower{GW170121_212536}}$}{GW170202_135657}{$\redshiftmedian{GW170202_135657}^{+\redshiftupper{GW170202_135657}}_{\redshiftlower{GW170202_135657}}$}{GW170304_163753}{$\redshiftmedian{GW170304_163753}^{+\redshiftupper{GW170304_163753}}_{\redshiftlower{GW170304_163753}}$}{GW170403_230611}{$\redshiftmedian{GW170403_230611}^{+\redshiftupper{GW170403_230611}}_{\redshiftlower{GW170403_230611}}$}{GW170425_055334}{$\redshiftmedian{GW170425_055334}^{+\redshiftupper{GW170425_055334}}_{\redshiftlower{GW170425_055334}}$}{GW170727_010430}{$\redshiftmedian{GW170727_010430}^{+\redshiftupper{GW170727_010430}}_{\redshiftlower{GW170727_010430}}$}{GW190426_082124}{$\redshiftmedian{GW190426_082124}^{+\redshiftupper{GW190426_082124}}_{\redshiftlower{GW190426_082124}}$}{GW190427_180650}{$\redshiftmedian{GW190427_180650}^{+\redshiftupper{GW190427_180650}}_{\redshiftlower{GW190427_180650}}$}{GW190511_125545}{$\redshiftmedian{GW190511_125545}^{+\redshiftupper{GW190511_125545}}_{\redshiftlower{GW190511_125545}}$}{GW190511_163209}{$\redshiftmedian{GW190511_163209}^{+\redshiftupper{GW190511_163209}}_{\redshiftlower{GW190511_163209}}$}{GW190514_065416}{$\redshiftmedian{GW190514_065416}^{+\redshiftupper{GW190514_065416}}_{\redshiftlower{GW190514_065416}}$}{GW190523_085933}{$\redshiftmedian{GW190523_085933}^{+\redshiftupper{GW190523_085933}}_{\redshiftlower{GW190523_085933}}$}{GW190524_134109}{$\redshiftmedian{GW190524_134109}^{+\redshiftupper{GW190524_134109}}_{\redshiftlower{GW190524_134109}}$}{GW190530_030659}{$\redshiftmedian{GW190530_030659}^{+\redshiftupper{GW190530_030659}}_{\redshiftlower{GW190530_030659}}$}{GW190530_133833}{$\redshiftmedian{GW190530_133833}^{+\redshiftupper{GW190530_133833}}_{\redshiftlower{GW190530_133833}}$}{GW190604_103812}{$\redshiftmedian{GW190604_103812}^{+\redshiftupper{GW190604_103812}}_{\redshiftlower{GW190604_103812}}$}{GW190605_025957}{$\redshiftmedian{GW190605_025957}^{+\redshiftupper{GW190605_025957}}_{\redshiftlower{GW190605_025957}}$}{GW190607_083827}{$\redshiftmedian{GW190607_083827}^{+\redshiftupper{GW190607_083827}}_{\redshiftlower{GW190607_083827}}$}{GW190614_134749}{$\redshiftmedian{GW190614_134749}^{+\redshiftupper{GW190614_134749}}_{\redshiftlower{GW190614_134749}}$}{GW190615_030234}{$\redshiftmedian{GW190615_030234}^{+\redshiftupper{GW190615_030234}}_{\redshiftlower{GW190615_030234}}$}{GW190705_164632}{$\redshiftmedian{GW190705_164632}^{+\redshiftupper{GW190705_164632}}_{\redshiftlower{GW190705_164632}}$}{GW190707_083226}{$\redshiftmedian{GW190707_083226}^{+\redshiftupper{GW190707_083226}}_{\redshiftlower{GW190707_083226}}$}{GW190711_030756}{$\redshiftmedian{GW190711_030756}^{+\redshiftupper{GW190711_030756}}_{\redshiftlower{GW190711_030756}}$}{GW190718_160159}{$\redshiftmedian{GW190718_160159}^{+\redshiftupper{GW190718_160159}}_{\redshiftlower{GW190718_160159}}$}{GW190725_174728}{$\redshiftmedian{GW190725_174728}^{+\redshiftupper{GW190725_174728}}_{\redshiftlower{GW190725_174728}}$}{GW190805_105432}{$\redshiftmedian{GW190805_105432}^{+\redshiftupper{GW190805_105432}}_{\redshiftlower{GW190805_105432}}$}{GW190806_033721}{$\redshiftmedian{GW190806_033721}^{+\redshiftupper{GW190806_033721}}_{\redshiftlower{GW190806_033721}}$}{GW190814_192009}{$\redshiftmedian{GW190814_192009}^{+\redshiftupper{GW190814_192009}}_{\redshiftlower{GW190814_192009}}$}{GW190818_232544}{$\redshiftmedian{GW190818_232544}^{+\redshiftupper{GW190818_232544}}_{\redshiftlower{GW190818_232544}}$}{GW190821_124821}{$\redshiftmedian{GW190821_124821}^{+\redshiftupper{GW190821_124821}}_{\redshiftlower{GW190821_124821}}$}{GW190904_104631}{$\redshiftmedian{GW190904_104631}^{+\redshiftupper{GW190904_104631}}_{\redshiftlower{GW190904_104631}}$}{GW190906_054335}{$\redshiftmedian{GW190906_054335}^{+\redshiftupper{GW190906_054335}}_{\redshiftlower{GW190906_054335}}$}{GW190910_012619}{$\redshiftmedian{GW190910_012619}^{+\redshiftupper{GW190910_012619}}_{\redshiftlower{GW190910_012619}}$}{GW190911_195101}{$\redshiftmedian{GW190911_195101}^{+\redshiftupper{GW190911_195101}}_{\redshiftlower{GW190911_195101}}$}{GW190916_200658}{$\redshiftmedian{GW190916_200658}^{+\redshiftupper{GW190916_200658}}_{\redshiftlower{GW190916_200658}}$}{GW190926_050336}{$\redshiftmedian{GW190926_050336}^{+\redshiftupper{GW190926_050336}}_{\redshiftlower{GW190926_050336}}$}{GW191113_103541}{$\redshiftmedian{GW191113_103541}^{+\redshiftupper{GW191113_103541}}_{\redshiftlower{GW191113_103541}}$}{GW191117_023843}{$\redshiftmedian{GW191117_023843}^{+\redshiftupper{GW191117_023843}}_{\redshiftlower{GW191117_023843}}$}{GW191127_050227}{$\redshiftmedian{GW191127_050227}^{+\redshiftupper{GW191127_050227}}_{\redshiftlower{GW191127_050227}}$}{GW191208_080334}{$\redshiftmedian{GW191208_080334}^{+\redshiftupper{GW191208_080334}}_{\redshiftlower{GW191208_080334}}$}{GW191224_043228}{$\redshiftmedian{GW191224_043228}^{+\redshiftupper{GW191224_043228}}_{\redshiftlower{GW191224_043228}}$}{GW191228_195619}{$\redshiftmedian{GW191228_195619}^{+\redshiftupper{GW191228_195619}}_{\redshiftlower{GW191228_195619}}$}{GW200106_134123}{$\redshiftmedian{GW200106_134123}^{+\redshiftupper{GW200106_134123}}_{\redshiftlower{GW200106_134123}}$}{GW200109_195634}{$\redshiftmedian{GW200109_195634}^{+\redshiftupper{GW200109_195634}}_{\redshiftlower{GW200109_195634}}$}{GW200129_114245}{$\redshiftmedian{GW200129_114245}^{+\redshiftupper{GW200129_114245}}_{\redshiftlower{GW200129_114245}}$}{GW200208_211609}{$\redshiftmedian{GW200208_211609}^{+\redshiftupper{GW200208_211609}}_{\redshiftlower{GW200208_211609}}$}{GW200210_005122}{$\redshiftmedian{GW200210_005122}^{+\redshiftupper{GW200210_005122}}_{\redshiftlower{GW200210_005122}}$}{GW200210_100022}{$\redshiftmedian{GW200210_100022}^{+\redshiftupper{GW200210_100022}}_{\redshiftlower{GW200210_100022}}$}{GW200220_124850}{$\redshiftmedian{GW200220_124850}^{+\redshiftupper{GW200220_124850}}_{\redshiftlower{GW200220_124850}}$}{GW200214_223307}{$\redshiftmedian{GW200214_223307}^{+\redshiftupper{GW200214_223307}}_{\redshiftlower{GW200214_223307}}$}{GW200225_075134}{$\redshiftmedian{GW200225_075134}^{+\redshiftupper{GW200225_075134}}_{\redshiftlower{GW200225_075134}}$}{GW200301_211019}{$\redshiftmedian{GW200301_211019}^{+\redshiftupper{GW200301_211019}}_{\redshiftlower{GW200301_211019}}$}{GW200304_172806}{$\redshiftmedian{GW200304_172806}^{+\redshiftupper{GW200304_172806}}_{\redshiftlower{GW200304_172806}}$}{GW200305_084739}{$\redshiftmedian{GW200305_084739}^{+\redshiftupper{GW200305_084739}}_{\redshiftlower{GW200305_084739}}$}{GW200318_191337}{$\redshiftmedian{GW200318_191337}^{+\redshiftupper{GW200318_191337}}_{\redshiftlower{GW200318_191337}}$}}}

\DeclareRobustCommand{\chirpmasssourcetwoOGCdivergence}[1]{\IfEqCase{#1}{{GW151205_195525}{0.010}{GW151216_092416}{0.016}{GW170121_212536}{0.018}{GW170202_135657}{0.045}{GW170304_163753}{0.001}{GW170403_230611}{0.017}{GW170425_055334}{0.001}{GW170727_010430}{0.005}{GW190426_082124}{--}{GW190427_180650}{--}{GW190511_125545}{--}{GW190511_163209}{--}{GW190514_065416}{--}{GW190523_085933}{--}{GW190524_134109}{--}{GW190530_030659}{--}{GW190530_133833}{--}{GW190604_103812}{--}{GW190605_025957}{--}{GW190607_083827}{--}{GW190614_134749}{--}{GW190615_030234}{--}{GW190705_164632}{--}{GW190707_083226}{--}{GW190711_030756}{--}{GW190718_160159}{--}{GW190725_174728}{--}{GW190805_105432}{--}{GW190806_033721}{--}{GW190814_192009}{--}{GW190818_232544}{--}{GW190821_124821}{--}{GW190904_104631}{--}{GW190906_054335}{--}{GW190910_012619}{--}{GW190911_195101}{--}{GW190916_200658}{--}{GW190926_050336}{--}{GW191113_103541}{--}{GW191117_023843}{--}{GW191127_050227}{--}{GW191208_080334}{--}{GW191224_043228}{--}{GW191228_195619}{--}{GW200106_134123}{--}{GW200109_195634}{--}{GW200129_114245}{--}{GW200208_211609}{--}{GW200210_005122}{--}{GW200210_100022}{--}{GW200220_124850}{--}{GW200214_223307}{--}{GW200225_075134}{--}{GW200301_211019}{--}{GW200304_172806}{--}{GW200305_084739}{--}{GW200318_191337}{--}}}
\DeclareRobustCommand{\massratiotwoOGCdivergence}[1]{\IfEqCase{#1}{{GW151205_195525}{0.004}{GW151216_092416}{0.032}{GW170121_212536}{0.002}{GW170202_135657}{0.089}{GW170304_163753}{0.004}{GW170403_230611}{0.020}{GW170425_055334}{0.015}{GW170727_010430}{0.005}{GW190426_082124}{--}{GW190427_180650}{--}{GW190511_125545}{--}{GW190511_163209}{--}{GW190514_065416}{--}{GW190523_085933}{--}{GW190524_134109}{--}{GW190530_030659}{--}{GW190530_133833}{--}{GW190604_103812}{--}{GW190605_025957}{--}{GW190607_083827}{--}{GW190614_134749}{--}{GW190615_030234}{--}{GW190705_164632}{--}{GW190707_083226}{--}{GW190711_030756}{--}{GW190718_160159}{--}{GW190725_174728}{--}{GW190805_105432}{--}{GW190806_033721}{--}{GW190814_192009}{--}{GW190818_232544}{--}{GW190821_124821}{--}{GW190904_104631}{--}{GW190906_054335}{--}{GW190910_012619}{--}{GW190911_195101}{--}{GW190916_200658}{--}{GW190926_050336}{--}{GW191113_103541}{--}{GW191117_023843}{--}{GW191127_050227}{--}{GW191208_080334}{--}{GW191224_043228}{--}{GW191228_195619}{--}{GW200106_134123}{--}{GW200109_195634}{--}{GW200129_114245}{--}{GW200208_211609}{--}{GW200210_005122}{--}{GW200210_100022}{--}{GW200220_124850}{--}{GW200214_223307}{--}{GW200225_075134}{--}{GW200301_211019}{--}{GW200304_172806}{--}{GW200305_084739}{--}{GW200318_191337}{--}}}
\DeclareRobustCommand{\chiefftwoOGCdivergence}[1]{\IfEqCase{#1}{{GW151205_195525}{0.001}{GW151216_092416}{0.010}{GW170121_212536}{0.036}{GW170202_135657}{0.008}{GW170304_163753}{0.002}{GW170403_230611}{0.004}{GW170425_055334}{0.001}{GW170727_010430}{0.001}{GW190426_082124}{--}{GW190427_180650}{--}{GW190511_125545}{--}{GW190511_163209}{--}{GW190514_065416}{--}{GW190523_085933}{--}{GW190524_134109}{--}{GW190530_030659}{--}{GW190530_133833}{--}{GW190604_103812}{--}{GW190605_025957}{--}{GW190607_083827}{--}{GW190614_134749}{--}{GW190615_030234}{--}{GW190705_164632}{--}{GW190707_083226}{--}{GW190711_030756}{--}{GW190718_160159}{--}{GW190725_174728}{--}{GW190805_105432}{--}{GW190806_033721}{--}{GW190814_192009}{--}{GW190818_232544}{--}{GW190821_124821}{--}{GW190904_104631}{--}{GW190906_054335}{--}{GW190910_012619}{--}{GW190911_195101}{--}{GW190916_200658}{--}{GW190926_050336}{--}{GW191113_103541}{--}{GW191117_023843}{--}{GW191127_050227}{--}{GW191208_080334}{--}{GW191224_043228}{--}{GW191228_195619}{--}{GW200106_134123}{--}{GW200109_195634}{--}{GW200129_114245}{--}{GW200208_211609}{--}{GW200210_005122}{--}{GW200210_100022}{--}{GW200220_124850}{--}{GW200214_223307}{--}{GW200225_075134}{--}{GW200301_211019}{--}{GW200304_172806}{--}{GW200305_084739}{--}{GW200318_191337}{--}}}
\DeclareRobustCommand{\chiptwoOGCdivergence}[1]{\IfEqCase{#1}{{GW151205_195525}{0.001}{GW151216_092416}{0.009}{GW170121_212536}{0.003}{GW170202_135657}{0.025}{GW170304_163753}{0.000}{GW170403_230611}{0.003}{GW170425_055334}{0.000}{GW170727_010430}{0.001}{GW190426_082124}{--}{GW190427_180650}{--}{GW190511_125545}{--}{GW190511_163209}{--}{GW190514_065416}{--}{GW190523_085933}{--}{GW190524_134109}{--}{GW190530_030659}{--}{GW190530_133833}{--}{GW190604_103812}{--}{GW190605_025957}{--}{GW190607_083827}{--}{GW190614_134749}{--}{GW190615_030234}{--}{GW190705_164632}{--}{GW190707_083226}{--}{GW190711_030756}{--}{GW190718_160159}{--}{GW190725_174728}{--}{GW190805_105432}{--}{GW190806_033721}{--}{GW190814_192009}{--}{GW190818_232544}{--}{GW190821_124821}{--}{GW190904_104631}{--}{GW190906_054335}{--}{GW190910_012619}{--}{GW190911_195101}{--}{GW190916_200658}{--}{GW190926_050336}{--}{GW191113_103541}{--}{GW191117_023843}{--}{GW191127_050227}{--}{GW191208_080334}{--}{GW191224_043228}{--}{GW191228_195619}{--}{GW200106_134123}{--}{GW200109_195634}{--}{GW200129_114245}{--}{GW200208_211609}{--}{GW200210_005122}{--}{GW200210_100022}{--}{GW200220_124850}{--}{GW200214_223307}{--}{GW200225_075134}{--}{GW200301_211019}{--}{GW200304_172806}{--}{GW200305_084739}{--}{GW200318_191337}{--}}}
\DeclareRobustCommand{\luminositydistancetwoOGCdivergence}[1]{\IfEqCase{#1}{{GW151205_195525}{0.008}{GW151216_092416}{0.042}{GW170121_212536}{0.007}{GW170202_135657}{0.026}{GW170304_163753}{0.011}{GW170403_230611}{0.033}{GW170425_055334}{0.003}{GW170727_010430}{0.008}{GW190426_082124}{--}{GW190427_180650}{--}{GW190511_125545}{--}{GW190511_163209}{--}{GW190514_065416}{--}{GW190523_085933}{--}{GW190524_134109}{--}{GW190530_030659}{--}{GW190530_133833}{--}{GW190604_103812}{--}{GW190605_025957}{--}{GW190607_083827}{--}{GW190614_134749}{--}{GW190615_030234}{--}{GW190705_164632}{--}{GW190707_083226}{--}{GW190711_030756}{--}{GW190718_160159}{--}{GW190725_174728}{--}{GW190805_105432}{--}{GW190806_033721}{--}{GW190814_192009}{--}{GW190818_232544}{--}{GW190821_124821}{--}{GW190904_104631}{--}{GW190906_054335}{--}{GW190910_012619}{--}{GW190911_195101}{--}{GW190916_200658}{--}{GW190926_050336}{--}{GW191113_103541}{--}{GW191117_023843}{--}{GW191127_050227}{--}{GW191208_080334}{--}{GW191224_043228}{--}{GW191228_195619}{--}{GW200106_134123}{--}{GW200109_195634}{--}{GW200129_114245}{--}{GW200208_211609}{--}{GW200210_005122}{--}{GW200210_100022}{--}{GW200220_124850}{--}{GW200214_223307}{--}{GW200225_075134}{--}{GW200301_211019}{--}{GW200304_172806}{--}{GW200305_084739}{--}{GW200318_191337}{--}}}
\DeclareRobustCommand{\chirpmasssourcefourOGCdivergence}[1]{\IfEqCase{#1}{{GW151205_195525}{--}{GW151216_092416}{--}{GW170121_212536}{0.008}{GW170202_135657}{0.046}{GW170304_163753}{0.003}{GW170403_230611}{0.004}{GW170425_055334}{--}{GW170727_010430}{0.004}{GW190426_082124}{--}{GW190427_180650}{0.040}{GW190511_125545}{--}{GW190511_163209}{--}{GW190514_065416}{--}{GW190523_085933}{--}{GW190524_134109}{--}{GW190530_030659}{--}{GW190530_133833}{--}{GW190604_103812}{--}{GW190605_025957}{--}{GW190607_083827}{--}{GW190614_134749}{--}{GW190615_030234}{--}{GW190705_164632}{--}{GW190707_083226}{--}{GW190711_030756}{--}{GW190718_160159}{--}{GW190725_174728}{--}{GW190805_105432}{0.008}{GW190806_033721}{--}{GW190814_192009}{--}{GW190818_232544}{--}{GW190821_124821}{--}{GW190904_104631}{--}{GW190906_054335}{--}{GW190910_012619}{--}{GW190911_195101}{--}{GW190916_200658}{--}{GW190926_050336}{--}{GW191113_103541}{--}{GW191117_023843}{--}{GW191127_050227}{--}{GW191208_080334}{--}{GW191224_043228}{0.023}{GW191228_195619}{--}{GW200106_134123}{0.003}{GW200109_195634}{--}{GW200129_114245}{0.676}{GW200208_211609}{--}{GW200210_005122}{0.053}{GW200210_100022}{--}{GW200220_124850}{--}{GW200214_223307}{0.007}{GW200225_075134}{--}{GW200301_211019}{--}{GW200304_172806}{--}{GW200305_084739}{0.004}{GW200318_191337}{0.015}}}
\DeclareRobustCommand{\massratiofourOGCdivergence}[1]{\IfEqCase{#1}{{GW151205_195525}{--}{GW151216_092416}{--}{GW170121_212536}{0.002}{GW170202_135657}{0.032}{GW170304_163753}{0.004}{GW170403_230611}{0.002}{GW170425_055334}{--}{GW170727_010430}{0.005}{GW190426_082124}{--}{GW190427_180650}{0.010}{GW190511_125545}{--}{GW190511_163209}{--}{GW190514_065416}{--}{GW190523_085933}{--}{GW190524_134109}{--}{GW190530_030659}{--}{GW190530_133833}{--}{GW190604_103812}{--}{GW190605_025957}{--}{GW190607_083827}{--}{GW190614_134749}{--}{GW190615_030234}{--}{GW190705_164632}{--}{GW190707_083226}{--}{GW190711_030756}{--}{GW190718_160159}{--}{GW190725_174728}{--}{GW190805_105432}{0.002}{GW190806_033721}{--}{GW190814_192009}{--}{GW190818_232544}{--}{GW190821_124821}{--}{GW190904_104631}{--}{GW190906_054335}{--}{GW190910_012619}{--}{GW190911_195101}{--}{GW190916_200658}{--}{GW190926_050336}{--}{GW191113_103541}{--}{GW191117_023843}{--}{GW191127_050227}{--}{GW191208_080334}{--}{GW191224_043228}{0.006}{GW191228_195619}{--}{GW200106_134123}{0.002}{GW200109_195634}{--}{GW200129_114245}{0.433}{GW200208_211609}{--}{GW200210_005122}{0.002}{GW200210_100022}{--}{GW200220_124850}{--}{GW200214_223307}{0.008}{GW200225_075134}{--}{GW200301_211019}{--}{GW200304_172806}{--}{GW200305_084739}{0.005}{GW200318_191337}{0.003}}}
\DeclareRobustCommand{\chiefffourOGCdivergence}[1]{\IfEqCase{#1}{{GW151205_195525}{--}{GW151216_092416}{--}{GW170121_212536}{0.033}{GW170202_135657}{0.002}{GW170304_163753}{0.003}{GW170403_230611}{0.004}{GW170425_055334}{--}{GW170727_010430}{0.003}{GW190426_082124}{--}{GW190427_180650}{0.022}{GW190511_125545}{--}{GW190511_163209}{--}{GW190514_065416}{--}{GW190523_085933}{--}{GW190524_134109}{--}{GW190530_030659}{--}{GW190530_133833}{--}{GW190604_103812}{--}{GW190605_025957}{--}{GW190607_083827}{--}{GW190614_134749}{--}{GW190615_030234}{--}{GW190705_164632}{--}{GW190707_083226}{--}{GW190711_030756}{--}{GW190718_160159}{--}{GW190725_174728}{--}{GW190805_105432}{0.001}{GW190806_033721}{--}{GW190814_192009}{--}{GW190818_232544}{--}{GW190821_124821}{--}{GW190904_104631}{--}{GW190906_054335}{--}{GW190910_012619}{--}{GW190911_195101}{--}{GW190916_200658}{--}{GW190926_050336}{--}{GW191113_103541}{--}{GW191117_023843}{--}{GW191127_050227}{--}{GW191208_080334}{--}{GW191224_043228}{0.005}{GW191228_195619}{--}{GW200106_134123}{0.002}{GW200109_195634}{--}{GW200129_114245}{0.158}{GW200208_211609}{--}{GW200210_005122}{0.002}{GW200210_100022}{--}{GW200220_124850}{--}{GW200214_223307}{0.005}{GW200225_075134}{--}{GW200301_211019}{--}{GW200304_172806}{--}{GW200305_084739}{0.006}{GW200318_191337}{0.006}}}
\DeclareRobustCommand{\chipfourOGCdivergence}[1]{\IfEqCase{#1}{{GW151205_195525}{--}{GW151216_092416}{--}{GW170121_212536}{0.001}{GW170202_135657}{0.003}{GW170304_163753}{0.002}{GW170403_230611}{0.001}{GW170425_055334}{--}{GW170727_010430}{0.003}{GW190426_082124}{--}{GW190427_180650}{0.029}{GW190511_125545}{--}{GW190511_163209}{--}{GW190514_065416}{--}{GW190523_085933}{--}{GW190524_134109}{--}{GW190530_030659}{--}{GW190530_133833}{--}{GW190604_103812}{--}{GW190605_025957}{--}{GW190607_083827}{--}{GW190614_134749}{--}{GW190615_030234}{--}{GW190705_164632}{--}{GW190707_083226}{--}{GW190711_030756}{--}{GW190718_160159}{--}{GW190725_174728}{--}{GW190805_105432}{0.003}{GW190806_033721}{--}{GW190814_192009}{--}{GW190818_232544}{--}{GW190821_124821}{--}{GW190904_104631}{--}{GW190906_054335}{--}{GW190910_012619}{--}{GW190911_195101}{--}{GW190916_200658}{--}{GW190926_050336}{--}{GW191113_103541}{--}{GW191117_023843}{--}{GW191127_050227}{--}{GW191208_080334}{--}{GW191224_043228}{0.009}{GW191228_195619}{--}{GW200106_134123}{0.003}{GW200109_195634}{--}{GW200129_114245}{0.128}{GW200208_211609}{--}{GW200210_005122}{0.002}{GW200210_100022}{--}{GW200220_124850}{--}{GW200214_223307}{0.007}{GW200225_075134}{--}{GW200301_211019}{--}{GW200304_172806}{--}{GW200305_084739}{0.001}{GW200318_191337}{0.001}}}
\DeclareRobustCommand{\luminositydistancefourOGCdivergence}[1]{\IfEqCase{#1}{{GW151205_195525}{--}{GW151216_092416}{--}{GW170121_212536}{0.029}{GW170202_135657}{0.014}{GW170304_163753}{0.009}{GW170403_230611}{0.007}{GW170425_055334}{--}{GW170727_010430}{0.004}{GW190426_082124}{--}{GW190427_180650}{0.024}{GW190511_125545}{--}{GW190511_163209}{--}{GW190514_065416}{--}{GW190523_085933}{--}{GW190524_134109}{--}{GW190530_030659}{--}{GW190530_133833}{--}{GW190604_103812}{--}{GW190605_025957}{--}{GW190607_083827}{--}{GW190614_134749}{--}{GW190615_030234}{--}{GW190705_164632}{--}{GW190707_083226}{--}{GW190711_030756}{--}{GW190718_160159}{--}{GW190725_174728}{--}{GW190805_105432}{0.009}{GW190806_033721}{--}{GW190814_192009}{--}{GW190818_232544}{--}{GW190821_124821}{--}{GW190904_104631}{--}{GW190906_054335}{--}{GW190910_012619}{--}{GW190911_195101}{--}{GW190916_200658}{--}{GW190926_050336}{--}{GW191113_103541}{--}{GW191117_023843}{--}{GW191127_050227}{--}{GW191208_080334}{--}{GW191224_043228}{0.042}{GW191228_195619}{--}{GW200106_134123}{0.005}{GW200109_195634}{--}{GW200129_114245}{0.409}{GW200208_211609}{--}{GW200210_005122}{0.083}{GW200210_100022}{--}{GW200220_124850}{--}{GW200214_223307}{0.022}{GW200225_075134}{--}{GW200301_211019}{--}{GW200304_172806}{--}{GW200305_084739}{0.010}{GW200318_191337}{0.031}}}
\DeclareRobustCommand{\chirpmasssourceIAStwodivergence}[1]{\IfEqCase{#1}{{GW151205_195525}{--}{GW151216_092416}{0.127}{GW170121_212536}{0.004}{GW170202_135657}{0.058}{GW170304_163753}{0.018}{GW170403_230611}{0.021}{GW170425_055334}{0.012}{GW170727_010430}{0.005}{GW190426_082124}{--}{GW190427_180650}{--}{GW190511_125545}{--}{GW190511_163209}{--}{GW190514_065416}{--}{GW190523_085933}{--}{GW190524_134109}{--}{GW190530_030659}{--}{GW190530_133833}{--}{GW190604_103812}{--}{GW190605_025957}{--}{GW190607_083827}{--}{GW190614_134749}{--}{GW190615_030234}{--}{GW190705_164632}{--}{GW190707_083226}{--}{GW190711_030756}{--}{GW190718_160159}{--}{GW190725_174728}{--}{GW190805_105432}{--}{GW190806_033721}{--}{GW190814_192009}{--}{GW190818_232544}{--}{GW190821_124821}{--}{GW190904_104631}{--}{GW190906_054335}{--}{GW190910_012619}{--}{GW190911_195101}{--}{GW190916_200658}{--}{GW190926_050336}{--}{GW191113_103541}{--}{GW191117_023843}{--}{GW191127_050227}{--}{GW191208_080334}{--}{GW191224_043228}{--}{GW191228_195619}{--}{GW200106_134123}{--}{GW200109_195634}{--}{GW200129_114245}{--}{GW200208_211609}{--}{GW200210_005122}{--}{GW200210_100022}{--}{GW200220_124850}{--}{GW200214_223307}{--}{GW200225_075134}{--}{GW200301_211019}{--}{GW200304_172806}{--}{GW200305_084739}{--}{GW200318_191337}{--}}}
\DeclareRobustCommand{\massratioIAStwodivergence}[1]{\IfEqCase{#1}{{GW151205_195525}{--}{GW151216_092416}{0.263}{GW170121_212536}{0.034}{GW170202_135657}{0.006}{GW170304_163753}{0.055}{GW170403_230611}{0.022}{GW170425_055334}{0.094}{GW170727_010430}{0.017}{GW190426_082124}{--}{GW190427_180650}{--}{GW190511_125545}{--}{GW190511_163209}{--}{GW190514_065416}{--}{GW190523_085933}{--}{GW190524_134109}{--}{GW190530_030659}{--}{GW190530_133833}{--}{GW190604_103812}{--}{GW190605_025957}{--}{GW190607_083827}{--}{GW190614_134749}{--}{GW190615_030234}{--}{GW190705_164632}{--}{GW190707_083226}{--}{GW190711_030756}{--}{GW190718_160159}{--}{GW190725_174728}{--}{GW190805_105432}{--}{GW190806_033721}{--}{GW190814_192009}{--}{GW190818_232544}{--}{GW190821_124821}{--}{GW190904_104631}{--}{GW190906_054335}{--}{GW190910_012619}{--}{GW190911_195101}{--}{GW190916_200658}{--}{GW190926_050336}{--}{GW191113_103541}{--}{GW191117_023843}{--}{GW191127_050227}{--}{GW191208_080334}{--}{GW191224_043228}{--}{GW191228_195619}{--}{GW200106_134123}{--}{GW200109_195634}{--}{GW200129_114245}{--}{GW200208_211609}{--}{GW200210_005122}{--}{GW200210_100022}{--}{GW200220_124850}{--}{GW200214_223307}{--}{GW200225_075134}{--}{GW200301_211019}{--}{GW200304_172806}{--}{GW200305_084739}{--}{GW200318_191337}{--}}}
\DeclareRobustCommand{\chieffIAStwodivergence}[1]{\IfEqCase{#1}{{GW151205_195525}{--}{GW151216_092416}{0.413}{GW170121_212536}{0.013}{GW170202_135657}{0.024}{GW170304_163753}{0.002}{GW170403_230611}{0.269}{GW170425_055334}{0.036}{GW170727_010430}{0.006}{GW190426_082124}{--}{GW190427_180650}{--}{GW190511_125545}{--}{GW190511_163209}{--}{GW190514_065416}{--}{GW190523_085933}{--}{GW190524_134109}{--}{GW190530_030659}{--}{GW190530_133833}{--}{GW190604_103812}{--}{GW190605_025957}{--}{GW190607_083827}{--}{GW190614_134749}{--}{GW190615_030234}{--}{GW190705_164632}{--}{GW190707_083226}{--}{GW190711_030756}{--}{GW190718_160159}{--}{GW190725_174728}{--}{GW190805_105432}{--}{GW190806_033721}{--}{GW190814_192009}{--}{GW190818_232544}{--}{GW190821_124821}{--}{GW190904_104631}{--}{GW190906_054335}{--}{GW190910_012619}{--}{GW190911_195101}{--}{GW190916_200658}{--}{GW190926_050336}{--}{GW191113_103541}{--}{GW191117_023843}{--}{GW191127_050227}{--}{GW191208_080334}{--}{GW191224_043228}{--}{GW191228_195619}{--}{GW200106_134123}{--}{GW200109_195634}{--}{GW200129_114245}{--}{GW200208_211609}{--}{GW200210_005122}{--}{GW200210_100022}{--}{GW200220_124850}{--}{GW200214_223307}{--}{GW200225_075134}{--}{GW200301_211019}{--}{GW200304_172806}{--}{GW200305_084739}{--}{GW200318_191337}{--}}}
\DeclareRobustCommand{\chipIAStwodivergence}[1]{\IfEqCase{#1}{{GW151205_195525}{--}{GW151216_092416}{--}{GW170121_212536}{--}{GW170202_135657}{--}{GW170304_163753}{--}{GW170403_230611}{--}{GW170425_055334}{--}{GW170727_010430}{--}{GW190426_082124}{--}{GW190427_180650}{--}{GW190511_125545}{--}{GW190511_163209}{--}{GW190514_065416}{--}{GW190523_085933}{--}{GW190524_134109}{--}{GW190530_030659}{--}{GW190530_133833}{--}{GW190604_103812}{--}{GW190605_025957}{--}{GW190607_083827}{--}{GW190614_134749}{--}{GW190615_030234}{--}{GW190705_164632}{--}{GW190707_083226}{--}{GW190711_030756}{--}{GW190718_160159}{--}{GW190725_174728}{--}{GW190805_105432}{--}{GW190806_033721}{--}{GW190814_192009}{--}{GW190818_232544}{--}{GW190821_124821}{--}{GW190904_104631}{--}{GW190906_054335}{--}{GW190910_012619}{--}{GW190911_195101}{--}{GW190916_200658}{--}{GW190926_050336}{--}{GW191113_103541}{--}{GW191117_023843}{--}{GW191127_050227}{--}{GW191208_080334}{--}{GW191224_043228}{--}{GW191228_195619}{--}{GW200106_134123}{--}{GW200109_195634}{--}{GW200129_114245}{--}{GW200208_211609}{--}{GW200210_005122}{--}{GW200210_100022}{--}{GW200220_124850}{--}{GW200214_223307}{--}{GW200225_075134}{--}{GW200301_211019}{--}{GW200304_172806}{--}{GW200305_084739}{--}{GW200318_191337}{--}}}
\DeclareRobustCommand{\luminositydistanceIAStwodivergence}[1]{\IfEqCase{#1}{{GW151205_195525}{--}{GW151216_092416}{0.050}{GW170121_212536}{0.012}{GW170202_135657}{0.028}{GW170304_163753}{0.032}{GW170403_230611}{0.045}{GW170425_055334}{0.069}{GW170727_010430}{0.007}{GW190426_082124}{--}{GW190427_180650}{--}{GW190511_125545}{--}{GW190511_163209}{--}{GW190514_065416}{--}{GW190523_085933}{--}{GW190524_134109}{--}{GW190530_030659}{--}{GW190530_133833}{--}{GW190604_103812}{--}{GW190605_025957}{--}{GW190607_083827}{--}{GW190614_134749}{--}{GW190615_030234}{--}{GW190705_164632}{--}{GW190707_083226}{--}{GW190711_030756}{--}{GW190718_160159}{--}{GW190725_174728}{--}{GW190805_105432}{--}{GW190806_033721}{--}{GW190814_192009}{--}{GW190818_232544}{--}{GW190821_124821}{--}{GW190904_104631}{--}{GW190906_054335}{--}{GW190910_012619}{--}{GW190911_195101}{--}{GW190916_200658}{--}{GW190926_050336}{--}{GW191113_103541}{--}{GW191117_023843}{--}{GW191127_050227}{--}{GW191208_080334}{--}{GW191224_043228}{--}{GW191228_195619}{--}{GW200106_134123}{--}{GW200109_195634}{--}{GW200129_114245}{--}{GW200208_211609}{--}{GW200210_005122}{--}{GW200210_100022}{--}{GW200220_124850}{--}{GW200214_223307}{--}{GW200225_075134}{--}{GW200301_211019}{--}{GW200304_172806}{--}{GW200305_084739}{--}{GW200318_191337}{--}}}
\DeclareRobustCommand{\chirpmasssourceIASthreedivergence}[1]{\IfEqCase{#1}{{GW151205_195525}{--}{GW151216_092416}{--}{GW170121_212536}{--}{GW170202_135657}{--}{GW170304_163753}{--}{GW170403_230611}{--}{GW170425_055334}{--}{GW170727_010430}{--}{GW190426_082124}{--}{GW190427_180650}{--}{GW190511_125545}{--}{GW190511_163209}{--}{GW190514_065416}{--}{GW190523_085933}{--}{GW190524_134109}{--}{GW190530_030659}{--}{GW190530_133833}{--}{GW190604_103812}{--}{GW190605_025957}{--}{GW190607_083827}{--}{GW190614_134749}{--}{GW190615_030234}{--}{GW190705_164632}{--}{GW190707_083226}{0.467}{GW190711_030756}{0.590}{GW190718_160159}{0.659}{GW190725_174728}{--}{GW190805_105432}{--}{GW190806_033721}{--}{GW190814_192009}{0.220}{GW190818_232544}{0.490}{GW190821_124821}{0.693}{GW190904_104631}{--}{GW190906_054335}{0.377}{GW190910_012619}{0.693}{GW190911_195101}{--}{GW190916_200658}{--}{GW190926_050336}{--}{GW191113_103541}{--}{GW191117_023843}{--}{GW191127_050227}{--}{GW191208_080334}{--}{GW191224_043228}{--}{GW191228_195619}{--}{GW200106_134123}{--}{GW200109_195634}{--}{GW200129_114245}{--}{GW200208_211609}{--}{GW200210_005122}{--}{GW200210_100022}{--}{GW200220_124850}{--}{GW200214_223307}{--}{GW200225_075134}{--}{GW200301_211019}{--}{GW200304_172806}{--}{GW200305_084739}{--}{GW200318_191337}{--}}}
\DeclareRobustCommand{\massratioIASthreedivergence}[1]{\IfEqCase{#1}{{GW151205_195525}{--}{GW151216_092416}{--}{GW170121_212536}{--}{GW170202_135657}{--}{GW170304_163753}{--}{GW170403_230611}{--}{GW170425_055334}{--}{GW170727_010430}{--}{GW190426_082124}{--}{GW190427_180650}{--}{GW190511_125545}{--}{GW190511_163209}{--}{GW190514_065416}{--}{GW190523_085933}{--}{GW190524_134109}{--}{GW190530_030659}{--}{GW190530_133833}{--}{GW190604_103812}{--}{GW190605_025957}{--}{GW190607_083827}{--}{GW190614_134749}{--}{GW190615_030234}{--}{GW190705_164632}{--}{GW190707_083226}{0.033}{GW190711_030756}{0.049}{GW190718_160159}{0.015}{GW190725_174728}{--}{GW190805_105432}{--}{GW190806_033721}{--}{GW190814_192009}{0.054}{GW190818_232544}{0.074}{GW190821_124821}{0.042}{GW190904_104631}{--}{GW190906_054335}{0.053}{GW190910_012619}{0.430}{GW190911_195101}{--}{GW190916_200658}{--}{GW190926_050336}{--}{GW191113_103541}{--}{GW191117_023843}{--}{GW191127_050227}{--}{GW191208_080334}{--}{GW191224_043228}{--}{GW191228_195619}{--}{GW200106_134123}{--}{GW200109_195634}{--}{GW200129_114245}{--}{GW200208_211609}{--}{GW200210_005122}{--}{GW200210_100022}{--}{GW200220_124850}{--}{GW200214_223307}{--}{GW200225_075134}{--}{GW200301_211019}{--}{GW200304_172806}{--}{GW200305_084739}{--}{GW200318_191337}{--}}}
\DeclareRobustCommand{\chieffIASthreedivergence}[1]{\IfEqCase{#1}{{GW151205_195525}{--}{GW151216_092416}{--}{GW170121_212536}{--}{GW170202_135657}{--}{GW170304_163753}{--}{GW170403_230611}{--}{GW170425_055334}{--}{GW170727_010430}{--}{GW190426_082124}{--}{GW190427_180650}{--}{GW190511_125545}{--}{GW190511_163209}{--}{GW190514_065416}{--}{GW190523_085933}{--}{GW190524_134109}{--}{GW190530_030659}{--}{GW190530_133833}{--}{GW190604_103812}{--}{GW190605_025957}{--}{GW190607_083827}{--}{GW190614_134749}{--}{GW190615_030234}{--}{GW190705_164632}{--}{GW190707_083226}{0.007}{GW190711_030756}{0.017}{GW190718_160159}{0.113}{GW190725_174728}{--}{GW190805_105432}{--}{GW190806_033721}{--}{GW190814_192009}{0.007}{GW190818_232544}{0.026}{GW190821_124821}{0.048}{GW190904_104631}{--}{GW190906_054335}{0.032}{GW190910_012619}{0.086}{GW190911_195101}{--}{GW190916_200658}{--}{GW190926_050336}{--}{GW191113_103541}{--}{GW191117_023843}{--}{GW191127_050227}{--}{GW191208_080334}{--}{GW191224_043228}{--}{GW191228_195619}{--}{GW200106_134123}{--}{GW200109_195634}{--}{GW200129_114245}{--}{GW200208_211609}{--}{GW200210_005122}{--}{GW200210_100022}{--}{GW200220_124850}{--}{GW200214_223307}{--}{GW200225_075134}{--}{GW200301_211019}{--}{GW200304_172806}{--}{GW200305_084739}{--}{GW200318_191337}{--}}}
\DeclareRobustCommand{\chipIASthreedivergence}[1]{\IfEqCase{#1}{{GW151205_195525}{--}{GW151216_092416}{--}{GW170121_212536}{--}{GW170202_135657}{--}{GW170304_163753}{--}{GW170403_230611}{--}{GW170425_055334}{--}{GW170727_010430}{--}{GW190426_082124}{--}{GW190427_180650}{--}{GW190511_125545}{--}{GW190511_163209}{--}{GW190514_065416}{--}{GW190523_085933}{--}{GW190524_134109}{--}{GW190530_030659}{--}{GW190530_133833}{--}{GW190604_103812}{--}{GW190605_025957}{--}{GW190607_083827}{--}{GW190614_134749}{--}{GW190615_030234}{--}{GW190705_164632}{--}{GW190707_083226}{0.013}{GW190711_030756}{0.009}{GW190718_160159}{0.016}{GW190725_174728}{--}{GW190805_105432}{--}{GW190806_033721}{--}{GW190814_192009}{0.004}{GW190818_232544}{0.011}{GW190821_124821}{0.015}{GW190904_104631}{--}{GW190906_054335}{0.005}{GW190910_012619}{0.012}{GW190911_195101}{--}{GW190916_200658}{--}{GW190926_050336}{--}{GW191113_103541}{--}{GW191117_023843}{--}{GW191127_050227}{--}{GW191208_080334}{--}{GW191224_043228}{--}{GW191228_195619}{--}{GW200106_134123}{--}{GW200109_195634}{--}{GW200129_114245}{--}{GW200208_211609}{--}{GW200210_005122}{--}{GW200210_100022}{--}{GW200220_124850}{--}{GW200214_223307}{--}{GW200225_075134}{--}{GW200301_211019}{--}{GW200304_172806}{--}{GW200305_084739}{--}{GW200318_191337}{--}}}
\DeclareRobustCommand{\luminositydistanceIASthreedivergence}[1]{\IfEqCase{#1}{{GW151205_195525}{--}{GW151216_092416}{--}{GW170121_212536}{--}{GW170202_135657}{--}{GW170304_163753}{--}{GW170403_230611}{--}{GW170425_055334}{--}{GW170727_010430}{--}{GW190426_082124}{--}{GW190427_180650}{--}{GW190511_125545}{--}{GW190511_163209}{--}{GW190514_065416}{--}{GW190523_085933}{--}{GW190524_134109}{--}{GW190530_030659}{--}{GW190530_133833}{--}{GW190604_103812}{--}{GW190605_025957}{--}{GW190607_083827}{--}{GW190614_134749}{--}{GW190615_030234}{--}{GW190705_164632}{--}{GW190707_083226}{0.017}{GW190711_030756}{0.009}{GW190718_160159}{0.052}{GW190725_174728}{--}{GW190805_105432}{--}{GW190806_033721}{--}{GW190814_192009}{0.039}{GW190818_232544}{0.062}{GW190821_124821}{0.007}{GW190904_104631}{--}{GW190906_054335}{0.057}{GW190910_012619}{0.175}{GW190911_195101}{--}{GW190916_200658}{--}{GW190926_050336}{--}{GW191113_103541}{--}{GW191117_023843}{--}{GW191127_050227}{--}{GW191208_080334}{--}{GW191224_043228}{--}{GW191228_195619}{--}{GW200106_134123}{--}{GW200109_195634}{--}{GW200129_114245}{--}{GW200208_211609}{--}{GW200210_005122}{--}{GW200210_100022}{--}{GW200220_124850}{--}{GW200214_223307}{--}{GW200225_075134}{--}{GW200301_211019}{--}{GW200304_172806}{--}{GW200305_084739}{--}{GW200318_191337}{--}}}
\DeclareRobustCommand{\chirpmasssourceIASHMdivergence}[1]{\IfEqCase{#1}{{GW151205_195525}{--}{GW151216_092416}{--}{GW170121_212536}{--}{GW170202_135657}{--}{GW170304_163753}{--}{GW170403_230611}{--}{GW170425_055334}{--}{GW170727_010430}{--}{GW190426_082124}{--}{GW190427_180650}{0.040}{GW190511_125545}{--}{GW190511_163209}{0.115}{GW190514_065416}{--}{GW190523_085933}{--}{GW190524_134109}{0.043}{GW190530_030659}{0.053}{GW190530_133833}{0.023}{GW190604_103812}{0.271}{GW190605_025957}{0.059}{GW190607_083827}{--}{GW190614_134749}{--}{GW190615_030234}{0.054}{GW190705_164632}{--}{GW190707_083226}{0.049}{GW190711_030756}{0.066}{GW190718_160159}{--}{GW190725_174728}{--}{GW190805_105432}{--}{GW190806_033721}{0.092}{GW190814_192009}{--}{GW190818_232544}{0.027}{GW190821_124821}{--}{GW190904_104631}{--}{GW190906_054335}{0.029}{GW190910_012619}{--}{GW190911_195101}{0.061}{GW190916_200658}{--}{GW190926_050336}{--}{GW191113_103541}{0.040}{GW191117_023843}{--}{GW191127_050227}{--}{GW191208_080334}{--}{GW191224_043228}{--}{GW191228_195619}{0.072}{GW200106_134123}{--}{GW200109_195634}{0.034}{GW200129_114245}{--}{GW200208_211609}{--}{GW200210_005122}{--}{GW200210_100022}{0.030}{GW200220_124850}{--}{GW200214_223307}{--}{GW200225_075134}{--}{GW200301_211019}{0.035}{GW200304_172806}{--}{GW200305_084739}{--}{GW200318_191337}{--}}}
\DeclareRobustCommand{\massratioIASHMdivergence}[1]{\IfEqCase{#1}{{GW151205_195525}{--}{GW151216_092416}{--}{GW170121_212536}{--}{GW170202_135657}{--}{GW170304_163753}{--}{GW170403_230611}{--}{GW170425_055334}{--}{GW170727_010430}{--}{GW190426_082124}{--}{GW190427_180650}{0.445}{GW190511_125545}{--}{GW190511_163209}{0.188}{GW190514_065416}{--}{GW190523_085933}{--}{GW190524_134109}{0.020}{GW190530_030659}{0.059}{GW190530_133833}{0.022}{GW190604_103812}{0.260}{GW190605_025957}{0.093}{GW190607_083827}{--}{GW190614_134749}{--}{GW190615_030234}{0.034}{GW190705_164632}{--}{GW190707_083226}{0.025}{GW190711_030756}{0.006}{GW190718_160159}{--}{GW190725_174728}{--}{GW190805_105432}{--}{GW190806_033721}{0.033}{GW190814_192009}{--}{GW190818_232544}{0.081}{GW190821_124821}{--}{GW190904_104631}{--}{GW190906_054335}{0.024}{GW190910_012619}{--}{GW190911_195101}{0.026}{GW190916_200658}{--}{GW190926_050336}{--}{GW191113_103541}{0.091}{GW191117_023843}{--}{GW191127_050227}{--}{GW191208_080334}{--}{GW191224_043228}{--}{GW191228_195619}{0.060}{GW200106_134123}{--}{GW200109_195634}{0.025}{GW200129_114245}{--}{GW200208_211609}{--}{GW200210_005122}{--}{GW200210_100022}{0.107}{GW200220_124850}{--}{GW200214_223307}{--}{GW200225_075134}{--}{GW200301_211019}{0.035}{GW200304_172806}{--}{GW200305_084739}{--}{GW200318_191337}{--}}}
\DeclareRobustCommand{\chieffIASHMdivergence}[1]{\IfEqCase{#1}{{GW151205_195525}{--}{GW151216_092416}{--}{GW170121_212536}{--}{GW170202_135657}{--}{GW170304_163753}{--}{GW170403_230611}{--}{GW170425_055334}{--}{GW170727_010430}{--}{GW190426_082124}{--}{GW190427_180650}{0.022}{GW190511_125545}{--}{GW190511_163209}{0.315}{GW190514_065416}{--}{GW190523_085933}{--}{GW190524_134109}{0.007}{GW190530_030659}{0.049}{GW190530_133833}{0.010}{GW190604_103812}{0.116}{GW190605_025957}{0.075}{GW190607_083827}{--}{GW190614_134749}{--}{GW190615_030234}{0.021}{GW190705_164632}{--}{GW190707_083226}{0.149}{GW190711_030756}{0.086}{GW190718_160159}{--}{GW190725_174728}{--}{GW190805_105432}{--}{GW190806_033721}{0.050}{GW190814_192009}{--}{GW190818_232544}{0.297}{GW190821_124821}{--}{GW190904_104631}{--}{GW190906_054335}{0.137}{GW190910_012619}{--}{GW190911_195101}{0.036}{GW190916_200658}{--}{GW190926_050336}{--}{GW191113_103541}{0.041}{GW191117_023843}{--}{GW191127_050227}{--}{GW191208_080334}{--}{GW191224_043228}{--}{GW191228_195619}{0.031}{GW200106_134123}{--}{GW200109_195634}{0.307}{GW200129_114245}{--}{GW200208_211609}{--}{GW200210_005122}{--}{GW200210_100022}{0.193}{GW200220_124850}{--}{GW200214_223307}{--}{GW200225_075134}{--}{GW200301_211019}{0.107}{GW200304_172806}{--}{GW200305_084739}{--}{GW200318_191337}{--}}}
\DeclareRobustCommand{\chipIASHMdivergence}[1]{\IfEqCase{#1}{{GW151205_195525}{--}{GW151216_092416}{--}{GW170121_212536}{--}{GW170202_135657}{--}{GW170304_163753}{--}{GW170403_230611}{--}{GW170425_055334}{--}{GW170727_010430}{--}{GW190426_082124}{--}{GW190427_180650}{0.029}{GW190511_125545}{--}{GW190511_163209}{0.042}{GW190514_065416}{--}{GW190523_085933}{--}{GW190524_134109}{0.001}{GW190530_030659}{0.006}{GW190530_133833}{0.000}{GW190604_103812}{0.217}{GW190605_025957}{0.050}{GW190607_083827}{--}{GW190614_134749}{--}{GW190615_030234}{0.000}{GW190705_164632}{--}{GW190707_083226}{0.114}{GW190711_030756}{0.141}{GW190718_160159}{--}{GW190725_174728}{--}{GW190805_105432}{--}{GW190806_033721}{0.006}{GW190814_192009}{--}{GW190818_232544}{0.021}{GW190821_124821}{--}{GW190904_104631}{--}{GW190906_054335}{0.051}{GW190910_012619}{--}{GW190911_195101}{0.001}{GW190916_200658}{--}{GW190926_050336}{--}{GW191113_103541}{0.008}{GW191117_023843}{--}{GW191127_050227}{--}{GW191208_080334}{--}{GW191224_043228}{--}{GW191228_195619}{0.014}{GW200106_134123}{--}{GW200109_195634}{0.012}{GW200129_114245}{--}{GW200208_211609}{--}{GW200210_005122}{--}{GW200210_100022}{0.070}{GW200220_124850}{--}{GW200214_223307}{--}{GW200225_075134}{--}{GW200301_211019}{0.105}{GW200304_172806}{--}{GW200305_084739}{--}{GW200318_191337}{--}}}
\DeclareRobustCommand{\luminositydistanceIASHMdivergence}[1]{\IfEqCase{#1}{{GW151205_195525}{--}{GW151216_092416}{--}{GW170121_212536}{--}{GW170202_135657}{--}{GW170304_163753}{--}{GW170403_230611}{--}{GW170425_055334}{--}{GW170727_010430}{--}{GW190426_082124}{--}{GW190427_180650}{0.024}{GW190511_125545}{--}{GW190511_163209}{0.067}{GW190514_065416}{--}{GW190523_085933}{--}{GW190524_134109}{0.027}{GW190530_030659}{0.033}{GW190530_133833}{0.010}{GW190604_103812}{0.131}{GW190605_025957}{0.013}{GW190607_083827}{--}{GW190614_134749}{--}{GW190615_030234}{0.053}{GW190705_164632}{--}{GW190707_083226}{0.103}{GW190711_030756}{0.034}{GW190718_160159}{--}{GW190725_174728}{--}{GW190805_105432}{--}{GW190806_033721}{0.030}{GW190814_192009}{--}{GW190818_232544}{0.165}{GW190821_124821}{--}{GW190904_104631}{--}{GW190906_054335}{0.141}{GW190910_012619}{--}{GW190911_195101}{0.062}{GW190916_200658}{--}{GW190926_050336}{--}{GW191113_103541}{0.098}{GW191117_023843}{--}{GW191127_050227}{--}{GW191208_080334}{--}{GW191224_043228}{--}{GW191228_195619}{0.027}{GW200106_134123}{--}{GW200109_195634}{0.099}{GW200129_114245}{--}{GW200208_211609}{--}{GW200210_005122}{--}{GW200210_100022}{0.201}{GW200220_124850}{--}{GW200214_223307}{--}{GW200225_075134}{--}{GW200301_211019}{0.009}{GW200304_172806}{--}{GW200305_084739}{--}{GW200318_191337}{--}}}
\DeclareRobustCommand{\chirpmasssourceIASfourdivergence}[1]{\IfEqCase{#1}{{GW151205_195525}{--}{GW151216_092416}{--}{GW170121_212536}{--}{GW170202_135657}{--}{GW170304_163753}{--}{GW170403_230611}{--}{GW170425_055334}{--}{GW170727_010430}{--}{GW190426_082124}{--}{GW190427_180650}{--}{GW190511_125545}{--}{GW190511_163209}{--}{GW190514_065416}{--}{GW190523_085933}{--}{GW190524_134109}{--}{GW190530_030659}{--}{GW190530_133833}{--}{GW190604_103812}{--}{GW190605_025957}{--}{GW190607_083827}{--}{GW190614_134749}{--}{GW190615_030234}{--}{GW190705_164632}{--}{GW190707_083226}{--}{GW190711_030756}{--}{GW190718_160159}{--}{GW190725_174728}{--}{GW190805_105432}{--}{GW190806_033721}{--}{GW190814_192009}{--}{GW190818_232544}{--}{GW190821_124821}{--}{GW190904_104631}{--}{GW190906_054335}{--}{GW190910_012619}{--}{GW190911_195101}{--}{GW190916_200658}{--}{GW190926_050336}{--}{GW191113_103541}{--}{GW191117_023843}{0.125}{GW191127_050227}{--}{GW191208_080334}{--}{GW191224_043228}{--}{GW191228_195619}{0.693}{GW200106_134123}{--}{GW200109_195634}{0.006}{GW200129_114245}{--}{GW200208_211609}{--}{GW200210_005122}{--}{GW200210_100022}{0.069}{GW200220_124850}{--}{GW200214_223307}{--}{GW200225_075134}{0.116}{GW200301_211019}{--}{GW200304_172806}{--}{GW200305_084739}{--}{GW200318_191337}{--}}}
\DeclareRobustCommand{\massratioIASfourdivergence}[1]{\IfEqCase{#1}{{GW151205_195525}{--}{GW151216_092416}{--}{GW170121_212536}{--}{GW170202_135657}{--}{GW170304_163753}{--}{GW170403_230611}{--}{GW170425_055334}{--}{GW170727_010430}{--}{GW190426_082124}{--}{GW190427_180650}{--}{GW190511_125545}{--}{GW190511_163209}{--}{GW190514_065416}{--}{GW190523_085933}{--}{GW190524_134109}{--}{GW190530_030659}{--}{GW190530_133833}{--}{GW190604_103812}{--}{GW190605_025957}{--}{GW190607_083827}{--}{GW190614_134749}{--}{GW190615_030234}{--}{GW190705_164632}{--}{GW190707_083226}{--}{GW190711_030756}{--}{GW190718_160159}{--}{GW190725_174728}{--}{GW190805_105432}{--}{GW190806_033721}{--}{GW190814_192009}{--}{GW190818_232544}{--}{GW190821_124821}{--}{GW190904_104631}{--}{GW190906_054335}{--}{GW190910_012619}{--}{GW190911_195101}{--}{GW190916_200658}{--}{GW190926_050336}{--}{GW191113_103541}{--}{GW191117_023843}{0.503}{GW191127_050227}{--}{GW191208_080334}{--}{GW191224_043228}{--}{GW191228_195619}{0.608}{GW200106_134123}{--}{GW200109_195634}{0.006}{GW200129_114245}{--}{GW200208_211609}{--}{GW200210_005122}{--}{GW200210_100022}{0.132}{GW200220_124850}{--}{GW200214_223307}{--}{GW200225_075134}{0.002}{GW200301_211019}{--}{GW200304_172806}{--}{GW200305_084739}{--}{GW200318_191337}{--}}}
\DeclareRobustCommand{\chieffIASfourdivergence}[1]{\IfEqCase{#1}{{GW151205_195525}{--}{GW151216_092416}{--}{GW170121_212536}{--}{GW170202_135657}{--}{GW170304_163753}{--}{GW170403_230611}{--}{GW170425_055334}{--}{GW170727_010430}{--}{GW190426_082124}{--}{GW190427_180650}{--}{GW190511_125545}{--}{GW190511_163209}{--}{GW190514_065416}{--}{GW190523_085933}{--}{GW190524_134109}{--}{GW190530_030659}{--}{GW190530_133833}{--}{GW190604_103812}{--}{GW190605_025957}{--}{GW190607_083827}{--}{GW190614_134749}{--}{GW190615_030234}{--}{GW190705_164632}{--}{GW190707_083226}{--}{GW190711_030756}{--}{GW190718_160159}{--}{GW190725_174728}{--}{GW190805_105432}{--}{GW190806_033721}{--}{GW190814_192009}{--}{GW190818_232544}{--}{GW190821_124821}{--}{GW190904_104631}{--}{GW190906_054335}{--}{GW190910_012619}{--}{GW190911_195101}{--}{GW190916_200658}{--}{GW190926_050336}{--}{GW191113_103541}{--}{GW191117_023843}{0.037}{GW191127_050227}{--}{GW191208_080334}{--}{GW191224_043228}{--}{GW191228_195619}{0.123}{GW200106_134123}{--}{GW200109_195634}{0.011}{GW200129_114245}{--}{GW200208_211609}{--}{GW200210_005122}{--}{GW200210_100022}{0.036}{GW200220_124850}{--}{GW200214_223307}{--}{GW200225_075134}{0.000}{GW200301_211019}{--}{GW200304_172806}{--}{GW200305_084739}{--}{GW200318_191337}{--}}}
\DeclareRobustCommand{\chipIASfourdivergence}[1]{\IfEqCase{#1}{{GW151205_195525}{--}{GW151216_092416}{--}{GW170121_212536}{--}{GW170202_135657}{--}{GW170304_163753}{--}{GW170403_230611}{--}{GW170425_055334}{--}{GW170727_010430}{--}{GW190426_082124}{--}{GW190427_180650}{--}{GW190511_125545}{--}{GW190511_163209}{--}{GW190514_065416}{--}{GW190523_085933}{--}{GW190524_134109}{--}{GW190530_030659}{--}{GW190530_133833}{--}{GW190604_103812}{--}{GW190605_025957}{--}{GW190607_083827}{--}{GW190614_134749}{--}{GW190615_030234}{--}{GW190705_164632}{--}{GW190707_083226}{--}{GW190711_030756}{--}{GW190718_160159}{--}{GW190725_174728}{--}{GW190805_105432}{--}{GW190806_033721}{--}{GW190814_192009}{--}{GW190818_232544}{--}{GW190821_124821}{--}{GW190904_104631}{--}{GW190906_054335}{--}{GW190910_012619}{--}{GW190911_195101}{--}{GW190916_200658}{--}{GW190926_050336}{--}{GW191113_103541}{--}{GW191117_023843}{0.017}{GW191127_050227}{--}{GW191208_080334}{--}{GW191224_043228}{--}{GW191228_195619}{0.240}{GW200106_134123}{--}{GW200109_195634}{0.003}{GW200129_114245}{--}{GW200208_211609}{--}{GW200210_005122}{--}{GW200210_100022}{0.003}{GW200220_124850}{--}{GW200214_223307}{--}{GW200225_075134}{0.001}{GW200301_211019}{--}{GW200304_172806}{--}{GW200305_084739}{--}{GW200318_191337}{--}}}
\DeclareRobustCommand{\luminositydistanceIASfourdivergence}[1]{\IfEqCase{#1}{{GW151205_195525}{--}{GW151216_092416}{--}{GW170121_212536}{--}{GW170202_135657}{--}{GW170304_163753}{--}{GW170403_230611}{--}{GW170425_055334}{--}{GW170727_010430}{--}{GW190426_082124}{--}{GW190427_180650}{--}{GW190511_125545}{--}{GW190511_163209}{--}{GW190514_065416}{--}{GW190523_085933}{--}{GW190524_134109}{--}{GW190530_030659}{--}{GW190530_133833}{--}{GW190604_103812}{--}{GW190605_025957}{--}{GW190607_083827}{--}{GW190614_134749}{--}{GW190615_030234}{--}{GW190705_164632}{--}{GW190707_083226}{--}{GW190711_030756}{--}{GW190718_160159}{--}{GW190725_174728}{--}{GW190805_105432}{--}{GW190806_033721}{--}{GW190814_192009}{--}{GW190818_232544}{--}{GW190821_124821}{--}{GW190904_104631}{--}{GW190906_054335}{--}{GW190910_012619}{--}{GW190911_195101}{--}{GW190916_200658}{--}{GW190926_050336}{--}{GW191113_103541}{--}{GW191117_023843}{0.170}{GW191127_050227}{--}{GW191208_080334}{--}{GW191224_043228}{--}{GW191228_195619}{0.552}{GW200106_134123}{--}{GW200109_195634}{0.014}{GW200129_114245}{--}{GW200208_211609}{--}{GW200210_005122}{--}{GW200210_100022}{0.123}{GW200220_124850}{--}{GW200214_223307}{--}{GW200225_075134}{0.066}{GW200301_211019}{--}{GW200304_172806}{--}{GW200305_084739}{--}{GW200318_191337}{--}}}

\def\eventname#1#2{GW$\,#1\_#2$}

\usepackage{soul}

\def\add#1{{\color{blue}#1}}
\def\delete#1{{\color{red}#1}}

\def\add#1{{#1}}
\def\delete#1{}

\usepackage{tabularx}
\begin{document}

\title[Beyond GWTC-3]{Beyond GWTC-3: Analysing and verifying new gravitational-wave events from community catalogues}

\author{Daniel Williams}

\address{SUPA, School of Physics \& Astronomy, University of Glasgow, G12 8QQ}
\vspace{10pt}
\begin{indented}
\item[]April 2025
\end{indented}

\begin{abstract}
The public release of data from the LIGO and Virgo detectors has enabled the identification of potential gravitational wave signals by independent teams using alternative methodologies.
In addition to the LIGO-Virgo-KAGRA (LVK) collaboration's GWTC-3 catalogue there have been several additional works claiming the detection of signals in the data from the first three observing runs.
In this paper we present an analysis of these new signals using the same analysis workflow which was used to generate the GWTC-2.1 and GWTC-3 catalogues published by the LVK, matching the analysis configuration as closely as possible, and we provide our parameter estimation results in a format comparable to those of the GWTC-3 data release \add{for 57 events not previously analysed in LVK analyses. We find our results to be broadly consistent with those published by other groups}.
We also include a discussion of the workflow developed for this analysis.
\end{abstract}

\section{Introduction}

As the result of rapid developments in detector technology and data analysis techniques, the field of gravitational wave detection has advanced from the first detection of a signal, GW$150914\_095045$~\cite{gw150914}, from a binary black hole, in 2015, to frequent and routine observation of binary black hole coalescence (BBH) signals. 
This has allowed the creation of catalogues of observed signals, with four of these catalogues having been produced to date by the LIGO-Virgo-KAGRA collaborations~(LVK)~\cite{catalog-gwtc-1,catalog-gwtc-2,catalog-gwtc-2d1,catalog-gwtc-3}.

Gravitational wave signals are detected in a low signal-to-noise regime, and signals are generally short compared to the total observing time.
The needle-in-a-haystack-like problem of detecting signals currently necessitates performing analysis of the data in at least two stages.
In the first of these stages signals are identified by \emph{searches} of the data, for example~\cite{search-pycbc,search-mbta,search-spiir,search-cwb}.
Searches identify candidate signals in the detector data, estimate the significance of the event, and provide an estimate of some of the astrophysical parameters of the system which produced the signal.
The significance can either be calculated statistically, as the false alarm rate of the candidate, or by attempting to estimate the probability that the candidate had an astrophysical origin~\cite{farr_counting_2015, guglielmetti_backgroundsource_2009}.

The second stage uses statistical inference techniques to perform model selection on signals identified in the search stage.
This stage is referred-to as parameter estimation (PE), and provides more detailed and comprehensive analysis of the signal, examining only the data in a \add{short} time window around the signal.
The PE stage is generally computationally intensive, using Markov chain Monte Carlo techniques to determine posterior distributions in a Bayesian framework such as LALInference~\cite{code-lalinference} or Bilby~\cite{code-bilby}, and may require thousands of core hours to complete.
\add{Alternative approaches to this problem are in active development.
These include approaches which reduce the complexity of the inference problem, for example Cogwheel~\cite{code-cogwheel}, as well as approaches which use machine learning, such as Vitamin~\cite{code-vitamin} and Dingo~\cite{code-dingo}.
}

The events reported in these catalogues therefore fall into two sets; a full list of possible events which have been identified by a search, which may \delete{list}\add{number} in the thousands of candidates; and a much shorter list of candidates which have been analysed in greater detail using PE techniques to determine the properties of the astrophysical source.
This shorter list is generally created by applying a threshold to some measure of the signals' significance.
The method used to perform the search can affect the calculation of the false alarm rate for each trigger, thus by either choosing a different cut, or using a different search methodology it is possible to produce different event lists for further analysis.

The most complete catalogue produced by the LVK to date is GWTC-3~\cite{catalog-gwtc-3}, which includes results from the first three observing runs of the ground-based detectors\add{~\footnote{\add{The authors note that the LVK's GWTC catalogues are cumulative, so GWTC-3 is considered to contain events from all three observing runs, and not just the most recent observing run, O3b; we have treated the OGC catalogues in the same way in this work, but the nature of IAS catalogues suggests these catalogues are not cumulative.}}}.
Data from these detectors are made publicly available at the end of a proprietary period, allowing both the reproduction of results presented in LVK publications, and the identification of signals using alternative techniques to those used by the LVK.
The Open Gravitational-Wave Catalog (OGC) series~\cite{catalog-1ogc,catalog-2ogc,catalog-3ogc,catalog-4ogc} and the Institute for Advanced Studies (IAS) series \cite{catalog-ias-1,catalog-ias-2,catalog-ias-3, catalog-ias-4, catalog-ias-hm} are the most extensive third-party catalogues, with each spanning the three observing runs for which data is publicly available.
There are also a number of catalogues which span only a subset of the total observing time of the detectors.
These catalogues have generally accompanied descriptions of either novel techniques for improving the sensitivity of existing search codes, for example pycbc-kde~\cite{catalog-pycbc-kde} and cWB~\cite{catalog-cwb}; or entirely novel search techniques \add{such as AresGW which make use of machine learning}~\cite{catalog-gwares}.
Each of these catalogues present a different set of candidate events (although there is considerable overlap between each).
The differences in the set of events identified can be a result of the efficiency at which a code can identify signals (especially at low signal-to-noise ratios), the significance statistic, and method used to calculate it for a given signal~\footnote{Prior to GWTC-3, LVK analyses used the false alarm rate to determine this cut~\cite{catalog-gwtc-1,catalog-gwtc-2,catalog-gwtc-2d1}; in GWTC-3~\cite{catalog-gwtc-3} the methodology was changed to use the probability of astrophysical origin.}, or a combination of the two.
As a result of these various differences 61 events have been reported in third-party catalogues which are not present in the LVK's GWTC-3 catalogue, or its related data releases.

This work aims to provide detailed PE results for 57 of these events in the same format as those provided in the GWTC-3 data release, and produced in a way which was consistent with analyses used for that catalogue, in order to produce a consistent dataset of as many events claimed to date across the literature as possible.
In preparing this work we used Asimov~\cite{code-asimov}, a workflow management tool used by the LVK to produce GWTC-3, to analyse the triggers presented in third-party catalogs which were not present in the GWTC-3 PE analysis dataset, and we provide an accompanying dataset of Asimov blueprints which can be used to both reproduce these analyses, and easily configure additional analyses at scale across the extended set of events. 
\add{These are further detailed in~\ref{sec:blueprints}}.
We were unable to produce well-converged posterior distributions using our technique for four events: \eventname{190704}{104834}, \eventname{190920}{113516}, \eventname{191228}{085854}, and \eventname{200316}{235947} (the former two were first published in IAS-3, the latter two in IAS-4). 
\add{In order to ensure reproducibility we use only publicly available software and data to perform our analysis, and we provide data in an accompanying data release to facilitate the straightforward reproduction of our results.}

\add{In section~\ref{sec:selection} we describe the approach taken to select events for consideration in this work.}

In section~\ref{sec:method} we provide a brief overview of the method used to configure and perform the new PE analyses.
We discuss specifically the differences between the analyses and datasets presented in this work and the original GWTC-3 analysis in section~\ref{sec:differences}.
We present the results of these analyses, and a discussion of the differences between our results and those presented in the various third-party catalogues in section~\ref{sec:pe}.
Where PE results were provided in the third-party catalogue's data release, we have compared our PE results to their results.
Finally, \add{tables of estimated source parameters for each of these events are provided in~\ref{sec:tables}, and} a technical description of the Asimov blueprints used for the analysis is contained in~\ref{sec:blueprints}.

\section{Event selection}
\label{sec:selection}

In this work we present PE results for all of the gravitational wave events which have been presented across various third-party catalogues~\cite{catalog-1ogc,catalog-2ogc,catalog-3ogc,catalog-4ogc,catalog-ias-1,catalog-ias-2,catalog-ias-3,catalog-ias-4,catalog-ias-hm,catalog-pycbc-kde,catalog-gwares,catalog-cwb}, but which were not present in GWTC-3~\cite{catalog-gwtc-3}\delete{~\footnote{\delete{The authors note that the LVK's GWTC catalogues are cumulative, so GWTC-3 is considered to contain events from all three observing runs, and not just O3b; we have treated the OGC catalogues in the same way in this work, but the nature of IAS catalogues suggests these catalogues are not cumulative.}}}, or which were not selected for further PE study in that work.
Events \add{identified in each catalogue's publication as being new detections} were selected \delete{from each catalogue's publication, selecting those events which were identified as new detections in that catalogue}\footnote{A number of events were reported first in 3-OGC which were later reported in GWTC-2.1; those events have not been included in this work, as they are contained within GWTC-3. Other events may appear in some editions of third party catalogues, but were removed from later editions; we have chosen to produce PE for the maximal set of events.}.

The events which we provide PE for are summarised in table~\ref{tab:event-list}, alongside a summary of the catalogues these are contained within. 
The pycbc-kde catalogue present two sets, labeled ``gold'' and ``silver'' in that work; we present PE for both sets.
We have attempted to perform PE on all of the events for which these additional catalogues provided PE (or all of the new significant events identified in the case of those publications which did not provide PE samples in a data release).

We did not conduct our own search as part of this work, so do not provide new estimates of the significance of the events for which we present PE.

\begin{table}[]
    \centering
    \scriptsize
    \begin{tabular}{c|cccc|ccccc|ccc}\toprule
    Event  & \multicolumn{4}{c|}{OGC} & \multicolumn{5}{c|}{IAS} & pycbc & AresGW & cwb \\ 
           & 1 & 2 & 3 & 4            & 1 & 2 & 3 & 4 & HM       &  kde  &      &     \\ \midrule
    \eventname{151205}{195525} & $\cdot$ & Y & $\cdot$ & $\cdot$ & Y & Y & Y & Y & $\cdot$ & $\cdot$ & $\cdot$ & $\cdot$ \\
\eventname{151216}{092416} & $\cdot$ & $\cdot$ & $\cdot$ & $\cdot$ & Y & Y & Y & Y & $\cdot$ & $\cdot$ & $\cdot$ & $\cdot$ \\
\eventname{170121}{212536} & $\cdot$ & Y & Y & Y & $\cdot$ & Y & Y & Y & $\cdot$ & $\cdot$ & $\cdot$ & $\cdot$ \\
\eventname{170202}{135657} & $\cdot$ & $\cdot$ & Y & Y & $\cdot$ & Y & Y & Y & $\cdot$ & $\cdot$ & $\cdot$ & $\cdot$ \\
\eventname{170304}{163753} & $\cdot$ & Y & Y & Y & $\cdot$ & Y & Y & Y & $\cdot$ & $\cdot$ & $\cdot$ & $\cdot$ \\
\eventname{170403}{230611} & $\cdot$ & $\cdot$ & Y & Y & $\cdot$ & Y & Y & Y & $\cdot$ & $\cdot$ & $\cdot$ & $\cdot$ \\
\eventname{170425}{055334} & $\cdot$ & $\cdot$ & $\cdot$ & $\cdot$ & $\cdot$ & Y & Y & Y & $\cdot$ & $\cdot$ & $\cdot$ & $\cdot$ \\
\eventname{170727}{010430} & $\cdot$ & Y & Y & Y & $\cdot$ & Y & Y & Y & $\cdot$ & $\cdot$ & $\cdot$ & $\cdot$ \\
\eventname{190426}{082124} & $\cdot$ & $\cdot$ & $\cdot$ & $\cdot$ & $\cdot$ & $\cdot$ & $\cdot$ & $\cdot$ & $\cdot$ & $\cdot$ & Y & $\cdot$ \\
\eventname{190427}{180650} & $\cdot$ & $\cdot$ & $\cdot$ & Y & $\cdot$ & $\cdot$ & $\cdot$ & $\cdot$ & $\cdot$ & $\cdot$ & $\cdot$ & $\cdot$ \\
\eventname{190511}{125545} & $\cdot$ & $\cdot$ & $\cdot$ & $\cdot$ & $\cdot$ & $\cdot$ & $\cdot$ & $\cdot$ & $\cdot$ & $\cdot$ & Y & $\cdot$ \\
\eventname{190511}{163209} & $\cdot$ & $\cdot$ & $\cdot$ & $\cdot$ & $\cdot$ & $\cdot$ & $\cdot$ & $\cdot$ & Y & $\cdot$ & $\cdot$ & $\cdot$ \\
\eventname{190514}{065416} & $\cdot$ & $\cdot$ & $\cdot$ & $\cdot$ & $\cdot$ & $\cdot$ & $\cdot$ & $\cdot$ & $\cdot$ & Y & $\cdot$ & $\cdot$ \\
\eventname{190523}{085933} & $\cdot$ & $\cdot$ & $\cdot$ & $\cdot$ & $\cdot$ & $\cdot$ & $\cdot$ & $\cdot$ & $\cdot$ & $\cdot$ & Y & $\cdot$ \\
\eventname{190524}{134109} & $\cdot$ & $\cdot$ & $\cdot$ & $\cdot$ & $\cdot$ & $\cdot$ & $\cdot$ & $\cdot$ & Y & $\cdot$ & $\cdot$ & $\cdot$ \\
\eventname{190530}{030659} & $\cdot$ & $\cdot$ & $\cdot$ & $\cdot$ & $\cdot$ & $\cdot$ & $\cdot$ & $\cdot$ & Y & $\cdot$ & $\cdot$ & $\cdot$ \\
\eventname{190530}{133833} & $\cdot$ & $\cdot$ & $\cdot$ & $\cdot$ & $\cdot$ & $\cdot$ & $\cdot$ & $\cdot$ & Y & $\cdot$ & $\cdot$ & $\cdot$ \\
\eventname{190604}{103812} & $\cdot$ & $\cdot$ & $\cdot$ & $\cdot$ & $\cdot$ & $\cdot$ & $\cdot$ & $\cdot$ & Y & $\cdot$ & $\cdot$ & $\cdot$ \\
\eventname{190605}{025957} & $\cdot$ & $\cdot$ & $\cdot$ & $\cdot$ & $\cdot$ & $\cdot$ & $\cdot$ & $\cdot$ & Y & $\cdot$ & $\cdot$ & $\cdot$ \\
\eventname{190607}{083827} & $\cdot$ & $\cdot$ & $\cdot$ & $\cdot$ & $\cdot$ & $\cdot$ & $\cdot$ & $\cdot$ & $\cdot$ & $\cdot$ & Y & Y  \\
\eventname{190614}{134749} & $\cdot$ & $\cdot$ & $\cdot$ & $\cdot$ & $\cdot$ & $\cdot$ & $\cdot$ & $\cdot$ & $\cdot$ & $\cdot$ & Y & $\cdot$ \\
\eventname{190615}{030234} & $\cdot$ & $\cdot$ & $\cdot$ & $\cdot$ & $\cdot$ & $\cdot$ & $\cdot$ & $\cdot$ & Y & $\cdot$ & $\cdot$ & $\cdot$ \\
\eventname{190705}{164632} & $\cdot$ & $\cdot$ & $\cdot$ & $\cdot$ & $\cdot$ & $\cdot$ & $\cdot$ & $\cdot$ & $\cdot$ & $\cdot$ & Y & $\cdot$ \\
\eventname{190707}{083226} & $\cdot$ & $\cdot$ & $\cdot$ & $\cdot$ & $\cdot$ & $\cdot$ & Y & $\cdot$ & Y & $\cdot$ & $\cdot$ & $\cdot$ \\
\eventname{190711}{030756} & $\cdot$ & $\cdot$ & $\cdot$ & $\cdot$ & $\cdot$ & $\cdot$ & Y & $\cdot$ & Y & $\cdot$ & $\cdot$ &  Y  \\
\eventname{190718}{160159} & $\cdot$ & $\cdot$ & $\cdot$ & $\cdot$ & $\cdot$ & $\cdot$ & Y & $\cdot$ & $\cdot$ & $\cdot$ & $\cdot$ & $\cdot$ \\
\eventname{190725}{174728} & $\cdot$ & $\cdot$ & $\cdot$ & $\cdot$ & $\cdot$ & $\cdot$ & $\cdot$ & $\cdot$ & $\cdot$ & Y & $\cdot$ & $\cdot$ \\
\eventname{190805}{105432} & $\cdot$ & $\cdot$ & $\cdot$ & Y & $\cdot$ & $\cdot$ & $\cdot$ & $\cdot$ & $\cdot$ & $\cdot$ & $\cdot$ & $\cdot$ \\
\eventname{190806}{033721} & $\cdot$ & $\cdot$ & $\cdot$ & $\cdot$ & $\cdot$ & $\cdot$ & $\cdot$ & $\cdot$ & Y & $\cdot$ & $\cdot$ & $\cdot$ \\
\eventname{190814}{192009} & $\cdot$ & $\cdot$ & $\cdot$ & $\cdot$ & $\cdot$ & $\cdot$ & Y & $\cdot$ & $\cdot$ & $\cdot$ & $\cdot$ & $\cdot$ \\
\eventname{190818}{232544} & $\cdot$ & $\cdot$ & $\cdot$ & $\cdot$ & $\cdot$ & $\cdot$ & Y & $\cdot$ & Y & $\cdot$ & $\cdot$ & $\cdot$ \\
\eventname{190821}{124821} & $\cdot$ & $\cdot$ & $\cdot$ & $\cdot$ & $\cdot$ & $\cdot$ & Y & $\cdot$ & $\cdot$ & $\cdot$ & $\cdot$ & $\cdot$ \\
\eventname{190904}{104631} & $\cdot$ & $\cdot$ & $\cdot$ & $\cdot$ & $\cdot$ & $\cdot$ & $\cdot$ & $\cdot$ & $\cdot$ & $\cdot$ & Y & $\cdot$ \\
\eventname{190906}{054335} & $\cdot$ & $\cdot$ & $\cdot$ & $\cdot$ & $\cdot$ & $\cdot$ & Y & $\cdot$ & Y & $\cdot$ & $\cdot$ & $\cdot$ \\
\eventname{190910}{012619} & $\cdot$ & $\cdot$ & $\cdot$ & $\cdot$ & $\cdot$ & $\cdot$ & Y & $\cdot$ & $\cdot$ & $\cdot$ & $\cdot$ & $\cdot$ \\
\eventname{190911}{195101} & $\cdot$ & $\cdot$ & $\cdot$ & $\cdot$ & $\cdot$ & $\cdot$ & $\cdot$ & $\cdot$ & Y & $\cdot$ & $\cdot$ & $\cdot$ \\
\eventname{190916}{200658} & $\cdot$ & $\cdot$ & $\cdot$ & $\cdot$ & $\cdot$ & $\cdot$ & $\cdot$ & $\cdot$ & $\cdot$ & Y & $\cdot$ & $\cdot$ \\
\eventname{190926}{050336} & $\cdot$ & $\cdot$ & $\cdot$ & $\cdot$ & $\cdot$ & $\cdot$ & $\cdot$ & $\cdot$ & $\cdot$ & Y & $\cdot$ & $\cdot$ \\
\eventname{191113}{103541} & $\cdot$ & $\cdot$ & $\cdot$ & $\cdot$ & $\cdot$ & $\cdot$ & $\cdot$ & $\cdot$ & Y & $\cdot$ & $\cdot$ & $\cdot$ \\
\eventname{191117}{023843} & $\cdot$ & $\cdot$ & $\cdot$ & $\cdot$ & $\cdot$ & $\cdot$ & $\cdot$ & Y & $\cdot$ & $\cdot$ & $\cdot$ & $\cdot$ \\
\eventname{191127}{050227} & $\cdot$ & $\cdot$ & $\cdot$ & $\cdot$ & $\cdot$ & $\cdot$ & $\cdot$ & $\cdot$ & $\cdot$ & Y & $\cdot$ & $\cdot$ \\
\eventname{191208}{080334} & $\cdot$ & $\cdot$ & $\cdot$ & $\cdot$ & $\cdot$ & $\cdot$ & $\cdot$ & $\cdot$ & $\cdot$ & Y & $\cdot$ & $\cdot$ \\
\eventname{191224}{043228} & $\cdot$ & $\cdot$ & $\cdot$ & Y & $\cdot$ & $\cdot$ & $\cdot$ & $\cdot$ & $\cdot$ & Y & $\cdot$ & $\cdot$ \\
\eventname{191228}{195619} & $\cdot$ & $\cdot$ & $\cdot$ & $\cdot$ & $\cdot$ & $\cdot$ & $\cdot$ & $\cdot$ & Y & $\cdot$ & $\cdot$ & $\cdot$ \\
\eventname{200106}{134123} & $\cdot$ & $\cdot$ & $\cdot$ & Y & $\cdot$ & $\cdot$ & $\cdot$ & $\cdot$ & $\cdot$ & $\cdot$ & $\cdot$ & $\cdot$ \\
\eventname{200109}{195634} & $\cdot$ & $\cdot$ & $\cdot$ & $\cdot$ & $\cdot$ & $\cdot$ & $\cdot$ & Y & $\cdot$ & $\cdot$ & $\cdot$ & $\cdot$ \\
\eventname{200129}{114245} & $\cdot$ & $\cdot$ & $\cdot$ & Y & $\cdot$ & $\cdot$ & $\cdot$ & $\cdot$ & $\cdot$ & $\cdot$ & $\cdot$ & $\cdot$ \\
\eventname{200208}{211609} & $\cdot$ & $\cdot$ & $\cdot$ & $\cdot$ & $\cdot$ & $\cdot$ & $\cdot$ & $\cdot$ & $\cdot$ & $\cdot$ & Y & $\cdot$ \\
\eventname{200210}{005122} & $\cdot$ & $\cdot$ & $\cdot$ & Y & $\cdot$ & $\cdot$ & $\cdot$ & $\cdot$ & $\cdot$ & $\cdot$ & $\cdot$ & $\cdot$ \\
\eventname{200210}{100022} & $\cdot$ & $\cdot$ & $\cdot$ & $\cdot$ & $\cdot$ & $\cdot$ & $\cdot$ & Y & Y & $\cdot$ & $\cdot$ & $\cdot$ \\
\eventname{200220}{124850} & $\cdot$ & $\cdot$ & $\cdot$ & $\cdot$ & $\cdot$ & $\cdot$ & $\cdot$ & $\cdot$ & $\cdot$ & Y & $\cdot$ & $\cdot$ \\
\eventname{200214}{223307} & $\cdot$ & $\cdot$ & $\cdot$ & $\cdot$ & $\cdot$ & $\cdot$ & $\cdot$ & $\cdot$ & $\cdot$ & $\cdot$ & $\cdot$ & $\cdot$ \\
\eventname{200225}{075134} & $\cdot$ & $\cdot$ & $\cdot$ & $\cdot$ & $\cdot$ & $\cdot$ & $\cdot$ & Y & $\cdot$ & $\cdot$ & $\cdot$ & $\cdot$ \\
\eventname{200301}{211019} & $\cdot$ & $\cdot$ & $\cdot$ & $\cdot$ & $\cdot$ & $\cdot$ & $\cdot$ & $\cdot$ & Y & Y & $\cdot$ & $\cdot$ \\
\eventname{200304}{172806} & $\cdot$ & $\cdot$ & $\cdot$ & $\cdot$ & $\cdot$ & $\cdot$ & $\cdot$ & $\cdot$ & Y & $\cdot$ & $\cdot$ & $\cdot$ \\
\eventname{200305}{084739} & $\cdot$ & $\cdot$ & $\cdot$ & Y & $\cdot$ & $\cdot$ & $\cdot$ & $\cdot$ & $\cdot$ & $\cdot$ & $\cdot$ & $\cdot$ \\
\eventname{200318}{191337} & $\cdot$ & $\cdot$ & $\cdot$ & Y & $\cdot$ & $\cdot$ & $\cdot$ & $\cdot$ & $\cdot$ & $\cdot$ & $\cdot$ &  Y  \\
    \\\bottomrule
    \end{tabular}
    \caption{Events for which parameter estimation (PE) is provided in this work, and the catalogues from which they originate.}
    \label{tab:event-list}
\end{table}

\section{Analysis Workflow}
\label{sec:method}
We modelled our analysis workflow closely on that presented in GWTC-3, but have made alterations where required in order to use only information which is available to the general public, and to reduce the computing resource requirements where necessary.
A full description of the differences between the workflow and data products used in this work compared to GWTC-3 is contained in section~\ref{sec:differences}.

\add{In performing our analysis we exclusively use open source software which is publicly available, in addition to publicly available data.
In addition we provide data files (in a format described in~\ref{sec:blueprints}) which allow the precise analysis workflow used to be replicated, including the acquisition of data and its analysis.
These blueprint files, alongside the Asimov software, and publicly available analysis codes, are sufficient for reconstructing the entire analysis pipeline with a small number of command-line actions on an appropriate computer system.
The resulting workflow will download the publicly available strain data, estimate the noise in the data segment, configure and run the parameter estimation process itself, and will then perform appropriate post-processing on results in order to create a dataset comparable to one from GWTC-3.
}
\delete{These analyses were made possible thanks to the Asimov software tool, which allowed the GWTC-3 workflow to be easily replicated with the changes required for this work.}
This abstracts the process of determining the running-order of each step in the workflow, and interacting with a high-throughput computing cluster running HTCondor.
This workflow management substantially reduces the amount of human intervention required in producing results from this multi-stage analyses such as the one we employed.

We used event times presented in each catalogue, or where relevant, its data release, to determine the event time at which to perform the analysis.
Additionally, we used the estimate of either component masses or chirp mass for each signal to determine both a reasonable prior range for the chirp mass parameter in the analysis, and the segment length which should be analysed, based on the (2,2)-mode of the IMRPhenomXPHM waveform.
\add{The use of the (2,2)-mode was made as this mode contributes the vast majority of power to the signal, and ensuring that the segment length is sufficient to contain the entire waveform in this mode will ensure that power is not lost from the detector band.
Higher modes can also contribute to the signal power, however they are only likely to be observed in a small number of binaries~\cite{mills_measuring_2021} while potentially adding considerably to the required segment length, and consequently the time required to complete the analysis.
Consequently, and given the low SNR of the signals being considered we elected to only guarantee the full resolution of the (2,2)-mode.}
Choosing the shortest segment length which will still allow full analysis of the signal will optimise analysis time, while choosing a suitable prior over mass ratio will prevent the requirement for excessively long data segment lengths to match the length of waveforms for very low-mass systems.

These quantities were used to prepare \emph{blueprint} files which describe each event for the Asimov workflow system.
The same workflow was then applied to each event:
\begin{enumerate}
    \item Gravitational wave strain data were then downloaded, by Asimov, from the Gravitational-wave Open Science Centre (GWOSC) in the form of frame files.
    \item Bayeswave~\cite{code-bayeswave} was run on the data surrounding the event time, and containing the signal, from the frame files in order to produce on-source PSDs for each detector.
    \add{On-source PSDs provide an estimate of the noise content of data by first removing non-Gaussian features in the data before computing the PSD.
    In contrast, off-source PSDs use data collected before and after the signal to estimate the noise.}
    \item Bilby~\cite{code-bilby} (via Bilby\_pipe~\cite{code-bilby-pipe}) was used to perform model selection with the Dynesty~\cite{code-dynesty} nested sampler and the IMRPhenomXPHM~\cite{waveform-imrphenomxphm} waveform.
    \item PESummary~\cite{code-pesummary} was used to perform post-processing on the posterior samples produced by the Bilby analysis, and produce samples for derived quantities.
\end{enumerate}
We then reviewed the results for each event using plots produced by PESummary in the final step of the workflow in order to scrutinise the convergence of the samples, and identify problems with the analysis, such as situations where the posteriors railed against the priors, or where prior dominance was observed in the posterior.
In cases where problems were identified a new blueprint was prepared to produce a new Bilby analysis, and then this was run using Asimov.

In order to produce the final data release, and to simplify the process of comparing our results to those of the other catalogues, we used PESummary's ability to compare sample sets, and produce ``metafiles'' containing multiple sets of samples.
In keeping with the GWTC-3 data release we have labelled samples from our analysis as \texttt{C01:IMRPhenomXPHM}, but we have labelled the samples from the other catalogues with their respective catalogue name as used in table~\ref{tab:event-list}.

\subsection{Differences compared to GWTC-3}
\label{sec:differences}

While we have endeavoured to make our analyses as similar as possible to those performed for GWTC-3 there are a number of \add{minor} differences\delete{, however these are mostly minor}.

In this work we have used frame files which were published via GWOSC to perform \add{the }analysis.
The GWTC-3 analysis was run on proprietary detector data prior to public release, using data retrieval systems which are not generally accessible.
The GWOSC data products were produced from the same data source, but accessing the data in this way necessitated a slight change to the workflow in order to first retrieve the data frames containing the released strain data.
The channel names in the GWOSC frame files are different to those used internally by the LVK, so this may be noted as a difference between configuration files provided in our data release compared to those in the GWTC-3 data release.

We have used a newer version of the Bayeswave package than was used in GWTC-3 which provides a substantial improvement in performance, while still using the same underlying approach to producing on-source PSD estimates.
We have also used a more recent version of the Bilby library (v2.1.1~\cite{code-bilby-2d1d1}) which \delete{includes a number of performance improvements}\add{provides better computational  performance} compared to the one used for GWTC-3.

The GWTC-3 analyses marginalised over the uncertainty in detector calibration.
The data to allow this marginalisation to be performed is not made publicly available by the \delete{Collaborations}\add{LVK}, and so we elected not to attempt to include this uncertainty.
LIGO and Virgo data in observing runs 1, 2, and 3 were re-calibrated prior to release, and we did not expect the calibration uncertainty to present a large additional source of uncertainty in our analyses, especially given the low SNR of the majority of events in this work.
The 4-OGC catalogue does attempt to include this uncertainty, but uses the uncertainty envelopes which were published to coincide with specific GWTC-3 events; 4-OGC events are then analysed using the calibration uncertainty for the closest published event in time, with the reasoning that the uncertainty is expected to evolve slowly over time.

For the majority of events we performed sampling directly over a cosmologically-motivated distance prior (that is, a prior on the \delete{luminosity distance}\add{redshift, $z$, which is uniform in comoving volume, $V_{\rm c}$ and source-frame time with the form
\[
p(z) \propto \frac{1}{1 + z}\frac{dV_{c}}{dz}
\]} in comparison \delete{to} GWTC-3 \delete{which} reweighted its samples to this prior, having sampled over a power-law prior \add{ of the form 
\[
p(x) = \frac{3 x^2}{x_{\rm max}^3 - x_{\rm min}^3}
\]
for $x_{\rm max}, x_{\rm min}$ respectively the maximum and minimum bound of the prior over the distance, $x$}.
By performing the sampling directly from the cosmological prior we were able to achieve higher sampling efficiency and reduced the complication of the post-processing required.
As a result, however, we do not provide a set of samples from a power-law prior in our data release, and our samples should be treated as comparable to the \add{samples labelled as }``cosmo'' \delete{samples} in the \add{LVK's} GWTC-3 release.
A small number of events were sampled with the power-law prior, but these have been reweighted to make them comparable to the other results in this work~\add{\footnote{\add{An earlier iteration of this work focused only on these events, which were events newly reported in 4-OGC. When expanding the scope of the work to include a much larger number of events with a greater distribution of SNRs we chose to perform subsequent analysis with the cosmologically-motivated prior and then reweighted the previous analyses to the new prior.}}}.

In order to determine the optimal prior ranges for the events in this work we often required multiple analyses, which would be reviewed, and lead to subsequent analyses which were configured to address any observed problems.
\delete{The iterative process was partly avoided during the production of GWTC-3, which relied on analysing earlier, rougher, analyses performed by a dedicated team of researchers, which were in turn analysed to determine suitable prior boundaries.}
\add{GWTC-3 avoided this iterative process in part as a result of the existence of \emph{preliminary} analyses which had been produced shortly after signals were detected.
These analyses were intended to provide rough estimates of the signals' parameters in order to inform subsequent follow-up.
Scrutinising the posterior distributions from these preliminary analyses allowed suitable prior ranges to be estimated.}
The software product for performing this analysis was not publicly available at the time that we started our analysis, and we did not have access to preliminary results to analyse with it.
The tool used to perform this analysis is now publicly available as PE-Configurator~\cite{code-peconfigurator}.

The largest single difference between our analysis and that of GWTC-3 is the choice of waveform models; in our work we elected to only perform analysis using the IMRPhenomXPHM model; GWTC-3 contains samples from both IMRPhenomXPHM and SEOBNRv4PHM~\cite{waveform-seobnrv4}.
This choice was born out of practicality; GWTC-3 used the RIFT~\cite{code-rift-0, code-rift-1, code-rift-2, code-rift-3, code-rift-asimov} pipeline to produce samples for SEOBNRv4PHM, which generally requires much more time to evaluate a single waveform than IMRPhenomXPHM.
This pipeline made substantial use of both CPU and GPU acceleration resources, which were not available to us when preparing this work using a small computing resource at the Institute for Gravitational Research at the University of Glasgow.
\add{These two waveforms make use of different techniques to model signals; IMRPhenomXPHM makes use of phenomenological fits to waveforms generated by numerical relativity simulation, whereas SEOBNRv4PHM uses an effective one-body approximation which is calibrated against simulations}.
As a result we are also unable to provide ``mixed'' samples for these events in a similar format to those in the \add{LVK's}  GWTC-3 data release.
\add{(Mixed samples are produced by combining a random selection of points from the posterior distributions from multiple analyses which have compatible prior distributions.
In the GWTC-3 data release these are constructed such that half of the points in the mixed sample set were produced using the IMRPhenomXPHM waveform, and half from the SEOBNRv4PHM model.)}

We used the Planck15 cosmology from astropy~\cite{software-astropy-3,cosmology-planck15} when determining source-frame quantities, where GWTC-3 use a slightly different set of values to form their cosmology based on values in LALInference~\cite{code-lalinference} \add{(we use $H_0=67.74 {\rm km/Mpc/s},\ \Omega_{\rm m}=0.3075$, whereas GWTC-3 use $H_0=67.90\,{\rm km/Mpc/s},\ \Omega_{\rm m}=0.3065$)}.

\section{Source properties}
\label{sec:pe}

Here we present the properties of the various events reported in the community catalogues, and compare them to the results which were presented in those catalogues.
Due to the large number of events, we have divided these between each of the three observing runs.

\begin{figure}
    \centering
    \includegraphics[width=1\textwidth]{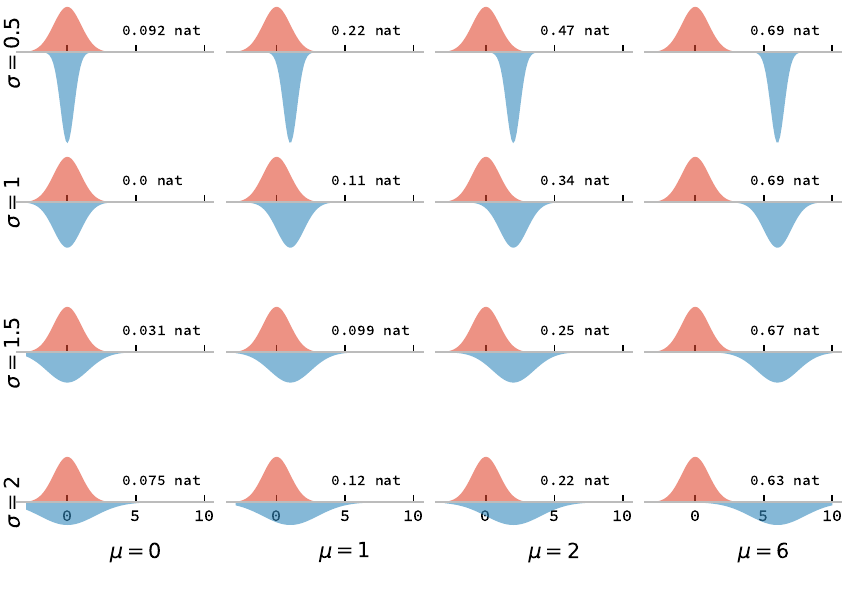}
    \caption{A selection of normal distributions with differing means ($\mu$) and standard deviations ($\sigma$). 
    In each case the red distribution above the $x$-axis has $\mu=0$ and $\sigma=1$.
    The blue distribution below the axis have $\mu = (0, 1, 2, 6)$ in each plot in the first, second, third, and fourth column respectively, counting from the left; and $\sigma = 0.5, 1, 1.5,$ and $2$ for the first, second, third, and fourth row, counting from the bottom.
    To the right of each pair of distributions the Jensen-Shannon Divergence between the distribution is shown in units of nat.
    }
    \label{fig:jsdivergence}
\end{figure}

\add{In order to quantitatively compare the posterior probability distributions between our analyses and those from the other catalogues we calculate the Jensen-Shannon (JS) divergence between the marginal distributions for a selection of source properties.
The JS divergence is derived from the Kullback-Leibler (KL) divergence, but is symmetrised and hence bounded.
It is defined
\begin{equation}
    \label{eq:jsd}
    {\rm JSD}(A || B) = \frac{{\rm KL}(A || M) + {\rm KL}(B || M)}{2}
\end{equation}
where $A$ and $B$ are probability distributions, 
$ M = (A + B)/2 $, and ${\rm KL}$ is the KL divergence~\cite{kullback_information_1951} between $A$ and $B$:
\begin{equation}
    \label{eq:kld}
    {\rm KL}(A||B) = \int_{-\infty}^{\infty} A(x) \log \frac{A(x)}{B(x)} {\rm d}x
\end{equation}
When the logarithm in equation \ref{eq:kld} is the natural logarithm the JSD will have a range of $[0\,{\rm nat}, \ln2\approx0.69\,{\rm nat}]$.
A value of $0$ implies the distributions are identical, while distributions with a JD divergence of $\ln 2\,{\rm nat}$ have no similarity. 
Examples of different distributions and their JS divergences are shown in figure~\ref{fig:jsdivergence}, including instances where the divergence is minimal and maximal.
Analyses do not directly produce posterior probability distributions, but rather produce samples drawn from the distribution.
In order to calculate the JS divergence we first calculate a Gaussian kernel density estimate from the samples which is used to evaluate equation~\ref{eq:jsd}.
}

\add{While the data release which accompanies this publication provides posterior probability distributions on all of the parameters included in GWTC-3 and its data release, we have chosen to show an illustrative subset of the source parameters in this work.}

\add{These are:
\begin{itemize}
    \item chirp mass, a reduced mass of the binary system, defined as 
\begin{equation*}
    \label{eq:chirp-mass}
    \mathcal{M}=\frac{(m_1 m_2)^{3/5}}{(m_1+m_2)^{1/5}}
\end{equation*}
for $m_1$ and $m_2$ the mass of the primary and secondary component of the binary respectively.
    \item mass ratio,
    $ q = m_2/m_1 $
    \item the effective spin parameter,
    \[
    \chi_{\rm eff} = \left[ \frac{m_1 \chi_{1,\perp} + m_2 \chi_{2,\perp}}{m_1 + m_2}    \right]
    \]
    where $\chi_i = cS_i/(Gm_i^2)$ is the dimensionless component spin for the $i$-th component, with $S_i$, the magnitude of the total spin vector of the $i$-th component, and $\chi_{i, \perp}$ is the component of this dimensionless component spin which is perpendicular to the direction of the Newtonian orbital angular momentum vector.
    A non-zero value for this parameter indicates support for spins being present within the system; if positive this indicates the components' spins are aligned, and if negative that they are anti-aligned.
    \item the precessing spin parameter, 
    \[ 
    \chi_{\rm p} = {\rm max} \left[\chi_{1,\perp}, \frac{q(4q+3)}{4+3q} \chi_{2, \perp} \right]
    \]
    This component provides a measure of the spin precession present in the binary.
\end{itemize}
In each case we quote values and present probability distributions for frame-dependent parameters in the source reference frame, indicated by a subscript ``src'' in the parameter symbol.
}

\subsection{Observing Run 1}

\begin{figure}
    \includegraphics[width=1\textwidth]{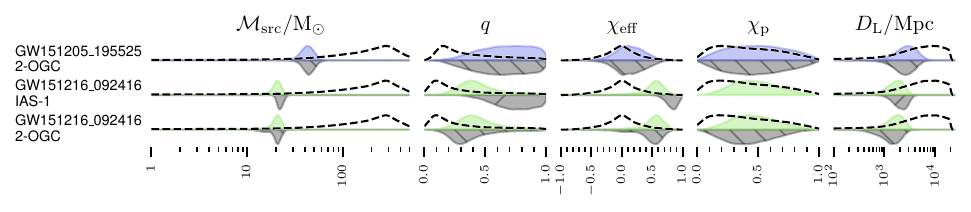}
    \caption{Violin plots showing the marginal probability distribution of a selection of inferred parameters for each signal from O1. From left to right these are the chirp mass, $\mathcal{M}_{\rm src}$ (in the source frame); the mass ratio; the effective dimensionless spin parameter, $\chi_{\rm eff}$; the precessing dimensionless spin parameter, $\chi_{\rm p}$; and the luminosity distance, $D_{\rm L}$.
    The coloured distributions above the $x$-axis represent the posterior probability distributions from our analysis, while those in grey below the $x$-axis are the posterior probability distributions reported in third-party catalogues. 
    \add{The distribution outlined by a dashed black line represents the prior probability distribution used in the analyses performed for this work.}
    Where PE samples are presented in multiple catalogues we present the comparison between our results and each catalogue's samples, with the catalogue noted beneath the name of the event, with our results repeated in each plot.
    \label{fig:violins:o1}}
\end{figure}

\begin{table}[]
    \footnotesize
    \begin{tabular}{c|ccccc|ccccc}\toprule
     & \multicolumn{5}{c|}{2-OGC} & \multicolumn{5}{c}{IAS-1} \\
Event & $\mathcal{M}_{\rm src}$ & $q$ & $\chi_{\rm eff}$ & $\chi_{\rm p}$ & $D_{\rm L}$ &$\mathcal{M}_{\rm src}$ & $q$ & $\chi_{\rm eff}$ & $\chi_{\rm p}$ & $D_{\rm L}$ \\
    \midrule
    \eventname{151205}{195525} & $\chirpmasssourcetwoOGCdivergence{GW151205_195525}$ & $\massratiotwoOGCdivergence{GW151205_195525}$ & $\chiefftwoOGCdivergence{GW151205_195525}$ & $\chiptwoOGCdivergence{GW151205_195525}$ & $\luminositydistancetwoOGCdivergence{GW151205_195525}$ & $\chirpmasssourceIAStwodivergence{GW151205_195525}$ & $\massratioIAStwodivergence{GW151205_195525}$ & $\chieffIAStwodivergence{GW151205_195525}$ & $\chipIAStwodivergence{GW151205_195525}$ & $\luminositydistanceIAStwodivergence{GW151205_195525}$ \\ 
\eventname{151216}{092416} & $\chirpmasssourcetwoOGCdivergence{GW151216_092416}$ & $\massratiotwoOGCdivergence{GW151216_092416}$ & $\chiefftwoOGCdivergence{GW151216_092416}$ & $\chiptwoOGCdivergence{GW151216_092416}$ & $\luminositydistancetwoOGCdivergence{GW151216_092416}$ & $\chirpmasssourceIAStwodivergence{GW151216_092416}$ & $\massratioIAStwodivergence{GW151216_092416}$ & $\chieffIAStwodivergence{GW151216_092416}$ & $\chipIAStwodivergence{GW151216_092416}$ & $\luminositydistanceIAStwodivergence{GW151216_092416}$
    \\\bottomrule
    \end{tabular}
     \caption{The Jensen-Shannon divergences, in nats, between the posteriors presented in this work and those from the 2-OGC and IAS-2 data releases for the marginal posterior distributions shown in figure \ref{fig:violins:o1}.}
    \label{tab:divergences:o1}
\end{table}

GWTC-3 contains three events from the first observing run, all of which are binary black hole (BBH) coalescences, and include the first observed gravitaional wave event, \eventname{150914}{095045}.
Two additional events have been claimed in the literature, \eventname{151205}{195525} was reported in 2-OGC, however it was later excluded from 3-OGC, and does not appear in subsequent OGC catalogues\add{, as changes to the search algorithm result in the signal no longer exceeding this catalogue's significance threshold}.
It does not appear in any of the other catalogues we considered in this work.

\eventname{151216}{092416} was reported in IAS-1, but has not been identified as a significant trigger by any of the other catalogues we considered in this work.
We note that the analyses presented in \cite{151216-paper} used an aligned-spin waveform model, so we are unable to compare posterior distributions on the precessing spin parameter.
This event has been the result of some discussion in the literature, with some authors noting \cite{lowest-snr-properties} that analysis of this event is strongly affected by the choice of prior in the analysis. 
The authors of \cite{astrophysical-gw151216} question the astrophysical nature of the signal, though \cite{improving-151216} support it.
This prior-dependence appears to be reflected in our analysis: our posterior distributions diverge substantially from those presented in \cite{catalog-ias-1}, especially in the effective spin.

We present the results of our analysis of these events for a subset of parameters in table~\ref{tab:source-properties}, and a summary plot of the marginal posteriors for a selection of source parameters of each of these O1 events in figure \ref{fig:violins:o1}; results from this work are shown as the distribution above the $x$-axis, and those from the IAS-2 and 2-OGC data releases below.
Additionally, table~\ref{tab:divergences:o1} shows the Jensen-Shannon divergence for each parameter depicted in this figure.

Results from both of these publications use a different waveform family to produce results from those used in this work. 
The results from 2-OGC were created with the IMRPhenomPv2~\cite{waveform-imrphenomp}, in contrast to those from both this work and 4-OGC which used IMRPhenomXPHM.
Further, the results from IAS-2 use the IMRPhenomD~\cite{waveform-imrphenomd} waveform, which is non-precessing. 
As a result we would anticipate some amount of divergence between these results and ours given the reduction in physics modelled in each of these waveforms.
Further, as inference performed with IMRPhenomD necessarily cannot measure precessing spin, posteriors for $\chi_{\rm p}$, the precessing dimensionless spin parameter, are not present for results derived from IMRPhenomD.
Results from IAS-2 were not sampled using a prior which was uniform in comoving volume as those from this work were, so we have reweighted these posteriors to represent draws with the same prior, and rejection-sampled the resulting weighted samples. 
This was performed using the PESummary library.

\subsection{Observing Run 2}

In addition to the eight gravitational-wave event candidates reported in GWTC-3, both IAS-2 and 2-OGC claim novel events, with additional events being identified in 3-OGC and 4-OGC.
We compare the results of our analyses to PE results available from each group for O2 events, 2-OGC, 4-OGC, and IAS-2; as with the events in O1, there is a difference in the waveforms used for sampling between each of these catalogues, and the IAS-2 results have been reweighted to account for the differing luminosity distance priors.
We present the median estimates of a subset of the source parameters in table~\ref{tab:source-properties}.

\begin{figure}
    \includegraphics[width=1\textwidth]{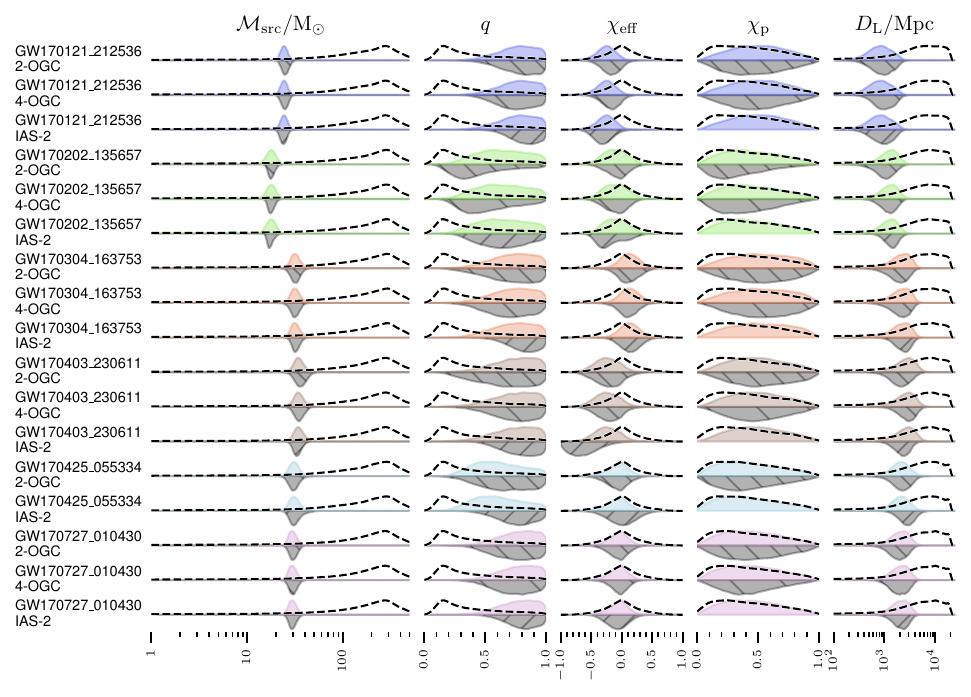}
    \caption{Violin plots showing the marginal probability distribution of a selection of inferred parameters for each signal from O2.
    The parameters displayed, and the prior and posterior quantities depicted in various parts of each plot are the same as in figure~\ref{fig:violins:o1}.
    Where PE samples are presented in multiple catalogues we present the comparison between our results and each catalogue's samples, with the catalogue noted beneath the name of the event.
    \label{fig:violins:o2}}
\end{figure}

In figure \ref{fig:violins:o2} we present marginal posteriors for the six O2 events from 4-OGC, and 11 from IAS-2.
The JS divergence between our posteriors and those in each data release are contained in table~\ref{tab:divergences:o2}.

The largest divergence appears in \eventname{170202}{135657} where the inferred source \add{chirp} mass\delete{es} \delete{have}\add{has} a divergence of \chirpmasssourceIAStwodivergence{GW170202_135657}\add{$\,{\rm nat}$}  between our results and the IAS-2 results.
The IAS-2 results show a degree of bimodality which is not seen in the 2-OGC, 4-OGC, or our own results.
Similarly, the IAS-2 posterior for \eventname{170403}{230611} shows support for extremal, negative $\chi_{\rm eff}$. 
Our results show some support for negative $\chi_{\rm eff}$ compared to our prior, this is much less pronounced, and is comparable to the results from the two OGC publications.
\add{It is unclear where the source of this difference arises. 
The primary difference between results produced here and in the OGC catalogues, and those produced in the IAS catalogues are the parameter estimation codes themselves.}

\begin{table}[]
    \footnotesize
    \begin{tabular}{c|ccccc|ccccc}\toprule
     & \multicolumn{5}{c|}{2-OGC} & \multicolumn{5}{c}{4-OGC} \\
Event & $\mathcal{M}_{\rm src}$ & $q$ & $\chi_{\rm eff}$ & $\chi_{\rm p}$ & $D_{\rm L}$ &$\mathcal{M}_{\rm src}$ & $q$ & $\chi_{\rm eff}$ & $\chi_{\rm p}$ & $D_{\rm L}$ \\
    \midrule
    \eventname{170121}{212536} & $\chirpmasssourcetwoOGCdivergence{GW170121_212536}$ & $\massratiotwoOGCdivergence{GW170121_212536}$ & $\chiefftwoOGCdivergence{GW170121_212536}$ & $\chiptwoOGCdivergence{GW170121_212536}$ & $\luminositydistancetwoOGCdivergence{GW170121_212536}$ & $\chirpmasssourcefourOGCdivergence{GW170121_212536}$ & $\massratiofourOGCdivergence{GW170121_212536}$ & $\chiefffourOGCdivergence{GW170121_212536}$ & $\chipfourOGCdivergence{GW170121_212536}$ & $\luminositydistancefourOGCdivergence{GW170121_212536}$ \\ 
\eventname{170202}{135657} & $\chirpmasssourcetwoOGCdivergence{GW170202_135657}$ & $\massratiotwoOGCdivergence{GW170202_135657}$ & $\chiefftwoOGCdivergence{GW170202_135657}$ & $\chiptwoOGCdivergence{GW170202_135657}$ & $\luminositydistancetwoOGCdivergence{GW170202_135657}$ & $\chirpmasssourcefourOGCdivergence{GW170202_135657}$ & $\massratiofourOGCdivergence{GW170202_135657}$ & $\chiefffourOGCdivergence{GW170202_135657}$ & $\chipfourOGCdivergence{GW170202_135657}$ & $\luminositydistancefourOGCdivergence{GW170202_135657}$ \\ 
\eventname{170304}{163753} & $\chirpmasssourcetwoOGCdivergence{GW170304_163753}$ & $\massratiotwoOGCdivergence{GW170304_163753}$ & $\chiefftwoOGCdivergence{GW170304_163753}$ & $\chiptwoOGCdivergence{GW170304_163753}$ & $\luminositydistancetwoOGCdivergence{GW170304_163753}$ & $\chirpmasssourcefourOGCdivergence{GW170304_163753}$ & $\massratiofourOGCdivergence{GW170304_163753}$ & $\chiefffourOGCdivergence{GW170304_163753}$ & $\chipfourOGCdivergence{GW170304_163753}$ & $\luminositydistancefourOGCdivergence{GW170304_163753}$ \\ 
\eventname{170403}{230611} & $\chirpmasssourcetwoOGCdivergence{GW170403_230611}$ & $\massratiotwoOGCdivergence{GW170403_230611}$ & $\chiefftwoOGCdivergence{GW170403_230611}$ & $\chiptwoOGCdivergence{GW170403_230611}$ & $\luminositydistancetwoOGCdivergence{GW170403_230611}$ & $\chirpmasssourcefourOGCdivergence{GW170403_230611}$ & $\massratiofourOGCdivergence{GW170403_230611}$ & $\chiefffourOGCdivergence{GW170403_230611}$ & $\chipfourOGCdivergence{GW170403_230611}$ & $\luminositydistancefourOGCdivergence{GW170403_230611}$ \\ 
\eventname{170425}{055334} & $\chirpmasssourcetwoOGCdivergence{GW170425_055334}$ & $\massratiotwoOGCdivergence{GW170425_055334}$ & $\chiefftwoOGCdivergence{GW170425_055334}$ & $\chiptwoOGCdivergence{GW170425_055334}$ & $\luminositydistancetwoOGCdivergence{GW170425_055334}$ & $\chirpmasssourcefourOGCdivergence{GW170425_055334}$ & $\massratiofourOGCdivergence{GW170425_055334}$ & $\chiefffourOGCdivergence{GW170425_055334}$ & $\chipfourOGCdivergence{GW170425_055334}$ & $\luminositydistancefourOGCdivergence{GW170425_055334}$ \\ 
\eventname{170727}{010430} & $\chirpmasssourcetwoOGCdivergence{GW170727_010430}$ & $\massratiotwoOGCdivergence{GW170727_010430}$ & $\chiefftwoOGCdivergence{GW170727_010430}$ & $\chiptwoOGCdivergence{GW170727_010430}$ & $\luminositydistancetwoOGCdivergence{GW170727_010430}$ & $\chirpmasssourcefourOGCdivergence{GW170727_010430}$ & $\massratiofourOGCdivergence{GW170727_010430}$ & $\chiefffourOGCdivergence{GW170727_010430}$ & $\chipfourOGCdivergence{GW170727_010430}$ & $\luminositydistancefourOGCdivergence{GW170727_010430}$
    \\\bottomrule
    \end{tabular}
    \begin{tabular}{c|ccccc}\toprule
     & \multicolumn{5}{c}{IAS-2} \\
Event & $\mathcal{M}_{\rm src}$ & $q$ & $\chi_{\rm eff}$ & $\chi_{\rm p}$ & $D_{\rm L}$ \\
    \midrule
    \eventname{170121}{212536} & $\chirpmasssourceIAStwodivergence{GW170121_212536}$ & $\massratioIAStwodivergence{GW170121_212536}$ & $\chieffIAStwodivergence{GW170121_212536}$ & $\chipIAStwodivergence{GW170121_212536}$ & $\luminositydistanceIAStwodivergence{GW170121_212536}$ \\ 
\eventname{170202}{135657} & $\chirpmasssourceIAStwodivergence{GW170202_135657}$ & $\massratioIAStwodivergence{GW170202_135657}$ & $\chieffIAStwodivergence{GW170202_135657}$ & $\chipIAStwodivergence{GW170202_135657}$ & $\luminositydistanceIAStwodivergence{GW170202_135657}$ \\ 
\eventname{170304}{163753} & $\chirpmasssourceIAStwodivergence{GW170304_163753}$ & $\massratioIAStwodivergence{GW170304_163753}$ & $\chieffIAStwodivergence{GW170304_163753}$ & $\chipIAStwodivergence{GW170304_163753}$ & $\luminositydistanceIAStwodivergence{GW170304_163753}$ \\ 
\eventname{170403}{230611} & $\chirpmasssourceIAStwodivergence{GW170403_230611}$ & $\massratioIAStwodivergence{GW170403_230611}$ & $\chieffIAStwodivergence{GW170403_230611}$ & $\chipIAStwodivergence{GW170403_230611}$ & $\luminositydistanceIAStwodivergence{GW170403_230611}$ \\ 
\eventname{170425}{055334} & $\chirpmasssourceIAStwodivergence{GW170425_055334}$ & $\massratioIAStwodivergence{GW170425_055334}$ & $\chieffIAStwodivergence{GW170425_055334}$ & $\chipIAStwodivergence{GW170425_055334}$ & $\luminositydistanceIAStwodivergence{GW170425_055334}$ \\ 
\eventname{170727}{010430} & $\chirpmasssourceIAStwodivergence{GW170727_010430}$ & $\massratioIAStwodivergence{GW170727_010430}$ & $\chieffIAStwodivergence{GW170727_010430}$ & $\chipIAStwodivergence{GW170727_010430}$ & $\luminositydistanceIAStwodivergence{GW170727_010430}$
    \\\bottomrule
    \end{tabular}
    \medskip
     \caption{The Jensen-Shannon divergences, in nats, between the posteriors presented in this work and those from the 2-OGC, 4-OGC, and IAS-2 data releases for the marginal posterior distributions shown in figure \ref{fig:violins:o2}.}
    \label{tab:divergences:o2}
\end{table}

\subsection{Observing Run 3}

The third observing run saw substantially improved sensitivity from all three gravitational detectors, and a much greater \add{network} duty cycle, \add{with one detector observing around 96\% of the time in O3~\cite{catalog-gwtc-3} compared to 42.8\% in O1 and 46\% in O2~\cite{collaboration_open_2021}}.
As a result an order of magnitude more events are reported in GWTC-3 from this period.
Similarly other catalogues which search and analyse this data find far more candidate events than in O1 and O2 combined.

As a result in the plots and tables which follow we divide results between the two parts of the observing run as defined by the LVK's GWTC-2.1 and GWTC-3 catalogues, into O3a and O3b, however we present the median inferred source parameters for the same subset of parameters as in table~\ref{tab:source-properties} for all O3 events in table~\ref{tab:source-properties-o3}.
A similar divide is present with some of the other catalogues; \add{events from} O3a \delete{is}\add{are described} by IAS-3, and O3b by IAS-4.
Both parts are considered in 4-OGC (which updates 3-OGC, which \delete{covered}\add{describes events from} only O3a), IAS-HM, pycbc-kde, AresGW, and cWB.
A substantially greater variety of catalogues and techniques cover this period, though only 3-OGC, 4-OGC, IAS-3, IAS-4, and IAS-HM provide publically available data releases of their PE results.
Given the similarity between the setup used in both 3-OGC and 4-OGC we have elected to consider only 4-OGC results, as these include all of the events contained within 3-OGC.

\begin{figure}
    \includegraphics[width=1\textwidth]{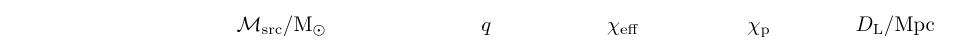}
    \includegraphics[width=1\textwidth]{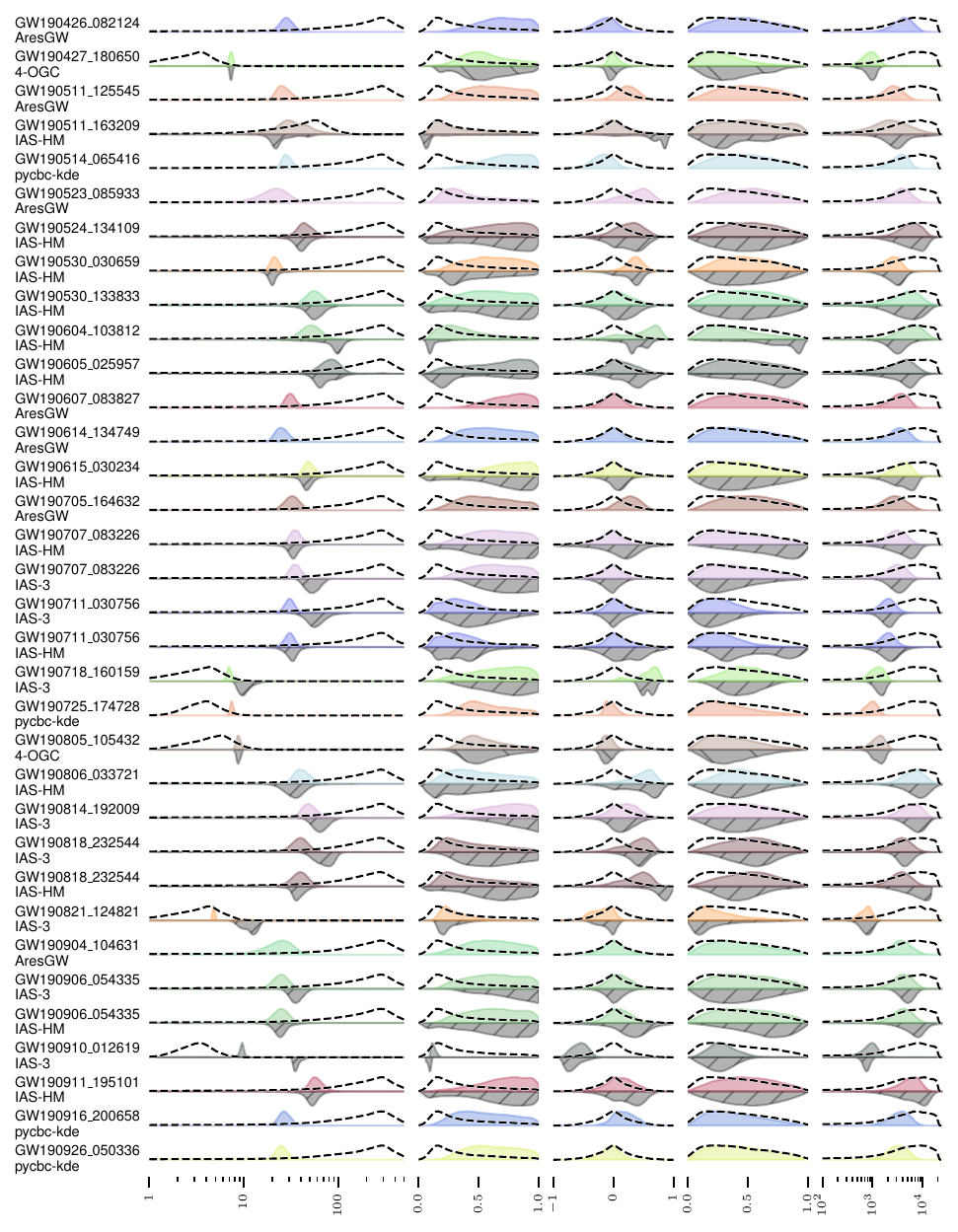}
    \caption{Violin plots showing the marginal probability distribution of a selection of inferred parameters for each signal from O3a. 
    The parameters displayed, and the prior and posterior quantities depicted in various parts of each plot are the same as in figure~\ref{fig:violins:o1}.
    Where PE samples are presented in multiple catalogues we present the comparison between our results and each catalogue's samples, with the catalogue noted beneath the name of the event.
    \label{fig:violins:o3a}}
\end{figure}

\begin{table}[]
    \footnotesize
    \begin{tabular}{c|ccccc|ccccc}\toprule
     & \multicolumn{5}{c|}{4-OGC} & \multicolumn{5}{c}{IAS-3} \\
Event & $\mathcal{M}_{\rm src}$ & $q$ & $\chi_{\rm eff}$ & $\chi_{\rm p}$ & $D_{\rm L}$ &$\mathcal{M}_{\rm src}$ & $q$ & $\chi_{\rm eff}$ & $\chi_{\rm p}$ & $D_{\rm L}$ \\
    \midrule
    \eventname{190427}{180650} & $\chirpmasssourcefourOGCdivergence{GW190427_180650}$ & $\massratiofourOGCdivergence{GW190427_180650}$ & $\chiefffourOGCdivergence{GW190427_180650}$ & $\chipfourOGCdivergence{GW190427_180650}$ & $\luminositydistancefourOGCdivergence{GW190427_180650}$ & $\chirpmasssourceIASthreedivergence{GW190427_180650}$ & $\massratioIASthreedivergence{GW190427_180650}$ & $\chieffIASthreedivergence{GW190427_180650}$ & $\chipIASthreedivergence{GW190427_180650}$ & $\luminositydistanceIASthreedivergence{GW190427_180650}$ \\ 
\eventname{190707}{083226} & $\chirpmasssourcefourOGCdivergence{GW190707_083226}$ & $\massratiofourOGCdivergence{GW190707_083226}$ & $\chiefffourOGCdivergence{GW190707_083226}$ & $\chipfourOGCdivergence{GW190707_083226}$ & $\luminositydistancefourOGCdivergence{GW190707_083226}$ & $\chirpmasssourceIASthreedivergence{GW190707_083226}$ & $\massratioIASthreedivergence{GW190707_083226}$ & $\chieffIASthreedivergence{GW190707_083226}$ & $\chipIASthreedivergence{GW190707_083226}$ & $\luminositydistanceIASthreedivergence{GW190707_083226}$ \\ 
\eventname{190711}{030756} & $\chirpmasssourcefourOGCdivergence{GW190711_030756}$ & $\massratiofourOGCdivergence{GW190711_030756}$ & $\chiefffourOGCdivergence{GW190711_030756}$ & $\chipfourOGCdivergence{GW190711_030756}$ & $\luminositydistancefourOGCdivergence{GW190711_030756}$ & $\chirpmasssourceIASthreedivergence{GW190711_030756}$ & $\massratioIASthreedivergence{GW190711_030756}$ & $\chieffIASthreedivergence{GW190711_030756}$ & $\chipIASthreedivergence{GW190711_030756}$ & $\luminositydistanceIASthreedivergence{GW190711_030756}$ \\ 
\eventname{190718}{160159} & $\chirpmasssourcefourOGCdivergence{GW190718_160159}$ & $\massratiofourOGCdivergence{GW190718_160159}$ & $\chiefffourOGCdivergence{GW190718_160159}$ & $\chipfourOGCdivergence{GW190718_160159}$ & $\luminositydistancefourOGCdivergence{GW190718_160159}$ & $\chirpmasssourceIASthreedivergence{GW190718_160159}$ & $\massratioIASthreedivergence{GW190718_160159}$ & $\chieffIASthreedivergence{GW190718_160159}$ & $\chipIASthreedivergence{GW190718_160159}$ & $\luminositydistanceIASthreedivergence{GW190718_160159}$ \\ 
\eventname{190805}{105432} & $\chirpmasssourcefourOGCdivergence{GW190805_105432}$ & $\massratiofourOGCdivergence{GW190805_105432}$ & $\chiefffourOGCdivergence{GW190805_105432}$ & $\chipfourOGCdivergence{GW190805_105432}$ & $\luminositydistancefourOGCdivergence{GW190805_105432}$ & $\chirpmasssourceIASthreedivergence{GW190805_105432}$ & $\massratioIASthreedivergence{GW190805_105432}$ & $\chieffIASthreedivergence{GW190805_105432}$ & $\chipIASthreedivergence{GW190805_105432}$ & $\luminositydistanceIASthreedivergence{GW190805_105432}$ \\ 
\eventname{190814}{192009} & $\chirpmasssourcefourOGCdivergence{GW190814_192009}$ & $\massratiofourOGCdivergence{GW190814_192009}$ & $\chiefffourOGCdivergence{GW190814_192009}$ & $\chipfourOGCdivergence{GW190814_192009}$ & $\luminositydistancefourOGCdivergence{GW190814_192009}$ & $\chirpmasssourceIASthreedivergence{GW190814_192009}$ & $\massratioIASthreedivergence{GW190814_192009}$ & $\chieffIASthreedivergence{GW190814_192009}$ & $\chipIASthreedivergence{GW190814_192009}$ & $\luminositydistanceIASthreedivergence{GW190814_192009}$ \\ 
\eventname{190818}{232544} & $\chirpmasssourcefourOGCdivergence{GW190818_232544}$ & $\massratiofourOGCdivergence{GW190818_232544}$ & $\chiefffourOGCdivergence{GW190818_232544}$ & $\chipfourOGCdivergence{GW190818_232544}$ & $\luminositydistancefourOGCdivergence{GW190818_232544}$ & $\chirpmasssourceIASthreedivergence{GW190818_232544}$ & $\massratioIASthreedivergence{GW190818_232544}$ & $\chieffIASthreedivergence{GW190818_232544}$ & $\chipIASthreedivergence{GW190818_232544}$ & $\luminositydistanceIASthreedivergence{GW190818_232544}$ \\ 
\eventname{190821}{124821} & $\chirpmasssourcefourOGCdivergence{GW190821_124821}$ & $\massratiofourOGCdivergence{GW190821_124821}$ & $\chiefffourOGCdivergence{GW190821_124821}$ & $\chipfourOGCdivergence{GW190821_124821}$ & $\luminositydistancefourOGCdivergence{GW190821_124821}$ & $\chirpmasssourceIASthreedivergence{GW190821_124821}$ & $\massratioIASthreedivergence{GW190821_124821}$ & $\chieffIASthreedivergence{GW190821_124821}$ & $\chipIASthreedivergence{GW190821_124821}$ & $\luminositydistanceIASthreedivergence{GW190821_124821}$ \\ 
\eventname{190906}{054335} & $\chirpmasssourcefourOGCdivergence{GW190906_054335}$ & $\massratiofourOGCdivergence{GW190906_054335}$ & $\chiefffourOGCdivergence{GW190906_054335}$ & $\chipfourOGCdivergence{GW190906_054335}$ & $\luminositydistancefourOGCdivergence{GW190906_054335}$ & $\chirpmasssourceIASthreedivergence{GW190906_054335}$ & $\massratioIASthreedivergence{GW190906_054335}$ & $\chieffIASthreedivergence{GW190906_054335}$ & $\chipIASthreedivergence{GW190906_054335}$ & $\luminositydistanceIASthreedivergence{GW190906_054335}$ \\ 
\eventname{190910}{012619} & $\chirpmasssourcefourOGCdivergence{GW190910_012619}$ & $\massratiofourOGCdivergence{GW190910_012619}$ & $\chiefffourOGCdivergence{GW190910_012619}$ & $\chipfourOGCdivergence{GW190910_012619}$ & $\luminositydistancefourOGCdivergence{GW190910_012619}$ & $\chirpmasssourceIASthreedivergence{GW190910_012619}$ & $\massratioIASthreedivergence{GW190910_012619}$ & $\chieffIASthreedivergence{GW190910_012619}$ & $\chipIASthreedivergence{GW190910_012619}$ & $\luminositydistanceIASthreedivergence{GW190910_012619}$ \\\bottomrule
    \end{tabular}
    \begin{tabular}{c|ccccc}
     & \multicolumn{5}{c}{IAS-HM} \\\toprule
Event & $\mathcal{M}_{\rm src}$ & $q$ & $\chi_{\rm eff}$ & $\chi_{\rm p}$ & $D_{\rm L}$ \\
    \midrule
\eventname{190427}{180650} & $\chirpmasssourceIASHMdivergence{GW190427_180650}$ & $\massratioIASHMdivergence{GW190427_180650}$ & $\chieffIASHMdivergence{GW190427_180650}$ & $\chipIASHMdivergence{GW190427_180650}$ & $\luminositydistanceIASHMdivergence{GW190427_180650}$ \\ 
\eventname{190511}{163209} & $\chirpmasssourceIASHMdivergence{GW190511_163209}$ & $\massratioIASHMdivergence{GW190511_163209}$ & $\chieffIASHMdivergence{GW190511_163209}$ & $\chipIASHMdivergence{GW190511_163209}$ & $\luminositydistanceIASHMdivergence{GW190511_163209}$ \\ 
\eventname{190524}{134109} & $\chirpmasssourceIASHMdivergence{GW190524_134109}$ & $\massratioIASHMdivergence{GW190524_134109}$ & $\chieffIASHMdivergence{GW190524_134109}$ & $\chipIASHMdivergence{GW190524_134109}$ & $\luminositydistanceIASHMdivergence{GW190524_134109}$ \\ 
\eventname{190530}{030659} & $\chirpmasssourceIASHMdivergence{GW190530_030659}$ & $\massratioIASHMdivergence{GW190530_030659}$ & $\chieffIASHMdivergence{GW190530_030659}$ & $\chipIASHMdivergence{GW190530_030659}$ & $\luminositydistanceIASHMdivergence{GW190530_030659}$ \\ 
\eventname{190530}{133833} & $\chirpmasssourceIASHMdivergence{GW190530_133833}$ & $\massratioIASHMdivergence{GW190530_133833}$ & $\chieffIASHMdivergence{GW190530_133833}$ & $\chipIASHMdivergence{GW190530_133833}$ & $\luminositydistanceIASHMdivergence{GW190530_133833}$ \\ 
\eventname{190604}{103812} & $\chirpmasssourceIASHMdivergence{GW190604_103812}$ & $\massratioIASHMdivergence{GW190604_103812}$ & $\chieffIASHMdivergence{GW190604_103812}$ & $\chipIASHMdivergence{GW190604_103812}$ & $\luminositydistanceIASHMdivergence{GW190604_103812}$ \\ 
\eventname{190605}{025957} & $\chirpmasssourceIASHMdivergence{GW190605_025957}$ & $\massratioIASHMdivergence{GW190605_025957}$ & $\chieffIASHMdivergence{GW190605_025957}$ & $\chipIASHMdivergence{GW190605_025957}$ & $\luminositydistanceIASHMdivergence{GW190605_025957}$ \\ 
\eventname{190615}{030234} & $\chirpmasssourceIASHMdivergence{GW190615_030234}$ & $\massratioIASHMdivergence{GW190615_030234}$ & $\chieffIASHMdivergence{GW190615_030234}$ & $\chipIASHMdivergence{GW190615_030234}$ & $\luminositydistanceIASHMdivergence{GW190615_030234}$ \\ 
\eventname{190707}{083226} & $\chirpmasssourceIASHMdivergence{GW190707_083226}$ & $\massratioIASHMdivergence{GW190707_083226}$ & $\chieffIASHMdivergence{GW190707_083226}$ & $\chipIASHMdivergence{GW190707_083226}$ & $\luminositydistanceIASHMdivergence{GW190707_083226}$ \\ 
\eventname{190711}{030756} & $\chirpmasssourceIASHMdivergence{GW190711_030756}$ & $\massratioIASHMdivergence{GW190711_030756}$ & $\chieffIASHMdivergence{GW190711_030756}$ & $\chipIASHMdivergence{GW190711_030756}$ & $\luminositydistanceIASHMdivergence{GW190711_030756}$ \\ 
\eventname{190806}{033721} & $\chirpmasssourceIASHMdivergence{GW190806_033721}$ & $\massratioIASHMdivergence{GW190806_033721}$ & $\chieffIASHMdivergence{GW190806_033721}$ & $\chipIASHMdivergence{GW190806_033721}$ & $\luminositydistanceIASHMdivergence{GW190806_033721}$ \\ 
\eventname{190818}{232544} & $\chirpmasssourceIASHMdivergence{GW190818_232544}$ & $\massratioIASHMdivergence{GW190818_232544}$ & $\chieffIASHMdivergence{GW190818_232544}$ & $\chipIASHMdivergence{GW190818_232544}$ & $\luminositydistanceIASHMdivergence{GW190818_232544}$ \\ 
\eventname{190906}{054335} & $\chirpmasssourceIASHMdivergence{GW190906_054335}$ & $\massratioIASHMdivergence{GW190906_054335}$ & $\chieffIASHMdivergence{GW190906_054335}$ & $\chipIASHMdivergence{GW190906_054335}$ & $\luminositydistanceIASHMdivergence{GW190906_054335}$ \\ 
\eventname{190911}{195101} & $\chirpmasssourceIASHMdivergence{GW190911_195101}$ & $\massratioIASHMdivergence{GW190911_195101}$ & $\chieffIASHMdivergence{GW190911_195101}$ & $\chipIASHMdivergence{GW190911_195101}$ & $\luminositydistanceIASHMdivergence{GW190911_195101}$ 
\\
    \bottomrule
    \end{tabular}
    \medskip
     \caption{The Jensen-Shannon divergences, in nats, between the posteriors presented in this work and those from the 4-OGC, IAS-3, and IAS-HM data releases for the marginal posterior distributions shown in figure \ref{fig:violins:o3a}.}
    \label{tab:divergences:o3a}
\end{table}

\begin{figure}
    \includegraphics[width=1\textwidth]{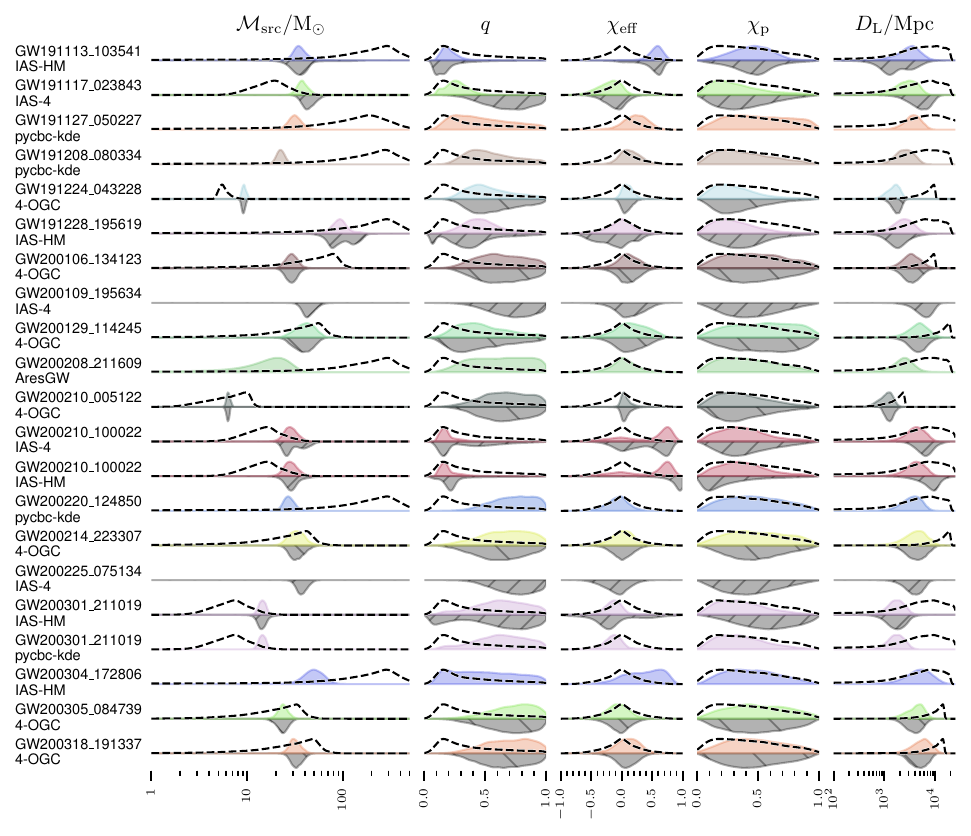}
    \caption{Violin plots showing the marginal probability distribution of a selection of inferred parameters for each signal from O3b. 
    The parameters displayed, and the prior and posterior quantities depicted in various parts of each plot are the same as in figure~\ref{fig:violins:o1}.
    Where PE samples are presented in multiple catalogues we present the comparison between our results and each catalogue's samples, with the catalogue noted beneath the name of the event.
    \label{fig:violins:o3b}}
\end{figure}

The results of the various O3a analyses are depicted in figure~\ref{fig:violins:o3a} \add{and those for O3b are shown in figure~\ref{fig:violins:o3b}}; events from pycbc-kde and AresGW show only our marginal posteriors\delete{, while those for O3b are shown in figure~\ref{fig:violins:o3b}}.

The cWB catalogue does not include any events which are not reported in another catalogue, so these are not depicted separately.
The JS divergences corresponding to the marginal posterior distributions shown in figure~\ref{fig:violins:o3a} are contained in table~\ref{tab:divergences:o3a}, while those for figure~\ref{fig:violins:o3b} are in table~\ref{tab:divergences:o3b}.

Similarly to results from O2, our results generally show less support for extremal binary configurations compared to results from the IAS-3 catalogue, but also show very large divergences between estimates of the chirp mass in the source frame.
There is considerable overlap between the event lists from GWTC-3 and 4-OGC in O3a, and as a result we compare results between only two events from this catalogue.
Neither show a large divergence between marginal posteriors.
There are a larger number of events not identified in GWTC-3 from O3b, and we find good agreement between our results and those in 4-OGC for all events \delete{except}\add{with} \eventname{200210}{005122} \add{having the largest disagreement}\delete{where there is considerable disagreement  between the inferred chirp masses} \add{as evidenced by the JS divergence of \chirpmasssourcefourOGCdivergence{GW200210_005122}~$\,{\rm nat}$ between the two source chirp mass posterior distributions, and \luminositydistancefourOGCdivergence{GW200210_005122}~$\,\rm{nat}$ between the luminosity distance posteriors indicating only a small difference between the two results}.

\delete{\eventname{191117}{023843} has the clearest divergence between our results and IAS-4, with the multimodality in the IAS-4 posterior not present in our results.
Other }\add{The} IAS-4 results are broadly comparable to ours, though \add{some} other differences \delete{also} stand out, such as the multimodality present in the IAS-4 estimate of \eventname{200210}{10022} which is not present in our \delete{samples}\add{posterior distributions}.

The IAS-HM catalogue present their PE results using the IMRPhenomXODE~\cite{waveform-imrphenomxode} waveform model, but also provide PE results using the IMRPhenomXPHM model.
We use these results from their data release produced using \delete{a similar} \add{the same} prior configuration to the one used here for the posteriors labelled as IAS-HM in figure~\ref{fig:violins:o3a} and table~\ref{tab:divergences:o3a}.
There is considerable variety in the divergences \delete{of}\add{between} our and their posteriors; \eventname{190511}{163209} shows perhaps the most dramatic difference, with the IAS-HM results favouring very high effective spin (though with a bimodal distribution), while our posteriors largely reproduce the prior distribution on this parameter.
\add{Again, we note that the primary difference between the means the two results were produced are the parameter estimation codes themselves.}

By construction, the events presented in this work are generally of low signficance, having failed to reach the significance threshold of at least GWTC-3.
Our confidence that the signals presented here are the result of astrophysical events should consider this.
However, if real, a number of these signals stand out as unusual.
\eventname{190523}{085933}, \eventname{190718}{160159}, \eventname{190718}{160159}, \eventname{191113}{103541}, and \eventname{200210}{100022} each show evidence for high positive $\chi_{\rm eff}$, while \eventname{190910}{012619} shows strong support for high negative $\chi_{\rm eff}$ indicating an anti-aligned spin configuration.

High positive $\chi_{\rm eff}$ has previously been reported for candidates in GWTC-3, e.g. \eventname{190517}{055101} and \eventname{190403}{051519}, but it does not contain any events with such clearly negative $\chi_{\rm eff}$.
\add{Large negative $\chi_{\rm eff}$ can be a strong indication that the binary system formed as the result of dynamical interaction~\cite{rodriguez_illuminating_2016}.
However, very high spin values are also inferred when analysing non-stationary noise artefacts (glitches) under the assumption that they represent compact binary signals~\cite{ashton_parameterised_2022}.
}

\begin{table}[]
    \footnotesize
    \begin{tabular}{c|ccccc|ccccc}\toprule
     & \multicolumn{5}{c|}{4-OGC} & \multicolumn{5}{c}{IAS-4} \\
Event & $\mathcal{M}_{\rm src}$ & $q$ & $\chi_{\rm eff}$ & $\chi_{\rm p}$ & $D_{\rm L}$ &$\mathcal{M}_{\rm src}$ & $q$ & $\chi_{\rm eff}$ & $\chi_{\rm p}$ & $D_{\rm L}$ \\
    \midrule
\eventname{191117}{023843} & $\chirpmasssourcefourOGCdivergence{GW191117_023843}$ & $\massratiofourOGCdivergence{GW191117_023843}$ & $\chiefffourOGCdivergence{GW191117_023843}$ & $\chipfourOGCdivergence{GW191117_023843}$ & $\luminositydistancefourOGCdivergence{GW191117_023843}$ & $\chirpmasssourceIASfourdivergence{GW191117_023843}$ & $\massratioIASfourdivergence{GW191117_023843}$ & $\chieffIASfourdivergence{GW191117_023843}$ & $\chipIASfourdivergence{GW191117_023843}$ & $\luminositydistanceIASfourdivergence{GW191117_023843}$ \\ 
\eventname{191224}{043228} & $\chirpmasssourcefourOGCdivergence{GW191224_043228}$ & $\massratiofourOGCdivergence{GW191224_043228}$ & $\chiefffourOGCdivergence{GW191224_043228}$ & $\chipfourOGCdivergence{GW191224_043228}$ & $\luminositydistancefourOGCdivergence{GW191224_043228}$ & $\chirpmasssourceIASfourdivergence{GW191224_043228}$ & $\massratioIASfourdivergence{GW191224_043228}$ & $\chieffIASfourdivergence{GW191224_043228}$ & $\chipIASfourdivergence{GW191224_043228}$ & $\luminositydistanceIASfourdivergence{GW191224_043228}$ \\ 
\eventname{200106}{134123} & $\chirpmasssourcefourOGCdivergence{GW200106_134123}$ & $\massratiofourOGCdivergence{GW200106_134123}$ & $\chiefffourOGCdivergence{GW200106_134123}$ & $\chipfourOGCdivergence{GW200106_134123}$ & $\luminositydistancefourOGCdivergence{GW200106_134123}$ & $\chirpmasssourceIASfourdivergence{GW200106_134123}$ & $\massratioIASfourdivergence{GW200106_134123}$ & $\chieffIASfourdivergence{GW200106_134123}$ & $\chipIASfourdivergence{GW200106_134123}$ & $\luminositydistanceIASfourdivergence{GW200106_134123}$ \\ 
\eventname{200109}{195634} & $\chirpmasssourcefourOGCdivergence{GW200109_195634}$ & $\massratiofourOGCdivergence{GW200109_195634}$ & $\chiefffourOGCdivergence{GW200109_195634}$ & $\chipfourOGCdivergence{GW200109_195634}$ & $\luminositydistancefourOGCdivergence{GW200109_195634}$ & $\chirpmasssourceIASfourdivergence{GW200109_195634}$ & $\massratioIASfourdivergence{GW200109_195634}$ & $\chieffIASfourdivergence{GW200109_195634}$ & $\chipIASfourdivergence{GW200109_195634}$ & $\luminositydistanceIASfourdivergence{GW200109_195634}$ \\ 
\eventname{200129}{114245} & $\chirpmasssourcefourOGCdivergence{GW200129_114245}$ & $\massratiofourOGCdivergence{GW200129_114245}$ & $\chiefffourOGCdivergence{GW200129_114245}$ & $\chipfourOGCdivergence{GW200129_114245}$ & $\luminositydistancefourOGCdivergence{GW200129_114245}$ & $\chirpmasssourceIASfourdivergence{GW200129_114245}$ & $\massratioIASfourdivergence{GW200129_114245}$ & $\chieffIASfourdivergence{GW200129_114245}$ & $\chipIASfourdivergence{GW200129_114245}$ & $\luminositydistanceIASfourdivergence{GW200129_114245}$ \\ 
\eventname{200210}{005122} & $\chirpmasssourcefourOGCdivergence{GW200210_005122}$ & $\massratiofourOGCdivergence{GW200210_005122}$ & $\chiefffourOGCdivergence{GW200210_005122}$ & $\chipfourOGCdivergence{GW200210_005122}$ & $\luminositydistancefourOGCdivergence{GW200210_005122}$ & $\chirpmasssourceIASfourdivergence{GW200210_005122}$ & $\massratioIASfourdivergence{GW200210_005122}$ & $\chieffIASfourdivergence{GW200210_005122}$ & $\chipIASfourdivergence{GW200210_005122}$ & $\luminositydistanceIASfourdivergence{GW200210_005122}$ \\ 
\eventname{200210}{100022} & $\chirpmasssourcefourOGCdivergence{GW200210_100022}$ & $\massratiofourOGCdivergence{GW200210_100022}$ & $\chiefffourOGCdivergence{GW200210_100022}$ & $\chipfourOGCdivergence{GW200210_100022}$ & $\luminositydistancefourOGCdivergence{GW200210_100022}$ & $\chirpmasssourceIASfourdivergence{GW200210_100022}$ & $\massratioIASfourdivergence{GW200210_100022}$ & $\chieffIASfourdivergence{GW200210_100022}$ & $\chipIASfourdivergence{GW200210_100022}$ & $\luminositydistanceIASfourdivergence{GW200210_100022}$ \\ 
\eventname{200214}{223307} & $\chirpmasssourcefourOGCdivergence{GW200214_223307}$ & $\massratiofourOGCdivergence{GW200214_223307}$ & $\chiefffourOGCdivergence{GW200214_223307}$ & $\chipfourOGCdivergence{GW200214_223307}$ & $\luminositydistancefourOGCdivergence{GW200214_223307}$ & $\chirpmasssourceIASfourdivergence{GW200214_223307}$ & $\massratioIASfourdivergence{GW200214_223307}$ & $\chieffIASfourdivergence{GW200214_223307}$ & $\chipIASfourdivergence{GW200214_223307}$ & $\luminositydistanceIASfourdivergence{GW200214_223307}$ \\ 
\eventname{200225}{075134} & $\chirpmasssourcefourOGCdivergence{GW200225_075134}$ & $\massratiofourOGCdivergence{GW200225_075134}$ & $\chiefffourOGCdivergence{GW200225_075134}$ & $\chipfourOGCdivergence{GW200225_075134}$ & $\luminositydistancefourOGCdivergence{GW200225_075134}$ & $\chirpmasssourceIASfourdivergence{GW200225_075134}$ & $\massratioIASfourdivergence{GW200225_075134}$ & $\chieffIASfourdivergence{GW200225_075134}$ & $\chipIASfourdivergence{GW200225_075134}$ & $\luminositydistanceIASfourdivergence{GW200225_075134}$ \\ 
\eventname{200305}{084739} & $\chirpmasssourcefourOGCdivergence{GW200305_084739}$ & $\massratiofourOGCdivergence{GW200305_084739}$ & $\chiefffourOGCdivergence{GW200305_084739}$ & $\chipfourOGCdivergence{GW200305_084739}$ & $\luminositydistancefourOGCdivergence{GW200305_084739}$ & $\chirpmasssourceIASfourdivergence{GW200305_084739}$ & $\massratioIASfourdivergence{GW200305_084739}$ & $\chieffIASfourdivergence{GW200305_084739}$ & $\chipIASfourdivergence{GW200305_084739}$ & $\luminositydistanceIASfourdivergence{GW200305_084739}$ \\ 
\eventname{200318}{191337} & $\chirpmasssourcefourOGCdivergence{GW200318_191337}$ & $\massratiofourOGCdivergence{GW200318_191337}$ & $\chiefffourOGCdivergence{GW200318_191337}$ & $\chipfourOGCdivergence{GW200318_191337}$ & $\luminositydistancefourOGCdivergence{GW200318_191337}$ & $\chirpmasssourceIASfourdivergence{GW200318_191337}$ & $\massratioIASfourdivergence{GW200318_191337}$ & $\chieffIASfourdivergence{GW200318_191337}$ & $\chipIASfourdivergence{GW200318_191337}$ & $\luminositydistanceIASfourdivergence{GW200318_191337}$ \\\bottomrule
    \end{tabular}
    \begin{tabular}{c|ccccc}
     & \multicolumn{5}{c}{IAS-HM} \\
Event & $\mathcal{M}_{\rm src}$ & $q$ & $\chi_{\rm eff}$ & $\chi_{\rm p}$ & $D_{\rm L}$ \\
    \midrule
    \eventname{191113}{103541} & $\chirpmasssourceIASHMdivergence{GW191113_103541}$ & $\massratioIASHMdivergence{GW191113_103541}$ & $\chieffIASHMdivergence{GW191113_103541}$ & $\chipIASHMdivergence{GW191113_103541}$ & $\luminositydistanceIASHMdivergence{GW191113_103541}$ \\ 
\eventname{191228}{195619} & $\chirpmasssourceIASHMdivergence{GW191228_195619}$ & $\massratioIASHMdivergence{GW191228_195619}$ & $\chieffIASHMdivergence{GW191228_195619}$ & $\chipIASHMdivergence{GW191228_195619}$ & $\luminositydistanceIASHMdivergence{GW191228_195619}$ \\ 
\eventname{200109}{195634} & $\chirpmasssourceIASHMdivergence{GW200109_195634}$ & $\massratioIASHMdivergence{GW200109_195634}$ & $\chieffIASHMdivergence{GW200109_195634}$ & $\chipIASHMdivergence{GW200109_195634}$ & $\luminositydistanceIASHMdivergence{GW200109_195634}$ \\ 
\eventname{200210}{100022} & $\chirpmasssourceIASHMdivergence{GW200210_100022}$ & $\massratioIASHMdivergence{GW200210_100022}$ & $\chieffIASHMdivergence{GW200210_100022}$ & $\chipIASHMdivergence{GW200210_100022}$ & $\luminositydistanceIASHMdivergence{GW200210_100022}$ \\ 
\eventname{200301}{211019} & $\chirpmasssourceIASHMdivergence{GW200301_211019}$ & $\massratioIASHMdivergence{GW200301_211019}$ & $\chieffIASHMdivergence{GW200301_211019}$ & $\chipIASHMdivergence{GW200301_211019}$ & $\luminositydistanceIASHMdivergence{GW200301_211019}$ 
 \\\bottomrule
    \end{tabular}
    \medskip
     \caption{The Jensen-Shannon divergences, in nats, between the posteriors presented in this work and those from the 4-OGC, IAS-4, and IAS-HM data releases for the marginal posterior distributions shown in figure \ref{fig:violins:o3b}.}
    \label{tab:divergences:o3b}
\end{table}

\section{Conclusion}

In this work we examined 57 gravitational wave events which have been reported in publications by authors \delete{outwith}\add{outside} the LIGO, Virgo, and KAGRA collaborations.
We have performed parameter estimation (PE) on each of these events using an analysis workflow which is as close as reasonably practicable to the one used in the preparation of the GWTC-3 catalogue, with the aim of providing a consistent dataset which encompasses all of these catalogues.


Many of the examined events were excluded from at least one other catalogue due to their not exceeding a significance threshold.
The large number of such events indicates that there is a genuine requirement for a means of determining event significance which is independent of the search pipeline identifying the event.
Previous approaches to this problem include~\cite{ashton_gravitational_2019} which was applied to \eventname{151216}{092416} in~\cite{ashton_astrophysical_2020}, but such a metric would need to be calculated for all events from PE in order to allow construction of a catalogue which is search agnostic.

We have identified divergence between the results \add{presented in the IAS catalogue publications which were} produced with the Cogwheel code~\add{~\cite{code-cogwheel}}, and ours, produced using Bilby.
We believe that these merit further investigation, as Cogwheel uses a different approach to calculating the signal likelihood (relative binning)~\add{~\cite{roulet_removing_2022}} \delete{to the approach}\add{compared to the one} used in Bilby which is based on \delete{that}\add{a method} developed in~\cite{code-lalinference}, and it is unclear from our investigations if this is the source of the difference, or if there is an additional difference we have not accounted for.
\add{Perhaps the most dramatic difference we observe is in the measurement of spin parameters by our approach and that employed by Cogwheel.
A clearer understanding of how these differences arise would be a valuable future investigation.}

The addition of 57 events to the 90 identified in GWTC-3 substantially increases the population of signals analysed using a GWTC-like analysis configuration, and the implications of these additional events to the understanding of compact binary populations also deserves future investigation.

With future developments in detector technology continually improving the sensitivity of the network of gravitational wave observatories we anticipate that the variety of search results from different groups and search pipelines will continue to grow, and so we present the use of Asimov as a means of creating consistent datasets for gravitational wave PE on public data.

\section{Data availability}
The data associated with this publication is available from Zenodo~\cite{data-release}.

\section{Acknowledgements}

The authors wish to thank John Veitch and Christopher Berry for their suggestions throughout the production of these analyses and the associated missive, as well as various discussions with members of the Institute for Gravitational Research as we developed the analysis. We also thank Michael Williams for useful suggestions on the text, and the two anonymous reviewers for their insightful and thorough critique of the manuscript.

This research has made use of data or software obtained from the Gravitational Wave Open Science Center (gwosc.org), a service of the LIGO Scientific Collaboration, the Virgo Collaboration, and KAGRA. This material is based upon work supported by NSF's LIGO Laboratory which is a major facility fully funded by the National Science Foundation, as well as the Science and Technology Facilities Council (STFC) of the United Kingdom, the Max-Planck-Society (MPS), and the State of Niedersachsen/Germany for support of the construction of Advanced LIGO and construction and operation of the GEO600 detector. Additional support for Advanced LIGO was provided by the Australian Research Council. Virgo is funded, through the European Gravitational Observatory (EGO), by the French Centre National de Recherche Scientifique (CNRS), the Italian Istituto Nazionale di Fisica Nucleare (INFN) and the Dutch Nikhef, with contributions by institutions from Belgium, Germany, Greece, Hungary, Ireland, Japan, Monaco, Poland, Portugal, Spain. KAGRA is supported by Ministry of Education, Culture, Sports, Science and Technology (MEXT), Japan Society for the Promotion of Science (JSPS) in Japan; National Research Foundation (NRF) and Ministry of Science and ICT (MSIT) in Korea; Academia Sinica (AS) and National Science and Technology Council (NSTC) in Taiwan.

DW is supported under STFC grant ST/V005634/1.

The analyses presented in this missive were run on the computing resources at the Institute for Gravitational Research at the University of Glasgow. We acknowledge the following software packages which were integral to this analysis: asimov~\cite{code-asimov}, Bilby~\cite{code-bilby}, Bilby\_pipe~\cite{code-bilby-pipe}, Bayeswave~\cite{code-bayeswave}, PESummary~\cite{code-pesummary}, Dynesty~\cite{code-dynesty}, numpy~\cite{software-numpy}, scipy\cite{software-scipy}, matplotlib\cite{software-matplotlib}, HTCondor~\cite{software-htcondor}, and astropy~\cite{software-astropy-1,software-astropy-2,software-astropy-3}.

\printbibliography

@ARTICLE{catalog-1ogc,
       author = {{Nitz}, Alexander H. and {Capano}, Collin and {Nielsen}, Alex B. and {Reyes}, Steven and {White}, Rebecca and {Brown}, Duncan A. and {Krishnan}, Badri},
        title = "{1-OGC: The First Open Gravitational-wave Catalog of Binary Mergers from Analysis of Public Advanced LIGO Data}",
      journal = {The Astrophysical Journal},
         year = 2019,
        month = feb,
       volume = {872},
       number = {2},
          eid = {195},
        pages = {195},
          doi = {10.3847/1538-4357/ab0108},
archivePrefix = {arXiv},
       eprint = {1811.01921},
 primaryClass = {gr-qc},
       adsurl = {https://ui.adsabs.harvard.edu/abs/2019ApJ...872..195N}
}

@ARTICLE{catalog-2ogc,
       author = {{Nitz}, Alexander H. and {Dent}, Thomas and {Davies}, Gareth S. and {Kumar}, Sumit and {Capano}, Collin D. and {Harry}, Ian and {Mozzon}, Simone and {Nuttall}, Laura and {Lundgren}, Andrew and {T{\'a}pai}, M{\'a}rton},
        title = "{2-OGC: Open Gravitational-wave Catalog of Binary Mergers from Analysis of Public Advanced LIGO and Virgo Data}",
      journal = {The Astrophysical Journal},
         year = 2020,
        month = mar,
       volume = {891},
       number = {2},
          eid = {123},
        pages = {123},
          doi = {10.3847/1538-4357/ab733f},
archivePrefix = {arXiv},
       eprint = {1910.05331},
 primaryClass = {astro-ph.HE},
       adsurl = {https://ui.adsabs.harvard.edu/abs/2020ApJ...891..123N}
}

@ARTICLE{catalog-3ogc,
       author = {{Nitz}, Alexander H. and {Capano}, Collin D. and {Kumar}, Sumit and {Wang}, Yi-Fan and {Kastha}, Shilpa and {Sch{\"a}fer}, Marlin and {Dhurkunde}, Rahul and {Cabero}, Miriam},
        title = "{3-OGC: Catalog of Gravitational Waves from Compact-binary Mergers}",
      journal = {The Astrophysical Journal},
         year = 2021,
        month = nov,
       volume = {922},
       number = {1},
          eid = {76},
        pages = {76},
          doi = {10.3847/1538-4357/ac1c03},
archivePrefix = {arXiv},
       eprint = {2105.09151},
 primaryClass = {astro-ph.HE},
       adsurl = {https://ui.adsabs.harvard.edu/abs/2021ApJ...922...76N}
}

@ARTICLE{catalog-4ogc,
       author = {{Nitz}, Alexander H. and {Kumar}, Sumit and {Wang}, Yi-Fan and {Kastha}, Shilpa and {Wu}, Shichao and {Sch{\"a}fer}, Marlin and {Dhurkunde}, Rahul and {Capano}, Collin D.},
        title = "{4-OGC: Catalog of Gravitational Waves from Compact Binary Mergers}",
      journal = {The Astrophysical Journal},
         year = 2023,
        month = apr,
       volume = {946},
       number = {2},
          eid = {59},
        pages = {59},
          doi = {10.3847/1538-4357/aca591},
       adsurl = {https://ui.adsabs.harvard.edu/abs/2023ApJ...946...59N}
}

@ARTICLE{code-dynesty,
       author = {{Speagle}, Joshua S.},
        title = "{DYNESTY: a dynamic nested sampling package for estimating Bayesian posteriors and evidences}",
      journal = {Monthly Notices of the Royal Astronomical Society},
         year = 2020,
        month = apr,
       volume = {493},
       number = {3},
        pages = {3132-3158},
          doi = {10.1093/mnras/staa278},
archivePrefix = {arXiv},
       eprint = {1904.02180},
 primaryClass = {astro-ph.IM},
       adsurl = {https://ui.adsabs.harvard.edu/abs/2020MNRAS.493.3132S}
}

@article{code-pesummary,
    author = "Hoy, Charlie and Raymond, Vivien",
    title = "{PESummary: the code agnostic Parameter Estimation Summary page builder}",
    eprint = "2006.06639",
    archivePrefix = "arXiv",
    primaryClass = "astro-ph.IM",
    reportNumber = "LIGO-P2000156",
    doi = "10.1016/j.softx.2021.100765",
    journal = "SoftwareX",
    volume = "15",
    pages = "100765",
    year = "2021"
}

@article{waveform-imrphenomxphm,
   title={Computationally efficient models for the dominant and subdominant harmonic modes of precessing binary black holes},
   volume={103},
   ISSN={2470-0029},
   url={http://dx.doi.org/10.1103/PhysRevD.103.104056},
   DOI={10.1103/physrevd.103.104056},
   number={10},
   journal={Physical Review D},
   publisher={American Physical Society (APS)},
   author={Pratten, Geraint and García-Quirós, Cecilio and Colleoni, Marta and Ramos-Buades, Antoni and Estellés, Héctor and Mateu-Lucena, Maite and Jaume, Rafel and Haney, Maria and Keitel, David and Thompson, Jonathan E. and Husa, Sascha},
   year={2021},
   month=may }

@article{waveform-seobnrv4,
   title={Multipolar effective-one-body waveforms for precessing binary black holes: Construction and validation},
   volume={102},
   ISSN={2470-0029},
   url={http://dx.doi.org/10.1103/PhysRevD.102.044055},
   DOI={10.1103/physrevd.102.044055},
   number={4},
   journal={Physical Review D},
   publisher={American Physical Society (APS)},
   author={Ossokine, Serguei and Buonanno, Alessandra and Marsat, Sylvain and Cotesta, Roberto and Babak, Stanislav and Dietrich, Tim and Haas, Roland and Hinder, Ian and Pfeiffer, Harald P. and Pürrer, Michael and Woodford, Charles J. and Boyle, Michael and Kidder, Lawrence E. and Scheel, Mark A. and Szilágyi, Béla},
   year={2020},
   month=aug }

@software{code-peconfigurator,
  author       = {Estellés Estrella, Héctor and
                  Ossokine, Serguei and
                  Williams, Daniel},
  title        = {pe-configurator},
  month        = nov,
  year         = 2023,
  publisher    = {Zenodo},
  version      = {1.0.1},
  doi          = {10.5281/zenodo.10133300},
  url          = {https://doi.org/10.5281/zenodo.10133300}
}

@article{code-bayeswave,
   title={Bayeswave: Bayesian inference for gravitational wave bursts and instrument glitches},
   volume={32},
   ISSN={1361-6382},
   url={http://dx.doi.org/10.1088/0264-9381/32/13/135012},
   DOI={10.1088/0264-9381/32/13/135012},
   number={13},
   journal={Classical and Quantum Gravity},
   publisher={IOP Publishing},
   author={Cornish, Neil J and Littenberg, Tyson B},
   year={2015},
   month=jun, pages={135012} }

@article{code-bilby,
   title={Bilby: A User-friendly Bayesian Inference Library for Gravitational-wave Astronomy},
   volume={241},
   ISSN={1538-4365},
   url={http://dx.doi.org/10.3847/1538-4365/ab06fc},
   DOI={10.3847/1538-4365/ab06fc},
   number={2},
   journal={The Astrophysical Journal Supplement Series},
   publisher={American Astronomical Society},
   author={Ashton, Gregory and Hübner, Moritz and Lasky, Paul D. and Talbot, Colm and Ackley, Kendall and Biscoveanu, Sylvia and Chu, Qi and Divakarla, Atul and Easter, Paul J. and Goncharov, Boris and Vivanco, Francisco Hernandez and Harms, Jan and Lower, Marcus E. and Meadors, Grant D. and Melchor, Denyz and Payne, Ethan and Pitkin, Matthew D. and Powell, Jade and Sarin, Nikhil and Smith, Rory J. E. and Thrane, Eric},
   year={2019},
   month=apr, pages={27} }

@dataset{data-release,
  author       = {Williams, Daniel},
  title        = {{Beyond GWTC-3: Analysing and verifying new 
                   gravitational-wave events from the 4-OGC Catalogue}},
  month        = jan,
  year         = 2025,
  publisher    = {Zenodo},
  version      = 2,
  doi          = {10.5281/zenodo.10479523},
  url          = {https://doi.org/10.5281/zenodo.10479523}
}

@article{gw150914,
   title={Observation of Gravitational Waves from a Binary Black Hole Merger},
   volume={116},
   ISSN={1079-7114},
   url={http://dx.doi.org/10.1103/PhysRevLett.116.061102},
   DOI={10.1103/physrevlett.116.061102},
   number={6},
   journal={Physical Review Letters},
   publisher={American Physical Society (APS)},
   author={Abbott, B. P. and Abbott, R. and Abbott, T. D. and Abernathy, M. R. and Acernese, F. and Ackley, K. and Adams, C. and Adams, T. and Addesso, P. and Adhikari, R. X. and Adya, V. B. and Affeldt, C. and Agathos, M. and Agatsuma, K. and Aggarwal, N. and Aguiar, O. D. and Aiello, L. and Ain, A. and Ajith, P. and Allen, B. and Allocca, A. and Altin, P. A. and Anderson, S. B. and Anderson, W. G. and Arai, K. and Arain, M. A. and Araya, M. C. and Arceneaux, C. C. and Areeda, J. S. and Arnaud, N. and Arun, K. G. and Ascenzi, S. and Ashton, G. and Ast, M. and Aston, S. M. and Astone, P. and Aufmuth, P. and Aulbert, C. and Babak, S. and Bacon, P. and Bader, M. K. M. and Baker, P. T. and Baldaccini, F. and Ballardin, G. and Ballmer, S. W. and Barayoga, J. C. and Barclay, S. E. and Barish, B. C. and Barker, D. and Barone, F. and Barr, B. and Barsotti, L. and Barsuglia, M. and Barta, D. and Bartlett, J. and Barton, M. A. and Bartos, I. and Bassiri, R. and Basti, A. and Batch, J. C. and Baune, C. and Bavigadda, V. and Bazzan, M. and Behnke, B. and Bejger, M. and Belczynski, C. and Bell, A. S. and Bell, C. J. and Berger, B. K. and Bergman, J. and Bergmann, G. and Berry, C. P. L. and Bersanetti, D. and Bertolini, A. and Betzwieser, J. and Bhagwat, S. and Bhandare, R. and Bilenko, I. A. and Billingsley, G. and Birch, J. and Birney, R. and Birnholtz, O. and Biscans, S. and Bisht, A. and Bitossi, M. and Biwer, C. and Bizouard, M. A. and Blackburn, J. K. and Blair, C. D. and Blair, D. G. and Blair, R. M. and Bloemen, S. and Bock, O. and Bodiya, T. P. and Boer, M. and Bogaert, G. and Bogan, C. and Bohe, A. and Bojtos, P. and Bond, C. and Bondu, F. and Bonnand, R. and Boom, B. A. and Bork, R. and Boschi, V. and Bose, S. and Bouffanais, Y. and Bozzi, A. and Bradaschia, C. and Brady, P. R. and Braginsky, V. B. and Branchesi, M. and Brau, J. E. and Briant, T. and Brillet, A. and Brinkmann, M. and Brisson, V. and Brockill, P. and Brooks, A. F. and Brown, D. A. and Brown, D. D. and Brown, N. M. and Buchanan, C. C. and Buikema, A. and Bulik, T. and Bulten, H. J. and Buonanno, A. and Buskulic, D. and Buy, C. and Byer, R. L. and Cabero, M. and Cadonati, L. and Cagnoli, G. and Cahillane, C. and Bustillo, J. Calderón and Callister, T. and Calloni, E. and Camp, J. B. and Cannon, K. C. and Cao, J. and Capano, C. D. and Capocasa, E. and Carbognani, F. and Caride, S. and Diaz, J. Casanueva and Casentini, C. and Caudill, S. and Cavaglià, M. and Cavalier, F. and Cavalieri, R. and Cella, G. and Cepeda, C. B. and Baiardi, L. Cerboni and Cerretani, G. and Cesarini, E. and Chakraborty, R. and Chalermsongsak, T. and Chamberlin, S. J. and Chan, M. and Chao, S. and Charlton, P. and Chassande-Mottin, E. and Chen, H. Y. and Chen, Y. and Cheng, C. and Chincarini, A. and Chiummo, A. and Cho, H. S. and Cho, M. and Chow, J. H. and Christensen, N. and Chu, Q. and Chua, S. and Chung, S. and Ciani, G. and Clara, F. and Clark, J. A. and Cleva, F. and Coccia, E. and Cohadon, P.-F. and Colla, A. and Collette, C. G. and Cominsky, L. and Constancio, M. and Conte, A. and Conti, L. and Cook, D. and Corbitt, T. R. and Cornish, N. and Corsi, A. and Cortese, S. and Costa, C. A. and Coughlin, M. W. and Coughlin, S. B. and Coulon, J.-P. and Countryman, S. T. and Couvares, P. and Cowan, E. E. and Coward, D. M. and Cowart, M. J. and Coyne, D. C. and Coyne, R. and Craig, K. and Creighton, J. D. E. and Creighton, T. D. and Cripe, J. and Crowder, S. G. and Cruise, A. M. and Cumming, A. and Cunningham, L. and Cuoco, E. and Canton, T. Dal and Danilishin, S. L. and D’Antonio, S. and Danzmann, K. and Darman, N. S. and Da Silva Costa, C. F. and Dattilo, V. and Dave, I. and Daveloza, H. P. and Davier, M. and Davies, G. S. and Daw, E. J. and Day, R. and De, S. and DeBra, D. and Debreczeni, G. and Degallaix, J. and De Laurentis, M. and Deléglise, S. and Del Pozzo, W. and Denker, T. and Dent, T. and Dereli, H. and Dergachev, V. and DeRosa, R. T. and De Rosa, R. and DeSalvo, R. and Dhurandhar, S. and Díaz, M. C. and Di Fiore, L. and Di Giovanni, M. and Di Lieto, A. and Di Pace, S. and Di Palma, I. and Di Virgilio, A. and Dojcinoski, G. and Dolique, V. and Donovan, F. and Dooley, K. L. and Doravari, S. and Douglas, R. and Downes, T. P. and Drago, M. and Drever, R. W. P. and Driggers, J. C. and Du, Z. and Ducrot, M. and Dwyer, S. E. and Edo, T. B. and Edwards, M. C. and Effler, A. and Eggenstein, H.-B. and Ehrens, P. and Eichholz, J. and Eikenberry, S. S. and Engels, W. and Essick, R. C. and Etzel, T. and Evans, M. and Evans, T. M. and Everett, R. and Factourovich, M. and Fafone, V. and Fair, H. and Fairhurst, S. and Fan, X. and Fang, Q. and Farinon, S. and Farr, B. and Farr, W. M. and Favata, M. and Fays, M. and Fehrmann, H. and Fejer, M. M. and Feldbaum, D. and Ferrante, I. and Ferreira, E. C. and Ferrini, F. and Fidecaro, F. and Finn, L. S. and Fiori, I. and Fiorucci, D. and Fisher, R. P. and Flaminio, R. and Fletcher, M. and Fong, H. and Fournier, J.-D. and Franco, S. and Frasca, S. and Frasconi, F. and Frede, M. and Frei, Z. and Freise, A. and Frey, R. and Frey, V. and Fricke, T. T. and Fritschel, P. and Frolov, V. V. and Fulda, P. and Fyffe, M. and Gabbard, H. A. G. and Gair, J. R. and Gammaitoni, L. and Gaonkar, S. G. and Garufi, F. and Gatto, A. and Gaur, G. and Gehrels, N. and Gemme, G. and Gendre, B. and Genin, E. and Gennai, A. and George, J. and Gergely, L. and Germain, V. and Ghosh, Abhirup and Ghosh, Archisman and Ghosh, S. and Giaime, J. A. and Giardina, K. D. and Giazotto, A. and Gill, K. and Glaefke, A. and Gleason, J. R. and Goetz, E. and Goetz, R. and Gondan, L. and González, G. and Castro, J. M. Gonzalez and Gopakumar, A. and Gordon, N. A. and Gorodetsky, M. L. and Gossan, S. E. and Gosselin, M. and Gouaty, R. and Graef, C. and Graff, P. B. and Granata, M. and Grant, A. and Gras, S. and Gray, C. and Greco, G. and Green, A. C. and Greenhalgh, R. J. S. and Groot, P. and Grote, H. and Grunewald, S. and Guidi, G. M. and Guo, X. and Gupta, A. and Gupta, M. K. and Gushwa, K. E. and Gustafson, E. K. and Gustafson, R. and Hacker, J. J. and Hall, B. R. and Hall, E. D. and Hammond, G. and Haney, M. and Hanke, M. M. and Hanks, J. and Hanna, C. and Hannam, M. D. and Hanson, J. and Hardwick, T. and Harms, J. and Harry, G. M. and Harry, I. W. and Hart, M. J. and Hartman, M. T. and Haster, C.-J. and Haughian, K. and Healy, J. and Heefner, J. and Heidmann, A. and Heintze, M. C. and Heinzel, G. and Heitmann, H. and Hello, P. and Hemming, G. and Hendry, M. and Heng, I. S. and Hennig, J. and Heptonstall, A. W. and Heurs, M. and Hild, S. and Hoak, D. and Hodge, K. A. and Hofman, D. and Hollitt, S. E. and Holt, K. and Holz, D. E. and Hopkins, P. and Hosken, D. J. and Hough, J. and Houston, E. A. and Howell, E. J. and Hu, Y. M. and Huang, S. and Huerta, E. A. and Huet, D. and Hughey, B. and Husa, S. and Huttner, S. H. and Huynh-Dinh, T. and Idrisy, A. and Indik, N. and Ingram, D. R. and Inta, R. and Isa, H. N. and Isac, J.-M. and Isi, M. and Islas, G. and Isogai, T. and Iyer, B. R. and Izumi, K. and Jacobson, M. B. and Jacqmin, T. and Jang, H. and Jani, K. and Jaranowski, P. and Jawahar, S. and Jiménez-Forteza, F. and Johnson, W. W. and Johnson-McDaniel, N. K. and Jones, D. I. and Jones, R. and Jonker, R. J. G. and Ju, L. and Haris, K. and Kalaghatgi, C. V. and Kalogera, V. and Kandhasamy, S. and Kang, G. and Kanner, J. B. and Karki, S. and Kasprzack, M. and Katsavounidis, E. and Katzman, W. and Kaufer, S. and Kaur, T. and Kawabe, K. and Kawazoe, F. and Kéfélian, F. and Kehl, M. S. and Keitel, D. and Kelley, D. B. and Kells, W. and Kennedy, R. and Keppel, D. G. and Key, J. S. and Khalaidovski, A. and Khalili, F. Y. and Khan, I. and Khan, S. and Khan, Z. and Khazanov, E. A. and Kijbunchoo, N. and Kim, C. and Kim, J. and Kim, K. and Kim, Nam-Gyu and Kim, Namjun and Kim, Y.-M. and King, E. J. and King, P. J. and Kinzel, D. L. and Kissel, J. S. and Kleybolte, L. and Klimenko, S. and Koehlenbeck, S. M. and Kokeyama, K. and Koley, S. and Kondrashov, V. and Kontos, A. and Koranda, S. and Korobko, M. and Korth, W. Z. and Kowalska, I. and Kozak, D. B. and Kringel, V. and Krishnan, B. and Królak, A. and Krueger, C. and Kuehn, G. and Kumar, P. and Kumar, R. and Kuo, L. and Kutynia, A. and Kwee, P. and Lackey, B. D. and Landry, M. and Lange, J. and Lantz, B. and Lasky, P. D. and Lazzarini, A. and Lazzaro, C. and Leaci, P. and Leavey, S. and Lebigot, E. O. and Lee, C. H. and Lee, H. K. and Lee, H. M. and Lee, K. and Lenon, A. and Leonardi, M. and Leong, J. R. and Leroy, N. and Letendre, N. and Levin, Y. and Levine, B. M. and Li, T. G. F. and Libson, A. and Littenberg, T. B. and Lockerbie, N. A. and Logue, J. and Lombardi, A. L. and London, L. T. and Lord, J. E. and Lorenzini, M. and Loriette, V. and Lormand, M. and Losurdo, G. and Lough, J. D. and Lousto, C. O. and Lovelace, G. and Lück, H. and Lundgren, A. P. and Luo, J. and Lynch, R. and Ma, Y. and MacDonald, T. and Machenschalk, B. and MacInnis, M. and Macleod, D. M. and Magaña-Sandoval, F. and Magee, R. M. and Mageswaran, M. and Majorana, E. and Maksimovic, I. and Malvezzi, V. and Man, N. and Mandel, I. and Mandic, V. and Mangano, V. and Mansell, G. L. and Manske, M. and Mantovani, M. and Marchesoni, F. and Marion, F. and Márka, S. and Márka, Z. and Markosyan, A. S. and Maros, E. and Martelli, F. and Martellini, L. and Martin, I. W. and Martin, R. M. and Martynov, D. V. and Marx, J. N. and Mason, K. and Masserot, A. and Massinger, T. J. and Masso-Reid, M. and Matichard, F. and Matone, L. and Mavalvala, N. and Mazumder, N. and Mazzolo, G. and McCarthy, R. and McClelland, D. E. and McCormick, S. and McGuire, S. C. and McIntyre, G. and McIver, J. and McManus, D. J. and McWilliams, S. T. and Meacher, D. and Meadors, G. D. and Meidam, J. and Melatos, A. and Mendell, G. and Mendoza-Gandara, D. and Mercer, R. A. and Merilh, E. and Merzougui, M. and Meshkov, S. and Messenger, C. and Messick, C. and Meyers, P. M. and Mezzani, F. and Miao, H. and Michel, C. and Middleton, H. and Mikhailov, E. E. and Milano, L. and Miller, J. and Millhouse, M. and Minenkov, Y. and Ming, J. and Mirshekari, S. and Mishra, C. and Mitra, S. and Mitrofanov, V. P. and Mitselmakher, G. and Mittleman, R. and Moggi, A. and Mohan, M. and Mohapatra, S. R. P. and Montani, M. and Moore, B. C. and Moore, C. J. and Moraru, D. and Moreno, G. and Morriss, S. R. and Mossavi, K. and Mours, B. and Mow-Lowry, C. M. and Mueller, C. L. and Mueller, G. and Muir, A. W. and Mukherjee, Arunava and Mukherjee, D. and Mukherjee, S. and Mukund, N. and Mullavey, A. and Munch, J. and Murphy, D. J. and Murray, P. G. and Mytidis, A. and Nardecchia, I. and Naticchioni, L. and Nayak, R. K. and Necula, V. and Nedkova, K. and Nelemans, G. and Neri, M. and Neunzert, A. and Newton, G. and Nguyen, T. T. and Nielsen, A. B. and Nissanke, S. and Nitz, A. and Nocera, F. and Nolting, D. and Normandin, M. E. N. and Nuttall, L. K. and Oberling, J. and Ochsner, E. and O’Dell, J. and Oelker, E. and Ogin, G. H. and Oh, J. J. and Oh, S. H. and Ohme, F. and Oliver, M. and Oppermann, P. and Oram, Richard J. and O’Reilly, B. and O’Shaughnessy, R. and Ott, C. D. and Ottaway, D. J. and Ottens, R. S. and Overmier, H. and Owen, B. J. and Pai, A. and Pai, S. A. and Palamos, J. R. and Palashov, O. and Palomba, C. and Pal-Singh, A. and Pan, H. and Pan, Y. and Pankow, C. and Pannarale, F. and Pant, B. C. and Paoletti, F. and Paoli, A. and Papa, M. A. and Paris, H. R. and Parker, W. and Pascucci, D. and Pasqualetti, A. and Passaquieti, R. and Passuello, D. and Patricelli, B. and Patrick, Z. and Pearlstone, B. L. and Pedraza, M. and Pedurand, R. and Pekowsky, L. and Pele, A. and Penn, S. and Perreca, A. and Pfeiffer, H. P. and Phelps, M. and Piccinni, O. and Pichot, M. and Pickenpack, M. and Piergiovanni, F. and Pierro, V. and Pillant, G. and Pinard, L. and Pinto, I. M. and Pitkin, M. and Poeld, J. H. and Poggiani, R. and Popolizio, P. and Post, A. and Powell, J. and Prasad, J. and Predoi, V. and Premachandra, S. S. and Prestegard, T. and Price, L. R. and Prijatelj, M. and Principe, M. and Privitera, S. and Prix, R. and Prodi, G. A. and Prokhorov, L. and Puncken, O. and Punturo, M. and Puppo, P. and Pürrer, M. and Qi, H. and Qin, J. and Quetschke, V. and Quintero, E. A. and Quitzow-James, R. and Raab, F. J. and Rabeling, D. S. and Radkins, H. and Raffai, P. and Raja, S. and Rakhmanov, M. and Ramet, C. R. and Rapagnani, P. and Raymond, V. and Razzano, M. and Re, V. and Read, J. and Reed, C. M. and Regimbau, T. and Rei, L. and Reid, S. and Reitze, D. H. and Rew, H. and Reyes, S. D. and Ricci, F. and Riles, K. and Robertson, N. A. and Robie, R. and Robinet, F. and Rocchi, A. and Rolland, L. and Rollins, J. G. and Roma, V. J. and Romano, J. D. and Romano, R. and Romanov, G. and Romie, J. H. and Rosińska, D. and Rowan, S. and Rüdiger, A. and Ruggi, P. and Ryan, K. and Sachdev, S. and Sadecki, T. and Sadeghian, L. and Salconi, L. and Saleem, M. and Salemi, F. and Samajdar, A. and Sammut, L. and Sampson, L. M. and Sanchez, E. J. and Sandberg, V. and Sandeen, B. and Sanders, G. H. and Sanders, J. R. and Sassolas, B. and Sathyaprakash, B. S. and Saulson, P. R. and Sauter, O. and Savage, R. L. and Sawadsky, A. and Schale, P. and Schilling, R. and Schmidt, J. and Schmidt, P. and Schnabel, R. and Schofield, R. M. S. and Schönbeck, A. and Schreiber, E. and Schuette, D. and Schutz, B. F. and Scott, J. and Scott, S. M. and Sellers, D. and Sengupta, A. S. and Sentenac, D. and Sequino, V. and Sergeev, A. and Serna, G. and Setyawati, Y. and Sevigny, A. and Shaddock, D. A. and Shaffer, T. and Shah, S. and Shahriar, M. S. and Shaltev, M. and Shao, Z. and Shapiro, B. and Shawhan, P. and Sheperd, A. and Shoemaker, D. H. and Shoemaker, D. M. and Siellez, K. and Siemens, X. and Sigg, D. and Silva, A. D. and Simakov, D. and Singer, A. and Singer, L. P. and Singh, A. and Singh, R. and Singhal, A. and Sintes, A. M. and Slagmolen, B. J. J. and Smith, J. R. and Smith, M. R. and Smith, N. D. and Smith, R. J. E. and Son, E. J. and Sorazu, B. and Sorrentino, F. and Souradeep, T. and Srivastava, A. K. and Staley, A. and Steinke, M. and Steinlechner, J. and Steinlechner, S. and Steinmeyer, D. and Stephens, B. C. and Stevenson, S. P. and Stone, R. and Strain, K. A. and Straniero, N. and Stratta, G. and Strauss, N. A. and Strigin, S. and Sturani, R. and Stuver, A. L. and Summerscales, T. Z. and Sun, L. and Sutton, P. J. and Swinkels, B. L. and Szczepańczyk, M. J. and Tacca, M. and Talukder, D. and Tanner, D. B. and Tápai, M. and Tarabrin, S. P. and Taracchini, A. and Taylor, R. and Theeg, T. and Thirugnanasambandam, M. P. and Thomas, E. G. and Thomas, M. and Thomas, P. and Thorne, K. A. and Thorne, K. S. and Thrane, E. and Tiwari, S. and Tiwari, V. and Tokmakov, K. V. and Tomlinson, C. and Tonelli, M. and Torres, C. V. and Torrie, C. I. and Töyrä, D. and Travasso, F. and Traylor, G. and Trifirò, D. and Tringali, M. C. and Trozzo, L. and Tse, M. and Turconi, M. and Tuyenbayev, D. and Ugolini, D. and Unnikrishnan, C. S. and Urban, A. L. and Usman, S. A. and Vahlbruch, H. and Vajente, G. and Valdes, G. and Vallisneri, M. and van Bakel, N. and van Beuzekom, M. and van den Brand, J. F. J. and Van Den Broeck, C. and Vander-Hyde, D. C. and van der Schaaf, L. and van Heijningen, J. V. and van Veggel, A. A. and Vardaro, M. and Vass, S. and Vasúth, M. and Vaulin, R. and Vecchio, A. and Vedovato, G. and Veitch, J. and Veitch, P. J. and Venkateswara, K. and Verkindt, D. and Vetrano, F. and Viceré, A. and Vinciguerra, S. and Vine, D. J. and Vinet, J.-Y. and Vitale, S. and Vo, T. and Vocca, H. and Vorvick, C. and Voss, D. and Vousden, W. D. and Vyatchanin, S. P. and Wade, A. R. and Wade, L. E. and Wade, M. and Waldman, S. J. and Walker, M. and Wallace, L. and Walsh, S. and Wang, G. and Wang, H. and Wang, M. and Wang, X. and Wang, Y. and Ward, H. and Ward, R. L. and Warner, J. and Was, M. and Weaver, B. and Wei, L.-W. and Weinert, M. and Weinstein, A. J. and Weiss, R. and Welborn, T. and Wen, L. and Weßels, P. and Westphal, T. and Wette, K. and Whelan, J. T. and Whitcomb, S. E. and White, D. J. and Whiting, B. F. and Wiesner, K. and Wilkinson, C. and Willems, P. A. and Williams, L. and Williams, R. D. and Williamson, A. R. and Willis, J. L. and Willke, B. and Wimmer, M. H. and Winkelmann, L. and Winkler, W. and Wipf, C. C. and Wiseman, A. G. and Wittel, H. and Woan, G. and Worden, J. and Wright, J. L. and Wu, G. and Yablon, J. and Yakushin, I. and Yam, W. and Yamamoto, H. and Yancey, C. C. and Yap, M. J. and Yu, H. and Yvert, M. and Zadrożny, A. and Zangrando, L. and Zanolin, M. and Zendri, J.-P. and Zevin, M. and Zhang, F. and Zhang, L. and Zhang, M. and Zhang, Y. and Zhao, C. and Zhou, M. and Zhou, Z. and Zhu, X. J. and Zucker, M. E. and Zuraw, S. E. and Zweizig, J.},
   year={2016},
   month=feb }

@article{code-bilby-pipe,
   title={Bayesian inference for compact binary coalescences with bilby: validation and application to the first LIGO–Virgo gravitational-wave transient catalogue},
   volume={499},
   ISSN={1365-2966},
   url={http://dx.doi.org/10.1093/mnras/staa2850},
   DOI={10.1093/mnras/staa2850},
   number={3},
   journal={Monthly Notices of the Royal Astronomical Society},
   publisher={Oxford University Press (OUP)},
   author={Romero-Shaw, I M and Talbot, C and Biscoveanu, S and D’Emilio, V and Ashton, G and Berry, C P L and Coughlin, S and Galaudage, S and Hoy, C and Hübner, M and Phukon, K S and Pitkin, M and Rizzo, M and Sarin, N and Smith, R and Stevenson, S and Vajpeyi, A and Arène, M and Athar, K and Banagiri, S and Bose, N and Carney, M and Chatziioannou, K and Clark, J A and Colleoni, M and Cotesta, R and Edelman, B and Estellés, H and García-Quirós, C and Ghosh, Abhirup and Green, R and Haster, C-J and Husa, S and Keitel, D and Kim, A X and Hernandez-Vivanco, F and Magaña Hernandez, I and Karathanasis, C and Lasky, P D and De Lillo, N and Lower, M E and Macleod, D and Mateu-Lucena, M and Miller, A and Millhouse, M and Morisaki, S and Oh, S H and Ossokine, S and Payne, E and Powell, J and Pratten, G and Pürrer, M and Ramos-Buades, A and Raymond, V and Thrane, E and Veitch, J and Williams, D and Williams, M J and Xiao, L},
   year={2020},
   month=sep, pages={3295–3319} }

@ARTICLE{astrophysical-gw151216,
       author = {{Ashton}, Gregory and {Thrane}, Eric},
        title = "{The astrophysical odds of GW151216}",
      journal = {Monthly Notices of the Royal Astronomical Society},
         year = 2020,
        month = oct,
       volume = {498},
       number = {2},
        pages = {1905-1910},
          doi = {10.1093/mnras/staa2332},
archivePrefix = {arXiv},
       eprint = {2006.05039},
 primaryClass = {astro-ph.HE},
       adsurl = {https://ui.adsabs.harvard.edu/abs/2020MNRAS.498.1905A}
}

@ARTICLE{improving-151216,
       author = {{Jadhav}, Shreejit and {Mukund}, Nikhil and {Gadre}, Bhooshan and {Mitra}, Sanjit and {Abraham}, Sheelu},
        title = "{Improving significance of binary black hole mergers in Advanced LIGO data using deep learning: Confirmation of GW151216}",
      journal = {Physical Review D},
         year = 2021,
        month = sep,
       volume = {104},
       number = {6},
          eid = {064051},
        pages = {064051},
          doi = {10.1103/PhysRevD.104.064051},
archivePrefix = {arXiv},
       eprint = {2010.08584},
 primaryClass = {gr-qc},
       adsurl = {https://ui.adsabs.harvard.edu/abs/2021PhRvD.104f4051J}
}

@ARTICLE{151216-paper,
       author = {{Zackay}, Barak and {Venumadhav}, Tejaswi and {Dai}, Liang and {Roulet}, Javier and {Zaldarriaga}, Matias},
        title = "{Highly spinning and aligned binary black hole merger in the Advanced LIGO first observing run}",
      journal = {Physical Review D},
         year = 2019,
        month = jul,
       volume = {100},
       number = {2},
          eid = {023007},
        pages = {023007},
          doi = {10.1103/PhysRevD.100.023007},
archivePrefix = {arXiv},
       eprint = {1902.10331},
 primaryClass = {astro-ph.HE},
       adsurl = {https://ui.adsabs.harvard.edu/abs/2019PhRvD.100b3007Z}
}

@ARTICLE{lowest-snr-properties,
       author = {{Huang}, Yiwen and {Haster}, Carl-Johan and {Roulet}, Javier and {Vitale}, Salvatore and {Zimmerman}, Aaron and {Venumadhav}, Tejaswi and {Zackay}, Barak and {Dai}, Liang and {Zaldarriaga}, Matias},
        title = "{Source properties of the lowest signal-to-noise-ratio binary black hole detections}",
      journal = {Physical Review D},
         year = 2020,
        month = nov,
       volume = {102},
       number = {10},
          eid = {103024},
        pages = {103024},
          doi = {10.1103/PhysRevD.102.103024},
archivePrefix = {arXiv},
       eprint = {2003.04513},
 primaryClass = {gr-qc},
       adsurl = {https://ui.adsabs.harvard.edu/abs/2020PhRvD.102j3024H}
}

@ARTICLE{catalog-pycbc-kde,
       author = {{Kumar}, Praveen and {Dent}, Thomas},
        title = "{Optimized search for a binary black hole merger population in LIGO-Virgo O3 data}",
      journal = {Physical Review D},
         year = 2024,
        month = aug,
       volume = {110},
       number = {4},
          eid = {043036},
        pages = {043036},
          doi = {10.1103/PhysRevD.110.043036},
archivePrefix = {arXiv},
       eprint = {2403.10439},
 primaryClass = {gr-qc},
       adsurl = {https://ui.adsabs.harvard.edu/abs/2024PhRvD.110d3036K}
}

@ARTICLE{code-dingo,
       author = {{Dax}, Maximilian and {Green}, Stephen R. and {Gair}, Jonathan and {Macke}, Jakob H. and {Buonanno}, Alessandra and {Sch{\"o}lkopf}, Bernhard},
        title = "{Real-Time Gravitational Wave Science with Neural Posterior Estimation}",
      journal = {Physical Review Letters},
         year = 2021,
        month = dec,
       volume = {127},
       number = {24},
          eid = {241103},
        pages = {241103},
          doi = {10.1103/PhysRevLett.127.241103},
archivePrefix = {arXiv},
       eprint = {2106.12594},
 primaryClass = {gr-qc},
       adsurl = {https://ui.adsabs.harvard.edu/abs/2021PhRvL.127x1103D}
}

@ARTICLE{code-vitamin,
       author = {{Gabbard}, Hunter and {Messenger}, Chris and {Heng}, Ik Siong and {Tonolini}, Francesco and {Murray-Smith}, Roderick},
        title = "{Bayesian parameter estimation using conditional variational autoencoders for gravitational-wave astronomy}",
      journal = {Nature Physics},
         year = 2022,
        month = jan,
       volume = {18},
       number = {1},
        pages = {112-117},
          doi = {10.1038/s41567-021-01425-7},
archivePrefix = {arXiv},
       eprint = {1909.06296},
 primaryClass = {astro-ph.IM},
       adsurl = {https://ui.adsabs.harvard.edu/abs/2022NatPh..18..112G}
}

@article{catalog-gwtc-2d1,
	title = {{GWTC}-2.1: {Deep} extended catalog of compact binary coalescences observed by {LIGO} and {Virgo} during the first half of the third observing run},
	volume = {109},
	issn = {2470-0010, 2470-0029},
	shorttitle = {{GWTC}-2.1},
	url = {https://link.aps.org/doi/10.1103/PhysRevD.109.022001},
	doi = {10.1103/PhysRevD.109.022001},
	language = {en},
	number = {2},
	urldate = {2025-04-03},
	journal = {Physical Review D},
	author = {Abbott, R. and Abbott, T. D. and Acernese, F. and Ackley, K. and Adams, C. and Adhikari, N. and Adhikari, R. X. and Adya, V. B. and Affeldt, C. and Agarwal, D. and Agathos, M. and Agatsuma, K. and Aggarwal, N. and Aguiar, O. D. and Aiello, L. and Ain, A. and Ajith, P. and Albanesi, S. and Allocca, A. and Altin, P. A. and Amato, A. and Anand, C. and Anand, S. and Ananyeva, A. and Anderson, S. B. and Anderson, W. G. and Andrade, T. and Andres, N. and Andrić, T. and Angelova, S. V. and Ansoldi, S. and Antelis, J. M. and Antier, S. and Appert, S. and Arai, K. and Araya, M. C. and Areeda, J. S. and Arène, M. and Arnaud, N. and Aronson, S. M. and Arun, K. G. and Asali, Y. and Ashton, G. and Assiduo, M. and Aston, S. M. and Astone, P. and Aubin, F. and Austin, C. and Babak, S. and Badaracco, F. and Bader, M. K. M. and Badger, C. and Bae, S. and Baer, A. M. and Bagnasco, S. and Bai, Y. and Baird, J. and Ball, M. and Ballardin, G. and Ballmer, S. W. and Balsamo, A. and Baltus, G. and Banagiri, S. and Bankar, D. and Barayoga, J. C. and Barbieri, C. and Barish, B. C. and Barker, D. and Barneo, P. and Barone, F. and Barr, B. and Barsotti, L. and Barsuglia, M. and Barta, D. and Bartlett, J. and Barton, M. A. and Bartos, I. and Bassiri, R. and Basti, A. and Bawaj, M. and Bayley, J. C. and Baylor, A. C. and Bazzan, M. and Bécsy, B. and Bedakihale, V. M. and Bejger, M. and Belahcene, I. and Benedetto, V. and Beniwal, D. and Bennett, T. F. and Bentley, J. D. and BenYaala, M. and Bergamin, F. and Berger, B. K. and Bernuzzi, S. and Berry, C. P. L. and Bersanetti, D. and Bertolini, A. and Betzwieser, J. and Beveridge, D. and Bhandare, R. and Bhardwaj, U. and Bhattacharjee, D. and Bhaumik, S. and Bilenko, I. A. and Billingsley, G. and Bini, S. and Birney, I. A. and Birnholtz, O. and Biscans, S. and Bischi, M. and Biscoveanu, S. and Bisht, A. and Biswas, B. and Bitossi, M. and Bizouard, M.-A. and Blackburn, J. K. and Blair, C. D. and Blair, D. G. and Blair, R. M. and Bobba, F. and Bode, N. and Boer, M. and Bogaert, G. and Boldrini, M. and Bonavena, L. D. and Bondu, F. and Bonilla, E. and Bonnand, R. and Booker, P. and Boom, B. A. and Bork, R. and Boschi, V. and Bose, N. and Bose, S. and Bossilkov, V. and Boudart, V. and Bouffanais, Y. and Bozzi, A. and Bradaschia, C. and Brady, P. R. and Bramley, A. and Branch, A. and Branchesi, M. and Brau, J. E. and Breschi, M. and Briant, T. and Briggs, J. H. and Brillet, A. and Brinkmann, M. and Brockill, P. and Brooks, A. F. and Brooks, J. and Brown, D. D. and Brunett, S. and Bruno, G. and Bruntz, R. and Bryant, J. and Bulik, T. and Bulten, H. J. and Buonanno, A. and Buscicchio, R. and Buskulic, D. and Buy, C. and Byer, R. L. and Cadonati, L. and Cagnoli, G. and Cahillane, C. and Bustillo, J. Calderón and Callaghan, J. D. and Callister, T. A. and Calloni, E. and Cameron, J. and Camp, J. B. and Canepa, M. and Canevarolo, S. and Cannavacciuolo, M. and Cannon, K. C. and Cao, H. and Capote, E. and Carapella, G. and Carbognani, F. and Carlin, J. B. and Carney, M. F. and Carpinelli, M. and Carrillo, G. and Carullo, G. and Carver, T. L. and Diaz, J. Casanueva and Casentini, C. and Castaldi, G. and Caudill, S. and Cavaglià, M. and Cavalier, F. and Cavalieri, R. and Ceasar, M. and Cella, G. and Cerdá-Durán, P. and Cesarini, E. and Chaibi, W. and Chakravarti, K. and Subrahmanya, S. Chalathadka and Champion, E. and Chan, C.-H. and Chan, C. and Chan, C. L. and Chan, K. and Chandra, K. and Chanial, P. and Chao, S. and Charlton, P. and Chase, E. A. and Chassande-Mottin, E. and Chatterjee, C. and Chatterjee, Debarati and Chatterjee, Deep and Chattopadhyay, D. and Chaturvedi, M. and Chaty, S. and Chatziioannou, K. and Chen, H. Y. and Chen, J. and Chen, X. and Chen, Y. and Chen, Z. and Cheng, H. and Cheong, C. K. and Cheung, H. Y. and Chia, H. Y. and Chiadini, F. and Chiarini, G. and Chierici, R. and Chincarini, A. and Chiofalo, M. L. and Chiummo, A. and Cho, G. and Cho, H. S. and Choudhary, R. K. and Choudhary, S. and Christensen, N. and Chu, Q. and Chua, S. and Chung, K. W. and Ciani, G. and Ciecielag, P. and Cieślar, M. and Cifaldi, M. and Ciobanu, A. A. and Ciolfi, R. and Cipriano, F. and Cirone, A. and Clara, F. and Clark, E. N. and Clark, J. A. and Clarke, L. and Clearwater, P. and Clesse, S. and Cleva, F. and Coccia, E. and Codazzo, E. and Cohadon, P.-F. and Cohen, D. E. and Cohen, L. and Colleoni, M. and Collette, C. G. and Colombo, A. and Colpi, M. and Compton, C. M. and Constancio, M. and Conti, L. and Cooper, S. J. and Corban, P. and Corbitt, T. R. and Cordero-Carrión, I. and Corezzi, S. and Corley, K. R. and Cornish, N. and Corre, D. and Corsi, A. and Cortese, S. and Costa, C. A. and Cotesta, R. and Coughlin, M. W. and Coulon, J.-P. and Countryman, S. T. and Cousins, B. and Couvares, P. and Coward, D. M. and Cowart, M. J. and Coyne, D. C. and Coyne, R. and Creighton, J. D. E. and Creighton, T. D. and Criswell, A. W. and Croquette, M. and Crowder, S. G. and Cudell, J. R. and Cullen, T. J. and Cumming, A. and Cummings, R. and Cunningham, L. and Cuoco, E. and Curyło, M. and Dabadie, P. and Canton, T. Dal and Dall’Osso, S. and Dálya, G. and Dana, A. and DaneshgaranBajastani, L. M. and D’Angelo, B. and Danila, B. and Danilishin, S. and D’Antonio, S. and Danzmann, K. and Darsow-Fromm, C. and Dasgupta, A. and Datrier, L. E. H. and Datta, S. and Dattilo, V. and Dave, I. and Davier, M. and Davies, G. S. and Davis, D. and Davis, M. C. and Daw, E. J. and Dean, R. and DeBra, D. and Deenadayalan, M. and Degallaix, J. and De Laurentis, M. and Deléglise, S. and Del Favero, V. and De Lillo, F. and De Lillo, N. and Del Pozzo, W. and DeMarchi, L. M. and De Matteis, F. and D’Emilio, V. and Demos, N. and Dent, T. and Depasse, A. and De Pietri, R. and De Rosa, R. and De Rossi, C. and DeSalvo, R. and De Simone, R. and Dhurandhar, S. and Díaz, M. C. and Diaz-Ortiz, M. and Didio, N. A. and Dietrich, T. and Di Fiore, L. and Di Fronzo, C. and Di Giorgio, C. and Di Giovanni, F. and Di Giovanni, M. and Di Girolamo, T. and Di Lieto, A. and Ding, B. and Di Pace, S. and Di Palma, I. and Di Renzo, F. and Divakarla, A. K. and {Divyajyoti} and Dmitriev, A. and Doctor, Z. and D’Onofrio, L. and Donovan, F. and Dooley, K. L. and Doravari, S. and Dorrington, I. and Drago, M. and Driggers, J. C. and Drori, Y. and Ducoin, J.-G. and Dupej, P. and Durante, O. and D’Urso, D. and Duverne, P.-A. and Dwyer, S. E. and Eassa, C. and Easter, P. J. and Ebersold, M. and Eckhardt, T. and Eddolls, G. and Edelman, B. and Edo, T. B. and Edy, O. and Effler, A. and Eichholz, J. and Eikenberry, S. S. and Eisenmann, M. and Eisenstein, R. A. and Ejlli, A. and Engelby, E. and Errico, L. and Essick, R. C. and Estellés, H. and Estevez, D. and Etienne, Z. and Etzel, T. and Evans, M. and Evans, T. M. and Ewing, B. E. and Fafone, V. and Fair, H. and Fairhurst, S. and Fanning, S. P. and Farah, A. M. and Farinon, S. and Farr, B. and Farr, W. M. and Farrow, N. W. and Fauchon-Jones, E. J. and Favaro, G. and Favata, M. and Fays, M. and Fazio, M. and Feicht, J. and Fejer, M. M. and Fenyvesi, E. and Ferguson, D. L. and Fernandez-Galiana, A. and Ferrante, I. and Ferreira, T. A. and Fidecaro, F. and Figura, P. and Fiori, I. and Fishbach, M. and Fisher, R. P. and Fittipaldi, R. and Fiumara, V. and Flaminio, R. and Floden, E. and Fong, H. and Font, J. A. and Fornal, B. and Forsyth, P. W. F. and Franke, A. and Frasca, S. and Frasconi, F. and Frederick, C. and Freed, J. P. and Frei, Z. and Freise, A. and Frey, R. and Fritschel, P. and Frolov, V. V. and Fronzé, G. G. and Fulda, P. and Fyffe, M. and Gabbard, H. A. and Gabella, W. and Gadre, B. U. and Gair, J. R. and Gais, J. and Galaudage, S. and Gamba, R. and Ganapathy, D. and Ganguly, A. and Gaonkar, S. G. and Garaventa, B. and García, F. and García-Núñez, C. and García-Quirós, C. and Garufi, F. and Gateley, B. and Gaudio, S. and Gayathri, V. and Gemme, G. and Gennai, A. and George, J. and George, R. N. and Gerberding, O. and Gergely, L. and Gewecke, P. and Ghonge, S. and Ghosh, Abhirup and Ghosh, Archisman and Ghosh, Shaon and Ghosh, Shrobana and Giacomazzo, B. and Giacoppo, L. and Giaime, J. A. and Giardina, K. D. and Gibson, D. R. and Gier, C. and Giesler, M. and Giri, P. and Gissi, F. and Glanzer, J. and Gleckl, A. E. and Godwin, P. and Goetz, E. and Goetz, R. and Gohlke, N. and Goncharov, B. and González, G. and Gopakumar, A. and Gosselin, M. and Gouaty, R. and Gould, D. W. and Grace, B. and Grado, A. and Granata, M. and Granata, V. and Grant, A. and Gras, S. and Grassia, P. and Gray, C. and Gray, R. and Greco, G. and Green, A. C. and Green, R. and Gretarsson, A. M. and Gretarsson, E. M. and Griffith, D. and Griffiths, W. and Griggs, H. L. and Grignani, G. and Grimaldi, A. and Grimm, S. J. and Grote, H. and Grunewald, S. and Gruning, P. and Guerra, D. and Guidi, G. M. and Guimaraes, A. R. and Guixé, G. and Gulati, H. K. and Guo, H.-K. and Guo, Y. and Gupta, Anchal and Gupta, Anuradha and Gupta, P. and Gustafson, E. K. and Gustafson, R. and Guzman, F. and Haegel, L. and Halim, O. and Hall, E. D. and Hamilton, E. Z. and Hammond, G. and Haney, M. and Hanks, J. and Hanna, C. and Hannam, M. D. and Hannuksela, O. and Hansen, H. and Hansen, T. J. and Hanson, J. and Harder, T. and Hardwick, T. and Haris, K. and Harms, J. and Harry, G. M. and Harry, I. W. and Hartwig, D. and Haskell, B. and Hasskew, R. K. and Haster, C.-J. and Haughian, K. and Hayes, F. J. and Healy, J. and Heidmann, A. and Heidt, A. and Heintze, M. C. and Heinze, J. and Heinzel, J. and Heitmann, H. and Hellman, F. and Hello, P. and Helmling-Cornell, A. F. and Hemming, G. and Hendry, M. and Heng, I. S. and Hennes, E. and Hennig, J. and Hennig, M. H. and Hernandez, A. G. and Vivanco, F. Hernandez and Heurs, M. and Hild, S. and Hill, P. and Hines, A. S. and Hochheim, S. and Hofman, D. and Hohmann, J. N. and Holcomb, D. G. and Holland, N. A. and Holley-Bockelmann, K. and Hollows, I. J. and Holmes, Z. J. and Holt, K. and Holz, D. E. and Hopkins, P. and Hough, J. and Hourihane, S. and Howell, E. J. and Hoy, C. G. and Hoyland, D. and Hreibi, A. and Hsu, Y. and Huang, Y. and Hübner, M. T. and Huddart, A. D. and Hughey, B. and Hui, V. and Husa, S. and Huttner, S. H. and Huxford, R. and Huynh-Dinh, T. and Idzkowski, B. and Iess, A. and Ingram, C. and Isi, M. and Isleif, K. and Iyer, B. R. and JaberianHamedan, V. and Jacqmin, T. and Jadhav, S. J. and Jadhav, S. P. and James, A. L. and Jan, A. Z. and Jani, K. and Janquart, J. and Janssens, K. and Janthalur, N. N. and Jaranowski, P. and Jariwala, D. and Jaume, R. and Jenkins, A. C. and Jenner, K. and Jeunon, M. and Jia, W. and Johns, G. R. and Johnson-McDaniel, N. K. and Jones, A. W. and Jones, D. I. and Jones, J. D. and Jones, P. and Jones, R. and Jonker, R. J. G. and Ju, L. and Junker, J. and Juste, V. and Kalaghatgi, C. V. and Kalogera, V. and Kamai, B. and Kandhasamy, S. and Kang, G. and Kanner, J. B. and Kao, Y. and Kapadia, S. J. and Kapasi, D. P. and Karat, S. and Karathanasis, C. and Karki, S. and Kashyap, R. and Kasprzack, M. and Kastaun, W. and Katsanevas, S. and Katsavounidis, E. and Katzman, W. and Kaur, T. and Kawabe, K. and Kéfélian, F. and Keitel, D. and Key, J. S. and Khadka, S. and Khalili, F. Y. and Khan, S. and Khazanov, E. A. and Khetan, N. and Khursheed, M. and Kijbunchoo, N. and Kim, C. and Kim, J. C. and Kim, K. and Kim, W. S. and Kim, Y.-M. and Kimball, C. and Kinley-Hanlon, M. and Kirchhoff, R. and Kissel, J. S. and Kleybolte, L. and Klimenko, S. and Knee, A. M. and Knowles, T. D. and Knyazev, E. and Koch, P. and Koekoek, G. and Koley, S. and Kolitsidou, P. and Kolstein, M. and Komori, K. and Kondrashov, V. and Kontos, A. and Koper, N. and Korobko, M. and Kovalam, M. and Kozak, D. B. and Kringel, V. and Krishnendu, N. V. and Królak, A. and Kuehn, G. and Kuei, F. and Kuijer, P. and Kulkarni, S. and Kumar, A. and Kumar, P. and Kumar, Rahul and Kumar, Rakesh and Kuns, K. and Kuwahara, S. and Lagabbe, P. and Laghi, D. and Lalande, E. and Lam, T. L. and Lamberts, A. and Landry, M. and Lane, B. B. and Lang, R. N. and Lange, J. and Lantz, B. and La Rosa, I. and Lartaux-Vollard, A. and Lasky, P. D. and Laxen, M. and Lazzarini, A. and Lazzaro, C. and Leaci, P. and Leavey, S. and Lecoeuche, Y. K. and Lee, H. M. and Lee, H. W. and Lee, J. and Lee, K. and Lehmann, J. and Lemaître, A. and Leroy, N. and Letendre, N. and Levesque, C. and Levin, Y. and Leviton, J. N. and Leyde, K. and Li, A. K. Y. and Li, B. and Li, J. and Li, T. G. F. and Li, X. and Linde, F. and Linker, S. D. and Linley, J. N. and Littenberg, T. B. and Liu, J. and Liu, K. and Liu, X. and Llamas, F. and Llorens-Monteagudo, M. and Lo, R. K. L. and Lockwood, A. and London, L. T. and Longo, A. and Lopez, D. and Portilla, M. Lopez and Lorenzini, M. and Loriette, V. and Lormand, M. and Losurdo, G. and Lott, T. P. and Lough, J. D. and Lousto, C. O. and Lovelace, G. and Lucaccioni, J. F. and Lück, H. and Lumaca, D. and Lundgren, A. P. and Lynam, J. E. and Macas, R. and MacInnis, M. and Macleod, D. M. and MacMillan, I. A. O. and Macquet, A. and Hernandez, I. Magaña and Magazzù, C. and Magee, R. M. and Maggiore, R. and Magnozzi, M. and Mahesh, S. and Majorana, E. and Makarem, C. and Maksimovic, I. and Maliakal, S. and Malik, A. and Man, N. and Mandic, V. and Mangano, V. and Mango, J. L. and Mansell, G. L. and Manske, M. and Mantovani, M. and Mapelli, M. and Marchesoni, F. and Marion, F. and Mark, Z. and Márka, S. and Márka, Z. and Markakis, C. and Markosyan, A. S. and Markowitz, A. and Maros, E. and Marquina, A. and Marsat, S. and Martelli, F. and Martin, I. W. and Martin, R. M. and Martinez, M. and Martinez, V. A. and Martinez, V. and Martinovic, K. and Martynov, D. V. and Marx, E. J. and Masalehdan, H. and Mason, K. and Massera, E. and Masserot, A. and Massinger, T. J. and Masso-Reid, M. and Mastrogiovanni, S. and Matas, A. and Mateu-Lucena, M. and Matichard, F. and Matiushechkina, M. and Mavalvala, N. and McCann, J. J. and McCarthy, R. and McClelland, D. E. and McClincy, P. K. and McCormick, S. and McCuller, L. and McGhee, G. I. and McGuire, S. C. and McIsaac, C. and McIver, J. and McRae, T. and McWilliams, S. T. and Meacher, D. and Mehmet, M. and Mehta, A. K. and Meijer, Q. and Melatos, A. and Melchor, D. A. and Mendell, G. and Menendez-Vazquez, A. and Menoni, C. S. and Mercer, R. A. and Mereni, L. and Merfeld, K. and Merilh, E. L. and Merritt, J. D. and Merzougui, M. and Meshkov, S. and Messenger, C. and Messick, C. and Meyers, P. M. and Meylahn, F. and Mhaske, A. and Miani, A. and Miao, H. and Michaloliakos, I. and Michel, C. and Middleton, H. and Milano, L. and Miller, A. and Miller, A. L. and Miller, B. and Millhouse, M. and Mills, J. C. and Milotti, E. and Minazzoli, O. and Minenkov, Y. and Mir, Ll. M. and Miravet-Tenés, M. and Mishra, C. and Mishra, T. and Mistry, T. and Mitra, S. and Mitrofanov, V. P. and Mitselmakher, G. and Mittleman, R. and Mo, Geoffrey and Moguel, E. and Mogushi, K. and Mohapatra, S. R. P. and Mohite, S. R. and Molina, I. and Molina-Ruiz, M. and Mondin, M. and Montani, M. and Moore, C. J. and Moraru, D. and Morawski, F. and More, A. and Moreno, C. and Moreno, G. and Morisaki, S. and Mours, B. and Mow-Lowry, C. M. and Mozzon, S. and Muciaccia, F. and Mukherjee, Arunava and Mukherjee, D. and Mukherjee, Soma and Mukherjee, Subroto and Mukherjee, Suvodip and Mukund, N. and Mullavey, A. and Munch, J. and Muñiz, E. A. and Murray, P. G. and Musenich, R. and Muusse, S. and Nadji, S. L. and Nagar, A. and Napolano, V. and Nardecchia, I. and Naticchioni, L. and Nayak, B. and Nayak, R. K. and Neil, B. F. and Neilson, J. and Nelemans, G. and Nelson, T. J. N. and Nery, M. and Neubauer, P. and Neunzert, A. and Ng, K. Y. and Ng, S. W. S. and Nguyen, C. and Nguyen, P. and Nguyen, T. and Nichols, S. A. and Nissanke, S. and Nitoglia, E. and Nocera, F. and Norman, M. and North, C. and Nuttall, L. K. and Oberling, J. and O’Brien, B. D. and O’Dell, J. and Oelker, E. and Oganesyan, G. and Oh, J. J. and Oh, S. H. and Ohme, F. and Ohta, H. and Okada, M. A. and Olivetto, C. and Oram, R. and O’Reilly, B. and Ormiston, R. G. and Ormsby, N. D. and Ortega, L. F. and O’Shaughnessy, R. and O’Shea, E. and Ossokine, S. and Osthelder, C. and Ottaway, D. J. and Overmier, H. and Pace, A. E. and Pagano, G. and Page, M. A. and Pagliaroli, G. and Pai, A. and Pai, S. A. and Palamos, J. R. and Palashov, O. and Palomba, C. and Pan, H. and Panda, P. K. and Pang, P. T. H. and Pankow, C. and Pannarale, F. and Pant, B. C. and Panther, F. H. and Paoletti, F. and Paoli, A. and Paolone, A. and Park, H. and Parker, W. and Pascucci, D. and Pasqualetti, A. and Passaquieti, R. and Passuello, D. and Patel, M. and Pathak, M. and Patricelli, B. and Patron, A. S. and Patrone, S. and Paul, S. and Payne, E. and Pedraza, M. and Pegoraro, M. and Pele, A. and Penn, S. and Perego, A. and Pereira, A. and Pereira, T. and Perez, C. J. and Périgois, C. and Perkins, C. C. and Perreca, A. and Perriès, S. and Petermann, J. and Petterson, D. and Pfeiffer, H. P. and Pham, K. A. and Phukon, K. S. and Piccinni, O. J. and Pichot, M. and Piendibene, M. and Piergiovanni, F. and Pierini, L. and Pierro, V. and Pillant, G. and Pillas, M. and Pilo, F. and Pinard, L. and Pinto, I. M. and Pinto, M. and Piotrzkowski, K. and Pirello, M. and Pitkin, M. D. and Placidi, E. and Planas, L. and Plastino, W. and Pluchar, C. and Poggiani, R. and Polini, E. and Pong, D. Y. T. and Ponrathnam, S. and Popolizio, P. and Porter, E. K. and Poulton, R. and Powell, J. and Pracchia, M. and Pradier, T. and Prajapati, A. K. and Prasai, K. and Prasanna, R. and Pratten, G. and Principe, M. and Prodi, G. A. and Prokhorov, L. and Prosposito, P. and Prudenzi, L. and Puecher, A. and Punturo, M. and Puosi, F. and Puppo, P. and Pürrer, M. and Qi, H. and Quetschke, V. and Quitzow-James, R. and Raab, F. J. and Raaijmakers, G. and Radkins, H. and Radulesco, N. and Raffai, P. and Rail, S. X. and Raja, S. and Rajan, C. and Ramirez, K. E. and Ramirez, T. D. and Ramos-Buades, A. and Rana, J. and Rapagnani, P. and Rapol, U. D. and Ray, A. and Raymond, V. and Raza, N. and Razzano, M. and Read, J. and Rees, L. A. and Regimbau, T. and Rei, L. and Reid, S. and Reid, S. W. and Reitze, D. H. and Relton, P. and Renzini, A. and Rettegno, P. and Reza, A. and Rezac, M. and Ricci, F. and Richards, D. and Richardson, J. W. and Richardson, L. and Riemenschneider, G. and Riles, K. and Rinaldi, S. and Rink, K. and Rizzo, M. and Robertson, N. A. and Robie, R. and Robinet, F. and Rocchi, A. and Rodriguez, S. and Rolland, L. and Rollins, J. G. and Romanelli, M. and Romano, R. and Romel, C. L. and Romero-Rodríguez, A. and Romero-Shaw, I. M. and Romie, J. H. and Ronchini, S. and Rosa, L. and Rose, C. A. and Rosell, M. J. B. and Rosińska, D. and Ross, M. P. and Rowan, S. and Rowlinson, S. J. and Roy, S. and Roy, Santosh and Roy, Soumen and Rozza, D. and Ruggi, P. and Ruiz-Rocha, K. and Ryan, K. and Sachdev, S. and Sadecki, T. and Sadiq, J. and Sakellariadou, M. and Salafia, O. S. and Salconi, L. and Saleem, M. and Salemi, F. and Samajdar, A. and Sanchez, E. J. and Sanchez, J. H. and Sanchez, L. E. and Sanchis-Gual, N. and Sanders, J. R. and Sanuy, A. and Saravanan, T. R. and Sarin, N. and Sassolas, B. and Satari, H. and Sauter, O. and Savage, R. L. and Sawant, D. and Sawant, H. L. and Sayah, S. and Schaetzl, D. and Scheel, M. and Scheuer, J. and Schiworski, M. and Schmidt, P. and Schmidt, S. and Schnabel, R. and Schneewind, M. and Schofield, R. M. S. and Schönbeck, A. and Schulte, B. W. and Schutz, B. F. and Schwartz, E. and Scott, J. and Scott, S. M. and Seglar-Arroyo, M. and Sellers, D. and Sengupta, A. S. and Sentenac, D. and Seo, E. G. and Sequino, V. and Sergeev, A. and Setyawati, Y. and Shaffer, T. and Shahriar, M. S. and Shams, B. and Sharma, A. and Sharma, P. and Shawhan, P. and Shcheblanov, N. S. and Shikauchi, M. and Shoemaker, D. H. and Shoemaker, D. M. and ShyamSundar, S. and Sieniawska, M. and Sigg, D. and Singer, L. P. and Singh, D. and Singh, N. and Singha, A. and Sintes, A. M. and Sipala, V. and Skliris, V. and Slagmolen, B. J. J. and Slaven-Blair, T. J. and Smetana, J. and Smith, J. R. and Smith, R. J. E. and Soldateschi, J. and Somala, S. N. and Son, E. J. and Soni, K. and Soni, S. and Sordini, V. and Sorrentino, F. and Sorrentino, N. and Soulard, R. and Souradeep, T. and Sowell, E. and Spagnuolo, V. and Spencer, A. P. and Spera, M. and Srinivasan, R. and Srivastava, A. K. and Srivastava, V. and Staats, K. and Stachie, C. and Steer, D. A. and Steinhoff, J. and Steinlechner, J. and Steinlechner, S. and Stevenson, S. and Stops, D. J. and Stover, M. and Strain, K. A. and Strang, L. C. and Stratta, G. and Strunk, A. and Sturani, R. and Stuver, A. L. and Sudhagar, S. and Sudhir, V. and Suh, H. G. and Summerscales, T. Z. and Sun, H. and Sun, L. and Sunil, S. and Sur, A. and Suresh, J. and Sutton, P. J. and Swinkels, B. L. and Szczepańczyk, M. J. and Szewczyk, P. and Tacca, M. and Tait, S. C. and Talbot, C. J. and Talbot, C. and Tanasijczuk, A. J. and Tanner, D. B. and Tao, D. and Tao, L. and Martín, E. N. Tapia San and Taranto, C. and Tasson, J. D. and Tenorio, R. and Terhune, J. E. and Terkowski, L. and Thirugnanasambandam, M. P. and Thomas, L. and Thomas, M. and Thomas, P. and Thompson, J. E. and Thondapu, S. R. and Thorne, K. A. and Thrane, E. and Tiwari, Shubhanshu and Tiwari, Srishti and Tiwari, V. and Toivonen, A. M. and Toland, K. and Tolley, A. E. and Tonelli, M. and Torres-Forné, A. and Torrie, C. I. and E Melo, I. Tosta and Töyrä, D. and Trapananti, A. and Travasso, F. and Traylor, G. and Trevor, M. and Tringali, M. C. and Tripathee, A. and Troiano, L. and Trovato, A. and Trozzo, L. and Trudeau, R. J. and Tsai, D. S. and Tsai, D. and Tsang, K. W. and Tse, M. and Tso, R. and Tsukada, L. and Tsuna, D. and Tsutsui, T. and Turbang, K. and Turconi, M. and Ubhi, A. S. and Udall, R. P. and Ueno, K. and Unnikrishnan, C. S. and Urban, A. L. and Utina, A. and Vahlbruch, H. and Vajente, G. and Vajpeyi, A. and Valdes, G. and Valentini, M. and Valsan, V. and Van Bakel, N. and Van Beuzekom, M. and Van Den Brand, J. F. J. and Van Den Broeck, C. and Vander-Hyde, D. C. and Van Der Schaaf, L. and Van Heijningen, J. V. and Vanosky, J. and Van Remortel, N. and Vardaro, M. and Vargas, A. F. and Varma, V. and Vasúth, M. and Vecchio, A. and Vedovato, G. and Veitch, J. and Veitch, P. J. and Venneberg, J. and Venugopalan, G. and Verkindt, D. and Verma, P. and Verma, Y. and Veske, D. and Vetrano, F. and Viceré, A. and Vidyant, S. and Viets, A. D. and Vijaykumar, A. and Villa-Ortega, V. and Vinet, J.-Y. and Virtuoso, A. and Vitale, S. and Vo, T. and Vocca, H. and Von Reis, E. R. G. and Von Wrangel, J. S. A. and Vorvick, C. and Vyatchanin, S. P. and Wade, L. E. and Wade, M. and Wagner, K. J. and Walet, R. C. and Walker, M. and Wallace, G. S. and Wallace, L. and Walsh, S. and Wang, J. Z. and Wang, W. H. and Ward, R. L. and Warner, J. and Was, M. and Washington, N. Y. and Watchi, J. and Weaver, B. and Webster, S. A. and Weinert, M. and Weinstein, A. J. and Weiss, R. and Weller, C. M. and Weller, R. and Wellmann, F. and Wen, L. and Weßels, P. and Wette, K. and Whelan, J. T. and White, D. D. and Whiting, B. F. and Whittle, C. and Wilken, D. and Williams, D. and Williams, M. J. and Williamson, A. R. and Willis, J. L. and Willke, B. and Wilson, D. J. and Winkler, W. and Wipf, C. C. and Wlodarczyk, T. and Woan, G. and Woehler, J. and Wofford, J. K. and Wong, I. C. F. and Wu, D. S. and Wysocki, D. M. and Xiao, L. and Yamamoto, H. and Yang, F. W. and Yang, L. and Yang, Yang and Yang, Z. and Yap, M. J. and Yeeles, D. W. and Yelikar, A. B. and Ying, M. and Yoo, J. and Yu, Hang and Yu, Haocun and Zadrożny, A. and Zanolin, M. and Zelenova, T. and Zendri, J.-P. and Zevin, M. and Zhang, J. and Zhang, L. and Zhang, T. and Zhang, Y. and Zhao, C. and Zhao, G. and Zhao, Yue and Zhou, R. and Zhou, Z. and Zhu, X. J. and Zimmerman, A. B. and Zlochower, Y. and Zucker, M. E. and Zweizig, J. and {The LIGO Scientific Collaboration and the Virgo Collaboration}},
	month = jan,
	year = {2024},
	note = {Version Number: 2},
	pages = {022001},
}

@article{catalog-gwtc-3,
	title = {{GWTC}-3: {Compact} {Binary} {Coalescences} {Observed} by {LIGO} and {Virgo} during the {Second} {Part} of the {Third} {Observing} {Run}},
	volume = {13},
	shorttitle = {{GWTC}-3},
	url = {https://ui.adsabs.harvard.edu/abs/2023PhRvX..13d1039A},
	doi = {10.1103/PhysRevX.13.041039},
	urldate = {2025-04-03},
	journal = {Physical Review X},
	author = {Abbott, R. and Abbott, T. D. and Acernese, F. and Ackley, K. and Adams, C. and Adhikari, N. and Adhikari, R. X. and Adya, V. B. and Affeldt, C. and Agarwal, D. and Agathos, M. and Agatsuma, K. and Aggarwal, N. and Aguiar, O. D. and Aiello, L. and Ain, A. and Ajith, P. and Akcay, S. and Akutsu, T. and Albanesi, S. and Allocca, A. and Altin, P. A. and Amato, A. and Anand, C. and Anand, S. and Ananyeva, A. and Anderson, S. B. and Anderson, W. G. and Ando, M. and Andrade, T. and Andres, N. and Andrić, T. and Angelova, S. V. and Ansoldi, S. and Antelis, J. M. and Antier, S. and Appert, S. and Arai, Koji and Arai, Koya and Arai, Y. and Araki, S. and Araya, A. and Araya, M. C. and Areeda, J. S. and Arène, M. and Aritomi, N. and Arnaud, N. and Arogeti, M. and Aronson, S. M. and Arun, K. G. and Asada, H. and Asali, Y. and Ashton, G. and Aso, Y. and Assiduo, M. and Aston, S. M. and Astone, P. and Aubin, F. and Austin, C. and Babak, S. and Badaracco, F. and Bader, M. K. M. and Badger, C. and Bae, S. and Bae, Y. and Baer, A. M. and Bagnasco, S. and Bai, Y. and Baiotti, L. and Baird, J. and Bajpai, R. and Ball, M. and Ballardin, G. and Ballmer, S. W. and Balsamo, A. and Baltus, G. and Banagiri, S. and Bankar, D. and Barayoga, J. C. and Barbieri, C. and Barish, B. C. and Barker, D. and Barneo, P. and Barone, F. and Barr, B. and Barsotti, L. and Barsuglia, M. and Barta, D. and Bartlett, J. and Barton, M. A. and Bartos, I. and Bassiri, R. and Basti, A. and Bawaj, M. and Bayley, J. C. and Baylor, A. C. and Bazzan, M. and Bécsy, B. and Bedakihale, V. M. and Bejger, M. and Belahcene, I. and Benedetto, V. and Beniwal, D. and Bennett, T. F. and Bentley, J. D. and Benyaala, M. and Bergamin, F. and Berger, B. K. and Bernuzzi, S. and Berry, C. P. L. and Bersanetti, D. and Bertolini, A. and Betzwieser, J. and Beveridge, D. and Bhandare, R. and Bhardwaj, U. and Bhattacharjee, D. and Bhaumik, S. and Bilenko, I. A. and Billingsley, G. and Bini, S. and Birney, R. and Birnholtz, O. and Biscans, S. and Bischi, M. and Biscoveanu, S. and Bisht, A. and Biswas, B. and Bitossi, M. and Bizouard, M. -A. and Blackburn, J. K. and Blair, C. D. and Blair, D. G. and Blair, R. M. and Bobba, F. and Bode, N. and Boer, M. and Bogaert, G. and Boldrini, M. and Bonavena, L. D. and Bondu, F. and Bonilla, E. and Bonnand, R. and Booker, P. and Boom, B. A. and Bork, R. and Boschi, V. and Bose, N. and Bose, S. and Bossilkov, V. and Boudart, V. and Bouffanais, Y. and Bozzi, A. and Bradaschia, C. and Brady, P. R. and Bramley, A. and Branch, A. and Branchesi, M. and Brandt, J. and Brau, J. E. and Breschi, M. and Briant, T. and Briggs, J. H. and Brillet, A. and Brinkmann, M. and Brockill, P. and Brooks, A. F. and Brooks, J. and Brown, D. D. and Brunett, S. and Bruno, G. and Bruntz, R. and Bryant, J. and Bulik, T. and Bulten, H. J. and Buonanno, A. and Buscicchio, R. and Buskulic, D. and Buy, C. and Byer, R. L. and Davies, G. S. Cabourn and Cadonati, L. and Cagnoli, G. and Cahillane, C. and Bustillo, J. Calderón and Callaghan, J. D. and Callister, T. A. and Calloni, E. and Cameron, J. and Camp, J. B. and Canepa, M. and Canevarolo, S. and Cannavacciuolo, M. and Cannon, K. C. and Cao, H. and Cao, Z. and Capocasa, E. and Capote, E. and Carapella, G. and Carbognani, F.},
	month = oct,
	year = {2023},
	note = {Publisher: APS
ADS Bibcode: 2023PhRvX..13d1039A},
	pages = {041039},
}

@misc{code-bilby-2d1d1,
	title = {bilby-dev/bilby: v2.1.1},
	url = {https://doi.org/10.5281/zenodo.14025640},
	publisher = {Zenodo},
	author = {Talbot, Colm and Ashton, Gregory and Hübner, Moritz and Pitkin, Matt and {plasky} and {asb5468} and Williams, Michael J. and Smith, Rory and Vijaykumar, Aditya and {SMorisaki} and Veitch, John and Sarin, Nikhil and Edelman, Bruce and Macleod, Duncan and Williams, Daniel and {MarcArene} and {JasperMartins} and Berry, C P L and Raymond, Vivien and Keitel, David and {Ceciliogq} and {oliviawilk} and {AlexandreGoettel} and {BenGPatterson} and Hoy, Charlie and Legred, Isaac and {IsobelMarguarethe} and {Jacopo} and Willis, Josh and Wette, Karl},
	month = nov,
	year = {2024},
}

@article{code-rift-1,
	title = {An architecture for efficient gravitational wave parameter estimation with multimodal linear surrogate models},
	volume = {34},
	issn = {0264-9381},
	url = {https://dx.doi.org/10.1088/1361-6382/aa7649},
	doi = {10.1088/1361-6382/aa7649},
	language = {en},
	number = {14},
	urldate = {2024-09-10},
	journal = {Classical and Quantum Gravity},
	author = {O’Shaughnessy, Richard and Blackman, Jonathan and Field, Scott E.},
	month = jun,
	year = {2017},
	note = {Publisher: IOP Publishing},
	pages = {144002},
}

@article{code-rift-2,
	title = {Parameter estimation method that directly compares gravitational wave observations to numerical relativity},
	volume = {96},
	url = {https://link.aps.org/doi/10.1103/PhysRevD.96.104041},
	doi = {10.1103/PhysRevD.96.104041},
	number = {10},
	urldate = {2025-04-03},
	journal = {Physical Review D},
	author = {Lange, J. and O’Shaughnessy, R. and Boyle, M. and Calderón Bustillo, J. and Campanelli, M. and Chu, T. and Clark, J. A. and Demos, N. and Fong, H. and Healy, J. and Hemberger, D. A. and Hinder, I. and Jani, K. and Khamesra, B. and Kidder, L. E. and Kumar, P. and Laguna, P. and Lousto, C. O. and Lovelace, G. and Ossokine, S. and Pfeiffer, H. and Scheel, M. A. and Shoemaker, D. M. and Szilagyi, B. and Teukolsky, S. and Zlochower, Y.},
	month = nov,
	year = {2017},
	note = {Publisher: American Physical Society},
	pages = {104041},
}

@article{code-rift-3,
	title = {Accelerating parameter inference with graphics processing units},
	volume = {99},
	url = {https://link.aps.org/doi/10.1103/PhysRevD.99.084026},
	doi = {10.1103/PhysRevD.99.084026},
	number = {8},
	urldate = {2025-04-03},
	journal = {Physical Review D},
	author = {Wysocki, D. and O’Shaughnessy, R. and Lange, Jacob and Fang, Yao-Lung L.},
	month = apr,
	year = {2019},
	note = {Publisher: American Physical Society},
	pages = {084026},
}

@article{code-rift-0,
	title = {Novel scheme for rapid parallel parameter estimation of gravitational waves from compact binary coalescences},
	volume = {92},
	url = {https://link.aps.org/doi/10.1103/PhysRevD.92.023002},
	doi = {10.1103/PhysRevD.92.023002},
	number = {2},
	urldate = {2025-04-03},
	journal = {Physical Review D},
	author = {Pankow, C. and Brady, P. and Ochsner, E. and O’Shaughnessy, R.},
	month = jul,
	year = {2015},
	note = {Publisher: American Physical Society},
	pages = {023002},
}

@article{rodriguez_illuminating_2016,
	title = {Illuminating {Black} {Hole} {Binary} {Formation} {Channels} with {Spins} in {Advanced} {Ligo}},
	volume = {832},
	issn = {2041-8205},
	url = {https://dx.doi.org/10.3847/2041-8205/832/1/L2},
	doi = {10.3847/2041-8205/832/1/L2},
	language = {en},
	number = {1},
	urldate = {2025-04-01},
	journal = {The Astrophysical Journal Letters},
	author = {Rodriguez, Carl L. and Zevin, Michael and Pankow, Chris and Kalogera, Vasilliki and Rasio, Frederic A.},
	month = nov,
	year = {2016},
	pages = {L2},
}

@article{catalog-cwb,
	title = {Gravitational waves detected by a burst search in {LIGO}/{Virgo}'s third observing run},
	volume = {111},
	issn = {1550-79980556-2821},
	url = {https://ui.adsabs.harvard.edu/abs/2025PhRvD.111b3054M},
	doi = {10.1103/PhysRevD.111.023054},
	urldate = {2025-04-03},
	journal = {Physical Review D},
	author = {Mishra, T. and Bhaumik, S. and Gayathri, V. and Szczepańczyk, Marek J. and Bartos, I. and Klimenko, S.},
	month = jan,
	year = {2025},
	note = {Publication Title: arXiv e-prints
Publisher: APS
ADS Bibcode: 2025PhRvD.111b3054M
ADS Bibcode: 2024arXiv241015191M},
	pages = {023054},
}

@article{catalog-ias-4,
	title = {New binary black hole mergers in the {LIGO}-{Virgo} {O3b} data},
	volume = {111},
	issn = {1550-79980556-2821},
	url = {https://ui.adsabs.harvard.edu/abs/2025PhRvD.111b4049M},
	doi = {10.1103/PhysRevD.111.024049},
	urldate = {2025-04-03},
	journal = {Physical Review D},
	author = {Mehta, Ajit Kumar and Olsen, Seth and Wadekar, Digvijay and Roulet, Javier and Venumadhav, Tejaswi and Mushkin, Jonathan and Zackay, Barak and Zaldarriaga, Matias},
	month = jan,
	year = {2025},
	note = {Publication Title: arXiv e-prints
Publisher: APS
ADS Bibcode: 2025PhRvD.111b4049M
ADS Bibcode: 2023arXiv231106061M},
	pages = {024049},
}

@article{search-pycbc,
	title = {The {PyCBC} search for gravitational waves from compact binary coalescence},
	volume = {33},
	issn = {0264-9381},
	url = {https://ui.adsabs.harvard.edu/abs/2016CQGra..33u5004U},
	doi = {10.1088/0264-9381/33/21/215004},
	urldate = {2024-08-14},
	journal = {Classical and Quantum Gravity},
	author = {Usman, Samantha A. and Nitz, Alexander H. and Harry, Ian W. and Biwer, Christopher M. and Brown, Duncan A. and Cabero, Miriam and Capano, Collin D. and Dal Canton, Tito and Dent, Thomas and Fairhurst, Stephen and Kehl, Marcel S. and Keppel, Drew and Krishnan, Badri and Lenon, Amber and Lundgren, Andrew and Nielsen, Alex B. and Pekowsky, Larne P. and Pfeiffer, Harald P. and Saulson, Peter R. and West, Matthew and Willis, Joshua L.},
	month = nov,
	year = {2016},
	pages = {215004},
}

@article{catalog-gwares,
	title = {New gravitational wave discoveries enabled by machine learning},
	volume = {6},
	url = {https://ui.adsabs.harvard.edu/abs/2025MLS&T...6a5054K},
	doi = {10.1088/2632-2153/adb5ed},
	urldate = {2025-04-03},
	journal = {Machine Learning: Science and Technology},
	author = {Koloniari, Alexandra E. and Koursoumpa, Evdokia C. and Nousi, Paraskevi and Lampropoulos, Paraskevas and Passalis, Nikolaos and Tefas, Anastasios and Stergioulas, Nikolaos},
	month = mar,
	year = {2025},
	note = {Publisher: IOP
ADS Bibcode: 2025MLS\&T...6a5054K},
	pages = {015054},
}

@article{ashton_parameterised_2022,
	title = {Parameterised population models of transient non-{Gaussian} noise in the {LIGO} gravitational-wave detectors},
	volume = {39},
	issn = {0264-9381, 1361-6382},
	url = {http://arxiv.org/abs/2110.02689},
	doi = {10.1088/1361-6382/ac8094},
	number = {17},
	urldate = {2025-04-01},
	journal = {Classical and Quantum Gravity},
	author = {Ashton, Gregory and Thiele, Sarah and Lecoeuche, Yannick and McIver, Jess and Nuttall, Laura K.},
	month = sep,
	year = {2022},
	note = {arXiv:2110.02689 [gr-qc]},
	pages = {175004},
}

@article{collaboration_open_2021,
	title = {Open data from the first and second observing runs of {Advanced} {LIGO} and {Advanced} {Virgo}},
	volume = {13},
	issn = {23527110},
	url = {http://arxiv.org/abs/1912.11716},
	doi = {10.1016/j.softx.2021.100658},
	urldate = {2025-03-21},
	journal = {SoftwareX},
	author = {Collaboration, The LIGO Scientific and Collaboration, the Virgo and Abbott, R. and Abbott, T. D. and Abraham, S. and Acernese, F. and Ackley, K. and Adams, C. and Adhikari, R. X. and Adya, V. B. and Affeldt, C. and Agathos, M. and Agatsuma, K. and Aggarwal, N. and Aguiar, O. D. and Aich, A. and Aiello, L. and Ain, A. and Ajith, P. and Allen, G. and Allocca, A. and Altin, P. A. and Amato, A. and Anand, S. and Ananyeva, A. and Anderson, S. B. and Anderson, W. G. and Angelova, S. V. and Ansoldi, S. and Antier, S. and Appert, S. and Arai, K. and Araya, M. C. and Areeda, J. S. and Arène, M. and Arnaud, N. and Aronson, S. M. and Arun, K. G. and Ascenzi, S. and Ashton, G. and Aston, S. M. and Astone, P. and Aubin, F. and Aufmuth, P. and AultONeal, K. and Austin, C. and Avendano, V. and Babak, S. and Bacon, P. and Badaracco, F. and Bader, M. K. M. and Bae, S. and Baer, A. M. and Baird, J. and Baldaccini, F. and Ballardin, G. and Ballmer, S. W. and Bals, A. and Balsamo, A. and Baltus, G. and Banagiri, S. and Bankar, D. and Bankar, R. S. and Barayoga, J. C. and Barbieri, C. and Barish, B. C. and Barker, D. and Barkett, K. and Barneo, P. and Barone, F. and Barr, B. and Barsotti, L. and Barsuglia, M. and Barta, D. and Bartlett, J. and Bartos, I. and Bassiri, R. and Basti, A. and Bawaj, M. and Bayley, J. C. and Bazzan, M. and Bécsy, B. and Bejger, M. and Belahcene, I. and Bell, A. S. and Beniwal, D. and Benjamin, M. G. and Bentley, J. D. and Bergamin, F. and Berger, B. K. and Bergmann, G. and Bernuzzi, S. and Berry, C. P. L. and Bersanetti, D. and Bertolini, A. and Betzwieser, J. and Bhandare, R. and Bhandari, A. V. and Bidler, J. and Biggs, E. and Bilenko, I. A. and Billingsley, G. and Birney, R. and Birnholtz, O. and Biscans, S. and Bischi, M. and Biscoveanu, S. and Bisht, A. and Bissenbayeva, G. and Bitossi, M. and Bizouard, M. A. and Blackburn, J. K. and Blackman, J. and Blair, C. D. and Blair, D. G. and Blair, R. M. and Bobba, F. and Bode, N. and Boer, M. and Boetzel, Y. and Bogaert, G. and Bondu, F. and Bonilla, E. and Bonnand, R. and Booker, P. and Boom, B. A. and Bork, R. and Boschi, V. and Bose, S. and Bossilkov, V. and Bosveld, J. and Bouffanais, Y. and Bozzi, A. and Bradaschia, C. and Brady, P. R. and Bramley, A. and Branchesi, M. and Brau, J. E. and Breschi, M. and Briant, T. and Briggs, J. H. and Brighenti, F. and Brillet, A. and Brinkmann, M. and Brockill, P. and Brooks, A. F. and Brooks, J. and Brown, D. D. and Brunett, S. and Bruno, G. and Bruntz, R. and Buikema, A. and Bulik, T. and Bulten, H. J. and Buonanno, A. and Buskulic, D. and Byer, R. L. and Cabero, M. and Cadonati, L. and Cagnoli, G. and Cahillane, C. and Bustillo, J. Calderón and Callaghan, J. D. and Callister, T. A. and Calloni, E. and Camp, J. B. and Canepa, M. and Cannon, K. C. and Cao, H. and Cao, J. and Carapella, G. and Carbognani, F. and Caride, S. and Carney, M. F. and Carullo, G. and Diaz, J. Casanueva and Casentini, C. and Castañeda, J. and Caudill, S. and Cavaglià, M. and Cavalier, F. and Cavalieri, R. and Cella, G. and Cerdá-Durán, P. and Cesarini, E. and Chaibi, O. and Chakravarti, K. and Chan, C. and Chan, M. and Chao, S. and Charlton, P. and Chase, E. A. and Chassande-Mottin, E. and Chatterjee, D. and Chaturvedi, M. and Chen, H. Y. and Chen, X. and Chen, Y. and Cheng, H.-P. and Cheong, C. K. and Chia, H. Y. and Chiadini, F. and Chierici, R. and Chincarini, A. and Chiummo, A. and Cho, G. and Cho, H. S. and Cho, M. and Christensen, N. and Chu, Q. and Chua, S. and Chung, K. W. and Chung, S. and Ciani, G. and Ciecielag, P. and Cie\{ś\}lar, M. and Ciobanu, A. A. and Ciolfi, R. and Cipriano, F. and Cirone, A. and Clara, F. and Clark, J. A. and Clearwater, P. and Clesse, S. and Cleva, F. and Coccia, E. and Cohadon, P.-F. and Cohen, D. and Colleoni, M. and Collette, C. G. and Collins, C. and Colpi, M. and Jr, M. Constancio and Conti, L. and Cooper, S. J. and Corban, P. and Corbitt, T. R. and Cordero-Carrión, I. and Corezzi, S. and Corley, K. R. and Cornish, N. and Corre, D. and Corsi, A. and Cortese, S. and Costa, C. A. and Cotesta, R. and Coughlin, M. W. and Coughlin, S. B. and Coulon, J.-P. and Countryman, S. T. and Couvares, P. and Covas, P. B. and Coward, D. M. and Cowart, M. J. and Coyne, D. C. and Coyne, R. and Creighton, J. D. E. and Creighton, T. D. and Cripe, J. and Croquette, M. and Crowder, S. G. and Cudell, J.-R. and Cullen, T. J. and Cumming, A. and Cummings, R. and Cunningham, L. and Cuoco, E. and Curylo, M. and Canton, T. Dal and Dálya, G. and Dana, A. and Daneshgaran-Bajastani, L. M. and D'Angelo, B. and Danilishin, S. L. and D'Antonio, S. and Danzmann, K. and Darsow-Fromm, C. and Dasgupta, A. and Datrier, L. E. H. and Dattilo, V. and Dave, I. and Davier, M. and Davies, G. S. and Davis, D. and Daw, E. J. and DeBra, D. and Deenadayalan, M. and Degallaix, J. and Laurentis, M. De and Deléglise, S. and Delfavero, M. and Lillo, N. De and Pozzo, W. Del and DeMarchi, L. M. and D'Emilio, V. and Demos, N. and Dent, T. and Pietri, R. De and Rosa, R. De and Rossi, C. De and DeSalvo, R. and Varona, O. de and Dhurandhar, S. and Díaz, M. C. and Jr, M. Diaz-Ortiz and Dietrich, T. and Fiore, L. Di and Fronzo, C. Di and Giorgio, C. Di and Giovanni, F. Di and Giovanni, M. Di and Girolamo, T. Di and Lieto, A. Di and Ding, B. and Pace, S. Di and Palma, I. Di and Renzo, F. Di and Divakarla, A. K. and Dmitriev, A. and Doctor, Z. and Donovan, F. and Dooley, K. L. and Doravari, S. and Dorrington, I. and Downes, T. P. and Drago, M. and Driggers, J. C. and Du, Z. and Ducoin, J.-G. and Dupej, P. and Durante, O. and D'Urso, D. and Dwyer, S. E. and Easter, P. J. and Eddolls, G. and Edelman, B. and Edo, T. B. and Edy, O. and Effler, A. and Ehrens, P. and Eichholz, J. and Eikenberry, S. S. and Eisenmann, M. and Eisenstein, R. A. and Ejlli, A. and Errico, L. and Essick, R. C. and Estelles, H. and Estevez, D. and Etienne, Z. B. and Etzel, T. and Evans, M. and Evans, T. M. and Ewing, B. E. and Fafone, V. and Fairhurst, S. and Fan, X. and Farinon, S. and Farr, B. and Farr, W. M. and Fauchon-Jones, E. J. and Favata, M. and Fays, M. and Fazio, M. and Feicht, J. and Fejer, M. M. and Feng, F. and Fenyvesi, E. and Ferguson, D. L. and Fernandez-Galiana, A. and Ferrante, I. and Ferreira, E. C. and Ferreira, T. A. and Fidecaro, F. and Fiori, I. and Fiorucci, D. and Fishbach, M. and Fisher, R. P. and Fittipaldi, R. and Fitz-Axen, M. and Fiumara, V. and Flaminio, R. and Floden, E. and Flynn, E. and Fong, H. and Font, J. A. and Forsyth, P. W. F. and Fournier, J.-D. and Frasca, S. and Frasconi, F. and Frei, Z. and Freise, A. and Frey, R. and Frey, V. and Fritschel, P. and Frolov, V. V. and Fronzè, G. and Fulda, P. and Fyffe, M. and Gabbard, H. A. and Gadre, B. U. and Gaebel, S. M. and Gair, J. R. and Galaudage, S. and Ganapathy, D. and Gaonkar, S. G. and García-Quirós, C. and Garufi, F. and Gateley, B. and Gaudio, S. and Gayathri, V. and Gemme, G. and Genin, E. and Gennai, A. and George, D. and George, J. and Gergely, L. and Ghonge, S. and Ghosh, Abhirup and Ghosh, Archisman and Ghosh, S. and Giacomazzo, B. and Giaime, J. A. and Giardina, K. D. and Gibson, D. R. and Gier, C. and Gill, K. and Glanzer, J. and Gniesmer, J. and Godwin, P. and Goetz, E. and Goetz, R. and Gohlke, N. and Goncharov, B. and González, G. and Gopakumar, A. and Gossan, S. E. and Gosselin, M. and Gouaty, R. and Grace, B. and Grado, A. and Granata, M. and Grant, A. and Gras, S. and Grassia, P. and Gray, C. and Gray, R. and Greco, G. and Green, A. C. and Green, R. and Gretarsson, E. M. and Griggs, H. L. and Grignani, G. and Grimaldi, A. and Grimm, S. J. and Grote, H. and Grunewald, S. and Gruning, P. and Guidi, G. M. and Guimaraes, A. R. and Guixé, G. and Gulati, H. K. and Guo, Y. and Gupta, A. and Gupta, Anchal and Gupta, P. and Gustafson, E. K. and Gustafson, R. and Haegel, L. and Halim, O. and Hall, E. D. and Hamilton, E. Z. and Hammond, G. and Haney, M. and Hanke, M. M. and Hanks, J. and Hanna, C. and Hannam, M. D. and Hannuksela, O. A. and Hansen, T. J. and Hanson, J. and Harder, T. and Hardwick, T. and Haris, K. and Harms, J. and Harry, G. M. and Harry, I. W. and Hasskew, R. K. and Haster, C.-J. and Haughian, K. and Hayes, F. J. and Healy, J. and Heidmann, A. and Heintze, M. C. and Heinze, J. and Heitmann, H. and Hellman, F. and Hello, P. and Hemming, G. and Hendry, M. and Heng, I. S. and Hennes, E. and Hennig, J. and Heurs, M. and Hild, S. and Hinderer, T. and Hoback, S. Y. and Hochheim, S. and Hofgard, E. and Hofman, D. and Holgado, A. M. and Holland, N. A. and Holt, K. and Holz, D. E. and Hopkins, P. and Horst, C. and Hough, J. and Howell, E. J. and Hoy, C. G. and Huang, Y. and Hübner, M. T. and Huerta, E. A. and Huet, D. and Hughey, B. and Hui, V. and Husa, S. and Huttner, S. H. and Huxford, R. and Huynh-Dinh, T. and Idzkowski, B. and Iess, A. and Inchauspe, H. and Ingram, C. and Intini, G. and Isac, J.-M. and Isi, M. and Iyer, B. R. and Jacqmin, T. and Jadhav, S. J. and Jadhav, S. P. and James, A. L. and Jani, K. and Janthalur, N. N. and Jaranowski, P. and Jariwala, D. and Jaume, R. and Jenkins, A. C. and Jiang, J. and Johns, G. R. and Jones, A. W. and Jones, D. I. and Jones, J. D. and Jones, P. and Jones, R. and Jonker, R. J. G. and Ju, L. and Junker, J. and Kalaghatgi, C. V. and Kalogera, V. and Kamai, B. and Kandhasamy, S. and Kang, G. and Kanner, J. B. and Kapadia, S. J. and Karki, S. and Kashyap, R. and Kasprzack, M. and Kastaun, W. and Katsanevas, S. and Katsavounidis, E. and Katzman, W. and Kaufer, S. and Kawabe, K. and Kéfélian, F. and Keitel, D. and Keivani, A. and Kennedy, R. and Key, J. S. and Khadka, S. and Khalili, F. Y. and Khan, I. and Khan, S. and Khan, Z. A. and Khazanov, E. A. and Khetan, N. and Khursheed, M. and Kijbunchoo, N. and Kim, Chunglee and Kim, G. J. and Kim, J. C. and Kim, K. and Kim, W. and Kim, W. S. and Kim, Y.-M. and Kimball, C. and King, P. J. and Kinley-Hanlon, M. and Kirchhoff, R. and Kissel, J. S. and Kleybolte, L. and Klimenko, S. and Knowles, T. D. and Koch, P. and Koehlenbeck, S. M. and Koekoek, G. and Koley, S. and Kondrashov, V. and Kontos, A. and Koper, N. and Korobko, M. and Korth, W. Z. and Kovalam, M. and Kozak, D. B. and Kringel, V. and Krishnendu, N. V. and Królak, A. and Krupinski, N. and Kuehn, G. and Kumar, A. and Kumar, P. and Kumar, Rahul and Kumar, Rakesh and Kumar, S. and Kuo, L. and Kutynia, A. and Lackey, B. D. and Laghi, D. and Lalande, E. and Lam, T. L. and Lamberts, A. and Landry, M. and Lane, B. B. and Lang, R. N. and Lange, J. and Lantz, B. and Lanza, R. K. and Rosa, I. La and Lartaux-Vollard, A. and Lasky, P. D. and Laxen, M. and Lazzarini, A. and Lazzaro, C. and Leaci, P. and Leavey, S. and Lecoeuche, Y. K. and Lee, C. H. and Lee, H. M. and Lee, H. W. and Lee, J. and Lee, K. and Lehmann, J. and Leroy, N. and Letendre, N. and Levin, Y. and Li, A. K. Y. and Li, J. and li, K. and Li, T. G. F. and Li, X. and Linde, F. and Linker, S. D. and Linley, J. N. and Littenberg, T. B. and Liu, J. and Liu, X. and Llorens-Monteagudo, M. and Lo, R. K. L. and Lockwood, A. and London, L. T. and Longo, A. and Lorenzini, M. and Loriette, V. and Lormand, M. and Losurdo, G. and Lough, J. D. and Lousto, C. O. and Lovelace, G. and Lück, H. and Lumaca, D. and Lundgren, A. P. and Ma, Y. and Macas, R. and Macfoy, S. and MacInnis, M. and Macleod, D. M. and MacMillan, I. A. O. and Macquet, A. and Hernandez, I. Magaña and Magaña-Sandoval, F. and Magee, R. M. and Majorana, E. and Maksimovic, I. and Malik, A. and Man, N. and Mandic, V. and Mangano, V. and Mansell, G. L. and Manske, M. and Mantovani, M. and Mapelli, M. and Marchesoni, F. and Marion, F. and Márka, S. and Márka, Z. and Markakis, C. and Markosyan, A. S. and Markowitz, A. and Maros, E. and Marquina, A. and Marsat, S. and Martelli, F. and Martin, I. W. and Martin, R. M. and Martinez, V. and Martynov, D. V. and Masalehdan, H. and Mason, K. and Massera, E. and Masserot, A. and Massinger, T. J. and Masso-Reid, M. and Mastrogiovanni, S. and Matas, A. and Matichard, F. and Mavalvala, N. and Maynard, E. and McCann, J. J. and McCarthy, R. and McClelland, D. E. and McCormick, S. and McCuller, L. and McGuire, S. C. and McIsaac, C. and McIver, J. and McManus, D. J. and McRae, T. and McWilliams, S. T. and Meacher, D. and Meadors, G. D. and Mehmet, M. and Mehta, A. K. and Villa, E. Mejuto and Melatos, A. and Mendell, G. and Mercer, R. A. and Mereni, L. and Merfeld, K. and Merilh, E. L. and Merritt, J. D. and Merzougui, M. and Meshkov, S. and Messenger, C. and Messick, C. and Metzdorff, R. and Meyers, P. M. and Meylahn, F. and Mhaske, A. and Miani, A. and Miao, H. and Michaloliakos, I. and Michel, C. and Middleton, H. and Milano, L. and Miller, A. L. and Millhouse, M. and Mills, J. C. and Milotti, E. and Milovich-Goff, M. C. and Minazzoli, O. and Minenkov, Y. and Mishkin, A. and Mishra, C. and Mistry, T. and Mitra, S. and Mitrofanov, V. P. and Mitselmakher, G. and Mittleman, R. and Mo, G. and Mogushi, K. and Mohapatra, S. R. P. and Mohite, S. R. and Molina-Ruiz, M. and Mondin, M. and Montani, M. and Moore, C. J. and Moraru, D. and Morawski, F. and Moreno, G. and Morisaki, S. and Mours, B. and Mow-Lowry, C. M. and Mozzon, S. and Muciaccia, F. and Mukherjee, Arunava and Mukherjee, D. and Mukherjee, S. and Mukherjee, Subroto and Mukund, N. and Mullavey, A. and Munch, J. and Muñiz, E. A. and Murray, P. G. and Nagar, A. and Nardecchia, I. and Naticchioni, L. and Nayak, R. K. and Neil, B. F. and Neilson, J. and Nelemans, G. and Nelson, T. J. N. and Nery, M. and Neunzert, A. and Ng, K. Y. and Ng, S. and Nguyen, C. and Nguyen, P. and Nichols, D. and Nichols, S. A. and Nissanke, S. and Nocera, F. and Noh, M. and North, C. and Nothard, D. and Nuttall, L. K. and Oberling, J. and O'Brien, B. D. and Oganesyan, G. and Ogin, G. H. and Oh, J. J. and Oh, S. H. and Ohme, F. and Ohta, H. and Okada, M. A. and Oliver, M. and Olivetto, C. and Oppermann, P. and Oram, Richard J. and O'Reilly, B. and Ormiston, R. G. and Ortega, L. F. and O'Shaughnessy, R. and Ossokine, S. and Osthelder, C. and Ottaway, D. J. and Overmier, H. and Owen, B. J. and Pace, A. E. and Pagano, G. and Page, M. A. and Pagliaroli, G. and Pai, A. and Pai, S. A. and Palamos, J. R. and Palashov, O. and Palomba, C. and Pan, H. and Panda, P. K. and Pang, P. T. H. and Pankow, C. and Pannarale, F. and Pant, B. C. and Paoletti, F. and Paoli, A. and Parida, A. and Parker, W. and Pascucci, D. and Pasqualetti, A. and Passaquieti, R. and Passuello, D. and Patricelli, B. and Payne, E. and Pearlstone, B. L. and Pechsiri, T. C. and Pedersen, A. J. and Pedraza, M. and Pele, A. and Penn, S. and Perego, A. and Perez, C. J. and Périgois, C. and Perreca, A. and Perriès, S. and Petermann, J. and Pfeiffer, H. P. and Phelps, M. and Phukon, K. S. and Piccinni, O. J. and Pichot, M. and Piendibene, M. and Piergiovanni, F. and Pierro, V. and Pillant, G. and Pinard, L. and Pinto, I. M. and Piotrzkowski, K. and Pirello, M. and Pitkin, M. and Plastino, W. and Poggiani, R. and Pong, D. Y. T. and Ponrathnam, S. and Popolizio, P. and Porter, E. K. and Powell, J. and Prajapati, A. K. and Prasai, K. and Prasanna, R. and Pratten, G. and Prestegard, T. and Principe, M. and Prodi, G. A. and Prokhorov, L. and Punturo, M. and Puppo, P. and Pürrer, M. and Qi, H. and Quetschke, V. and Quinonez, P. J. and Raab, F. J. and Raaijmakers, G. and Radkins, H. and Radulesco, N. and Raffai, P. and Rafferty, H. and Raja, S. and Rajan, C. and Rajbhandari, B. and Rakhmanov, M. and Ramirez, K. E. and Ramos-Buades, A. and Rana, Javed and Rao, K. and Rapagnani, P. and Raymond, V. and Razzano, M. and Read, J. and Regimbau, T. and Rei, L. and Reid, S. and Reitze, D. H. and Rettegno, P. and Ricci, F. and Richardson, C. J. and Richardson, J. W. and Ricker, P. M. and Riemenschneider, G. and Riles, K. and Rizzo, M. and Robertson, N. A. and Robinet, F. and Rocchi, A. and Rodriguez-Soto, R. D. and Rolland, L. and Rollins, J. G. and Roma, V. J. and Romanelli, M. and Romano, R. and Romel, C. L. and Romero-Shaw, I. M. and Romie, J. H. and Rose, C. A. and Rose, D. and Rose, K. and Rosińska, D. and Rosofsky, S. G. and Ross, M. P. and Rowan, S. and Rowlinson, S. J. and Roy, P. K. and Roy, Santosh and Roy, Soumen and Ruggi, P. and Rutins, G. and Ryan, K. and Sachdev, S. and Sadecki, T. and Sakellariadou, M. and Salafia, O. S. and Salconi, L. and Saleem, M. and Samajdar, A. and Sanchez, E. J. and Sanchez, L. E. and Sanchis-Gual, N. and Sanders, J. R. and Santiago, K. A. and Santos, E. and Sarin, N. and Sassolas, B. and Sathyaprakash, B. S. and Sauter, O. and Savage, R. L. and Savant, V. and Sawant, D. and Sayah, S. and Schaetzl, D. and Schale, P. and Scheel, M. and Scheuer, J. and Schmidt, P. and Schnabel, R. and Schofield, R. M. S. and Schönbeck, A. and Schreiber, E. and Schulte, B. W. and Schutz, B. F. and Schwarm, O. and Schwartz, E. and Scott, J. and Scott, S. M. and Seidel, E. and Sellers, D. and Sengupta, A. S. and Sennett, N. and Sentenac, D. and Sequino, V. and Sergeev, A. and Setyawati, Y. and Shaddock, D. A. and Shaffer, T. and Shahriar, M. S. and Sharma, A. and Sharma, P. and Shawhan, P. and Shen, H. and Shikauchi, M. and Shink, R. and Shoemaker, D. H. and Shoemaker, D. M. and Shukla, K. and ShyamSundar, S. and Siellez, K. and Sieniawska, M. and Sigg, D. and Singer, L. P. and Singh, D. and Singh, N. and Singha, A. and Singhal, A. and Sintes, A. M. and Sipala, V. and Skliris, V. and Slagmolen, B. J. J. and Slaven-Blair, T. J. and Smetana, J. and Smith, J. R. and Smith, R. J. E. and Somala, S. and Son, E. J. and Soni, S. and Sorazu, B. and Sordini, V. and Sorrentino, F. and Souradeep, T. and Sowell, E. and Spencer, A. P. and Spera, M. and Srivastava, A. K. and Srivastava, V. and Staats, K. and Stachie, C. and Standke, M. and Steer, D. A. and Steinke, M. and Steinlechner, J. and Steinlechner, S. and Steinmeyer, D. and Stocks, D. and Stops, D. J. and Stover, M. and Strain, K. A. and Stratta, G. and Strunk, A. and Sturani, R. and Stuver, A. L. and Sudhagar, S. and Sudhir, V. and Summerscales, T. Z. and Sun, L. and Sunil, S. and Sur, A. and Suresh, J. and Sutton, P. J. and Swinkels, B. L. and Szczepańczyk, M. J. and Tacca, M. and Tait, S. C. and Talbot, C. and Tanasijczuk, A. J. and Tanner, D. B. and Tao, D. and Tápai, M. and Tapia, A. and Martin, E. N. Tapia San and Tasson, J. D. and Taylor, R. and Tenorio, R. and Terkowski, L. and Thirugnanasambandam, M. P. and Thomas, M. and Thomas, P. and Thompson, J. E. and Thondapu, S. R. and Thorne, K. A. and Thrane, E. and Tinsman, C. L. and Saravanan, T. R. and Tiwari, Shubhanshu and Tiwari, S. and Tiwari, V. and Toland, K. and Tonelli, M. and Tornasi, Z. and Torres-Forné, A. and Torrie, C. I. and Melo, I. Tosta e and Töyrä, D. and Trail, E. A. and Travasso, F. and Traylor, G. and Tringali, M. C. and Tripathee, A. and Trovato, A. and Trudeau, R. J. and Tsang, K. W. and Tse, M. and Tso, R. and Tsukada, L. and Tsuna, D. and Tsutsui, T. and Turconi, M. and Ubhi, A. S. and Ueno, K. and Ugolini, D. and Unnikrishnan, C. S. and Urban, A. L. and Usman, S. A. and Utina, A. C. and Vahlbruch, H. and Vajente, G. and Valdes, G. and Valentini, M. and Vallisneri, M. and Bakel, N. van and Beuzekom, M. van and Brand, J. F. J. van den and Broeck, C. Van Den and Vander-Hyde, D. C. and Schaaf, L. van der and Heijningen, J. V. Van and Veggel, A. A. van and Vardaro, M. and Varma, V. and Vass, S. and Vasúth, M. and Vecchio, A. and Vedovato, G. and Veitch, J. and Veitch, P. J. and Venkateswara, K. and Venugopalan, G. and Verkindt, D. and Veske, D. and Vetrano, F. and Viceré, A. and Viets, A. D. and Vinciguerra, S. and Vine, D. J. and Vinet, J.-Y. and Vitale, S. and Vivanco, Francisco Hernandez and Vo, T. and Vocca, H. and Vorvick, C. and Vyatchanin, S. P. and Wade, A. R. and Wade, L. E. and Wade, M. and Walet, R. and Walker, M. and Wallace, G. S. and Wallace, L. and Walsh, S. and Wang, J. Z. and Wang, S. and Wang, W. H. and Wang, Y. F. and Ward, R. L. and Warden, Z. A. and Warner, J. and Was, M. and Watchi, J. and Weaver, B. and Wei, L.-W. and Weinert, M. and Weinstein, A. J. and Weiss, R. and Wellmann, F. and Wen, L. and Weßels, P. and Westhouse, J. W. and Wette, K. and Whelan, J. T. and Whiting, B. F. and Whittle, C. and Wilken, D. M. and Williams, D. and Williams, R. D. and Willis, J. L. and Willke, B. and Winkler, W. and Wipf, C. C. and Wittel, H. and Woan, G. and Woehler, J. and Wofford, J. K. and Wong, C. and Wright, J. L. and Wu, D. S. and Wysocki, D. M. and Xiao, L. and Yamamoto, H. and Yang, L. and Yang, Y. and Yang, Z. and Yap, M. J. and Yazback, M. and Yeeles, D. W. and Yu, Hang and Yu, Haocun and Yuen, S. H. R. and Zadrożny, A. K. and Zadrożny, A. and Zanolin, M. and Zelenova, T. and Zendri, J.-P. and Zevin, M. and Zhang, J. and Zhang, L. and Zhang, T. and Zhao, C. and Zhao, G. and Zhou, M. and Zhou, Z. and Zhu, X. J. and Zimmerman, A. B. and Zucker, M. E. and Zweizig, J.},
	month = jan,
	year = {2021},
	note = {arXiv:1912.11716 [gr-qc]},
	pages = {100658},
}

@article{kullback_information_1951,
	title = {On {Information} and {Sufficiency}},
	volume = {22},
	issn = {0003-4851, 2168-8990},
	url = {https://projecteuclid.org/journals/annals-of-mathematical-statistics/volume-22/issue-1/On-Information-and-Sufficiency/10.1214/aoms/1177729694.full},
	doi = {10.1214/aoms/1177729694},
	number = {1},
	urldate = {2025-03-17},
	journal = {The Annals of Mathematical Statistics},
	author = {Kullback, S. and Leibler, R. A.},
	month = mar,
	year = {1951},
	note = {Publisher: Institute of Mathematical Statistics},
	pages = {79--86},
}

@article{mills_measuring_2021,
	title = {Measuring gravitational-wave higher-order multipoles},
	volume = {103},
	issn = {1550-79980556-2821},
	url = {https://ui.adsabs.harvard.edu/abs/2021PhRvD.103b4042M},
	doi = {10.1103/PhysRevD.103.024042},
	urldate = {2025-03-14},
	journal = {Physical Review D},
	author = {Mills, Cameron and Fairhurst, Stephen},
	month = jan,
	year = {2021},
	note = {Publisher: APS
ADS Bibcode: 2021PhRvD.103b4042M},
	pages = {024042},
}

@misc{catalog-ias-hm,
	title = {New black hole mergers in the {LIGO}-{Virgo} {O3} data from a gravitational wave search including higher-order harmonics},
	url = {https://ui.adsabs.harvard.edu/abs/2023arXiv231206631W},
	doi = {10.48550/arXiv.2312.06631},
	urldate = {2024-08-14},
	author = {Wadekar, Digvijay and Roulet, Javier and Venumadhav, Tejaswi and Mehta, Ajit Kumar and Zackay, Barak and Mushkin, Jonathan and Olsen, Seth and Zaldarriaga, Matias},
	month = dec,
	year = {2023},
	note = {Publication Title: arXiv e-prints
ADS Bibcode: 2023arXiv231206631W},
}

@article{roulet_removing_2022,
	title = {Removing degeneracy and multimodality in gravitational wave source parameters},
	volume = {106},
	issn = {1550-79980556-2821},
	url = {https://ui.adsabs.harvard.edu/abs/2022PhRvD.106l3015R},
	doi = {10.1103/PhysRevD.106.123015},
	urldate = {2024-12-12},
	journal = {Physical Review D},
	author = {Roulet, Javier and Olsen, Seth and Mushkin, Jonathan and Islam, Tousif and Venumadhav, Tejaswi and Zackay, Barak and Zaldarriaga, Matias},
	month = dec,
	year = {2022},
	note = {Publisher: APS
ADS Bibcode: 2022PhRvD.106l3015R},
	pages = {123015},
}

@article{software-astropy-1,
	title = {Astropy: {A} community {Python} package for astronomy},
	volume = {558},
	issn = {0004-6361},
	shorttitle = {Astropy},
	url = {https://ui.adsabs.harvard.edu/abs/2013A&A...558A..33A},
	doi = {10.1051/0004-6361/201322068},
	urldate = {2024-12-11},
	journal = {Astronomy and Astrophysics},
	author = {{Astropy Collaboration} and Robitaille, Thomas P. and Tollerud, Erik J. and Greenfield, Perry and Droettboom, Michael and Bray, Erik and Aldcroft, Tom and Davis, Matt and Ginsburg, Adam and Price-Whelan, Adrian M. and Kerzendorf, Wolfgang E. and Conley, Alexander and Crighton, Neil and Barbary, Kyle and Muna, Demitri and Ferguson, Henry and Grollier, Frédéric and Parikh, Madhura M. and Nair, Prasanth H. and Unther, Hans M. and Deil, Christoph and Woillez, Julien and Conseil, Simon and Kramer, Roban and Turner, James E. H. and Singer, Leo and Fox, Ryan and Weaver, Benjamin A. and Zabalza, Victor and Edwards, Zachary I. and Azalee Bostroem, K. and Burke, D. J. and Casey, Andrew R. and Crawford, Steven M. and Dencheva, Nadia and Ely, Justin and Jenness, Tim and Labrie, Kathleen and Lim, Pey Lian and Pierfederici, Francesco and Pontzen, Andrew and Ptak, Andy and Refsdal, Brian and Servillat, Mathieu and Streicher, Ole},
	month = oct,
	year = {2013},
	note = {ADS Bibcode: 2013A\&A...558A..33A},
	pages = {A33},
}

@article{software-astropy-2,
	title = {The {Astropy} {Project}: {Building} an {Open}-science {Project} and {Status} of the v2.0 {Core} {Package}},
	volume = {156},
	issn = {0004-6256},
	shorttitle = {The {Astropy} {Project}},
	url = {https://ui.adsabs.harvard.edu/abs/2018AJ....156..123A},
	doi = {10.3847/1538-3881/aabc4f},
	urldate = {2024-12-11},
	journal = {The Astronomical Journal},
	author = {{Astropy Collaboration} and Price-Whelan, A. M. and Sipőcz, B. M. and Günther, H. M. and Lim, P. L. and Crawford, S. M. and Conseil, S. and Shupe, D. L. and Craig, M. W. and Dencheva, N. and Ginsburg, A. and VanderPlas, J. T. and Bradley, L. D. and Pérez-Suárez, D. and de Val-Borro, M. and Aldcroft, T. L. and Cruz, K. L. and Robitaille, T. P. and Tollerud, E. J. and Ardelean, C. and Babej, T. and Bach, Y. P. and Bachetti, M. and Bakanov, A. V. and Bamford, S. P. and Barentsen, G. and Barmby, P. and Baumbach, A. and Berry, K. L. and Biscani, F. and Boquien, M. and Bostroem, K. A. and Bouma, L. G. and Brammer, G. B. and Bray, E. M. and Breytenbach, H. and Buddelmeijer, H. and Burke, D. J. and Calderone, G. and Cano Rodríguez, J. L. and Cara, M. and Cardoso, J. V. M. and Cheedella, S. and Copin, Y. and Corrales, L. and Crichton, D. and D'Avella, D. and Deil, C. and Depagne, É. and Dietrich, J. P. and Donath, A. and Droettboom, M. and Earl, N. and Erben, T. and Fabbro, S. and Ferreira, L. A. and Finethy, T. and Fox, R. T. and Garrison, L. H. and Gibbons, S. L. J. and Goldstein, D. A. and Gommers, R. and Greco, J. P. and Greenfield, P. and Groener, A. M. and Grollier, F. and Hagen, A. and Hirst, P. and Homeier, D. and Horton, A. J. and Hosseinzadeh, G. and Hu, L. and Hunkeler, J. S. and Ivezić, Ž. and Jain, A. and Jenness, T. and Kanarek, G. and Kendrew, S. and Kern, N. S. and Kerzendorf, W. E. and Khvalko, A. and King, J. and Kirkby, D. and Kulkarni, A. M. and Kumar, A. and Lee, A. and Lenz, D. and Littlefair, S. P. and Ma, Z. and Macleod, D. M. and Mastropietro, M. and McCully, C. and Montagnac, S. and Morris, B. M. and Mueller, M. and Mumford, S. J. and Muna, D. and Murphy, N. A. and Nelson, S. and Nguyen, G. H. and Ninan, J. P. and Nöthe, M. and Ogaz, S. and Oh, S. and Parejko, J. K. and Parley, N. and Pascual, S. and Patil, R. and Patil, A. A. and Plunkett, A. L. and Prochaska, J. X. and Rastogi, T. and Reddy Janga, V. and Sabater, J. and Sakurikar, P. and Seifert, M. and Sherbert, L. E. and Sherwood-Taylor, H. and Shih, A. Y. and Sick, J. and Silbiger, M. T. and Singanamalla, S. and Singer, L. P. and Sladen, P. H. and Sooley, K. A. and Sornarajah, S. and Streicher, O. and Teuben, P. and Thomas, S. W. and Tremblay, G. R. and Turner, J. E. H. and Terrón, V. and van Kerkwijk, M. H. and de la Vega, A. and Watkins, L. L. and Weaver, B. A. and Whitmore, J. B. and Woillez, J. and Zabalza, V. and {Astropy Contributors}},
	month = sep,
	year = {2018},
	note = {Publisher: IOP
ADS Bibcode: 2018AJ....156..123A},
	pages = {123},
}

@article{software-astropy-3,
	title = {The {Astropy} {Project}: {Sustaining} and {Growing} a {Community}-oriented {Open}-source {Project} and the {Latest} {Major} {Release} (v5.0) of the {Core} {Package}},
	volume = {935},
	issn = {0004-637X, 1538-4357},
	shorttitle = {The {Astropy} {Project}},
	url = {http://arxiv.org/abs/2206.14220},
	doi = {10.3847/1538-4357/ac7c74},
	number = {2},
	urldate = {2024-12-11},
	journal = {The Astrophysical Journal},
	author = {Collaboration, The Astropy and Price-Whelan, Adrian M. and Lim, Pey Lian and Earl, Nicholas and Starkman, Nathaniel and Bradley, Larry and Shupe, David L. and Patil, Aarya A. and Corrales, Lia and Brasseur, C. E. and Nöthe, Maximilian and Donath, Axel and Tollerud, Erik and Morris, Brett M. and Ginsburg, Adam and Vaher, Eero and Weaver, Benjamin A. and Tocknell, James and Jamieson, William and Kerkwijk, Marten H. van and Robitaille, Thomas P. and Merry, Bruce and Bachetti, Matteo and Günther, H. Moritz and Aldcroft, Thomas L. and Alvarado-Montes, Jaime A. and Archibald, Anne M. and Bódi, Attila and Bapat, Shreyas and Barentsen, Geert and Bazán, Juanjo and Biswas, Manish and Boquien, Médéric and Burke, D. J. and Cara, Daria and Cara, Mihai and Conroy, Kyle E. and Conseil, Simon and Craig, Matthew W. and Cross, Robert M. and Cruz, Kelle L. and D'Eugenio, Francesco and Dencheva, Nadia and Devillepoix, Hadrien A. R. and Dietrich, Jörg P. and Eigenbrot, Arthur Davis and Erben, Thomas and Ferreira, Leonardo and Foreman-Mackey, Daniel and Fox, Ryan and Freij, Nabil and Garg, Suyog and Geda, Robel and Glattly, Lauren and Gondhalekar, Yash and Gordon, Karl D. and Grant, David and Greenfield, Perry and Groener, Austen M. and Guest, Steve and Gurovich, Sebastian and Handberg, Rasmus and Hart, Akeem and Hatfield-Dodds, Zac and Homeier, Derek and Hosseinzadeh, Griffin and Jenness, Tim and Jones, Craig K. and Joseph, Prajwel and Kalmbach, J. Bryce and Karamehmetoglu, Emir and Kałuszyński, Mikołaj and Kelley, Michael S. P. and Kern, Nicholas and Kerzendorf, Wolfgang E. and Koch, Eric W. and Kulumani, Shankar and Lee, Antony and Ly, Chun and Ma, Zhiyuan and MacBride, Conor and Maljaars, Jakob M. and Muna, Demitri and Murphy, N. A. and Norman, Henrik and O'Steen, Richard and Oman, Kyle A. and Pacifici, Camilla and Pascual, Sergio and Pascual-Granado, J. and Patil, Rohit R. and Perren, Gabriel I. and Pickering, Timothy E. and Rastogi, Tanuj and Roulston, Benjamin R. and Ryan, Daniel F. and Rykoff, Eli S. and Sabater, Jose and Sakurikar, Parikshit and Salgado, Jesús and Sanghi, Aniket and Saunders, Nicholas and Savchenko, Volodymyr and Schwardt, Ludwig and Seifert-Eckert, Michael and Shih, Albert Y. and Jain, Anany Shrey and Shukla, Gyanendra and Sick, Jonathan and Simpson, Chris and Singanamalla, Sudheesh and Singer, Leo P. and Singhal, Jaladh and Sinha, Manodeep and Sipőcz, Brigitta M. and Spitler, Lee R. and Stansby, David and Streicher, Ole and Šumak, Jani and Swinbank, John D. and Taranu, Dan S. and Tewary, Nikita and Tremblay, Grant R. and Val-Borro, Miguel de and Kooten, Samuel J. Van and Vasović, Zlatan and Verma, Shresth and Cardoso, José Vinícius de Miranda and Williams, Peter K. G. and Wilson, Tom J. and Winkel, Benjamin and Wood-Vasey, W. M. and Xue, Rui and Yoachim, Peter and ZHANG, Chen and Zonca, Andrea},
	month = aug,
	year = {2022},
	note = {arXiv:2206.14220 [astro-ph]},
	pages = {167},
}

@article{cosmology-planck15,
	title = {Planck 2015 results - {XIII}. {Cosmological} parameters},
	volume = {594},
	copyright = {© ESO, 2016},
	issn = {0004-6361, 1432-0746},
	url = {https://www.aanda.org/articles/aa/abs/2016/10/aa25830-15/aa25830-15.html},
	doi = {10.1051/0004-6361/201525830},
	language = {en},
	urldate = {2024-12-11},
	journal = {Astronomy \& Astrophysics},
	author = {Ade, P. a. R. and Aghanim, N. and Arnaud, M. and Ashdown, M. and Aumont, J. and Baccigalupi, C. and Banday, A. J. and Barreiro, R. B. and Bartlett, J. G. and Bartolo, N. and Battaner, E. and Battye, R. and Benabed, K. and Benoît, A. and Benoit-Lévy, A. and Bernard, J.-P. and Bersanelli, M. and Bielewicz, P. and Bock, J. J. and Bonaldi, A. and Bonavera, L. and Bond, J. R. and Borrill, J. and Bouchet, F. R. and Boulanger, F. and Bucher, M. and Burigana, C. and Butler, R. C. and Calabrese, E. and Cardoso, J.-F. and Catalano, A. and Challinor, A. and Chamballu, A. and Chary, R.-R. and Chiang, H. C. and Chluba, J. and Christensen, P. R. and Church, S. and Clements, D. L. and Colombi, S. and Colombo, L. P. L. and Combet, C. and Coulais, A. and Crill, B. P. and Curto, A. and Cuttaia, F. and Danese, L. and Davies, R. D. and Davis, R. J. and Bernardis, P. de and Rosa, A. de and Zotti, G. de and Delabrouille, J. and Désert, F.-X. and Valentino, E. Di and Dickinson, C. and Diego, J. M. and Dolag, K. and Dole, H. and Donzelli, S. and Doré, O. and Douspis, M. and Ducout, A. and Dunkley, J. and Dupac, X. and Efstathiou, G. and Elsner, F. and Enßlin, T. A. and Eriksen, H. K. and Farhang, M. and Fergusson, J. and Finelli, F. and Forni, O. and Frailis, M. and Fraisse, A. A. and Franceschi, E. and Frejsel, A. and Galeotta, S. and Galli, S. and Ganga, K. and Gauthier, C. and Gerbino, M. and Ghosh, T. and Giard, M. and Giraud-Héraud, Y. and Giusarma, E. and Gjerløw, E. and González-Nuevo, J. and Górski, K. M. and Gratton, S. and Gregorio, A. and Gruppuso, A. and Gudmundsson, J. E. and Hamann, J. and Hansen, F. K. and Hanson, D. and Harrison, D. L. and Helou, G. and Henrot-Versillé, S. and Hernández-Monteagudo, C. and Herranz, D. and Hildebrandt, S. R. and Hivon, E. and Hobson, M. and Holmes, W. A. and Hornstrup, A. and Hovest, W. and Huang, Z. and Huffenberger, K. M. and Hurier, G. and Jaffe, A. H. and Jaffe, T. R. and Jones, W. C. and Juvela, M. and Keihänen, E. and Keskitalo, R. and Kisner, T. S. and Kneissl, R. and Knoche, J. and Knox, L. and Kunz, M. and Kurki-Suonio, H. and Lagache, G. and Lähteenmäki, A. and Lamarre, J.-M. and Lasenby, A. and Lattanzi, M. and Lawrence, C. R. and Leahy, J. P. and Leonardi, R. and Lesgourgues, J. and Levrier, F. and Lewis, A. and Liguori, M. and Lilje, P. B. and Linden-Vørnle, M. and López-Caniego, M. and Lubin, P. M. and Macías-Pérez, J. F. and Maggio, G. and Maino, D. and Mandolesi, N. and Mangilli, A. and Marchini, A. and Maris, M. and Martin, P. G. and Martinelli, M. and Martínez-González, E. and Masi, S. and Matarrese, S. and McGehee, P. and Meinhold, P. R. and Melchiorri, A. and Melin, J.-B. and Mendes, L. and Mennella, A. and Migliaccio, M. and Millea, M. and Mitra, S. and Miville-Deschênes, M.-A. and Moneti, A. and Montier, L. and Morgante, G. and Mortlock, D. and Moss, A. and Munshi, D. and Murphy, J. A. and Naselsky, P. and Nati, F. and Natoli, P. and Netterfield, C. B. and Nørgaard-Nielsen, H. U. and Noviello, F. and Novikov, D. and Novikov, I. and Oxborrow, C. A. and Paci, F. and Pagano, L. and Pajot, F. and Paladini, R. and Paoletti, D. and Partridge, B. and Pasian, F. and Patanchon, G. and Pearson, T. J. and Perdereau, O. and Perotto, L. and Perrotta, F. and Pettorino, V. and Piacentini, F. and Piat, M. and Pierpaoli, E. and Pietrobon, D. and Plaszczynski, S. and Pointecouteau, E. and Polenta, G. and Popa, L. and Pratt, G. W. and Prézeau, G. and Prunet, S. and Puget, J.-L. and Rachen, J. P. and Reach, W. T. and Rebolo, R. and Reinecke, M. and Remazeilles, M. and Renault, C. and Renzi, A. and Ristorcelli, I. and Rocha, G. and Rosset, C. and Rossetti, M. and Roudier, G. and d’Orfeuil, B. Rouillé and Rowan-Robinson, M. and Rubiño-Martín, J. A. and Rusholme, B. and Said, N. and Salvatelli, V. and Salvati, L. and Sandri, M. and Santos, D. and Savelainen, M. and Savini, G. and Scott, D. and Seiffert, M. D. and Serra, P. and Shellard, E. P. S. and Spencer, L. D. and Spinelli, M. and Stolyarov, V. and Stompor, R. and Sudiwala, R. and Sunyaev, R. and Sutton, D. and Suur-Uski, A.-S. and Sygnet, J.-F. and Tauber, J. A. and Terenzi, L. and Toffolatti, L. and Tomasi, M. and Tristram, M. and Trombetti, T. and Tucci, M. and Tuovinen, J. and Türler, M. and Umana, G. and Valenziano, L. and Valiviita, J. and Tent, F. Van and Vielva, P. and Villa, F. and Wade, L. A. and Wandelt, B. D. and Wehus, I. K. and White, M. and White, S. D. M. and Wilkinson, A. and Yvon, D. and Zacchei, A. and Zonca, A.},
	month = oct,
	year = {2016},
	note = {Publisher: EDP Sciences},
	pages = {A13},
}

@misc{docs-asimov,
	title = {Asimov documentation},
	url = {https://asimov.docs.ligo.org/asimov/master/index.html},
	urldate = {2024-12-10},
}

@misc{code-rift-asimov,
	title = {Efficient reanalysis of events from {GWTC}-3 with {RIFT} and asimov},
	url = {http://arxiv.org/abs/2412.02999},
	doi = {10.48550/arXiv.2412.02999},
	urldate = {2024-12-05},
	publisher = {arXiv},
	author = {Fernando, D. and O'Shaughnessy, R. and Williams, D.},
	month = dec,
	year = {2024},
	note = {arXiv:2412.02999 [astro-ph]},
}

@article{software-scipy,
	title = {{SciPy} 1.0: fundamental algorithms for scientific computing in {Python}},
	volume = {17},
	shorttitle = {{SciPy} 1.0},
	url = {https://ui.adsabs.harvard.edu/abs/2020NatMe..17..261V},
	doi = {10.1038/s41592-019-0686-2},
	urldate = {2024-12-01},
	journal = {Nature Methods},
	author = {Virtanen, Pauli and Gommers, Ralf and Oliphant, Travis E. and Haberland, Matt and Reddy, Tyler and Cournapeau, David and Burovski, Evgeni and Peterson, Pearu and Weckesser, Warren and Bright, Jonathan and van der Walt, Stéfan J. and Brett, Matthew and Wilson, Joshua and Millman, K. Jarrod and Mayorov, Nikolay and Nelson, Andrew R. J. and Jones, Eric and Kern, Robert and Larson, Eric and Carey, C. J. and Polat, İlhan and Feng, Yu and Moore, Eric W. and VanderPlas, Jake and Laxalde, Denis and Perktold, Josef and Cimrman, Robert and Henriksen, Ian and Quintero, E. A. and Harris, Charles R. and Archibald, Anne M. and Ribeiro, Antônio H. and Pedregosa, Fabian and van Mulbregt, Paul and {SciPy 1. 0 Contributors}},
	month = feb,
	year = {2020},
	note = {ADS Bibcode: 2020NatMe..17..261V},
	pages = {261--272},
}

@article{software-numpy,
	title = {Array programming with {NumPy}},
	volume = {585},
	copyright = {2020 The Author(s)},
	issn = {1476-4687},
	url = {https://www.nature.com/articles/s41586-020-2649-2},
	doi = {10.1038/s41586-020-2649-2},
	language = {en},
	number = {7825},
	urldate = {2024-12-10},
	journal = {Nature},
	author = {Harris, Charles R. and Millman, K. Jarrod and van der Walt, Stéfan J. and Gommers, Ralf and Virtanen, Pauli and Cournapeau, David and Wieser, Eric and Taylor, Julian and Berg, Sebastian and Smith, Nathaniel J. and Kern, Robert and Picus, Matti and Hoyer, Stephan and van Kerkwijk, Marten H. and Brett, Matthew and Haldane, Allan and del Río, Jaime Fernández and Wiebe, Mark and Peterson, Pearu and Gérard-Marchant, Pierre and Sheppard, Kevin and Reddy, Tyler and Weckesser, Warren and Abbasi, Hameer and Gohlke, Christoph and Oliphant, Travis E.},
	month = sep,
	year = {2020},
	note = {Publisher: Nature Publishing Group},
	pages = {357--362},
}

@article{waveform-imrphenomxode,
	title = {Accurate and efficient waveform model for precessing binary black holes},
	volume = {108},
	issn = {1550-79980556-2821},
	url = {https://ui.adsabs.harvard.edu/abs/2023PhRvD.108f4059Y},
	doi = {10.1103/PhysRevD.108.064059},
	urldate = {2024-12-09},
	journal = {Physical Review D},
	author = {Yu, Hang and Roulet, Javier and Venumadhav, Tejaswi and Zackay, Barak and Zaldarriaga, Matias},
	month = sep,
	year = {2023},
	note = {Publisher: APS
ADS Bibcode: 2023PhRvD.108f4059Y},
	pages = {064059},
}

@article{waveform-imrphenomd,
	title = {Frequency-domain gravitational waves from nonprecessing black-hole binaries. {I}. {New} numerical waveforms and anatomy of the signal},
	volume = {93},
	issn = {1550-79980556-2821},
	url = {https://ui.adsabs.harvard.edu/abs/2016PhRvD..93d4006H},
	doi = {10.1103/PhysRevD.93.044006},
	urldate = {2024-12-09},
	journal = {Physical Review D},
	author = {Husa, Sascha and Khan, Sebastian and Hannam, Mark and Pürrer, Michael and Ohme, Frank and Forteza, Xisco Jiménez and Bohé, Alejandro},
	month = feb,
	year = {2016},
	note = {Publisher: APS
ADS Bibcode: 2016PhRvD..93d4006H},
	pages = {044006},
}

@article{waveform-imrphenomp,
	title = {Frequency-domain gravitational waves from nonprecessing black-hole binaries. {II}. {A} phenomenological model for the advanced detector era},
	volume = {93},
	issn = {1550-79980556-2821},
	url = {https://ui.adsabs.harvard.edu/abs/2016PhRvD..93d4007K},
	doi = {10.1103/PhysRevD.93.044007},
	urldate = {2024-12-09},
	journal = {Physical Review D},
	author = {Khan, Sebastian and Husa, Sascha and Hannam, Mark and Ohme, Frank and Pürrer, Michael and Forteza, Xisco Jiménez and Bohé, Alejandro},
	month = feb,
	year = {2016},
	note = {Publisher: APS
ADS Bibcode: 2016PhRvD..93d4007K},
	pages = {044007},
}

@article{code-cogwheel,
	title = {Fast marginalization algorithm for optimizing gravitational wave detection, parameter estimation, and sky localization},
	volume = {110},
	issn = {1550-79980556-2821},
	url = {https://ui.adsabs.harvard.edu/abs/2024PhRvD.110d4010R},
	doi = {10.1103/PhysRevD.110.044010},
	urldate = {2024-11-19},
	journal = {Physical Review D},
	author = {Roulet, Javier and Mushkin, Jonathan and Wadekar, Digvijay and Venumadhav, Tejaswi and Zackay, Barak and Zaldarriaga, Matias},
	month = aug,
	year = {2024},
	note = {Publisher: APS
ADS Bibcode: 2024PhRvD.110d4010R},
	pages = {044010},
}

@article{catalog-ias-1,
	title = {New search pipeline for compact binary mergers: {Results} for binary black holes in the first observing run of {Advanced} {LIGO}},
	volume = {100},
	issn = {1550-79980556-2821},
	shorttitle = {New search pipeline for compact binary mergers},
	url = {https://ui.adsabs.harvard.edu/abs/2019PhRvD.100b3011V},
	doi = {10.1103/PhysRevD.100.023011},
	urldate = {2024-08-14},
	journal = {Physical Review D},
	author = {Venumadhav, Tejaswi and Zackay, Barak and Roulet, Javier and Dai, Liang and Zaldarriaga, Matias},
	month = jul,
	year = {2019},
	note = {Publisher: APS
ADS Bibcode: 2019PhRvD.100b3011V},
	pages = {023011},
}

@article{catalog-ias-2,
	title = {New binary black hole mergers in the second observing run of {Advanced} {LIGO} and {Advanced} {Virgo}},
	volume = {101},
	issn = {1550-79980556-2821},
	url = {https://ui.adsabs.harvard.edu/abs/2020PhRvD.101h3030V},
	doi = {10.1103/PhysRevD.101.083030},
	urldate = {2024-08-14},
	journal = {Physical Review D},
	author = {Venumadhav, Tejaswi and Zackay, Barak and Roulet, Javier and Dai, Liang and Zaldarriaga, Matias},
	month = apr,
	year = {2020},
	note = {Publisher: APS
ADS Bibcode: 2020PhRvD.101h3030V},
	pages = {083030},
}

@article{catalog-ias-3,
	title = {New binary black hole mergers in the {LIGO}-{Virgo} {O3a} data},
	volume = {106},
	issn = {1550-79980556-2821},
	url = {https://ui.adsabs.harvard.edu/abs/2022PhRvD.106d3009O},
	doi = {10.1103/PhysRevD.106.043009},
	urldate = {2024-08-14},
	journal = {Physical Review D},
	author = {Olsen, Seth and Venumadhav, Tejaswi and Mushkin, Jonathan and Roulet, Javier and Zackay, Barak and Zaldarriaga, Matias},
	month = aug,
	year = {2022},
	note = {Publisher: APS
ADS Bibcode: 2022PhRvD.106d3009O},
	pages = {043009},
}

@article{ashton_gravitational_2019,
	title = {Gravitational wave detection without boot straps: {A} {Bayesian} approach},
	volume = {100},
	issn = {1550-79980556-2821},
	shorttitle = {Gravitational wave detection without boot straps},
	url = {https://ui.adsabs.harvard.edu/abs/2019PhRvD.100l3018A},
	doi = {10.1103/PhysRevD.100.123018},
	urldate = {2024-12-08},
	journal = {Physical Review D},
	author = {Ashton, Gregory and Thrane, Eric and Smith, Rory J. E.},
	month = dec,
	year = {2019},
	note = {Publisher: APS
ADS Bibcode: 2019PhRvD.100l3018A},
	pages = {123018},
}

@article{ashton_astrophysical_2020,
	title = {The astrophysical odds of {GW151216}},
	volume = {498},
	issn = {0035-8711},
	url = {https://ui.adsabs.harvard.edu/abs/2020MNRAS.498.1905A},
	doi = {10.1093/mnras/staa2332},
	urldate = {2024-12-08},
	journal = {Monthly Notices of the Royal Astronomical Society},
	author = {Ashton, Gregory and Thrane, Eric},
	month = oct,
	year = {2020},
	note = {Publisher: OUP
ADS Bibcode: 2020MNRAS.498.1905A},
	pages = {1905--1910},
}

@article{catalog-gwtc-1,
	title = {{GWTC}-1: {A} {Gravitational}-{Wave} {Transient} {Catalog} of {Compact} {Binary} {Mergers} {Observed} by {LIGO} and {Virgo} during the {First} and {Second} {Observing} {Runs}},
	volume = {9},
	shorttitle = {{GWTC}-1},
	url = {https://ui.adsabs.harvard.edu/abs/2019PhRvX...9c1040A},
	doi = {10.1103/PhysRevX.9.031040},
	urldate = {2024-08-14},
	journal = {Physical Review X},
	author = {Abbott, B. P. and Abbott, R. and Abbott, T. D. and Abraham, S. and Acernese, F. and Ackley, K. and Adams, C. and Adhikari, R. X. and Adya, V. B. and Affeldt, C. and Agathos, M. and Agatsuma, K. and Aggarwal, N. and Aguiar, O. D. and Aiello, L. and Ain, A. and Ajith, P. and Allen, G. and Allocca, A. and Aloy, M. A. and Altin, P. A. and Amato, A. and Ananyeva, A. and Anderson, S. B. and Anderson, W. G. and Angelova, S. V. and Antier, S. and Appert, S. and Arai, K. and Araya, M. C. and Areeda, J. S. and Arène, M. and Arnaud, N. and Arun, K. G. and Ascenzi, S. and Ashton, G. and Aston, S. M. and Astone, P. and Aubin, F. and Aufmuth, P. and AultONeal, K. and Austin, C. and Avendano, V. and Avila-Alvarez, A. and Babak, S. and Bacon, P. and Badaracco, F. and Bader, M. K. M. and Bae, S. and Baker, P. T. and Baldaccini, F. and Ballardin, G. and Ballmer, S. W. and Banagiri, S. and Barayoga, J. C. and Barclay, S. E. and Barish, B. C. and Barker, D. and Barkett, K. and Barnum, S. and Barone, F. and Barr, B. and Barsotti, L. and Barsuglia, M. and Barta, D. and Bartlett, J. and Bartos, I. and Bassiri, R. and Basti, A. and Bawaj, M. and Bayley, J. C. and Bazzan, M. and Bécsy, B. and Bejger, M. and Belahcene, I. and Bell, A. S. and Beniwal, D. and Berger, B. K. and Bergmann, G. and Bernuzzi, S. and Bero, J. J. and Berry, C. P. L. and Bersanetti, D. and Bertolini, A. and Betzwieser, J. and Bhandare, R. and Bidler, J. and Bilenko, I. A. and Bilgili, S. A. and Billingsley, G. and Birch, J. and Birney, R. and Birnholtz, O. and Biscans, S. and Biscoveanu, S. and Bisht, A. and Bitossi, M. and Bizouard, M. A. and Blackburn, J. K. and Blackman, J. and Blair, C. D. and Blair, D. G. and Blair, R. M. and Bloemen, S. and Bode, N. and Boer, M. and Boetzel, Y. and Bogaert, G. and Bondu, F. and Bonilla, E. and Bonnand, R. and Booker, P. and Boom, B. A. and Booth, C. D. and Bork, R. and Boschi, V. and Bose, S. and Bossie, K. and Bossilkov, V. and Bosveld, J. and Bouffanais, Y. and Bozzi, A. and Bradaschia, C. and Brady, P. R. and Bramley, A. and Branchesi, M. and Brau, J. E. and Briant, T. and Briggs, J. H. and Brighenti, F. and Brillet, A. and Brinkmann, M. and Brisson, V. and Brockill, P. and Brooks, A. F. and Brown, D. D. and Brunett, S. and Buikema, A. and Bulik, T. and Bulten, H. J. and Buonanno, A. and Buskulic, D. and Bustamante Rosell, M. J. and Buy, C. and Byer, R. L. and Cabero, M. and Cadonati, L. and Cagnoli, G. and Cahillane, C. and Calderón Bustillo, J. and Callister, T. A. and Calloni, E. and Camp, J. B. and Campbell, W. A. and Canepa, M. and Cannon, K. C. and Cao, H. and Cao, J. and Capocasa, E. and Carbognani, F. and Caride, S. and Carney, M. F. and Carullo, G. and Casanueva Diaz, J. and Casentini, C. and Caudill, S. and Cavaglià, M. and Cavalier, F. and Cavalieri, R. and Cella, G. and Cerdá-Durán, P. and Cerretani, G. and Cesarini, E. and Chaibi, O. and Chakravarti, K. and Chamberlin, S. J. and Chan, M. and Chao, S. and Charlton, P. and Chase, E. A. and Chassande-Mottin, E. and Chatterjee, D. and Chaturvedi, M. and Chatziioannou, K. and Cheeseboro, B. D. and Chen, H. Y. and Chen, X. and Chen, Y. and Cheng, H. -P. and Cheong, C. K. and Chia, H. Y. and Chincarini, A. and Chiummo, A. and Cho, G. and Cho, H. S. and Cho, M. and Christensen, N. and Chu, Q. and Chua, S. and Chung, K. W. and Chung, S. and Ciani, G. and Ciobanu, A. A. and Ciolfi, R. and Cipriano, F. and Cirone, A. and Clara, F. and Clark, J. A. and Clearwater, P. and Cleva, F. and Cocchieri, C. and Coccia, E. and Cohadon, P. -F. and Cohen, D. and Colgan, R. and Colleoni, M. and Collette, C. G. and Collins, C. and Cominsky, L. R. and Constancio, M. and Conti, L. and Cooper, S. J. and Corban, P. and Corbitt, T. R. and Cordero-Carrión, I. and Corley, K. R. and Cornish, N. and Corsi, A. and Cortese, S. and Costa, C. A. and Cotesta, R. and Coughlin, M. W. and Coughlin, S. B. and Coulon, J. -P. and Countryman, S. T. and Couvares, P. and Covas, P. B. and Cowan, E. E. and Coward, D. M. and Cowart, M. J. and Coyne, D. C. and Coyne, R. and Creighton, J. D. E. and Creighton, T. D. and Cripe, J. and Croquette, M. and Crowder, S. G. and Cullen, T. J. and Cumming, A. and Cunningham, L. and Cuoco, E. and Canton, T. Dal and Dálya, G. and Danilishin, S. L. and D'Antonio, S. and Danzmann, K. and Dasgupta, A. and Da Silva Costa, C. F. and Datrier, L. E. H. and Dattilo, V. and Dave, I. and Davier, M. and Davis, D. and Daw, E. J. and DeBra, D. and Deenadayalan, M. and Degallaix, J. and De Laurentis, M. and Deléglise, S. and Del Pozzo, W. and DeMarchi, L. M. and Demos, N. and Dent, T. and De Pietri, R. and Derby, J. and De Rosa, R. and De Rossi, C. and DeSalvo, R. and de Varona, O. and Dhurandhar, S. and Díaz, M. C. and Dietrich, T. and Di Fiore, L. and Di Giovanni, M. and Di Girolamo, T. and Di Lieto, A. and Ding, B. and Di Pace, S. and Di Palma, I. and Di Renzo, F. and Dmitriev, A. and Doctor, Z. and Donovan, F. and Dooley, K. L. and Doravari, S. and Dorrington, I. and Downes, T. P. and Drago, M. and Driggers, J. C. and Du, Z. and Ducoin, J. -G. and Dupej, P. and Dwyer, S. E. and Easter, P. J. and Edo, T. B. and Edwards, M. C. and Effler, A. and Ehrens, P. and Eichholz, J. and Eikenberry, S. S. and Eisenmann, M. and Eisenstein, R. A. and Essick, R. C. and Estelles, H. and Estevez, D. and Etienne, Z. B. and Etzel, T. and Evans, M. and Evans, T. M. and Fafone, V. and Fair, H. and Fairhurst, S. and Fan, X. and Farinon, S. and Farr, B. and Farr, W. M. and Fauchon-Jones, E. J. and Favata, M. and Fays, M. and Fazio, M. and Fee, C. and Feicht, J. and Fejer, M. M. and Feng, F. and Fernandez-Galiana, A. and Ferrante, I. and Ferreira, E. C. and Ferreira, T. A. and Ferrini, F. and Fidecaro, F. and Fiori, I. and Fiorucci, D. and Fishbach, M. and Fisher, R. P. and Fishner, J. M. and Fitz-Axen, M. and Flaminio, R. and Fletcher, M. and Flynn, E. and Fong, H. and Font, J. A. and Forsyth, P. W. F. and Fournier, J. -D. and Frasca, S. and Frasconi, F. and Frei, Z. and Freise, A. and Frey, R. and Frey, V. and Fritschel, P. and Frolov, V. V. and Fulda, P. and Fyffe, M. and Gabbard, H. A. and Gadre, B. U. and Gaebel, S. M. and Gair, J. R. and Gammaitoni, L. and Ganija, M. R. and Gaonkar, S. G. and Garcia, A. and García-Quirós, C. and Garufi, F. and Gateley, B. and Gaudio, S. and Gaur, G. and Gayathri, V. and Gemme, G. and Genin, E. and Gennai, A. and George, D. and George, J. and Gergely, L. and Germain, V. and Ghonge, S. and Ghosh, Abhirup and Ghosh, Archisman and Ghosh, S. and Giacomazzo, B. and Giaime, J. A. and Giardina, K. D. and Giazotto, A. and Gill, K. and Giordano, G. and Glover, L. and Godwin, P. and Goetz, E. and Goetz, R. and Goncharov, B. and González, G. and Gonzalez Castro, J. M. and Gopakumar, A. and Gorodetsky, M. L. and Gossan, S. E. and Gosselin, M. and Gouaty, R. and Grado, A. and Graef, C. and Granata, M. and Grant, A. and Gras, S. and Grassia, P. and Gray, C. and Gray, R. and Greco, G. and Green, A. C. and Green, R. and Gretarsson, E. M. and Groot, P. and Grote, H. and Grunewald, S. and Gruning, P. and Guidi, G. M. and Gulati, H. K. and Guo, Y. and Gupta, A. and Gupta, M. K. and Gustafson, E. K. and Gustafson, R. and Haegel, L. and Halim, O. and Hall, B. R. and Hall, E. D. and Hamilton, E. Z. and Hammond, G. and Haney, M. and Hanke, M. M. and Hanks, J. and Hanna, C. and Hannam, M. D. and Hannuksela, O. A. and Hanson, J. and Hardwick, T. and Haris, K. and Harms, J. and Harry, G. M. and Harry, I. W. and Haster, C. -J. and Haughian, K. and Hayes, F. J. and Healy, J. and Heidmann, A. and Heintze, M. C. and Heitmann, H. and Hello, P. and Hemming, G. and Hendry, M. and Heng, I. S. and Hennig, J. and Heptonstall, A. W. and Hernandez Vivanco, Francisco and Heurs, M. and Hild, S. and Hinderer, T. and Hoak, D. and Hochheim, S. and Hofman, D. and Holgado, A. M. and Holland, N. A. and Holt, K. and Holz, D. E. and Hopkins, P. and Horst, C. and Hough, J. and Howell, E. J. and Hoy, C. G. and Hreibi, A. and Huang, Y. and Huerta, E. A. and Huet, D. and Hughey, B. and Hulko, M. and Husa, S. and Huttner, S. H. and Huynh-Dinh, T. and Idzkowski, B. and Iess, A. and Ingram, C. and Inta, R. and Intini, G. and Irwin, B. and Isa, H. N. and Isac, J. -M. and Isi, M. and Iyer, B. R. and Izumi, K. and Jacqmin, T. and Jadhav, S. J. and Jani, K. and Janthalur, N. N. and Jaranowski, P. and Jenkins, A. C. and Jiang, J. and Johnson, D. S. and Johnson-McDaniel, N. K. and Jones, A. W. and Jones, D. I. and Jones, R. and Jonker, R. J. G. and Ju, L. and Junker, J. and Kalaghatgi, C. V. and Kalogera, V. and Kamai, B. and Kandhasamy, S. and Kang, G. and Kanner, J. B. and Kapadia, S. J. and Karki, S. and Karvinen, K. S. and Kashyap, R. and Kasprzack, M. and Katsanevas, S. and Katsavounidis, E. and Katzman, W. and Kaufer, S. and Kawabe, K. and Keerthana, N. V. and Kéfélian, F. and Keitel, D. and Kennedy, R. and Key, J. S. and Khalili, F. Y. and Khan, H. and Khan, I. and Khan, S. and Khan, Z. and Khazanov, E. A. and Khursheed, M. and Kijbunchoo, N. and Kim, Chunglee and Kim, J. C. and Kim, K. and Kim, W. and Kim, W. S. and Kim, Y. -M. and Kimball, C. and King, E. J. and King, P. J. and Kinley-Hanlon, M. and Kirchhoff, R. and Kissel, J. S. and Kleybolte, L. and Klika, J. H. and Klimenko, S. and Knowles, T. D. and Koch, P. and Koehlenbeck, S. M. and Koekoek, G. and Koley, S. and Kondrashov, V. and Kontos, A. and Koper, N. and Korobko, M. and Korth, W. Z. and Kowalska, I. and Kozak, D. B. and Kringel, V. and Krishnendu, N. and Królak, A. and Kuehn, G. and Kumar, A. and Kumar, P. and Kumar, R. and Kumar, S. and Kuo, L. and Kutynia, A. and Kwang, S. and Lackey, B. D. and Lai, K. H. and Lam, T. L. and Landry, M. and Lane, B. B. and Lang, R. N. and Lange, J. and Lantz, B. and Lanza, R. K. and Lartaux-Vollard, A. and Lasky, P. D. and Laxen, M. and Lazzarini, A. and Lazzaro, C. and Leaci, P. and Leavey, S. and Lecoeuche, Y. K. and Lee, C. H. and Lee, H. K. and Lee, H. M. and Lee, H. W. and Lee, J. and Lee, K. and Lehmann, J. and Lenon, A. and Leroy, N. and Letendre, N. and Levin, Y. and Li, J. and Li, K. J. L. and Li, T. G. F. and Li, X. and Lin, F. and Linde, F. and Linker, S. D. and Littenberg, T. B. and Liu, J. and Liu, X. and Lo, R. K. L. and Lockerbie, N. A. and London, L. T. and Longo, A. and Lorenzini, M. and Loriette, V. and Lormand, M. and Losurdo, G. and Lough, J. D. and Lousto, C. O. and Lovelace, G. and Lower, M. E. and Lück, H. and Lumaca, D. and Lundgren, A. P. and Lynch, R. and Ma, Y. and Macas, R. and Macfoy, S. and MacInnis, M. and Macleod, D. M. and Macquet, A. and Magaña-Sandoval, F. and Magaña Zertuche, L. and Magee, R. M. and Majorana, E. and Maksimovic, I. and Malik, A. and Man, N. and Mandic, V. and Mangano, V. and Mansell, G. L. and Manske, M. and Mantovani, M. and Marchesoni, F. and Marion, F. and Márka, S. and Márka, Z. and Markakis, C. and Markosyan, A. S. and Markowitz, A. and Maros, E. and Marquina, A. and Marsat, S. and Martelli, F. and Martin, I. W. and Martin, R. M. and Martynov, D. V. and Mason, K. and Massera, E. and Masserot, A. and Massinger, T. J. and Masso-Reid, M. and Mastrogiovanni, S. and Matas, A. and Matichard, F. and Matone, L. and Mavalvala, N. and Mazumder, N. and McCann, J. J. and McCarthy, R. and McClelland, D. E. and McCormick, S. and McCuller, L. and McGuire, S. C. and McIver, J. and McManus, D. J. and McRae, T. and McWilliams, S. T. and Meacher, D. and Meadors, G. D. and Mehmet, M. and Mehta, A. K. and Meidam, J. and Melatos, A. and Mendell, G. and Mercer, R. A. and Mereni, L. and Merilh, E. L. and Merzougui, M. and Meshkov, S. and Messenger, C. and Messick, C. and Metzdorff, R. and Meyers, P. M. and Miao, H. and Michel, C. and Middleton, H. and Mikhailov, E. E. and Milano, L. and Miller, A. L. and Miller, A. and Millhouse, M. and Mills, J. C. and Milovich-Goff, M. C. and Minazzoli, O. and Minenkov, Y. and Mishkin, A. and Mishra, C. and Mistry, T. and Mitra, S. and Mitrofanov, V. P. and Mitselmakher, G. and Mittleman, R. and Mo, G. and Moffa, D. and Mogushi, K. and Mohapatra, S. R. P. and Montani, M. and Moore, C. J. and Moraru, D. and Moreno, G. and Morisaki, S. and Mours, B. and Mow-Lowry, C. M. and Mukherjee, Arunava and Mukherjee, D. and Mukherjee, S. and Mukund, N. and Mullavey, A. and Munch, J. and Muñiz, E. A. and Muratore, M. and Murray, P. G. and Nagar, A. and Nardecchia, I. and Naticchioni, L. and Nayak, R. K. and Neilson, J. and Nelemans, G. and Nelson, T. J. N. and Nery, M. and Neunzert, A. and Ng, K. Y. and Ng, S. and Nguyen, P. and Nichols, D. and Nielsen, A. B. and Nissanke, S. and Nitz, A. and Nocera, F. and North, C. and Nuttall, L. K. and Obergaulinger, M. and Oberling, J. and O'Brien, B. D. and O'Dea, G. D. and Ogin, G. H. and Oh, J. J. and Oh, S. H. and Ohme, F. and Ohta, H. and Okada, M. A. and Oliver, M. and Oppermann, P. and Oram, Richard J. and O'Reilly, B. and Ormiston, R. G. and Ortega, L. F. and O'Shaughnessy, R. and Ossokine, S. and Ottaway, D. J. and Overmier, H. and Owen, B. J. and Pace, A. E. and Pagano, G. and Page, M. A. and Pai, A. and Pai, S. A. and Palamos, J. R. and Palashov, O. and Palomba, C. and Pal-Singh, A. and Pan, Huang-Wei and Pang, B. and Pang, P. T. H. and Pankow, C. and Pannarale, F. and Pant, B. C. and Paoletti, F. and Paoli, A. and Papa, M. A. and Parida, A. and Parker, W. and Pascucci, D. and Pasqualetti, A. and Passaquieti, R. and Passuello, D. and Patil, M. and Patricelli, B. and Pearlstone, B. L. and Pedersen, C. and Pedraza, M. and Pedurand, R. and Pele, A. and Penn, S. and Perego, A. and Perez, C. J. and Perreca, A. and Pfeiffer, H. P. and Phelps, M. and Phukon, K. S. and Piccinni, O. J. and Pichot, M. and Piergiovanni, F. and Pillant, G. and Pinard, L. and Pirello, M. and Pitkin, M. and Poggiani, R. and Pong, D. Y. T. and Ponrathnam, S. and Popolizio, P. and Porter, E. K. and Powell, J. and Prajapati, A. K. and Prasad, J. and Prasai, K. and Prasanna, R. and Pratten, G. and Prestegard, T. and Privitera, S. and Prodi, G. A. and Prokhorov, L. G. and Puncken, O. and Punturo, M. and Puppo, P. and Pürrer, M. and Qi, H. and Quetschke, V. and Quinonez, P. J. and Quintero, E. A. and Quitzow-James, R. and Raab, F. J. and Radkins, H. and Radulescu, N. and Raffai, P. and Raja, S. and Rajan, C. and Rajbhandari, B. and Rakhmanov, M. and Ramirez, K. E. and Ramos-Buades, A. and Rana, Javed and Rao, K. and Rapagnani, P. and Raymond, V. and Razzano, M. and Read, J. and Regimbau, T. and Rei, L. and Reid, S. and Reitze, D. H. and Ren, W. and Ricci, F. and Richardson, C. J. and Richardson, J. W. and Ricker, P. M. and Riemenschneider, G. M. and Riles, K. and Rizzo, M. and Robertson, N. A. and Robie, R. and Robinet, F. and Rocchi, A. and Rolland, L. and Rollins, J. G. and Roma, V. J. and Romanelli, M. and Romano, R. and Romel, C. L. and Romie, J. H. and Rose, K. and Rosińska, D. and Rosofsky, S. G. and Ross, M. P. and Rowan, S. and Rüdiger, A. and Ruggi, P. and Rutins, G. and Ryan, K. and Sachdev, S. and Sadecki, T. and Sakellariadou, M. and Salafia, O. and Salconi, L. and Saleem, M. and Salemi, F. and Samajdar, A. and Sammut, L. and Sanchez, E. J. and Sanchez, L. E. and Sanchis-Gual, N. and Sandberg, V. and Sanders, J. R. and Santiago, K. A. and Sarin, N. and Sassolas, B. and Sathyaprakash, B. S. and Saulson, P. R. and Sauter, O. and Savage, R. L. and Schale, P. and Scheel, M. and Scheuer, J. and Schmidt, P. and Schnabel, R. and Schofield, R. M. S. and Schönbeck, A. and Schreiber, E. and Schulte, B. W. and Schutz, B. F. and Schwalbe, S. G. and Scott, J. and Scott, S. M. and Seidel, E. and Sellers, D. and Sengupta, A. S. and Sennett, N. and Sentenac, D. and Sequino, V. and Sergeev, A. and Setyawati, Y. and Shaddock, D. A. and Shaffer, T. and Shahriar, M. S. and Shaner, M. B. and Shao, L. and Sharma, P. and Shawhan, P. and Shen, H. and Shink, R. and Shoemaker, D. H. and Shoemaker, D. M. and ShyamSundar, S. and Siellez, K. and Sieniawska, M. and Sigg, D. and Silva, A. D. and Singer, L. P. and Singh, N. and Singhal, A. and Sintes, A. M. and Sitmukhambetov, S. and Skliris, V. and Slagmolen, B. J. J. and Slaven-Blair, T. J. and Smith, J. R. and Smith, R. J. E. and Somala, S. and Son, E. J. and Sorazu, B. and Sorrentino, F. and Souradeep, T. and Sowell, E. and Spencer, A. P. and Srivastava, A. K. and Srivastava, V. and Staats, K. and Stachie, C. and Standke, M. and Steer, D. A. and Steinke, M. and Steinlechner, J. and Steinlechner, S. and Steinmeyer, D. and Stevenson, S. P. and Stocks, D. and Stone, R. and Stops, D. J. and Strain, K. A. and Stratta, G. and Strigin, S. E. and Strunk, A. and Sturani, R. and Stuver, A. L. and Sudhir, V. and Summerscales, T. Z. and Sun, L. and Sunil, S. and Suresh, J. and Sutton, P. J. and Swinkels, B. L. and Szczepańczyk, M. J. and Tacca, M. and Tait, S. C. and Talbot, C. and Talukder, D. and Tanner, D. B. and Tápai, M. and Taracchini, A. and Tasson, J. D. and Taylor, R. and Thies, F. and Thomas, M. and Thomas, P. and Thondapu, S. R. and Thorne, K. A. and Thrane, E. and Tiwari, Shubhanshu and Tiwari, Srishti and Tiwari, V. and Toland, K. and Tonelli, M. and Tornasi, Z. and Torres-Forné, A. and Torrie, C. I. and Töyrä, D. and Travasso, F. and Traylor, G. and Tringali, M. C. and Trovato, A. and Trozzo, L. and Trudeau, R. and Tsang, K. W. and Tse, M. and Tso, R. and Tsukada, L. and Tsuna, D. and Tuyenbayev, D. and Ueno, K. and Ugolini, D. and Unnikrishnan, C. S. and Urban, A. L. and Usman, S. A. and Vahlbruch, H. and Vajente, G. and Valdes, G. and van Bakel, N. and van Beuzekom, M. and van den Brand, J. F. J. and Van Den Broeck, C. and Vander-Hyde, D. C. and van Heijningen, J. V. and van der Schaaf, L. and van Veggel, A. A. and Vardaro, M. and Varma, V. and Vass, S. and Vasúth, M. and Vecchio, A. and Vedovato, G. and Veitch, J. and Veitch, P. J. and Venkateswara, K. and Venugopalan, G. and Verkindt, D. and Vetrano, F. and Viceré, A. and Viets, A. D. and Vine, D. J. and Vinet, J. -Y. and Vitale, S. and Vo, T. and Vocca, H. and Vorvick, C. and Vyatchanin, S. P. and Wade, A. R. and Wade, L. E. and Wade, M. and Walet, R. and Walker, M. and Wallace, L. and Walsh, S. and Wang, G. and Wang, H. and Wang, J. Z. and Wang, W. H. and Wang, Y. F. and Ward, R. L. and Warden, Z. A. and Warner, J. and Was, M. and Watchi, J. and Weaver, B. and Wei, L. -W. and Weinert, M. and Weinstein, A. J. and Weiss, R. and Wellmann, F. and Wen, L. and Wessel, E. K. and Weßels, P. and Westhouse, J. W. and Wette, K. and Whelan, J. T. and White, L. V. and Whiting, B. F. and Whittle, C. and Wilken, D. M. and Williams, D. and Williamson, A. R. and Willis, J. L. and Willke, B. and Wimmer, M. H. and Winkler, W. and Wipf, C. C. and Wittel, H. and Woan, G. and Woehler, J. and Wofford, J. K. and Worden, J. and Wright, J. L. and Wu, D. S. and Wysocki, D. M. and Xiao, L. and Yamamoto, H. and Yancey, C. C. and Yang, L. and Yap, M. J. and Yazback, M. and Yeeles, D. W. and Yu, Hang and Yu, Haocun and Yuen, S. H. R. and Yvert, M. and ZadroŻny, A. K. and Zanolin, M. and Zappa, F. and Zelenova, T. and Zendri, J. -P. and Zevin, M. and Zhang, J. and Zhang, L. and Zhang, T. and Zhao, C. and Zhou, M. and Zhou, Z. and Zhu, X. J. and Zimmerman, A. B. and Zlochower, Y. and Zucker, M. E. and Zweizig, J. and {LIGO Scientific Collaboration} and {Virgo Collaboration}},
	month = jul,
	year = {2019},
	note = {Publisher: APS
ADS Bibcode: 2019PhRvX...9c1040A},
	pages = {031040},
}

@article{catalog-gwtc-2,
	title = {{GWTC}-2: {Compact} {Binary} {Coalescences} {Observed} by {LIGO} and {Virgo} during the {First} {Half} of the {Third} {Observing} {Run}},
	volume = {11},
	shorttitle = {{GWTC}-2},
	url = {https://ui.adsabs.harvard.edu/abs/2021PhRvX..11b1053A},
	doi = {10.1103/PhysRevX.11.021053},
	urldate = {2024-08-14},
	journal = {Physical Review X},
	author = {Abbott, R. and Abbott, T. D. and Abraham, S. and Acernese, F. and Ackley, K. and Adams, A. and Adams, C. and Adhikari, R. X. and Adya, V. B. and Affeldt, C. and Agathos, M. and Agatsuma, K. and Aggarwal, N. and Aguiar, O. D. and Aiello, L. and Ain, A. and Ajith, P. and Akcay, S. and Allen, G. and Allocca, A. and Altin, P. A. and Amato, A. and Anand, S. and Ananyeva, A. and Anderson, S. B. and Anderson, W. G. and Angelova, S. V. and Ansoldi, S. and Antelis, J. M. and Antier, S. and Appert, S. and Arai, K. and Araya, M. C. and Areeda, J. S. and Arène, M. and Arnaud, N. and Aronson, S. M. and Arun, K. G. and Asali, Y. and Ascenzi, S. and Ashton, G. and Aston, S. M. and Astone, P. and Aubin, F. and Aufmuth, P. and AultONeal, K. and Austin, C. and Avendano, V. and Babak, S. and Badaracco, F. and Bader, M. K. M. and Bae, S. and Baer, A. M. and Bagnasco, S. and Baird, J. and Ball, M. and Ballardin, G. and Ballmer, S. W. and Bals, A. and Balsamo, A. and Baltus, G. and Banagiri, S. and Bankar, D. and Bankar, R. S. and Barayoga, J. C. and Barbieri, C. and Barish, B. C. and Barker, D. and Barneo, P. and Barnum, S. and Barone, F. and Barr, B. and Barsotti, L. and Barsuglia, M. and Barta, D. and Bartlett, J. and Bartos, I. and Bassiri, R. and Basti, A. and Bawaj, M. and Bayley, J. C. and Bazzan, M. and Becher, B. R. and Bécsy, B. and Bedakihale, V. M. and Bejger, M. and Belahcene, I. and Beniwal, D. and Benjamin, M. G. and Bennett, T. F. and Bentley, J. D. and Bergamin, F. and Berger, B. K. and Bergmann, G. and Bernuzzi, S. and Berry, C. P. L. and Bersanetti, D. and Bertolini, A. and Betzwieser, J. and Bhandare, R. and Bhandari, A. V. and Bhattacharjee, D. and Bidler, J. and Bilenko, I. A. and Billingsley, G. and Birney, R. and Birnholtz, O. and Biscans, S. and Bischi, M. and Biscoveanu, S. and Bisht, A. and Bitossi, M. and Bizouard, M. -A. and Blackburn, J. K. and Blackman, J. and Blair, C. D. and Blair, D. G. and Blair, R. M. and Blanch, O. and Bobba, F. and Bode, N. and Boer, M. and Boetzel, Y. and Bogaert, G. and Boldrini, M. and Bondu, F. and Bonilla, E. and Bonnand, R. and Booker, P. and Boom, B. A. and Bork, R. and Boschi, V. and Bose, S. and Bossilkov, V. and Boudart, V. and Bouffanais, Y. and Bozzi, A. and Bradaschia, C. and Brady, P. R. and Bramley, A. and Branchesi, M. and Brau, J. E. and Breschi, M. and Briant, T. and Briggs, J. H. and Brighenti, F. and Brillet, A. and Brinkmann, M. and Brockill, P. and Brooks, A. F. and Brooks, J. and Brown, D. D. and Brunett, S. and Bruno, G. and Bruntz, R. and Buikema, A. and Bulik, T. and Bulten, H. J. and Buonanno, A. and Buscicchio, R. and Buskulic, D. and Byer, R. L. and Cabero, M. and Cadonati, L. and Caesar, M. and Cagnoli, G. and Cahillane, C. and Calderón Bustillo, J. and Callaghan, J. D. and Callister, T. A. and Calloni, E. and Camp, J. B. and Canepa, M. and Cannon, K. C. and Cao, H. and Cao, J. and Carapella, G. and Carbognani, F. and Carney, M. F. and Carpinelli, M. and Carullo, G. and Carver, T. L. and Casanueva Diaz, J. and Casentini, C. and Caudill, S. and Cavaglià, M. and Cavalier, F. and Cavalieri, R. and Cella, G. and Cerdá-Durán, P. and Cesarini, E. and Chaibi, W. and Chakravarti, K. and Chan, C. -L. and Chan, C. and Chandra, K. and Chanial, P. and Chao, S. and Charlton, P. and Chase, E. A. and Chassande-Mottin, E. and Chatterjee, D. and Chattopadhyay, D. and Chaturvedi, M. and Chatziioannou, K. and Chen, A. and Chen, H. Y. and Chen, X. and Chen, Y. and Cheng, H. -P. and Cheong, C. K. and Chia, H. Y. and Chiadini, F. and Chierici, R. and Chincarini, A. and Chiummo, A. and Cho, G. and Cho, H. S. and Cho, M. and Choate, S. and Christensen, N. and Chu, Q. and Chua, S. and Chung, K. W. and Chung, S. and Ciani, G. and Ciecielag, P. and Cieślar, M. and Cifaldi, M. and Ciobanu, A. A. and Ciolfi, R. and Cipriano, F. and Cirone, A. and Clara, F. and Clark, E. N. and Clark, J. A. and Clarke, L. and Clearwater, P. and Clesse, S. and Cleva, F. and Coccia, E. and Cohadon, P. -F. and Cohen, D. E. and Colleoni, M. and Collette, C. G. and Collins, C. and Colpi, M. and Constancio, M. and Conti, L. and Cooper, S. J. and Corban, P. and Corbitt, T. R. and Cordero-Carrión, I. and Corezzi, S. and Corley, K. R. and Cornish, N. and Corre, D. and Corsi, A. and Cortese, S. and Costa, C. A. and Cotesta, R. and Coughlin, M. W. and Coughlin, S. B. and Coulon, J. -P. and Countryman, S. T. and Cousins, B. and Couvares, P. and Covas, P. B. and Coward, D. M. and Cowart, M. J. and Coyne, D. C. and Coyne, R. and Creighton, J. D. E. and Creighton, T. D. and Croquette, M. and Crowder, S. G. and Cudell, J. R. and Cullen, T. J. and Cumming, A. and Cummings, R. and Cunningham, L. and Cuoco, E. and Curyło, M. and Canton, T. Dal and Dálya, G. and Dana, A. and DaneshgaranBajastani, L. M. and D'Angelo, B. and Danila, B. and Danilishin, S. L. and D'Antonio, S. and Danzmann, K. and Darsow-Fromm, C. and Dasgupta, A. and Datrier, L. E. H. and Dattilo, V. and Dave, I. and Davier, M. and Davies, G. S. and Davis, D. and Daw, E. J. and Dean, R. and DeBra, D. and Deenadayalan, M. and Degallaix, J. and De Laurentis, M. and Deléglise, S. and Del Favero, V. and De Lillo, F. and De Lillo, N. and Del Pozzo, W. and DeMarchi, L. M. and De Matteis, F. and D'Emilio, V. and Demos, N. and Denker, T. and Dent, T. and Depasse, A. and De Pietri, R. and De Rosa, R. and De Rossi, C. and DeSalvo, R. and de Varona, O. and Dhurandhar, S. and Díaz, M. C. and Diaz-Ortiz, M. and Didio, N. A. and Dietrich, T. and Di Fiore, L. and DiFronzo, C. and Di Giorgio, C. and Di Giovanni, F. and Di Giovanni, M. and Di Girolamo, T. and Di Lieto, A. and Ding, B. and Di Pace, S. and Di Palma, I. and Di Renzo, F. and Divakarla, A. K. and Dmitriev, A. and Doctor, Z. and D'Onofrio, L. and Donovan, F. and Dooley, K. L. and Doravari, S. and Dorrington, I. and Downes, T. P. and Drago, M. and Driggers, J. C. and Du, Z. and Ducoin, J. -G. and Dupej, P. and Durante, O. and D'Urso, D. and Duverne, P. -A. and Dwyer, S. E. and Easter, P. J. and Eddolls, G. and Edelman, B. and Edo, T. B. and Edy, O. and Effler, A. and Eichholz, J. and Eikenberry, S. S. and Eisenmann, M. and Eisenstein, R. A. and Ejlli, A. and Errico, L. and Essick, R. C. and Estellés, H. and Estevez, D. and Etienne, Z. B. and Etzel, T. and Evans, M. and Evans, T. M. and Ewing, B. E. and Fafone, V. and Fair, H. and Fairhurst, S. and Fan, X. and Farah, A. M. and Farinon, S. and Farr, B. and Farr, W. M. and Fauchon-Jones, E. J. and Favata, M. and Fays, M. and Fazio, M. and Feicht, J. and Fejer, M. M. and Feng, F. and Fenyvesi, E. and Ferguson, D. L. and Fernandez-Galiana, A. and Ferrante, I. and Ferreira, T. A. and Fidecaro, F. and Figura, P. and Fiori, I. and Fiorucci, D. and Fishbach, M. and Fisher, R. P. and Fishner, J. M. and Fittipaldi, R. and Fitz-Axen, M. and Fiumara, V. and Flaminio, R. and Floden, E. and Flynn, E. and Fong, H. and Font, J. A. and Forsyth, P. W. F. and Fournier, J. -D. and Frasca, S. and Frasconi, F. and Frei, Z. and Freise, A. and Frey, R. and Frey, V. and Fritschel, P. and Frolov, V. V. and Fronzé, G. G. and Fulda, P. and Fyffe, M. and Gabbard, H. A. and Gadre, B. U. and Gaebel, S. M. and Gair, J. R. and Gais, J. and Galaudage, S. and Gamba, R. and Ganapathy, D. and Ganguly, A. and Gaonkar, S. G. and Garaventa, B. and García-Quirós, C. and Garufi, F. and Gateley, B. and Gaudio, S. and Gayathri, V. and Gemme, G. and Gennai, A. and George, D. and George, J. and George, R. N. and Gergely, L. and Ghonge, S. and Ghosh, Abhirup and Ghosh, Archisman and Ghosh, S. and Giacomazzo, B. and Giacoppo, L. and Giaime, J. A. and Giardina, K. D. and Gibson, D. R. and Gier, C. and Gill, K. and Giri, P. and Glanzer, J. and Gleckl, A. E. and Godwin, P. and Goetz, E. and Goetz, R. and Gohlke, N. and Goncharov, B. and González, G. and Gopakumar, A. and Gossan, S. E. and Gosselin, M. and Gouaty, R. and Grace, B. and Grado, A. and Granata, M. and Granata, V. and Grant, A. and Gras, S. and Grassia, P. and Gray, C. and Gray, R. and Greco, G. and Green, A. C. and Green, R. and Gretarsson, E. M. and Griggs, H. L. and Grignani, G. and Grimaldi, A. and Grimes, E. and Grimm, S. J. and Grote, H. and Grunewald, S. and Gruning, P. and Guerrero, J. G. and Guidi, G. M. and Guimaraes, A. R. and Guixé, G. and Gulati, H. K. and Guo, Y. and Gupta, Anchal and Gupta, Anuradha and Gupta, P. and Gustafson, E. K. and Gustafson, R. and Guzman, F. and Haegel, L. and Halim, O. and Hall, E. D. and Hamilton, E. Z. and Hammond, G. and Haney, M. and Hanke, M. M. and Hanks, J. and Hanna, C. and Hannam, M. D. and Hannuksela, O. A. and Hannuksela, O. and Hansen, H. and Hansen, T. J. and Hanson, J. and Harder, T. and Hardwick, T. and Haris, K. and Harms, J. and Harry, G. M. and Harry, I. W. and Hartwig, D. and Hasskew, R. K. and Haster, C. -J. and Haughian, K. and Hayes, F. J. and Healy, J. and Heidmann, A. and Heintze, M. C. and Heinze, J. and Heinzel, J. and Heitmann, H. and Hellman, F. and Hello, P. and Helmling-Cornell, A. F. and Hemming, G. and Hendry, M. and Heng, I. S. and Hennes, E. and Hennig, J. and Hennig, M. H. and Hernandez Vivanco, F. and Heurs, M. and Hild, S. and Hill, P. and Hines, A. S. and Hochheim, S. and Hofgard, E. and Hofman, D. and Hohmann, J. N. and Holgado, A. M. and Holland, N. A. and Hollows, I. J. and Holmes, Z. J. and Holt, K. and Holz, D. E. and Hopkins, P. and Horst, C. and Hough, J. and Howell, E. J. and Hoy, C. G. and Hoyland, D. and Huang, Y. and Hübner, M. T. and Huddart, A. D. and Huerta, E. A. and Hughey, B. and Hui, V. and Husa, S. and Huttner, S. H. and Hutzler, B. M. and Huxford, R. and Huynh-Dinh, T. and Idzkowski, B. and Iess, A. and Imperato, S. and Inchauspe, H. and Ingram, C. and Intini, G. and Isi, M. and Iyer, B. R. and JaberianHamedan, V. and Jacqmin, T. and Jadhav, S. J. and Jadhav, S. P. and James, A. L. and Jani, K. and Janssens, K. and Janthalur, N. N. and Jaranowski, P. and Jariwala, D. and Jaume, R. and Jenkins, A. C. and Jeunon, M. and Jiang, J. and Johns, G. R. and Johnson-McDaniel, N. K. and Jones, A. W. and Jones, D. I. and Jones, J. D. and Jones, P. and Jones, R. and Jonker, R. J. G. and Ju, L. and Junker, J. and Kalaghatgi, C. V. and Kalogera, V. and Kamai, B. and Kandhasamy, S. and Kang, G. and Kanner, J. B. and Kapadia, S. J. and Kapasi, D. P. and Karathanasis, C. and Karki, S. and Kashyap, R. and Kasprzack, M. and Kastaun, W. and Katsanevas, S. and Katsavounidis, E. and Katzman, W. and Kawabe, K. and Kéfélian, F. and Keitel, D. and Key, J. S. and Khadka, S. and Khalili, F. Y. and Khan, I. and Khan, S. and Khazanov, E. A. and Khetan, N. and Khursheed, M. and Kijbunchoo, N. and Kim, C. and Kim, G. J. and Kim, J. C. and Kim, K. and Kim, W. S. and Kim, Y. -M. and Kimball, C. and King, P. J. and Kinley-Hanlon, M. and Kirchhoff, R. and Kissel, J. S. and Kleybolte, L. and Klimenko, S. and Knowles, T. D. and Knyazev, E. and Koch, P. and Koehlenbeck, S. M. and Koekoek, G. and Koley, S. and Kolstein, M. and Komori, K. and Kondrashov, V. and Kontos, A. and Koper, N. and Korobko, M. and Korth, W. Z. and Kovalam, M. and Kozak, D. B. and Krämer, C. and Kringel, V. and Krishnendu, N. V. and Królak, A. and Kuehn, G. and Kumar, A. and Kumar, P. and Kumar, Rahul and Kumar, Rakesh and Kuns, K. and Kwang, S. and Lackey, B. D. and Laghi, D. and Lalande, E. and Lam, T. L. and Lamberts, A. and Landry, M. and Lane, B. B. and Lang, R. N. and Lange, J. and Lantz, B. and Lanza, R. K. and La Rosa, I. and Lartaux-Vollard, A. and Lasky, P. D. and Laxen, M. and Lazzarini, A. and Lazzaro, C. and Leaci, P. and Leavey, S. and Lecoeuche, Y. K. and Lee, H. M. and Lee, H. W. and Lee, J. and Lee, K. and Lehmann, J. and Leon, E. and Leroy, N. and Letendre, N. and Levin, Y. and Li, A. and Li, J. and Li, K. J. L. and Li, T. G. F. and Li, X. and Linde, F. and Linker, S. D. and Linley, J. N. and Littenberg, T. B. and Liu, J. and Liu, X. and Llorens-Monteagudo, M. and Lo, R. K. L. and Lockwood, A. and London, L. T. and Longo, A. and Lorenzini, M. and Loriette, V. and Lormand, M. and Losurdo, G. and Lough, J. D. and Lousto, C. O. and Lovelace, G. and Lück, H. and Lumaca, D. and Lundgren, A. P. and Ma, Y. and Macas, R. and MacInnis, M. and Macleod, D. M. and MacMillan, I. A. O. and Macquet, A. and Magaña Hernandez, I. and Magaña-Sandoval, F. and Magazzù, C. and Magee, R. M. and Majorana, E. and Maksimovic, I. and Maliakal, S. and Malik, A. and Man, N. and Mandic, V. and Mangano, V. and Mansell, G. L. and Manske, M. and Mantovani, M. and Mapelli, M. and Marchesoni, F. and Marion, F. and Márka, S. and Márka, Z. and Markakis, C. and Markosyan, A. S. and Markowitz, A. and Maros, E. and Marquina, A. and Marsat, S. and Martelli, F. and Martin, I. W. and Martin, R. M. and Martinez, M. and Martinez, V. and Martynov, D. V. and Masalehdan, H. and Mason, K. and Massera, E. and Masserot, A. and Massinger, T. J. and Masso-Reid, M. and Mastrogiovanni, S. and Matas, A. and Mateu-Lucena, M. and Matichard, F. and Matiushechkina, M. and Mavalvala, N. and Maynard, E. and McCann, J. J. and McCarthy, R. and McClelland, D. E. and McCormick, S. and McCuller, L. and McGuire, S. C. and McIsaac, C. and McIver, J. and McManus, D. J. and McRae, T. and McWilliams, S. T. and Meacher, D. and Meadors, G. D. and Mehmet, M. and Mehta, A. K. and Melatos, A. and Melchor, D. A. and Mendell, G. and Menendez-Vazquez, A. and Mercer, R. A. and Mereni, L. and Merfeld, K. and Merilh, E. L. and Merritt, J. D. and Merzougui, M. and Meshkov, S. and Messenger, C. and Messick, C. and Metzdorff, R. and Meyers, P. M. and Meylahn, F. and Mhaske, A. and Miani, A. and Miao, H. and Michaloliakos, I. and Michel, C. and Middleton, H. and Milano, L. and Miller, A. L. and Millhouse, M. and Mills, J. C. and Milotti, E. and Milovich-Goff, M. C. and Minazzoli, O. and Minenkov, Y. and Mir, Ll. M. and Mishkin, A. and Mishra, C. and Mistry, T. and Mitra, S. and Mitrofanov, V. P. and Mitselmakher, G. and Mittleman, R. and Mo, G. and Mogushi, K. and Mohapatra, S. R. P. and Mohite, S. R. and Molina, I. and Molina-Ruiz, M. and Mondin, M. and Montani, M. and Moore, C. J. and Moraru, D. and Morawski, F. and Moreno, G. and Morisaki, S. and Mours, B. and Mow-Lowry, C. M. and Mozzon, S. and Muciaccia, F. and Mukherjee, Arunava and Mukherjee, D. and Mukherjee, Soma and Mukherjee, Subroto and Mukund, N. and Mullavey, A. and Munch, J. and Muñiz, E. A. and Murray, P. G. and Nadji, S. L. and Nagar, A. and Nardecchia, I. and Naticchioni, L. and Nayak, R. K. and Neil, B. F. and Neilson, J. and Nelemans, G. and Nelson, T. J. N. and Nery, M. and Neunzert, A. and Nitz, A. H. and Ng, K. Y. and Ng, S. and Nguyen, C. and Nguyen, P. and Nguyen, T. and Nichols, S. A. and Nissanke, S. and Nocera, F. and Noh, M. and North, C. and Nothard, D. and Nuttall, L. K. and Oberling, J. and O'Brien, B. D. and O'Dell, J. and Oganesyan, G. and Ogin, G. H. and Oh, J. J. and Oh, S. H. and Ohme, F. and Ohta, H. and Okada, M. A. and Olivetto, C. and Oppermann, P. and Oram, R. J. and O'Reilly, B. and Ormiston, R. G. and Ortega, L. F. and O'Shaughnessy, R. and Ossokine, S. and Osthelder, C. and Ottaway, D. J. and Overmier, H. and Owen, B. J. and Pace, A. E. and Pagano, G. and Page, M. A. and Pagliaroli, G. and Pai, A. and Pai, S. A. and Palamos, J. R. and Palashov, O. and Palomba, C. and Pan, H. and Panda, P. K. and Pang, T. H. and Pankow, C. and Pannarale, F. and Pant, B. C. and Paoletti, F. and Paoli, A. and Paolone, A. and Parker, W. and Pascucci, D. and Pasqualetti, A. and Passaquieti, R. and Passuello, D. and Patel, M. and Patricelli, B. and Payne, E. and Pechsiri, T. C. and Pedraza, M. and Pegoraro, M. and Pele, A. and Penn, S. and Perego, A. and Perez, C. J. and Périgois, C. and Perreca, A. and Perriès, S. and Petermann, J. and Petterson, D. and Pfeiffer, H. P. and Pham, K. A. and Phukon, K. S. and Piccinni, O. J. and Pichot, M. and Piendibene, M. and Piergiovanni, F. and Pierini, L. and Pierro, V. and Pillant, G. and Pilo, F. and Pinard, L. and Pinto, I. M. and Piotrzkowski, K. and Pirello, M. and Pitkin, M. and Placidi, E. and Plastino, W. and Pluchar, C. and Poggiani, R. and Polini, E. and Pong, D. Y. T. and Ponrathnam, S. and Popolizio, P. and Porter, E. K. and Poverman, A. and Powell, J. and Pracchia, M. and Prajapati, A. K. and Prasai, K. and Prasanna, R. and Pratten, G. and Prestegard, T. and Principe, M. and Prodi, G. A. and Prokhorov, L. and Prosposito, P. and Prudenzi, L. and Puecher, A. and Punturo, M. and Puosi, F. and Puppo, P. and Pürrer, M. and Qi, H. and Quetschke, V. and Quinonez, P. J. and Quitzow-James, R. and Raab, F. J. and Raaijmakers, G. and Radkins, H. and Radulesco, N. and Raffai, P. and Rafferty, H. and Rail, S. X. and Raja, S. and Rajan, C. and Rajbhandari, B. and Rakhmanov, M. and Ramirez, K. E. and Ramirez, T. D. and Ramos-Buades, A. and Rana, J. and Rao, K. and Rapagnani, P. and Rapol, U. D. and Ratto, B. and Raymond, V. and Razzano, M. and Read, J. and Regimbau, T. and Rei, L. and Reid, S. and Reitze, D. H. and Rettegno, P. and Ricci, F. and Richardson, C. J. and Richardson, J. W. and Richardson, L. and Ricker, P. M. and Riemenschneider, G. and Riles, K. and Rizzo, M. and Robertson, N. A. and Robinet, F. and Rocchi, A. and Rocha, J. A. and Rodriguez, S. and Rodriguez-Soto, R. D. and Rolland, L. and Rollins, J. G. and Roma, V. J. and Romanelli, M. and Romano, R. and Romel, C. L. and Romero, A. and Romero-Shaw, I. M. and Romie, J. H. and Ronchini, S. and Rose, C. A. and Rose, D. and Rose, K. and Rosell, M. J. B. and Rosińska, D. and Rosofsky, S. G. and Ross, M. P. and Rowan, S. and Rowlinson, S. J. and Roy, Santosh and Roy, Soumen and Ruggi, P. and Ryan, K. and Sachdev, S. and Sadecki, T. and Sadiq, J. and Sakellariadou, M. and Salafia, O. S. and Salconi, L. and Saleem, M. and Samajdar, A. and Sanchez, E. J. and Sanchez, J. H. and Sanchez, L. E. and Sanchis-Gual, N. and Sanders, J. R. and Sandles, L. and Santiago, K. A. and Santos, E. and Saravanan, T. R. and Sarin, N. and Sassolas, B. and Sathyaprakash, B. S. and Sauter, O. and Savage, R. L. and Savant, V. and Sawant, D. and Sayah, S. and Schaetzl, D. and Schale, P. and Scheel, M. and Scheuer, J. and Schindler-Tyka, A. and Schmidt, P. and Schnabel, R. and Schofield, R. M. S. and Schönbeck, A. and Schreiber, E. and Schulte, B. W. and Schutz, B. F. and Schwarm, O. and Schwartz, E. and Scott, J. and Scott, S. M. and Seglar-Arroyo, M. and Seidel, E. and Sellers, D. and Sengupta, A. S. and Sennett, N. and Sentenac, D. and Sequino, V. and Sergeev, A. and Setyawati, Y. and Shaffer, T. and Shahriar, M. S. and Sharifi, S. and Sharma, A. and Sharma, P. and Shawhan, P. and Shen, H. and Shikauchi, M. and Shink, R. and Shoemaker, D. H. and Shoemaker, D. M. and Shukla, K. and ShyamSundar, S. and Sieniawska, M. and Sigg, D. and Singer, L. P. and Singh, D. and Singh, N. and Singha, A. and Singhal, A. and Sintes, A. M. and Sipala, V. and Skliris, V. and Slagmolen, B. J. J. and Slaven-Blair, T. J. and Smetana, J. and Smith, J. R. and Smith, R. J. E. and Somala, S. N. and Son, E. J. and Soni, K. and Soni, S. and Sorazu, B. and Sordini, V. and Sorrentino, F. and Sorrentino, N. and Soulard, R. and Souradeep, T. and Sowell, E. and Spencer, A. P. and Spera, M. and Srivastava, A. K. and Srivastava, V. and Staats, K. and Stachie, C. and Steer, D. A. and Steinhoff, J. and Steinke, M. and Steinlechner, J. and Steinlechner, S. and Steinmeyer, D. and Stevenson, S. P. and Stolle-McAllister, G. and Stops, D. J. and Stover, M. and Strain, K. A. and Stratta, G. and Strunk, A. and Sturani, R. and Stuver, A. L. and Südbeck, J. and Sudhagar, S. and Sudhir, V. and Suh, H. G. and Summerscales, T. Z. and Sun, H. and Sun, L. and Sunil, S. and Sur, A. and Suresh, J. and Sutton, P. J. and Swinkels, B. L. and Szczepańczyk, M. J. and Tacca, M. and Tait, S. C. and Talbot, C. and Tanasijczuk, A. J. and Tanner, D. B. and Tao, D. and Tapia, A. and Tapia San Martin, E. N. and Tasson, J. D. and Taylor, R. and Tenorio, R. and Terkowski, L. and Thirugnanasambandam, M. P. and Thomas, L. M. and Thomas, M. and Thomas, P. and Thompson, J. E. and Thondapu, S. R. and Thorne, K. A. and Thrane, E. and Tiwari, Shubhanshu and Tiwari, Srishti and Tiwari, V. and Toland, K. and Tolley, A. E. and Tonelli, M. and Tornasi, Z. and Torres-Forné, A. and Torrie, C. I. and e Melo, I. Tosta and Töyrä, D. and Tran, A. T. and Trapananti, A. and Travasso, F. and Traylor, G. and Tringali, M. C. and Tripathee, A. and Trovato, A. and Trudeau, R. J. and Tsai, D. S. and Tsang, K. W. and Tse, M. and Tso, R. and Tsukada, L. and Tsuna, D. and Tsutsui, T. and Turconi, M. and Ubhi, A. S. and Udall, R. P. and Ueno, K. and Ugolini, D. and Unnikrishnan, C. S. and Urban, A. L. and Usman, S. A. and Utina, A. C. and Vahlbruch, H. and Vajente, G. and Vajpeyi, A. and Valdes, G. and Valentini, M. and Valsan, V. and van Bakel, N. and van Beuzekom, M. and van den Brand, J. F. J. and Van Den Broeck, C. and Vander-Hyde, D. C. and van der Schaaf, L. and van Heijningen, J. V. and Vardaro, M. and Vargas, A. F. and Varma, V. and Vass, S. and Vasúth, M. and Vecchio, A. and Vedovato, G. and Veitch, J. and Veitch, P. J. and Venkateswara, K. and Venneberg, J. and Venugopalan, G. and Verkindt, D. and Verma, Y. and Veske, D. and Vetrano, F. and Viceré, A. and Viets, A. D. and Vijaykumar, A. and Villa-Ortega, V. and Vinet, J. -Y. and Vitale, S. and Vo, T. and Vocca, H. and Vorvick, C. and Vyatchanin, S. P. and Wade, A. R. and Wade, L. E. and Wade, M. and Walet, R. C. and Walker, M. and Wallace, G. S. and Wallace, L. and Walsh, S. and Wang, J. Z. and Wang, S. and Wang, W. H. and Wang, Y. F. and Ward, R. L. and Warner, J. and Was, M. and Washington, N. Y. and Watchi, J. and Weaver, B. and Wei, L. and Weinert, M. and Weinstein, A. J. and Weiss, R. and Wellmann, F. and Wen, L. and Weßels, P. and Westhouse, J. W. and Wette, K. and Whelan, J. T. and White, D. D. and White, L. V. and Whiting, B. F. and Whittle, C. and Wilken, D. M. and Williams, D. and Williams, M. J. and Williamson, A. R. and Willis, J. L. and Willke, B. and Wilson, D. J. and Wimmer, M. H. and Winkler, W. and Wipf, C. C. and Woan, G. and Woehler, J. and Wofford, J. K. and Wong, I. C. F. and Wrangel, J. and Wright, J. L. and Wu, D. S. and Wysocki, D. M. and Xiao, L. and Yamamoto, H. and Yang, L. and Yang, Y. and Yang, Z. and Yap, M. J. and Yeeles, D. W. and Yoon, A. and Yu, Hang and Yu, Haocun and Yuen, S. H. R. and ZadroŻny, A. and Zanolin, M. and Zelenova, T. and Zendri, J. -P. and Zevin, M. and Zhang, J. and Zhang, L. and Zhang, R. and Zhang, T. and Zhao, C. and Zhao, G. and Zheng, Y. and Zhou, M. and Zhou, Z. and Zhu, X. J. and Zimmerman, A. B. and Zlochower, Y. and Zucker, M. E. and Zweizig, J. and {LIGO Scientific Collaboration} and {Virgo Collaboration}},
	month = apr,
	year = {2021},
	note = {Publisher: APS
ADS Bibcode: 2021PhRvX..11b1053A},
	pages = {021053},
}

@article{guglielmetti_backgroundsource_2009,
	title = {Background–source separation in astronomical images with {Bayesian} probability theory – {I}. {The} method},
	volume = {396},
	issn = {0035-8711},
	url = {https://doi.org/10.1111/j.1365-2966.2009.14739.x},
	doi = {10.1111/j.1365-2966.2009.14739.x},
	number = {1},
	urldate = {2024-12-01},
	journal = {Monthly Notices of the Royal Astronomical Society},
	author = {Guglielmetti, F. and Fischer, R. and Dose, V.},
	month = jun,
	year = {2009},
	pages = {165--190},
}

@article{farr_counting_2015,
	title = {Counting and confusion: {Bayesian} rate estimation with multiple populations},
	volume = {91},
	shorttitle = {Counting and confusion},
	url = {https://link.aps.org/doi/10.1103/PhysRevD.91.023005},
	doi = {10.1103/PhysRevD.91.023005},
	number = {2},
	urldate = {2024-12-01},
	journal = {Physical Review D},
	author = {Farr, Will M. and Gair, Jonathan R. and Mandel, Ilya and Cutler, Curt},
	month = jan,
	year = {2015},
	note = {Publisher: American Physical Society},
	pages = {023005},
}

@article{software-htcondor,
	title = {Distributed computing in practice: the {Condor} experience},
	volume = {17},
	copyright = {Copyright © 2005 John Wiley \& Sons, Ltd.},
	issn = {1532-0634},
	shorttitle = {Distributed computing in practice},
	url = {https://onlinelibrary.wiley.com/doi/abs/10.1002/cpe.938},
	doi = {10.1002/cpe.938},
	language = {en},
	number = {2-4},
	urldate = {2024-12-01},
	journal = {Concurrency and Computation: Practice and Experience},
	author = {Thain, Douglas and Tannenbaum, Todd and Livny, Miron},
	year = {2005},
	note = {\_eprint: https://onlinelibrary.wiley.com/doi/pdf/10.1002/cpe.938},
	pages = {323--356},
}

@article{software-matplotlib,
	title = {Matplotlib: {A} {2D} {Graphics} {Environment}},
	volume = {9},
	shorttitle = {Matplotlib},
	url = {https://ui.adsabs.harvard.edu/abs/2007CSE.....9...90H},
	doi = {10.1109/MCSE.2007.55},
	urldate = {2024-12-01},
	journal = {Computing in Science and Engineering},
	author = {Hunter, John D.},
	month = may,
	year = {2007},
	note = {ADS Bibcode: 2007CSE.....9...90H},
	pages = {90--95},
}

@article{code-asimov,
	title = {Asimov: {A} framework for coordinating parameter estimation workflows},
	volume = {8},
	issn = {2475-9066},
	shorttitle = {Asimov},
	url = {https://joss.theoj.org/papers/10.21105/joss.04170},
	doi = {10.21105/joss.04170},
	language = {en},
	number = {84},
	urldate = {2024-11-29},
	journal = {Journal of Open Source Software},
	author = {Williams, Daniel and Veitch, John and Chiofalo, Maria Luisa and Schmidt, Patricia and Udall, Rhiannon P. and Vajpeji, Avi and Hoy, Charlie},
	month = apr,
	year = {2023},
	pages = {4170},
}

@article{code-lalinference,
	title = {Parameter estimation for compact binaries with ground-based gravitational-wave observations using the {LALInference} software library},
	volume = {91},
	issn = {1550-79980556-2821},
	url = {https://ui.adsabs.harvard.edu/abs/2015PhRvD..91d2003V},
	doi = {10.1103/PhysRevD.91.042003},
	urldate = {2024-11-29},
	journal = {Physical Review D},
	author = {Veitch, J. and Raymond, V. and Farr, B. and Farr, W. and Graff, P. and Vitale, S. and Aylott, B. and Blackburn, K. and Christensen, N. and Coughlin, M. and Del Pozzo, W. and Feroz, F. and Gair, J. and Haster, C. -J. and Kalogera, V. and Littenberg, T. and Mandel, I. and O'Shaughnessy, R. and Pitkin, M. and Rodriguez, C. and Röver, C. and Sidery, T. and Smith, R. and Van Der Sluys, M. and Vecchio, A. and Vousden, W. and Wade, L.},
	month = feb,
	year = {2015},
	note = {Publisher: APS
ADS Bibcode: 2015PhRvD..91d2003V},
	pages = {042003},
}

@article{search-spiir,
	title = {{SPIIR} online coherent pipeline to search for gravitational waves from compact binary coalescences},
	volume = {105},
	url = {https://link.aps.org/doi/10.1103/PhysRevD.105.024023},
	doi = {10.1103/PhysRevD.105.024023},
	number = {2},
	urldate = {2024-11-29},
	journal = {Physical Review D},
	author = {Chu, Qi and Kovalam, Manoj and Wen, Linqing and Slaven-Blair, Teresa and Bosveld, Joel and Chen, Yanbei and Clearwater, Patrick and Codoreanu, Alex and Du, Zhihui and Guo, Xiangyu and Guo, Xiaoyang and Kim, Kyungmin and Li, Tjonnie G. F. and Oloworaran, Victor and Panther, Fiona and Powell, Jade and Sengupta, Anand S. and Wette, Karl and Zhu, Xingjiang},
	month = jan,
	year = {2022},
	note = {Publisher: American Physical Society},
	pages = {024023},
}

@article{search-cwb,
	title = {Method for detection and reconstruction of gravitational wave transients with networks of advanced detectors},
	volume = {93},
	issn = {1550-79980556-2821},
	url = {https://ui.adsabs.harvard.edu/abs/2016PhRvD..93d2004K},
	doi = {10.1103/PhysRevD.93.042004},
	urldate = {2024-11-29},
	journal = {Physical Review D},
	author = {Klimenko, S. and Vedovato, G. and Drago, M. and Salemi, F. and Tiwari, V. and Prodi, G. A. and Lazzaro, C. and Ackley, K. and Tiwari, S. and Da Silva, C. F. and Mitselmakher, G.},
	month = feb,
	year = {2016},
	note = {Publisher: APS
ADS Bibcode: 2016PhRvD..93d2004K},
	pages = {042004},
}

@article{search-mbta,
	title = {Low-latency analysis pipeline for compact binary coalescences in the advanced gravitational wave detector era},
	volume = {33},
	issn = {0264-9381},
	url = {https://ui.adsabs.harvard.edu/abs/2016CQGra..33q5012A},
	doi = {10.1088/0264-9381/33/17/175012},
	urldate = {2024-11-29},
	journal = {Classical and Quantum Gravity},
	author = {Adams, T. and Buskulic, D. and Germain, V. and Guidi, G. M. and Marion, F. and Montani, M. and Mours, B. and Piergiovanni, F. and Wang, G.},
	month = sep,
	year = {2016},
	note = {Publisher: IOP
ADS Bibcode: 2016CQGra..33q5012A},
	pages = {175012},
}

\appendix 

\section{Source properties -- tables}
\label{sec:tables}

In this section we present the source properties and their associated uncertainties for the events from O1 and O2 (table~\ref{tab:source-properties}) and O3 (table~\ref{tab:source-properties-o3}).

\begin{table}[]
    \centering
    \scriptsize
    \begin{tabularx}{\linewidth}{lrrrrrrr}\toprule
    Event & $M_{\rm src}  / {\rm M}_{\odot} $ & $\mathcal{M}_{\rm src} / {\rm M}_{\odot} $ & $m_{1,{\rm src}}  / {\rm M}_{\odot} $ & $m_{2,{\rm src}}  / {\rm M}_{\odot} $ & $\chi_{\rm eff}$ & $D_{\rm L} / {\rm Mpc}$ & $z$ \\
    \midrule
    \eventname{151205}{195525} & \totalmasssource{GW151205_195525} & \chirpmasssource{GW151205_195525} & \massonesource{GW151205_195525} & \masstwosource{GW151205_195525} & \chieff{GW151205_195525} & \luminositydistance{GW151205_195525} & \redshift{GW151205_195525} \\ 
\eventname{151216}{092416} & \totalmasssource{GW151216_092416} & \chirpmasssource{GW151216_092416} & \massonesource{GW151216_092416} & \masstwosource{GW151216_092416} & \chieff{GW151216_092416} & \luminositydistance{GW151216_092416} & \redshift{GW151216_092416} \\ \midrule
\eventname{170121}{212536} & \totalmasssource{GW170121_212536} & \chirpmasssource{GW170121_212536} & \massonesource{GW170121_212536} & \masstwosource{GW170121_212536} & \chieff{GW170121_212536} & \luminositydistance{GW170121_212536} & \redshift{GW170121_212536} \\ 
\eventname{170202}{135657} & \totalmasssource{GW170202_135657} & \chirpmasssource{GW170202_135657} & \massonesource{GW170202_135657} & \masstwosource{GW170202_135657} & \chieff{GW170202_135657} & \luminositydistance{GW170202_135657} & \redshift{GW170202_135657} \\ 
\eventname{170304}{163753} & \totalmasssource{GW170304_163753} & \chirpmasssource{GW170304_163753} & \massonesource{GW170304_163753} & \masstwosource{GW170304_163753} & \chieff{GW170304_163753} & \luminositydistance{GW170304_163753} & \redshift{GW170304_163753} \\ 
\eventname{170403}{230611} & \totalmasssource{GW170403_230611} & \chirpmasssource{GW170403_230611} & \massonesource{GW170403_230611} & \masstwosource{GW170403_230611} & \chieff{GW170403_230611} & \luminositydistance{GW170403_230611} & \redshift{GW170403_230611} \\ 
\eventname{170425}{055334} & \totalmasssource{GW170425_055334} & \chirpmasssource{GW170425_055334} & \massonesource{GW170425_055334} & \masstwosource{GW170425_055334} & \chieff{GW170425_055334} & \luminositydistance{GW170425_055334} & \redshift{GW170425_055334} \\ 
\eventname{170727}{010430} & \totalmasssource{GW170727_010430} & \chirpmasssource{GW170727_010430} & \massonesource{GW170727_010430} & \masstwosource{GW170727_010430} & \chieff{GW170727_010430} & \luminositydistance{GW170727_010430} & \redshift{GW170727_010430} \\ 
\eventname{190426}{082124} & \totalmasssource{GW190426_082124} & \chirpmasssource{GW190426_082124} & \massonesource{GW190426_082124} & \masstwosource{GW190426_082124} & \chieff{GW190426_082124} & \luminositydistance{GW190426_082124} & \redshift{GW190426_082124} \\ 
\eventname{190427}{180650} & \totalmasssource{GW190427_180650} & \chirpmasssource{GW190427_180650} & \massonesource{GW190427_180650} & \masstwosource{GW190427_180650} & \chieff{GW190427_180650} & \luminositydistance{GW190427_180650} & \redshift{GW190427_180650}

    \\\bottomrule
    \end{tabularx}
    \medskip
    \caption{
        The median, and 90\% highest probability density intervals \add{(that is, the narrowest region containing 90\% of the posterior probability)} for the selected inferred source properties of the signals from O1 and O2 analysed in this work.
        The columns show the total mass of the binary $M_{\rm src}$; the chirp mass $\mathcal{M}_{\rm src}$; the component objects' masses $m_{\rm 1, src}$, and $m_{\rm 2, src}$; the effective inspiral spin; $\chi_{\rm eff}$; the luminosity distance $D_{\rm L}$; and the redshift $z$. All masses are quoted in the source frame, and are derived from inference using the IMRPhenomXPHM BBH waveform model.
    }
    \label{tab:source-properties}
\end{table}

\begin{table}[]
    \centering
    \scriptsize
    \begin{tabularx}{\linewidth}{lrrrrrrr}\toprule
    Event & $M  / {\rm M}_{\odot} $ & $\mathcal{M} / {\rm M}_{\odot} $ & $m_1  / {\rm M}_{\odot} $ & $m_2  / {\rm M}_{\odot} $ & $\chi_{\rm eff}$ & $D_{\rm L} / {\rm Mpc}$ & $z$ \\
    \midrule
    \eventname{190511}{125545} & \totalmasssource{GW190511_125545} & \chirpmasssource{GW190511_125545} & \massonesource{GW190511_125545} & \masstwosource{GW190511_125545} & \chieff{GW190511_125545} & \luminositydistance{GW190511_125545} & \redshift{GW190511_125545} \\ 
\eventname{190511}{163209} & \totalmasssource{GW190511_163209} & \chirpmasssource{GW190511_163209} & \massonesource{GW190511_163209} & \masstwosource{GW190511_163209} & \chieff{GW190511_163209} & \luminositydistance{GW190511_163209} & \redshift{GW190511_163209} \\ 
\eventname{190514}{065416} & \totalmasssource{GW190514_065416} & \chirpmasssource{GW190514_065416} & \massonesource{GW190514_065416} & \masstwosource{GW190514_065416} & \chieff{GW190514_065416} & \luminositydistance{GW190514_065416} & \redshift{GW190514_065416} \\ 
\eventname{190523}{085933} & \totalmasssource{GW190523_085933} & \chirpmasssource{GW190523_085933} & \massonesource{GW190523_085933} & \masstwosource{GW190523_085933} & \chieff{GW190523_085933} & \luminositydistance{GW190523_085933} & \redshift{GW190523_085933} \\ 
\eventname{190524}{134109} & \totalmasssource{GW190524_134109} & \chirpmasssource{GW190524_134109} & \massonesource{GW190524_134109} & \masstwosource{GW190524_134109} & \chieff{GW190524_134109} & \luminositydistance{GW190524_134109} & \redshift{GW190524_134109} \\ 
\eventname{190530}{030659} & \totalmasssource{GW190530_030659} & \chirpmasssource{GW190530_030659} & \massonesource{GW190530_030659} & \masstwosource{GW190530_030659} & \chieff{GW190530_030659} & \luminositydistance{GW190530_030659} & \redshift{GW190530_030659} \\ 
\eventname{190530}{133833} & \totalmasssource{GW190530_133833} & \chirpmasssource{GW190530_133833} & \massonesource{GW190530_133833} & \masstwosource{GW190530_133833} & \chieff{GW190530_133833} & \luminositydistance{GW190530_133833} & \redshift{GW190530_133833} \\ 
\eventname{190604}{103812} & \totalmasssource{GW190604_103812} & \chirpmasssource{GW190604_103812} & \massonesource{GW190604_103812} & \masstwosource{GW190604_103812} & \chieff{GW190604_103812} & \luminositydistance{GW190604_103812} & \redshift{GW190604_103812} \\ 
\eventname{190605}{025957} & \totalmasssource{GW190605_025957} & \chirpmasssource{GW190605_025957} & \massonesource{GW190605_025957} & \masstwosource{GW190605_025957} & \chieff{GW190605_025957} & \luminositydistance{GW190605_025957} & \redshift{GW190605_025957} \\ 
\eventname{190607}{083827} & \totalmasssource{GW190607_083827} & \chirpmasssource{GW190607_083827} & \massonesource{GW190607_083827} & \masstwosource{GW190607_083827} & \chieff{GW190607_083827} & \luminositydistance{GW190607_083827} & \redshift{GW190607_083827} \\ 
\eventname{190614}{134749} & \totalmasssource{GW190614_134749} & \chirpmasssource{GW190614_134749} & \massonesource{GW190614_134749} & \masstwosource{GW190614_134749} & \chieff{GW190614_134749} & \luminositydistance{GW190614_134749} & \redshift{GW190614_134749} \\ 
\eventname{190615}{030234} & \totalmasssource{GW190615_030234} & \chirpmasssource{GW190615_030234} & \massonesource{GW190615_030234} & \masstwosource{GW190615_030234} & \chieff{GW190615_030234} & \luminositydistance{GW190615_030234} & \redshift{GW190615_030234} \\ 
\eventname{190705}{164632} & \totalmasssource{GW190705_164632} & \chirpmasssource{GW190705_164632} & \massonesource{GW190705_164632} & \masstwosource{GW190705_164632} & \chieff{GW190705_164632} & \luminositydistance{GW190705_164632} & \redshift{GW190705_164632} \\ 
\eventname{190707}{083226} & \totalmasssource{GW190707_083226} & \chirpmasssource{GW190707_083226} & \massonesource{GW190707_083226} & \masstwosource{GW190707_083226} & \chieff{GW190707_083226} & \luminositydistance{GW190707_083226} & \redshift{GW190707_083226} \\ 
\eventname{190711}{030756} & \totalmasssource{GW190711_030756} & \chirpmasssource{GW190711_030756} & \massonesource{GW190711_030756} & \masstwosource{GW190711_030756} & \chieff{GW190711_030756} & \luminositydistance{GW190711_030756} & \redshift{GW190711_030756} \\ 
\eventname{190718}{160159} & \totalmasssource{GW190718_160159} & \chirpmasssource{GW190718_160159} & \massonesource{GW190718_160159} & \masstwosource{GW190718_160159} & \chieff{GW190718_160159} & \luminositydistance{GW190718_160159} & \redshift{GW190718_160159} \\ 
\eventname{190725}{174728} & \totalmasssource{GW190725_174728} & \chirpmasssource{GW190725_174728} & \massonesource{GW190725_174728} & \masstwosource{GW190725_174728} & \chieff{GW190725_174728} & \luminositydistance{GW190725_174728} & \redshift{GW190725_174728} \\ 
\eventname{190805}{105432} & \totalmasssource{GW190805_105432} & \chirpmasssource{GW190805_105432} & \massonesource{GW190805_105432} & \masstwosource{GW190805_105432} & \chieff{GW190805_105432} & \luminositydistance{GW190805_105432} & \redshift{GW190805_105432} \\ 
\eventname{190806}{033721} & \totalmasssource{GW190806_033721} & \chirpmasssource{GW190806_033721} & \massonesource{GW190806_033721} & \masstwosource{GW190806_033721} & \chieff{GW190806_033721} & \luminositydistance{GW190806_033721} & \redshift{GW190806_033721} \\ 
\eventname{190814}{192009} & \totalmasssource{GW190814_192009} & \chirpmasssource{GW190814_192009} & \massonesource{GW190814_192009} & \masstwosource{GW190814_192009} & \chieff{GW190814_192009} & \luminositydistance{GW190814_192009} & \redshift{GW190814_192009} \\ 
\eventname{190818}{232544} & \totalmasssource{GW190818_232544} & \chirpmasssource{GW190818_232544} & \massonesource{GW190818_232544} & \masstwosource{GW190818_232544} & \chieff{GW190818_232544} & \luminositydistance{GW190818_232544} & \redshift{GW190818_232544} \\ 
\eventname{190821}{124821} & \totalmasssource{GW190821_124821} & \chirpmasssource{GW190821_124821} & \massonesource{GW190821_124821} & \masstwosource{GW190821_124821} & \chieff{GW190821_124821} & \luminositydistance{GW190821_124821} & \redshift{GW190821_124821} \\ 
\eventname{190904}{104631} & \totalmasssource{GW190904_104631} & \chirpmasssource{GW190904_104631} & \massonesource{GW190904_104631} & \masstwosource{GW190904_104631} & \chieff{GW190904_104631} & \luminositydistance{GW190904_104631} & \redshift{GW190904_104631} \\ 
\eventname{190906}{054335} & \totalmasssource{GW190906_054335} & \chirpmasssource{GW190906_054335} & \massonesource{GW190906_054335} & \masstwosource{GW190906_054335} & \chieff{GW190906_054335} & \luminositydistance{GW190906_054335} & \redshift{GW190906_054335} \\ 
\eventname{190910}{012619} & \totalmasssource{GW190910_012619} & \chirpmasssource{GW190910_012619} & \massonesource{GW190910_012619} & \masstwosource{GW190910_012619} & \chieff{GW190910_012619} & \luminositydistance{GW190910_012619} & \redshift{GW190910_012619} \\ 
\eventname{190911}{195101} & \totalmasssource{GW190911_195101} & \chirpmasssource{GW190911_195101} & \massonesource{GW190911_195101} & \masstwosource{GW190911_195101} & \chieff{GW190911_195101} & \luminositydistance{GW190911_195101} & \redshift{GW190911_195101} \\ 
\eventname{190916}{200658} & \totalmasssource{GW190916_200658} & \chirpmasssource{GW190916_200658} & \massonesource{GW190916_200658} & \masstwosource{GW190916_200658} & \chieff{GW190916_200658} & \luminositydistance{GW190916_200658} & \redshift{GW190916_200658} \\ 
\eventname{190926}{050336} & \totalmasssource{GW190926_050336} & \chirpmasssource{GW190926_050336} & \massonesource{GW190926_050336} & \masstwosource{GW190926_050336} & \chieff{GW190926_050336} & \luminositydistance{GW190926_050336} & \redshift{GW190926_050336} \\ 
\eventname{191113}{103541} & \totalmasssource{GW191113_103541} & \chirpmasssource{GW191113_103541} & \massonesource{GW191113_103541} & \masstwosource{GW191113_103541} & \chieff{GW191113_103541} & \luminositydistance{GW191113_103541} & \redshift{GW191113_103541} \\ 
\eventname{191117}{023843} & \totalmasssource{GW191117_023843} & \chirpmasssource{GW191117_023843} & \massonesource{GW191117_023843} & \masstwosource{GW191117_023843} & \chieff{GW191117_023843} & \luminositydistance{GW191117_023843} & \redshift{GW191117_023843} \\ 
\eventname{191127}{050227} & \totalmasssource{GW191127_050227} & \chirpmasssource{GW191127_050227} & \massonesource{GW191127_050227} & \masstwosource{GW191127_050227} & \chieff{GW191127_050227} & \luminositydistance{GW191127_050227} & \redshift{GW191127_050227} \\ 
\eventname{191208}{080334} & \totalmasssource{GW191208_080334} & \chirpmasssource{GW191208_080334} & \massonesource{GW191208_080334} & \masstwosource{GW191208_080334} & \chieff{GW191208_080334} & \luminositydistance{GW191208_080334} & \redshift{GW191208_080334} \\ 
\eventname{191224}{043228} & \totalmasssource{GW191224_043228} & \chirpmasssource{GW191224_043228} & \massonesource{GW191224_043228} & \masstwosource{GW191224_043228} & \chieff{GW191224_043228} & \luminositydistance{GW191224_043228} & \redshift{GW191224_043228} \\ 
\eventname{191228}{195619} & \totalmasssource{GW191228_195619} & \chirpmasssource{GW191228_195619} & \massonesource{GW191228_195619} & \masstwosource{GW191228_195619} & \chieff{GW191228_195619} & \luminositydistance{GW191228_195619} & \redshift{GW191228_195619} \\ 
\eventname{200106}{134123} & \totalmasssource{GW200106_134123} & \chirpmasssource{GW200106_134123} & \massonesource{GW200106_134123} & \masstwosource{GW200106_134123} & \chieff{GW200106_134123} & \luminositydistance{GW200106_134123} & \redshift{GW200106_134123} \\ 
\eventname{200109}{195634} & \totalmasssource{GW200109_195634} & \chirpmasssource{GW200109_195634} & \massonesource{GW200109_195634} & \masstwosource{GW200109_195634} & \chieff{GW200109_195634} & \luminositydistance{GW200109_195634} & \redshift{GW200109_195634} \\ 
\eventname{200129}{114245} & \totalmasssource{GW200129_114245} & \chirpmasssource{GW200129_114245} & \massonesource{GW200129_114245} & \masstwosource{GW200129_114245} & \chieff{GW200129_114245} & \luminositydistance{GW200129_114245} & \redshift{GW200129_114245} \\ 
\eventname{200208}{211609} & \totalmasssource{GW200208_211609} & \chirpmasssource{GW200208_211609} & \massonesource{GW200208_211609} & \masstwosource{GW200208_211609} & \chieff{GW200208_211609} & \luminositydistance{GW200208_211609} & \redshift{GW200208_211609} \\ 
\eventname{200210}{005122} & \totalmasssource{GW200210_005122} & \chirpmasssource{GW200210_005122} & \massonesource{GW200210_005122} & \masstwosource{GW200210_005122} & \chieff{GW200210_005122} & \luminositydistance{GW200210_005122} & \redshift{GW200210_005122} \\ 
\eventname{200210}{100022} & \totalmasssource{GW200210_100022} & \chirpmasssource{GW200210_100022} & \massonesource{GW200210_100022} & \masstwosource{GW200210_100022} & \chieff{GW200210_100022} & \luminositydistance{GW200210_100022} & \redshift{GW200210_100022} \\ 
\eventname{200220}{124850} & \totalmasssource{GW200220_124850} & \chirpmasssource{GW200220_124850} & \massonesource{GW200220_124850} & \masstwosource{GW200220_124850} & \chieff{GW200220_124850} & \luminositydistance{GW200220_124850} & \redshift{GW200220_124850} \\ 
\eventname{200214}{223307} & \totalmasssource{GW200214_223307} & \chirpmasssource{GW200214_223307} & \massonesource{GW200214_223307} & \masstwosource{GW200214_223307} & \chieff{GW200214_223307} & \luminositydistance{GW200214_223307} & \redshift{GW200214_223307} \\ 
\eventname{200225}{075134} & \totalmasssource{GW200225_075134} & \chirpmasssource{GW200225_075134} & \massonesource{GW200225_075134} & \masstwosource{GW200225_075134} & \chieff{GW200225_075134} & \luminositydistance{GW200225_075134} & \redshift{GW200225_075134} \\ 
\eventname{200301}{211019} & \totalmasssource{GW200301_211019} & \chirpmasssource{GW200301_211019} & \massonesource{GW200301_211019} & \masstwosource{GW200301_211019} & \chieff{GW200301_211019} & \luminositydistance{GW200301_211019} & \redshift{GW200301_211019} \\ 
\eventname{200304}{172806} & \totalmasssource{GW200304_172806} & \chirpmasssource{GW200304_172806} & \massonesource{GW200304_172806} & \masstwosource{GW200304_172806} & \chieff{GW200304_172806} & \luminositydistance{GW200304_172806} & \redshift{GW200304_172806} \\ 
\eventname{200305}{084739} & \totalmasssource{GW200305_084739} & \chirpmasssource{GW200305_084739} & \massonesource{GW200305_084739} & \masstwosource{GW200305_084739} & \chieff{GW200305_084739} & \luminositydistance{GW200305_084739} & \redshift{GW200305_084739} \\ 
\eventname{200318}{191337} & \totalmasssource{GW200318_191337} & \chirpmasssource{GW200318_191337} & \massonesource{GW200318_191337} & \masstwosource{GW200318_191337} & \chieff{GW200318_191337} & \luminositydistance{GW200318_191337} & \redshift{GW200318_191337} 
    \\\bottomrule
    \end{tabularx}
    \medskip
    \caption{
        The median, and 90\% highest probability density intervals \add{(that is, the narrowest region containing 90\% of the posterior probability)} for the selected inferred source properties of the signals from O3 analysed in this work.
        The columns show the total mass of the binary $M$; the chirp mass $\mathcal{M}$; the component objects' masses $m_1$, and $m_2$; the effective inspiral spin; $\chi_{\rm eff}$; the luminosity distance $D_{\rm L}$; and the redshift $z$. All masses are quoted in the source frame, and are derived from inference using the IMRPhenomXPHM BBH waveform model.
    }
    \label{tab:source-properties-o3}
\end{table}

\section{Asimov blueprints}
\label{sec:blueprints}

This section provides a short technical overview of Asimov's "blueprint" files which are used to configure analyses in a consistent way across different analysis pipelines and computing facilities while abstracting as much of that configuration as possible.
Blueprints are intended to provide a uniform configuration interface for a number of different analysis codes, allowing consistent and comparable analyses to be performed in a straightforward manner.
In this work we perform parameter estimation (PE) analyses only using the Bilby PE library, but within our workflow we also ran a Bayeswave analysis; Asimov blueprints allow this workflow to be specified with minimal repetition or reconfiguration between events.
Asimov also makes it possible to replace parts of the workflow with minimal additional work, for example the Bilby analyses could be replaced by a LALInference~\cite{code-lalinference} or RIFT~\cite{code-rift-0, code-rift-1, code-rift-2,code-rift-asimov} analysis by changing a single line in a blueprint, where previously the various configuration parameters would need to be mapped between the Bilby and other code's input format.

To achieve this, it is normal to configure an Asimov \emph{project} with a series of different settings; these are then treated hierarchically by Asimov when constructing an analysis, with settings specified with higher precedence over-riding those from below.
This allows almost total flexibility when creating an individual analysis, while generally ensuring consistency.
In sequence of increasing precedence these are:
\begin{enumerate}
    \item \textbf{Project-wide defaults}: for example default priors which should normally be applied to all events, information which should be passed to the batch scheduling system being used to perform computation, and information about default.
    \item \textbf{Pipeline defaults}: these apply to all analyses for a given analysis code or pipeline, and will often include sampling settings which are specific to a given analysis code.
    \item \textbf{Event/subject settings} these apply to a single event (or subject) which is to be analysed by Asimov, and will typically include information such as the event time; priors, such as those on the chirp mass which are specific to a given event; and the list of observatories which detected the signal.
    \item \textbf{Analysis-specific settings} which apply only to a single analysis: these will typically include specific setup information, such as the name of the pipeline to be used in this analysis, and the waveform model to be used.
\end{enumerate}

The hierarchical structure allows a substantial reduction in the number of settings which must be curated for each analysis.
For example, in this project the majority of analyses were configured using the same analysis-specific blueprint, with only the event-specific blueprints varying substantially between each event.

Asimov blueprints use the YAML serialisation format, and represent a development in the approach used for configuring analyses compared to the one described previously~\cite{code-asimov}, as the method in that work was non-portable and did not allow easy distribution of the configuration data required to reproduce analyses in the manner that blueprint files do.
Fuller documentation of blueprint files and their usage can be found in the Asimov documentation~\cite{docs-asimov}, along with a number of tutorials.

\subsection{Blueprints used in this work}

As a concrete example, we display extracts of the blueprints used in this work in this section.
In order to improve the clarity of presentation we have removed some information from these, though the full files are available in the accompanying data release~\cite{data-release}.

The first example is our project-wide blueprint.
This was applied to our project once, as we were setting it up, and before any events or analyses had been added to the project.

\begin{verbatim}
kind: configuration
data:
  channels:
    H1: H1:GWOSC-16KHZ_R1_STRAIN
    L1: L1:GWOSC-16KHZ_R1_STRAIN
    V1: V1:GWOSC-16KHZ_R1_STRAIN
  frame types:
    H1: H1:H1_GWOSC_O3b_16KHZ_R1
    L1: L1:L1_GWOSC_O3b_16KHZ_R1
    V1: V1:V1_GWOSC_O3b_16KHZ_R1
likelihood:
  roll off time: 0.4
postprocessing:
  pesummary:
    cosmology: Planck15
    evolve spins: forwards
    multiprocess: 4
    redshift: exact
    calculate:
    - precessing snr
    skymap samples: 2000
quality:
  state vector:
    L1: L1:GDS-CALIB_STATE_VECTOR_AR
    H1: H1:GDS-CALIB_STATE_VECTOR_AR
    V1: V1:DQ_ANALYSIS_STATE_VECTOR
  minimum frequency:
    H1: 20
    L1: 20
    V1: 20
\end{verbatim}

This file contains settings which are not expected to vary between events or analyses, such as the default frame types, and the setup for postprocessing.

Our pipeline specific settings are similarly not expected to change between analyses, but can be used to configure parameters which will be required only for analyses using a specific pipeline; in this instance Bilby and Bayeswave have differing configuration requirements and so are specified as pipeline-specific configurations.

\begin{verbatim}
kind: configuration
pipelines:

  bilby:
    sampler:
      sampler: dynesty
      kwargs:
        nlive: 1000
        naccept: 60
        sample: acceptance-walk
        check_point_plot: True
        maxmcmc: 100000
      parallel jobs: 3
    scheduler:
      request cpus: 16
      request disk: 8 #GB
      osg: True
      copy frames: True
    likelihood:
      roll off time: 1
      marginalization:
        time: False
        distance: True

  bayeswave:
    scheduler:
      request memory: 1024
      request post memory: 16384
      copy frames: True
      request disk: 3000000
      osg: True
    likelihood:
      roll off time: 1
      iterations: 100000
      chains: 8
      threads: 4
\end{verbatim}

Event-specific settings can then be added in a third blueprint.
Unlike the previous two blueprints, which are used once per-project, a different blueprint should be created for each event.

An example blueprint from this project, for \eventname{200210}{005122}, is shown below.

\begin{verbatim}
data:
  segment length: 256
event time: 1265331100.7424316
interferometers: ['H1', 'L1']
kind: event
likelihood:
  psd length: 256
  reference frequency: 20
  sample rate: 8192
  window length: 256
name: GW200210_005122
priors:
  chirp mass:
    maximum: 12
    minimum: 4

\end{verbatim}

This blueprint needs to specify the event time, as well as other settings specific to that event, such as the required segment length (which relates to the length of the signal), the chirp mass prior, and the list of interferometers which observed the signal.
An event blueprint can typically be created directly from search results, and these can be produced programmatically from the event tables which are published with most catalogues.

Finally, each analysis may be specified in a blueprint.
The two main analysis steps in our workflow can be specified in the following minimal blueprints (note that these may be contained in the same file, separated by three hyphen characters).
In addition we include the blueprint required to allow Asimov to acquire public data frames as the first "analysis" in the workflow.

\begin{verbatim}
kind: analysis
name: get-data
pipeline: gwdata
file length: 4096
download:
  - frames
scheduler:
  request memory: 1024
  request post memory: 16384
---
kind: analysis
name: generate-psd
pipeline: bayeswave
comment: Bayeswave on-source PSD estimation process
needs:
  - get-data
---
kind: analysis
name: bilby-IMRPhenomXPHM
pipeline: bilby
waveform:
  approximant: IMRPhenomXPHM

comment: PE job using IMRPhenomXPHM and bilby
needs:
    - generate-psd
\end{verbatim}

The dependency structure can be specified by listing the names of analyses which must have completed on a given event in the `needs` section of the blueprint.
The minimal structure of these files also allows alternative analyses to be created.
For example, if an additional run using a second waveform was desired, a copy of the last blueprint could be created, with only the change of waveform approximant name required to configure the extra analysis.

\end{document}